\documentclass[a4paper,11pt]{article}
\usepackage{jheppub}
\usepackage[utf8]{inputenc}

\usepackage{pdflscape}
\usepackage{tikz}
\usepackage{tikz-3dplot}
\usetikzlibrary{positioning}
\usetikzlibrary{arrows}
\usetikzlibrary{decorations.pathreplacing}
\usetikzlibrary{shapes.geometric}
\usetikzlibrary{calc}
\usepackage{todonotes}
\usepackage{enumitem}
\usepackage{subcaption}
\usepackage{afterpage}

\usepackage{mathtools}

\usepackage{MnSymbol}

\tikzset{hasse/.style={circle, fill,inner sep=2pt}}
\tikzset{hasseb/.style={circle, fill=black!40,inner sep=4pt}}
\tikzset{hassebr/.style={circle, fill=red!40,inner sep=4pt}}
\tikzset{hassegray/.style={circle, fill=black!10,inner sep=2pt}}
\tikzset{linegray/.style={black!10}}
\tikzset{gauge/.style={inner sep=1mm,draw=none,fill=white,minimum size=2mm,circle, draw}}
\tikzset{gauger/.style={inner sep=1mm,draw=none,fill=goodred,minimum size=2mm,circle, draw}}
\tikzset{gaugeo/.style={inner sep=1mm,draw=none,fill=goodorange,minimum size=2mm,circle, draw}}
\tikzset{gaugeb/.style={inner sep=1mm,draw=none,fill=blue,minimum size=2mm,circle, draw}}
\tikzset{gaugeg/.style={inner sep=1mm,draw=none,fill=goodgreen,minimum size=2mm,circle, draw}}
\tikzset{gaugec/.style={inner sep=1mm,draw=none,fill=cyan,minimum size=2mm,circle, draw}}
\tikzset{gaugem/.style={inner sep=1mm,draw=none,fill=goodmagenta,minimum size=2mm,circle, draw}}
\tikzset{flavour/.style={draw=none,minimum size=0.3mm,fill=white, regular polygon,regular polygon sides=4,draw}}
\tikzset{sev/.style={inner sep=1mm,draw=none,fill=white,minimum size=4mm,circle, draw}}
\tikzset{sevR/.style={inner sep=1mm,draw=none,fill=white,minimum size=4mm,circle, draw, red,fill=white}}
\tikzset{sevB/.style={inner sep=1mm,draw=none,fill=white,minimum size=4mm,circle, draw, blue,fill=white}}
\tikzset{sevG/.style={inner sep=1mm,draw=none,fill=white,minimum size=4mm,circle, draw, goodgreen,fill=white}}
\tikzset{NS/.style={circle, fill=red,inner sep=3pt}}
\tikzset{rec/.style={draw, rectangle, minimum size=2mm,align=center}}
\tikzset{quadruple/.style={double distance=#1-\pgflinewidth,thick,
        postaction={draw,thick,double distance=#1/3-\pgflinewidth}},
    quadruple/.default=.5em}
\tikzset{doublea/.style={double distance=1/4*(#1-\pgflinewidth),thick},
    doublea/.default=.5em}
\colorlet{c1}{white}
\colorlet{c3}{orange!10}
\colorlet{c6}{red!20}
\newcommand{\smallhasse}[1]{
\raisebox{-.5\height}{\scalebox{.5}{\begin{tikzpicture}[scale=0.7]
    #1
    \end{tikzpicture}}}}
    
\newcommand{\HS}{\mathrm{HS}}
\newcommand{\HV}{\mathcal{H}\mathcal{V}}
\newcommand{\HSCH}{\mathcal{H}\mathcal{S}}
\newcommand{\pal}{\ldots \text{palindrome}\ldots}

\definecolor{goodgreen}{RGB}{55,169,49}
\definecolor{goodred}{RGB}{230,0,0}
\definecolor{goodorange}{RGB}{255,200,0}
\definecolor{goodmagenta}{RGB}{255,150,255}
\definecolor{goodyellow}{RGB}{160,160,0}

\newcommand{\convexpath}[2]{
  [   
  create hullcoords/.code={
    \global\edef\namelist{#1}
    \foreach [count=\counter] \nodename in \namelist {
      \global\edef\numberofnodes{\counter}
      \coordinate (hullcoord\counter) at (\nodename);
    }
    \coordinate (hullcoord0) at (hullcoord\numberofnodes);
    \pgfmathtruncatemacro\lastnumber{\numberofnodes+1}
    \coordinate (hullcoord\lastnumber) at (hullcoord1);
  },
  create hullcoords
  ]
  ($(hullcoord1)!#2!-90:(hullcoord0)$)
  \foreach [
  evaluate=\currentnode as \previousnode using \currentnode-1,
  evaluate=\currentnode as \nextnode using \currentnode+1
  ] \currentnode in {1,...,\numberofnodes} {
    let \p1 = ($(hullcoord\currentnode) - (hullcoord\previousnode)$),
    \n1 = {atan2(\y1,\x1) + 90},
    \p2 = ($(hullcoord\nextnode) - (hullcoord\currentnode)$),
    \n2 = {atan2(\y2,\x2) + 90},
    \n{delta} = {Mod(\n2-\n1,360) - 360}
    in 
    {arc [start angle=\n1, delta angle=\n{delta}, radius=#2]}
    -- ($(hullcoord\nextnode)!#2!-90:(hullcoord\currentnode)$) 
  }
}

\usepackage[makeroom]{cancel}

\preprint{Imperial/TP/23/AH/01}

\title{A Tale of N Cones}

\author[a,b]{Antoine Bourget,}
\author[c]{Julius F. Grimminger,}
\author[c]{Amihay Hanany,}
\author[c]{Rudolph Kalveks,}
\author[d]{Marcus Sperling,}
\author[c,e]{and Zhenghao Zhong}
\affiliation[a]{Université Paris-Saclay, CEA, CNRS, Institut de physique théorique, 91191, Gif-sur-Yvette, France}
\affiliation[b]{Laboratoire de Physique de l'\'Ecole Normale Sup\'erieure, PSL University, \\ 24 rue Lhomond, 75005 Paris, France}
\affiliation[c]{Theoretical Physics Group, The Blackett Laboratory, Imperial College London, Prince Consort Road
London, SW7 2AZ, UK}
\affiliation[d]{Shing-Tung Yau Center, Southeast University \\
 Xuanwu District, Nanjing, Jiangsu, 210096, China}
\affiliation[e]{Mathematical Institute, University of Oxford,\\
Andrew Wiles Building, Woodstock Road, Oxford, OX2 6GG, UK}
\emailAdd{antoine.bourget@polytechnique.org}
\emailAdd{julius.grimminger17@imperial.ac.uk}
\emailAdd{a.hanany@imperial.ac.uk}
\emailAdd{rudolph.kalveks09@imperial.ac.uk}
\emailAdd{msperling@seu.edu.cn}
\emailAdd{zhenghao.zhong@maths.ox.ac.uk}

\abstract{We study particular families of bad 3d $\mathcal{N}=4$ quiver gauge theories, whose Higgs branches consist of many cones. We show the role of a novel brane configuration in realizing the Higgs moduli for each distinct cone. Through brane constructions, magnetic quivers, Hasse diagrams, and Hilbert series computations we study the intricate structure of the classical Higgs branches. These Higgs branches are both non-normal (since they consist of multiple cones) and non-reduced (due to the presence of nilpotent operators in the chiral ring). Applying the principle of \emph{inversion} to the classical Higgs branch Hasse diagrams, we conjecture the quantum Coulomb branch Hasse diagrams. These Coulomb branches have several most singular loci, corresponding to the several cones in the Higgs branch. We propose the Hasse diagrams of the full quantum moduli spaces of our theories. The quivers we study can be taken to be 5d effective gauge theories living on brane webs. Their infinite coupling theories have Higgs branches which also consist of multiple cones. Some of these cones have \emph{decorated} magnetic quivers, whose 3d Coulomb branches remain elusive.
}

\begin{document}

\maketitle

\clearpage

\section{Introduction and Summary}

Supersymmetric field theories provide a rich playground to explore properties of quantum field theory in a framework where exact computations are possible. In particular, the high degree of symmetry often implies the existence of a moduli space of vacua, which can be seen as a crude observable that partially characterizes the theory under consideration, and reveals, for instance, certain aspects of its symmetries. In this work, we consider theories with 8 supercharges in space-time dimension $3 \leq d \leq 5$. We focus mainly on the Higgs branch, which is a singular hyper-K\"ahler space. 

For a large class of theories, the Higgs branch is a normal hyper-K\"ahler cone, with symplectic singularities \cite{beauville2000symplectic} -- conjecturally, this class contains all 4d $\mathcal{N}=2$ SCFTs \cite{Beem:2017ooy,Bourget:2021csg}. However, it has been observed that certain theories have a classical Higgs branch made up of two cones that can intersect at their common tip, or along a sub-cone of positive dimension. A well known example is the theory with gauge group $\mathrm{SU}(2)$ and $N_f = 2$ fundamental hypermultiplets in 4d $\mathcal{N}=2$ \cite{Seiberg:1994aj} and 3d $\mathcal{N}=4$ \cite{Seiberg:1996nz} whose classical Higgs branch is the union of two singularities $\mathbb{C}^2 / \mathbb{Z}_2$ meeting at the origin. This has been generalized to the theories with gauge group $\mathrm{Sp}(k)$ and $2k$ fundamental hypermultiplets, for which the Higgs branch is the closure of the very even nilpotent orbit $\overline{\mathcal{O}}_{\mathrm{O}}^{[2^{2k}]} (\mathfrak{so}(4k))$ \cite{Ferlito:2016grh}. This is reviewed below in Section \ref{sec:2cones}. Another well-known example is $\mathrm{SU}(k)$ SQCD with $k \leq N_f \leq 2k-2$ fundamental hypermultiplets, which exhibits a mesonic branch and a baryonic branch \cite{Argyres:1996eh,Bourget:2019rtl}. Multiple cones in the Higgs branch were also observed for quiver gauge theories with underbalanced unitary and special unitary nodes \cite{Bourget:2021jwo}. Generically the picture that emerges is that cone multiplicity can arise for 3d $\mathcal{N}=4$ bad theories \cite{Gaiotto:2008ak}, and 4d $\mathcal{N}=2$ asymptotically free theories. Note that these theories are not conformal, thus there is no conflict with the claim above. In all cases which were studied in the literature, the presence of multiple cones in the classical Higgs branch of a 3d $\mathcal{N}=4$ or 4d $\mathcal{N}=2$ theory implies that there are multiple most singular loci in the quantum Coulomb branch of the theory, where the individual cones of the classical Higgs branch emanate \cite{Seiberg:1994aj,Seiberg:1996nz,Argyres:1996eh,Dey:2017fqs,Assel:2018exy,Akhond:2022jts}, and the cones in the classical Higgs branch are \emph{split} along the quantum Coulomb branch. See Appendix \ref{app:badCB} for a more detailed discussion including mixed branches. An excellent tool to study the intricate singularity structure of moduli spaces is the Hasse diagram \cite{Bourget:2019aer}. Based on \emph{inversion} \cite{Grimminger:2020dmg} of the classical Higgs branch Hasse diagram, we conjecture the quantum Coulomb branch and full moduli space Hasse diagrams for the 3d $\mathcal{N}=4$ theories studied in this paper.

In the case of 5d $\mathcal{N}=1$ SCFTs the Higgs branch can consist of multiple intricately intersecting cones, all emanating from the origin of the Coulomb branch, where the SCFT is realized \cite{Seiberg:1996bd,Cremonesi:2015lsa,Ferlito:2017xdq,Cabrera:2018jxt}. It turns out that identifying the different Higgs phases correctly from the underlying brane web may require one to use the recently introduced concept of \emph{decorated magnetic quiver} \cite{Bourget:2022ehw} see especially \cite[Appendix B]{Bourget:2022tmw}. For those quivers, we only know of a brane configuration description, while a precise definition of the 3d Coulomb branch, along with the ability to compute the corresponding Hilbert series, is still missing.

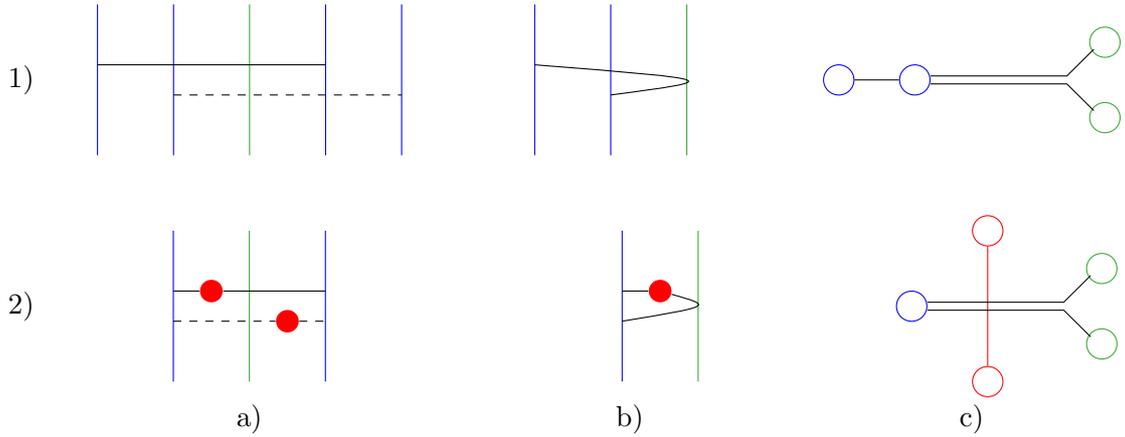
\begin{figure}
    \centering
    \begin{tikzpicture}
        \node at (0,0) {$\begin{tikzpicture}
            \draw[blue] (0,-1)--(0,1) (1,-1)--(1,1) (3,-1)--(3,1) (4,-1)--(4,1);
            \draw[goodgreen] (2,-1)--(2,1);
            \draw (0,0.2)--(3,0.2);
            \draw[dashed] (1,-0.2)--(4,-0.2);
        \end{tikzpicture}$};
        \node at (5,0) {$\begin{tikzpicture}
            \draw[blue] (0,-1)--(0,1) (1,-1)--(1,1);
            \draw[goodgreen] (2,-1)--(2,1);
            \draw (0,0.2) .. controls (2.5,0) .. (1,-0.2);
        \end{tikzpicture}$};
        \node at (9.5,0) {$\begin{tikzpicture}
            \node[sevB] (1) at (0,0) {};
            \node[sevB] (2) at (1,0) {};
            \node[sevG] (4u) at (3.5,0.5) {};
            \node[sevG] (4d) at (3.5,-0.5) {};
            \draw (1)--(2);
            \draw[transform canvas={yshift=1.5pt}] (2)--(3,0)--(4u);
            \draw[transform canvas={yshift=-1.5pt}] (2)--(3,0)--(4d);
        \end{tikzpicture}$};
        \node at (0,-3) {$\begin{tikzpicture}
            \node at (0,0) {};
            \node at (4,0) {};
            \draw[blue] (1,-1)--(1,1) (3,-1)--(3,1);
            \draw[goodgreen] (2,-1)--(2,1);
            \node[NS] (ns) at (1.5,0.2) {};
            \node[NS] (nsm) at (2.5,-0.2) {};
            \draw (1,0.2)--(ns)--(3,0.2);
            \draw[dashed] (1,-0.2)--(nsm)--(3,-0.2);
        \end{tikzpicture}$};
        \node at (5,-3) {$\begin{tikzpicture}
            \draw[white] (0,-1)--(0,1);
            \draw[blue] (1,-1)--(1,1);
            \draw[goodgreen] (2,-1)--(2,1);
            \node[NS] (ns) at (1.5,0.2) {};
            \draw (1,0.2)--(ns);
            \draw (ns) .. controls (2.2,0) .. (1,-0.2);
        \end{tikzpicture}$};
        \node at (9.5,-3) {$\begin{tikzpicture}
            \node at (0,0) {};
            \node[sevB] (2) at (1,0) {};
            \node[sevR] (3u) at (2,1) {};
            \node[sevR] (3d) at (2,-1) {};
            \node[sevG] (4u) at (3.5,0.5) {};
            \node[sevG] (4d) at (3.5,-0.5) {};
            \draw[red] (3u)--(3d);
            \draw[transform canvas={yshift=1.5pt}] (2)--(3,0)--(4u);
            \draw[transform canvas={yshift=-1.5pt}] (2)--(3,0)--(4d);
        \end{tikzpicture}$};
        \node at (-3,0) {1)};
        \node at (-3,-3) {2)};
        \node at (0,-4.5) {a)};
        \node at (5,-4.5) {b)};
        \node at (9.5,-4.5) {c)};
    \end{tikzpicture}
    \caption{Two supersymmetric brane configurations involving an O$p^-$ orientifold plane. a) Double cover of 3d brane system (NS5 branes in red, D3 branes in black, D5 branes in blue, O5$^-$ plane in green; for clarity the mirror D3 brane is depicted as a dashed line). b) Physical 3d brane system. c) T-dual (resolved) 5d brane web (NS5 branes in red, other 5-branes in black, D7 branes in blue, resolved O7${}^-$ in green; colors do not denote a brane web decomposition). In presence of an O5${}^-$ plane a D3 brane cannot end on a D5 brane and its image. One can allow for D3 branes by either adding an extra D5 brane or adding an extra NS5 brane. While configuration 1) is well known in the literature, configuration 2) has not been appreciated in the 3d setting, and allows to realize novel Higgs branch moduli. This should be thought of as an extension of \cite{Sen:1998rg} whereby the stable non-BPS configuration where a D3 connects a D5 and its image becomes BPS in the presence of an NS5 brane. Configuration 2c) is a known supersymmetric brane web configuration, see for example \cite{Mikhailov:1998bx,Bergman:1998ej}.}
    \label{fig:S-ruleConfig}
\end{figure}

In this paper, we demonstrate that there is no limit to the number of cones that can make up Higgs branches of theories with 8 supercharges. We do so by explicitly constructing families of theories labeled by an integer $p \in \mathbb{N}$ with classical Higgs branch consisting of exactly $p+1$ cones. This can be seen equivalently as a 4d $\mathcal{N}=2$ or as a 3d $\mathcal{N}=4$ statement. 

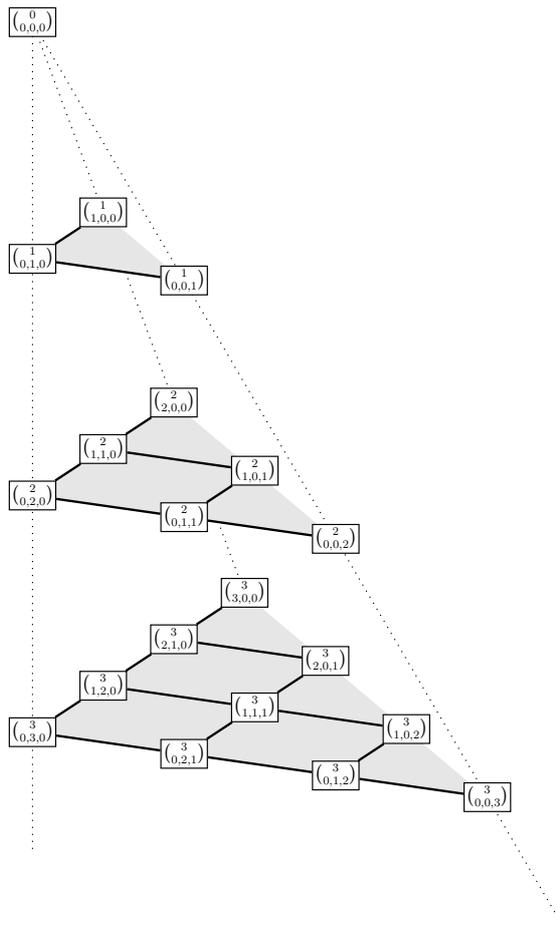
\begin{figure}
\centering
\tdplotsetmaincoords{-108}{65}
\scalebox{1.1}{\begin{tikzpicture}[tdplot_main_coords]
\newcommand{\z}{3}
\newcommand{\y}{2}
\newcommand{\x}{2}
\draw[dotted] (0,0,0)--(0*\x,0*\y,3.5*\z);
\draw[dotted] (0,0,0)--(0*\x,3.5*\y,3.5*\z);
\draw[dotted] (0,0,0)--(3.5*\x,0*\y,3.5*\z);
\draw[white,fill=black!10] (0*\x,0*\y,1*\z)--(0*\x,1*\y,1*\z)--(1*\x,0*\y,1*\z)--(0*\x,0*\y,1*\z);
\draw[white,fill=black!10] (0*\x,0*\y,2*\z)--(0*\x,2*\y,2*\z)--(2*\x,0*\y,2*\z)--(0*\x,0*\y,2*\z);
\draw[white,fill=black!10] (0*\x,0*\y,3*\z)--(0*\x,3*\y,3*\z)--(3*\x,0*\y,3*\z)--(0*\x,0*\y,3*\z);
\node[draw,fill=white,inner sep=1pt] (000) at (0,0,0) {\scalebox{.5}{$\binom{0}{0,0,0}$}};
\node[draw,fill=white,inner sep=1pt] (001) at (0*\x,0*\y,1*\z) {\scalebox{.5}{$\binom{1}{0,1,0}$}};
\node[draw,fill=white,inner sep=1pt] (101) at (1*\x,0*\y,1*\z) {\scalebox{.5}{$\binom{1}{1,0,0}$}};
\node[draw,fill=white,inner sep=1pt] (011) at (0*\x,1*\y,1*\z) {\scalebox{.5}{$\binom{1}{0,0,1}$}};
\node[draw,fill=white,inner sep=1pt] (002) at (0*\x,0*\y,2*\z) {\scalebox{.5}{$\binom{2}{0,2,0}$}};
\node[draw,fill=white,inner sep=1pt] (102) at (1*\x,0*\y,2*\z) {\scalebox{.5}{$\binom{2}{1,1,0}$}};
\node[draw,fill=white,inner sep=1pt] (012) at (0*\x,1*\y,2*\z) {\scalebox{.5}{$\binom{2}{0,1,1}$}};
\node[draw,fill=white,inner sep=1pt] (202) at (2*\x,0*\y,2*\z) {\scalebox{.5}{$\binom{2}{2,0,0}$}};
\node[draw,fill=white,inner sep=1pt] (112) at (1*\x,1*\y,2*\z) {\scalebox{.5}{$\binom{2}{1,0,1}$}};
\node[draw,fill=white,inner sep=1pt] (022) at (0*\x,2*\y,2*\z) {\scalebox{.5}{$\binom{2}{0,0,2}$}};
\node[draw,fill=white,inner sep=1pt] (003) at (0*\x,0*\y,3*\z) {\scalebox{.5}{$\binom{3}{0,3,0}$}};
\node[draw,fill=white,inner sep=1pt] (103) at (1*\x,0*\y,3*\z) {\scalebox{.5}{$\binom{3}{1,2,0}$}};
\node[draw,fill=white,inner sep=1pt] (013) at (0*\x,1*\y,3*\z) {\scalebox{.5}{$\binom{3}{0,2,1}$}};
\node[draw,fill=white,inner sep=1pt] (203) at (2*\x,0*\y,3*\z) {\scalebox{.5}{$\binom{3}{2,1,0}$}};
\node[draw,fill=white,inner sep=1pt] (113) at (1*\x,1*\y,3*\z) {\scalebox{.5}{$\binom{3}{1,1,1}$}};
\node[draw,fill=white,inner sep=1pt] (023) at (0*\x,2*\y,3*\z) {\scalebox{.5}{$\binom{3}{0,1,2}$}};
\node[draw,fill=white,inner sep=1pt] (303) at (3*\x,0*\y,3*\z) {\scalebox{.5}{$\binom{3}{3,0,0}$}};
\node[draw,fill=white,inner sep=1pt] (213) at (2*\x,1*\y,3*\z) {\scalebox{.5}{$\binom{3}{2,0,1}$}};
\node[draw,fill=white,inner sep=1pt] (123) at (1*\x,2*\y,3*\z) {\scalebox{.5}{$\binom{3}{1,0,2}$}};
\node[draw,fill=white,inner sep=1pt] (033) at (0*\x,3*\y,3*\z) {\scalebox{.5}{$\binom{3}{0,0,3}$}};
\draw[thick] (101)--(001)--(011);
\draw[thick] (202)--(102)--(112)--(012)--(022) (102)--(002)--(012);
\draw[thick] (303)--(203)--(213)--(113)--(123)--(023)--(033) (203)--(103)--(113)--(013)--(023) (103)--(003)--(013);
\end{tikzpicture}}
\caption{This figure depicts the Hasse diagrams for the Higgs branches of \eqref{eq:quivUneq2} for $p=0,1,2,3$ from top to bottom (thick black lines). The bottom leaves of the diagrams are on the left, and the top leaves are on the right. All the transitions are Klein singularities of type $A$, see \eqref{Hasse:3d-nis2}. The diagram at floor $p$ has $p+1$ maxima for the $p+1$ cones in the Higgs branch. The number in boxes give the multiplicities of nilpotent elements on each leaf. The entries $\binom{j+k+l}{j,k,l}$ can be evaluated as integers, in which case they reproduce Pascal's pyramid (each entry is the sum of the numbers above it), or as $t^2$-analogs \eqref{eq:binomialAnalog}, in which case they correspond (up to a factor $(-1)^k t^{k(k-1)}$) to the multiplicity factor of the leaf closure in the Higgs scheme. }
\label{fig:pyramid}
\end{figure}

A novel ingredient that makes our construction possible is a new brane configuration in the D3-D5-NS5 system with an O5${}^-$ plane, see Figure \ref{fig:S-ruleConfig}-2b). 
An example of such a family is the quiver theory 
\begin{equation}
    \raisebox{-.5\height}{\begin{tikzpicture}
    \node[gauge,label=below:{$\mathrm{U}(2)$}] (1) at (0,0) {};
    \node[gauge,label=below:{$\mathrm{U}(4)$}] (2) at (2,0) {};
    \node (d) at (3,0) {$\cdots$};
    \node[gauge,label=below:{$\mathrm{U}(2p-2)$}] (3) at (4,0) {};
    \node[gauge,label=below:{$\mathrm{USp}(2p)$}] (4) at (6,0) {};
    \node[flavour,label=below:{$D_2$}] (5) at (8,0) {};
    \draw (1)--(2)--(d)--(3)--(4)--(5);
    \end{tikzpicture}} \;   
    \label{eq:quivUneq2}
\end{equation}
which is directly inspired from the brane construction. Out of the $p+1$ cones in its Higgs branch, $p$ are realized by the new brane configuration. 
The Hasse diagrams of its Higgs branches are represented schematically in Figure \ref{fig:pyramid}, for values of $p = 0,1,2,3$. 
Note that here we only consider the classical Higgs branch, i.e.\ the hyper-K\"ahler quotient, and do not worry how the Higgs branch actually appears in the full moduli space of the theory. We also study the Higgs branch of \eqref{eq:quivUneq2} with unitary nodes replaced by special unitary ones in Section \ref{sec:5d}. In this case we may take the quiver to describe an effective 5d $\mathcal{N}=1$ theory and also consider the Higgs branch after taking various couplings to infinity.

\begin{figure}
    \centering
    \begin{tikzpicture}
        \node (C) at (-7,1) {$\begin{tikzpicture}
            \draw[blue] (0,-1)--(0,1) (1,-1)--(1,1);
            \draw[goodgreen] (2,-1)--(2,1);
            \node[NS] (ns) at (-1,0) {};
            \draw[transform canvas={yshift=1.5pt}] (ns)--(2,0);
            \draw[transform canvas={yshift=-1.5pt}] (ns)--(2,0);
        \end{tikzpicture}$};
        \node at (-7,-0.5) {USp$(2)-[D_2]$};
        \node (O) at (0,0) {$\begin{tikzpicture}
            \draw[blue] (0,-1)--(0,1) (1,-1)--(1,1);
            \draw[goodgreen] (2,-1)--(2,1);
            \node[NS] (ns) at (1.5,0) {};
            \draw (0,0)--(ns);
            \draw (ns) .. controls (2.2,-0.2) .. (1,-0.4);
        \end{tikzpicture}$};
        \node (L) at (-2,3) {$\begin{tikzpicture}
            \draw[blue] (0,-1)--(0,1) (1,-1)--(1,1);
            \draw[goodgreen] (2,-1)--(2,1);
            \node[NS] (ns) at (1.5,0) {};
            \draw (0,-0.4) .. controls (2.5,-0.6) .. (1,-0.8);
        \end{tikzpicture}$};
        \node (R) at (2,3) {$\begin{tikzpicture}
            \draw[blue] (0,-1)--(0,1) (1,-1)--(1,1);
            \draw[goodgreen] (2,-1)--(2,1);
            \node[NS] (ns) at (1.5,0) {};
            \draw (0,-0.8)--(1,-0.8) (1,0)--(ns);
            \draw (ns) .. controls (2.2,-0.2) .. (1,-0.4);
        \end{tikzpicture}$};
        \draw (O)--(L) (O)--(R);
        \node at (-1.5,1.5) {$A_1$};
        \node at (1.5,1.5) {$A_1$};
    \end{tikzpicture}
    \caption{Left: The brane system for a USp$(2)$ theory with 2 hypermultiplets in the fundamental representation.
    Right: The various Higgs phases. At the bottom we depict the origin of the Higgs branch, going up to the left we depict the cone which was found in \cite{Ferlito:2016grh}, going up to the right we depict the cone which was not identified from the brane system in \cite{Ferlito:2016grh}. In this basic example the two cones are isomorphic. When more than one NS5 brane is involved we find non-isomorphic cones, and realizing the various different cones in the brane system is instrumental for computing the various distinct magnetic quivers.}
    \label{fig:Intro_Moduli}
\end{figure}
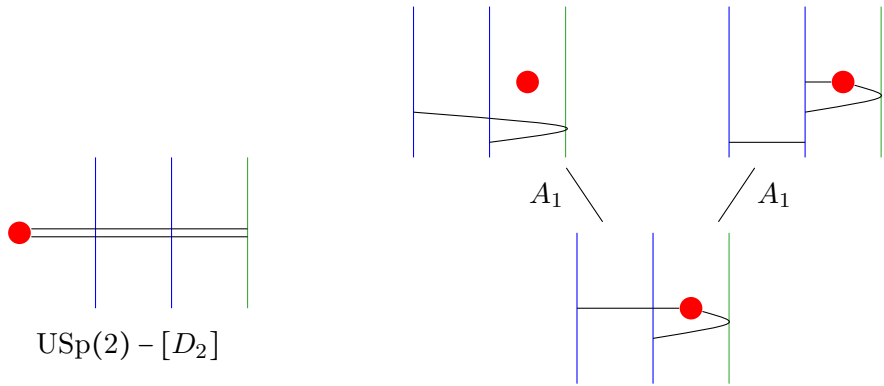

As is well known, in presence of an O5${}^-$ plane, a D3 brane suspended between a D5 brane and its image does not lead to a BPS state. However we can have a D3 brane end on one D5 and the image of a different D5 (Figure \ref{fig:S-ruleConfig}-1). When there is an NS5 brane right next to the O5${}^-$ plane a D3 brane spanning from one D5 brane can end on the NS5 and then continue to the image of the same D5 brane (Figure \ref{fig:S-ruleConfig}-2). This way one can realize different Higgs branch moduli as shown in Figure \ref{fig:Intro_Moduli} for Sp$(1)$ with $2$ fundamental hypermultiplets. This is T-dual to a brane web with inequivalent maximal decompositions. 
In this paper, we show how both brane systems are used to derive the intricate structure of Higgs branches. The multiple cones correspond to multiple highest leaves in the Hasse diagram. The correspondence between these maxima and the brane systems is made explicit using magnetic quivers \cite{Hanany:1996ie,DelZotto:2014kka,Ferlito:2017xdq,Hanany:2018uhm,Cabrera:2018jxt,Cabrera:2019izd,Cabrera:2019dob,Bourget:2019rtl,Bourget:2020asf,Bourget:2020gzi,Beratto:2020wmn,Closset:2020scj,Akhond:2020vhc,Bourget:2020mez,vanBeest:2020kou,Giacomelli:2020gee,Giacomelli:2020ryy,VanBeest:2020kxw,Closset:2020afy,Akhond:2021knl,Martone:2021ixp,Arias-Tamargo:2021ppf,Carta:2021whq,Bourget:2021xex,vanBeest:2021xyt,Carta:2021dyx,Xie:2021ewm,Sperling:2021fcf,Nawata:2021nse,Closset:2021lwy,Bhardwaj:2021mzl,Akhond:2022jts,Carta:2022spy,Carta:2022fxc,Kang:2022zsl,Bertolini:2022osy,Giacomelli:2022drw,Hanany:2022itc,Fazzi:2022yca,Fazzi:2022hal,Nawata:2023rdx,Bourget:2023uhe}. This allows to compute Hilbert series for the corresponding varieties, thereby providing a confirmation for our claims.

As observed in \cite{Bourget:2019rtl}, it is important to distinguish the Higgs variety $\HV$ and the Higgs scheme $\HSCH$. The latter is the affine scheme that is associated to the Higgs chiral ring, which may contain nilpotent operators, while the former is the underlying algebraic variety (reduced scheme, i.e.\ containing no nilpotent operators). Nilpotent operators show up prominently in Higgs branches of 5d $\mathcal{N}=1$ SCFTs \cite{Cremonesi:2015lsa,Ferlito:2017xdq}. In 4d / 3d however, as far as we know, these nilpotent operators only show up for asymptotically free / bad theories. The simplest example of a non-reduced scheme, which does show up as moduli space of supersymmetric theories, is the spectrum of the ring $\mathbb{C}[x]/(x^n)$, which contains an operator $x$ of nilpotency degree $n$. When combined with the Hasse diagram formalism, the presence of nilpotent operators in $\HSCH$ can be seen as an additional structure on $\HV$ whereby leaves acquire multiplicities, which roughly correspond to the order of nilpotency of nilpotent operators on their closures. There is however more structure than just a number: as the nilpotent operators carry R-charge, the multiplicity is given by $q$-analogs of integers and products thereof.\footnote{We use $q = t^2$ for consistency with previous literature using Hilbert series. So the Hilbert series for $\mathrm{Spec} (\mathbb{C}[x]/(x^n))$ is $1 + t^2 + \dots + t^{2n-2} \coloneqq  [n]_{t^2}$. } For the example \eqref{eq:quivUneq2}, one gets the series of diagrams shown in Figure \ref{fig:pyramid}, where on each leaf we have added  the multiplicity of nilpotent elements. We refer the reader to Section \ref{sec:scheme} for more details. It is an open problem to identify nilpotent operators from a brane system \cite{Bourget:2019rtl}.

\subsection*{Plan of the paper}

The paper is organized as follows. In Section \ref{sec:3dd}, we consider classical Higgs branches of unitary-orthosymplectic quivers realized on generalizations of the brane system of Figure \ref{fig:Intro_Moduli}. In Section \ref{sec:5d}, we turn our attention to special unitary-orthosymplectic quivers, using instead brane web techniques, and in Section \ref{sec:5dinf} we analyze how the multiple cones behave under taking various gauge parameters to infinity. 

Two appendices complement the main text. Appendix \ref{app:badCB} collects details on bad Coulomb branches and inversion of Hasse diagrams. Appendix \ref{app:eightcones} contains computational details.

\section{$\mathrm{U}(n)-\mathrm{U}(2n)-\cdots-\mathrm{U}((p-1)n)-\mathrm{USp}(pn)-[D_n]$ on HW Brane Systems}
\label{sec:3dd}
In this section we study theories living on Hanany-Witten brane systems in the presence of an O$5^-$ plane.

\subsection{A Tale of Two Cones}
\label{sec:2cones}

In this section we review a 1-parameter family of quiver gauge theories \cite{Ferlito:2016grh}
\begin{equation}
 \raisebox{-.5\height}{
    \begin{tikzpicture}
    \node[gauge,label=right:{USp$(2k)$}] (1) at (0,0) {};
    \node[flavour,label=right:{$D_{2k}$}] (2) at (0,1) {};
    \draw (1)--(2);
    \end{tikzpicture}
    }
    \label{eq:TwoConeQuiver}
\end{equation}
whose Higgs branch is a union of two cones.
The classical Higgs branch of this theory is straightforward to compute. It is the closure of the maximal height two nilpotent orbit in $\mathfrak{so}(4k)$ under the full orthogonal group $\mathrm{O}(4k)$, which we denote as $\mathcal{O}^{[2^{2k}]}_{\mathrm{O}}$. Nilpotent orbits and the Hasse diagrams of their closures were analyzed in \cite{kraft1980minimal,Kraft1982}. The orbit $\mathcal{O}^{[2^{2k}]}_{\mathrm{O}}$ is a so called \emph{very even} orbit, and as such it is a disjoint union of two isomorphic orbits under the special orthogonal group $\mathrm{SO}(4k)$, which we shall call $\mathcal{O}^{[2^{2k}]I}_{\mathrm{SO}}$ and $\mathcal{O}^{[2^{2k}]II}_{\mathrm{SO}}$.
Its closure, denoted $\overline{\mathcal{O}}^{[2^{2k}]}_{\mathrm{O}}$, is a union of two cones, i.e.\ the closures of the orbits $\mathcal{O}^{[2^{2k}]I}_{\mathrm{SO}}$ and $\mathcal{O}^{[2^{2k}]II}_{\mathrm{SO}}$:
\begin{equation}
\label{eq:veryEven}
    \overline{\mathcal{O}}^{[2^{2k}]}_{\mathrm{O}}=\overline{\mathcal{O}}^{[2^{2k}]I}_{\mathrm{SO}}\cup\overline{\mathcal{O}}^{[2^{2k}]II}_{\mathrm{SO}}\;.
\end{equation}
The intersection of the cones is:
\begin{equation}
    \overline{\mathcal{O}}^{[2^{2k}]I}_{\mathrm{SO}}\cap\overline{\mathcal{O}}^{[2^{2k}]II}_{\mathrm{SO}}=\overline{\mathcal{O}}^{[2^{2k-2},1^4]}_{\mathrm{O}}\;.
    \label{eq:KPInt}
\end{equation}
The Hasse diagram of $\overline{\mathcal{O}}^{[2^{2k}]}_{\mathrm{O}}$ follows from \cite{Kraft1982}:
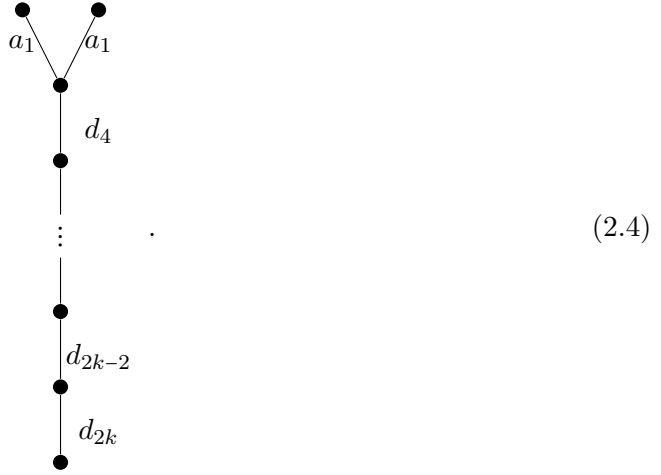
\begin{equation}
    \raisebox{-.5\height}{\begin{tikzpicture}
    \node[hasse,label={[shift={(0.5,0)}]$d_{2k}$}] (1) at (0,0) {};
    \node[hasse,label={[shift={(0.5,0)}]$d_{2k-2}$}] (2) at (0,1) {};
    \node[hasse] (3) at (0,2) {};
    \node (4) at (0,3) {$\vdots$};
    \node[hasse,label={[shift={(0.5,0)}]$d_{4}$}] (5) at (0,4) {};
    \node[hasse] (6) at (0,5) {};
    \node[hasse,label={[shift={(0,-0.8)}]$a_1$}] (7a) at (-0.5,6) {};
    \node[hasse,label={[shift={(0,-0.8)}]$a_1$}] (7b) at (0.5,6) {};
    \draw (1)--(2)--(3)--(4)--(5)--(6)--(7a) (7b)--(6);
    \end{tikzpicture}}\;.
    \label{hasse:Sp(k)-[D2k]}
\end{equation}

The theory \eqref{eq:TwoConeQuiver} does not have a well defined notion of a 3d mirror, as there is no unique interacting SCFT in the IR. In fact, there are two singularities in the Coulomb branch of \eqref{eq:TwoConeQuiver}, cf.\ \cite{Assel:2018exy}. Emanating from each singularity is one of the two cones that make up the classical Higgs branch.\footnote{This phenomenon also happens in 4d $\mathcal{N}=2$ theories \cite{Argyres:1996eh}.} One can still find a set of magnetic quivers, one for each cone in the Higgs branch. This can be thought of as a generalization of the notion of 3d mirror symmetry. The Coulomb branches of the magnetic quivers correspond to the individual cones in the Higgs branch of our theory. The Higgs branches of the magnetic quivers describe the local geometry of the Coulomb branch of the theory close to the different singularities \cite[Section 5]{Assel:2018exy}, as explained in Appendix \ref{app:badCB}. The Higgs branches of magnetic quivers \eqref{eq:KPC1} and \eqref{eq:KPC2} are intersections of Slodowy slices with closures of nilpotent orbits, as discussed in \cite{Cabrera:2018ldc,Hanany:2019tji}. The full moduli space is discussed in detail in Section \ref{sec:3dFullMod}.

In \cite{Ferlito:2016grh} a magnetic quiver was derived from a brane construction to represent one of the cones in the Higgs branch of theory \eqref{eq:TwoConeQuiver}. The magnetic quiver is:
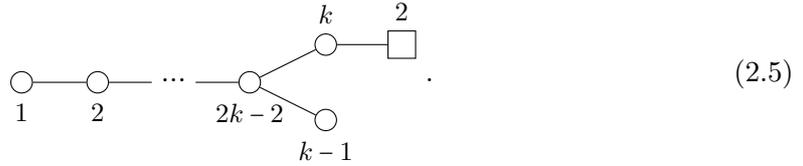
\begin{equation}
    \raisebox{-.5\height}{\begin{tikzpicture}
    \node[gauge,label=below:{\small$1$}] (1) at (0,0) {};
    \node[gauge,label=below:{\small$2$}] (2) at (1,0) {};
    \node (3) at (2,0) {$\cdots$};
    \node[gauge,label=below:{\small$2k-2$}] (4) at (3,0) {};
    \node[gauge,label=above:{\small$k$}] (5a) at (4,0.5) {};
    \node[gauge,label=below:{\small$k-1$}] (5b) at (4,-0.5) {};
    \node[flavour,label=above:{\small$2$}] (6a) at (5,0.5) {};
    \draw (1)--(2)--(3)--(4)--(5a)--(6a) (5b)--(4);
    \end{tikzpicture}}\;.
    \label{eq:KPC1}
\end{equation}
It is straightforward to determine from the algebraic fact \eqref{eq:veryEven} that the magnetic quiver for the other (isomorphic) cone should be:
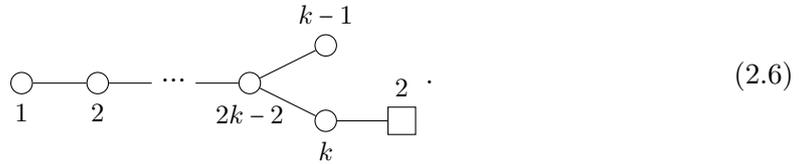
\begin{equation}
    \raisebox{-.5\height}{\begin{tikzpicture}
    \node[gauge,label=below:{\small$1$}] (1) at (0,0) {};
    \node[gauge,label=below:{\small$2$}] (2) at (1,0) {};
    \node (3) at (2,0) {$\cdots$};
    \node[gauge,label=below:{\small$2k-2$}] (4) at (3,0) {};
    \node[gauge,label=above:{\small$k-1$}] (5a) at (4,0.5) {};
    \node[gauge,label=below:{\small$k$}] (5b) at (4,-0.5) {};
    \node[flavour,label=above:{\small$2$}] (6b) at (5,-0.5) {};
    \draw (1)--(2)--(3)--(4)--(5a) (6b)--(5b)--(4);
    \end{tikzpicture}}\;.
    \label{eq:KPC2}
\end{equation}
Let us now derive both of these quivers from a brane construction. The 3d $\mathcal{N}=4$ theory \eqref{eq:TwoConeQuiver} may be realized in a Hanany-Witten brane system through two different constructions:
\begin{enumerate}
    \item Using an $O3^+$ orientifold plane parallel to the gauge D3 branes.
    \item Using an $O5^-$ orientifold plane parallel to the flavor D5 branes.
\end{enumerate}
While the former construction yields orthosymplectic magnetic quivers, the latter construction yields unitary magnetic quivers. As argued in \cite[Section 4.5]{Bourget:2020gzi}, one can employ the second construction, using an $O5^-$ plane, to identify the two cones.\footnote{In \cite{Bourget:2020gzi} 5d $\mathcal{N}=1$ theories on brane webs are studied, but the discussion about two cones is essentially the same in the case presented in this paper.} We are not able to see the two cones with the O3 plane construction -- to date this is still an open question and we encourage the reader to find the second cone. \\

The brane construction representing \eqref{eq:TwoConeQuiver} at the origin\footnote{The notion of origin only makes sense for the classical moduli space, as the quantum moduli space has multiple most singular points.} of its moduli space is:
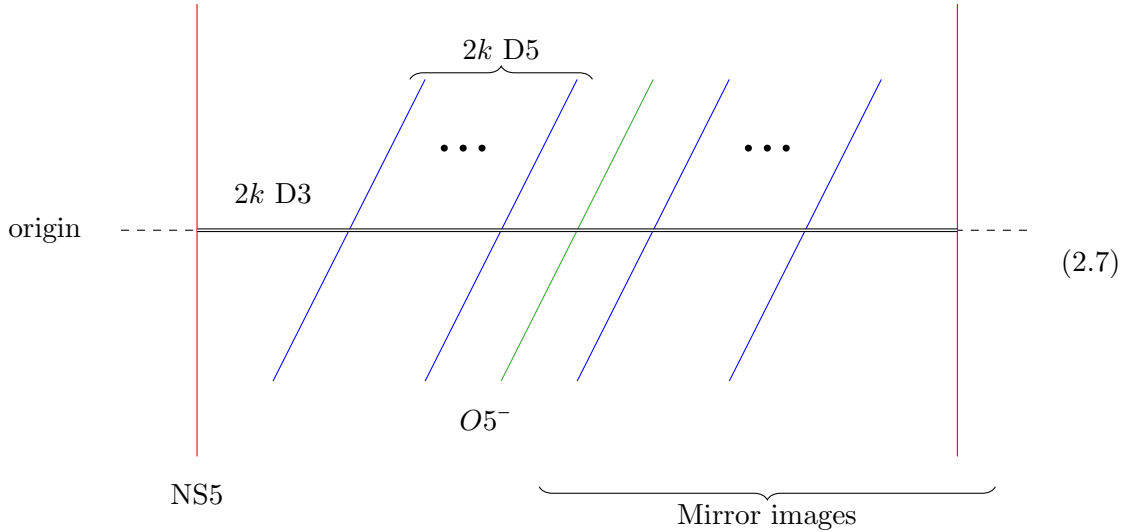
\begin{equation}
    \raisebox{-.5\height}{\begin{tikzpicture}
    \draw[goodred] (0,-3)--(0,3);
    \draw[blue] (1,-2)--(3,2) (3,-2)--(5,2);
    \node at (3.5,1) {\Huge$\cdots$};
    \draw[goodgreen] (4,-2)--(6,2);
    \draw[blue] (5,-2)--(7,2) (7,-2)--(9,2);
    \node at (7.5,1) {\Huge$\cdots$};
    \draw[goodred] (10,-3)--(10,3);
    \draw[dashed] (-1,0)--(11,0);
    \draw[double] (0,0)--(10,0);
    \node at (3.8,-2.5) {$O5^-$};
    \node at (1,0.5) {$2k$ D3};
    \node at (0,-3.5) {NS5};
    \draw [decorate,decoration={brace,amplitude=5pt},xshift=0pt,yshift=-0.2cm]
    (2.8,2.2)--(5.2,2.2) node [black,midway,yshift=0.4cm] {$2k$ D5};
    \draw [decorate,decoration={brace,amplitude=5pt},xshift=0pt,yshift=-0.2cm]
    (10.5,-3.2)--(4.5,-3.2) node [black,midway,yshift=-0.4cm] {Mirror images};
    \node at (-2,0) {origin};
    \end{tikzpicture}}
    \label{branes:Sp(k)-[D2k]full}
\end{equation}
Since we are interested in the Higgs branch, we have to focus on the moduli of the D3 branes moving along the D5 branes. Hence it is enough to depict the brane system in two dimensions, looking at \eqref{branes:Sp(k)-[D2k]full} ``from above'', and we only focus on the physical relevant space, without drawing the mirror images:
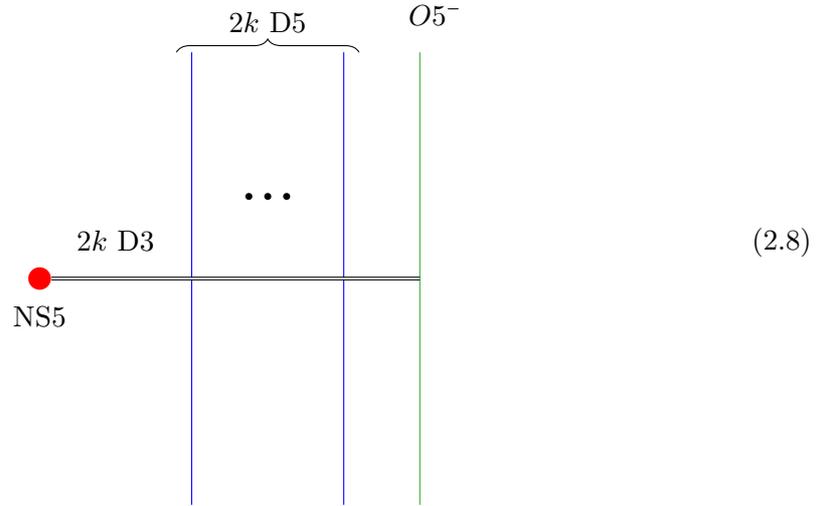
\begin{equation}
    \raisebox{-.5\height}{\begin{tikzpicture}
    \node[NS] (1) at (0,0) {};
    \draw[blue] (2,-3)--(2,3) (4,-3)--(4,3);
    \draw[goodgreen] (5,-3)--(5,3);
    \draw[double] (1)--(5,0);
    \node at (3,1) {\Huge$\cdots$};
    \node at (5.2,3.5) {$O5^-$};
    \node at (1,0.5) {$2k$ D3};
    \node at (0,-0.5) {NS5};
    \draw [decorate,decoration={brace,amplitude=5pt},xshift=0pt,yshift=-0.2cm]
    (1.8,3.2)--(4.2,3.2) node [black,midway,yshift=0.4cm] {$2k$ D5};
    \end{tikzpicture}}
    \label{branes:Sp(k)-[D2k]HiggsHalf}
\end{equation}
From now on we suppress the labels for the branes and simply write the number of D3 branes. In order to read magnetic quivers in a straightforward manner, we perform a series of Hanany-Witten transitions to obtain the brane system:
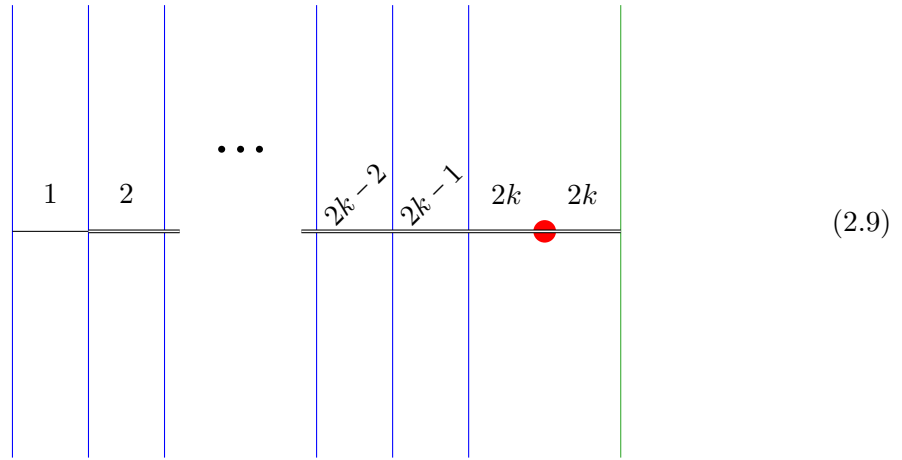
\begin{equation}
    \raisebox{-.5\height}{\begin{tikzpicture}
    \draw[blue] (0,-3)--(0,3) (1,-3)--(1,3) (2,-3)--(2,3) (4,-3)--(4,3) (5,-3)--(5,3) (6,-3)--(6,3);
    \draw[goodgreen] (8,-3)--(8,3);
    \node at (3,1) {\Huge$\cdots$};
    \node[NS] at (7,0) {};
    \draw (0,0)--(1,0);
    \draw[double] (1,0)--(2.2,0) (3.8,0)--(8,0);
    \node at (0.5,0.5) {$1$};
    \node at (1.5,0.5) {$2$};
    \node[rotate=45] at (4.5,0.5) {$2k-2$};
    \node[rotate=45] at (5.5,0.5) {$2k-1$};
    \node at (6.5,0.5) {$2k$};
    \node at (7.5,0.5) {$2k$};
    \end{tikzpicture}}
    \label{branes:Sp(k)-[D2k]HW}
\end{equation}
Now we can break the D3 branes along the D5 branes and move them along the Higgs branch, keeping in mind that we are dealing with an $O5^-$ plane:
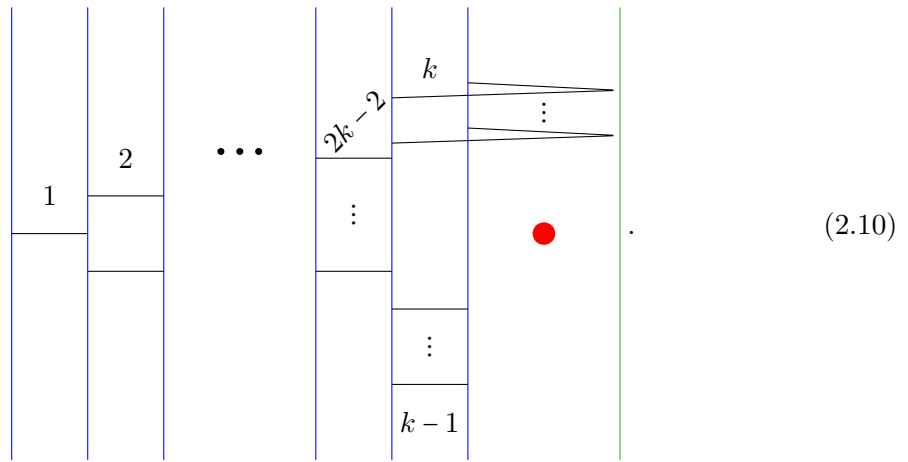
\begin{equation}
    \raisebox{-.5\height}{\begin{tikzpicture}
    \draw[blue] (0,-3)--(0,3) (1,-3)--(1,3) (2,-3)--(2,3) (4,-3)--(4,3) (5,-3)--(5,3) (6,-3)--(6,3);
    \draw[goodgreen] (8,-3)--(8,3);
    \node at (3,1) {\Huge$\cdots$};
    \node[NS] at (7,0) {};
    \draw (0,0)--(1,0);
    \node at (0.5,0.5) {$1$};
    \node at (1.5,1) {$2$};
    \draw (1,0.5)--(2,0.5) (1,-0.5)--(2,-0.5);
    \node[rotate=45] at (4.5,1.5) {$2k-2$};
    \draw (4,1)--(5,1) (4,-0.5)--(5,-0.5);
    \node at (4.5,0.25) {$\vdots$};
    \node at (5.5,-2.5) {$k-1$};
    \draw (5,-1)--(6,-1) (5,-2)--(6,-2);
    \node at (5.5,-1.5) {$\vdots$};
    \node at (5.5,2.2) {$k$};
    \draw[rounded corners] (6,2)--(8,1.9)--(5,1.8);
    \node at (7,1.6) {$\vdots$};
    \draw[rounded corners] (6,1.4)--(8,1.3)--(5,1.2);
    \end{tikzpicture}}\;.
    \label{branes:Sp(k)-[D2k]C1}
\end{equation}
We can identify the magnetic quiver \eqref{eq:KPC1}. This is the analysis presented in \cite{Ferlito:2016grh},\footnote{We do not perform an S-duality as it is not necessary for reading magnetic quivers.} and matches \eqref{eq:KPC1}.

\subsection{New Higgs Phases from Old Brane Systems}
\label{sec:NewBraneSys}
In order to identify the second magnetic quiver let us turn to a different type of brane system, in fact a brane web, which supports the 5d $\mathcal{N}=1$ version of the theory. The classical Higgs branch of the theory in dimensions $3-6$ is the same, so the magnetic quivers read from the brane web apply to the 3d theory just as well. The relevant brane web, after resolving an O7$^-$ plane \cite{Sen:1996vd,Zafrir:2015rga} and suitable Hanany-Witten transitions, is:
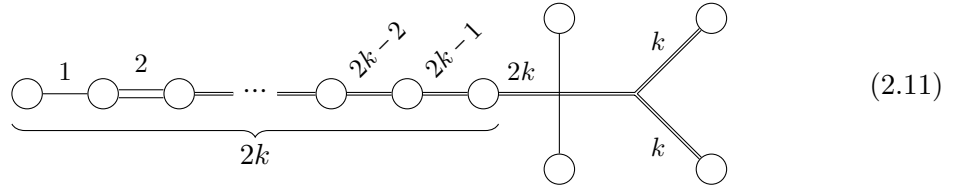
\begin{equation}
    \raisebox{-.5\height}{\begin{tikzpicture}
    \node[sev] (1) at (0,0) {};
    \node[sev] (2) at (1,0) {};
    \node[sev] (3) at (2,0) {};
    \node (4) at (3,0) {$\cdots$};
    \node[sev] (5) at (4,0) {};
    \node[sev] (6) at (5,0) {};
    \node[sev] (7) at (6,0) {};
    \node[sev] (8a) at (7,1) {};
    \node[sev] (8b) at (7,-1) {};
    \node[sev] (9a) at (9,1) {};
    \node[sev] (9b) at (9,-1) {};
    \draw[transform canvas={yshift=-1.5pt}] (2)--(3);
    \draw[transform canvas={yshift=1.5pt}] (2)--(3);
    \draw[double] (3)--(4)--(5)--(6)--(7)--(8,0)--(9a) (9b)--(8,0);
    \draw (1)--(2) (8a)--(8b);
    \draw [decorate,decoration={brace,amplitude=5pt},xshift=0pt,yshift=-0.2cm]
    (6.2,-0.2)--(-0.2,-0.2) node [black,midway,yshift=-0.4cm] {$2k$};
    \node at (0.5,0.3) {\small$1$};
    \node at (1.5,0.4) {\small$2$};
    \node[rotate=45] at (4.6,0.6) {\small$2k-2$};
    \node[rotate=45] at (5.6,0.6) {\small$2k-1$};
    \node at (6.5,0.3) {\small$2k$};
    \node at (8.3,0.7) {\small$k$};
    \node at (8.3,-0.7) {\small$k$};
    \end{tikzpicture}}
    \label{branes:twocones5d}
\end{equation}
We can use the rules developed in \cite{Cabrera:2018jxt} to read off the magnetic quivers from maximal subdivisions of the web. Keeping in mind the S-rule, one maximal subdivision is:

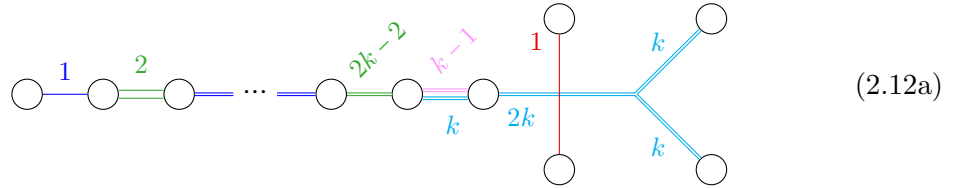
\begin{subequations}
\begin{equation}
    \raisebox{-.5\height}{\begin{tikzpicture}
    \node[sev] (1) at (0,0) {};
    \node[sev] (2) at (1,0) {};
    \node[sev] (3) at (2,0) {};
    \node (4) at (3,0) {$\cdots$};
    \node[sev] (5) at (4,0) {};
    \node[sev] (6) at (5,0) {};
    \node[sev] (7) at (6,0) {};
    \node[sev] (8a) at (7,1) {};
    \node[sev] (8b) at (7,-1) {};
    \node[sev] (9a) at (9,1) {};
    \node[sev] (9b) at (9,-1) {};
    \draw[blue] (1)--(2);
    \draw[goodgreen,transform canvas={yshift=-1.5pt}] (2)--(3);
    \draw[goodgreen,transform canvas={yshift=1.5pt}] (2)--(3);
    \draw[blue,double] (3)--(4)--(5);
    \draw[goodgreen,double] (5)--(6);
    \draw[goodmagenta,double,transform canvas={yshift=1.5pt}] (6)--(7);
    \draw[cyan,double,transform canvas={yshift=-1.5pt}] (6)--(7);
    \draw[cyan,double] (7)--(8,0)--(9a) (9b)--(8,0);
    \draw[goodred] (8a)--(8b);
    \node at (0.5,0.3) {\color{blue}\small$1$};
    \node at (1.5,0.4) {\color{goodgreen}\small$2$};
    \node[rotate=45] at (4.6,0.6) {\color{goodgreen}\small$2k-2$};
    \node[rotate=45] at (5.6,0.6) {\color{goodmagenta}\small$k-1$};
    \node at (6.7,0.7) {\color{goodred}\small$1$};
    \node at (5.6,-0.4) {\color{cyan}\small$k$};
    \node at (6.5,-0.3) {\color{cyan}\small$2k$};
    \node at (8.3,0.7) {\color{cyan}\small$k$};
    \node at (8.3,-0.7) {\color{cyan}\small$k$};
    \end{tikzpicture}}
    \label{branes:twocones5d1}
\end{equation}
Alternatively, we can combine the {\color{goodred}red} NS5 brane with two parallel {\color{cyan}cyan} D5 branes and obtain the second maximal subdivision:
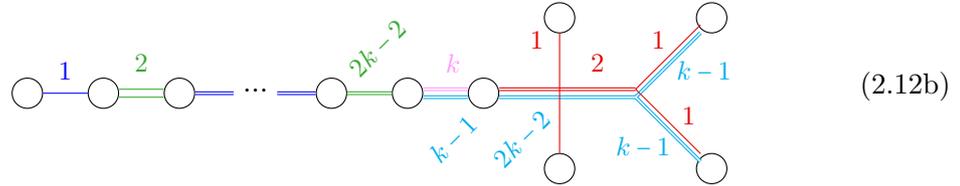
\begin{equation}
    \raisebox{-.5\height}{\begin{tikzpicture}
    \node[sev] (1) at (0,0) {};
    \node[sev] (2) at (1,0) {};
    \node[sev] (3) at (2,0) {};
    \node (4) at (3,0) {$\cdots$};
    \node[sev] (5) at (4,0) {};
    \node[sev] (6) at (5,0) {};
    \node[sev] (7) at (6,0) {};
    \node[sev] (8a) at (7,1) {};
    \node[sev] (8b) at (7,-1) {};
    \node[sev] (9a) at (9,1) {};
    \node[sev] (9b) at (9,-1) {};
    \draw[blue] (1)--(2);
    \draw[goodgreen,transform canvas={yshift=-1.5pt}] (2)--(3);
    \draw[goodgreen,transform canvas={yshift=1.5pt}] (2)--(3);
    \draw[blue,double] (3)--(4)--(5);
    \draw[goodgreen,double] (5)--(6);
    \draw[goodmagenta,double,transform canvas={yshift=1.5pt}] (6)--(7);
    \draw[cyan,double,transform canvas={yshift=-1.5pt}] (6)--(7);
    \draw[cyan,double,transform canvas={yshift=-1.5pt}] (7)--(8,0)--(9a) (9b)--(8,0);
    \draw[goodred,transform canvas={yshift=1.5pt}] (8,0)--(9a) (9b)--(8,0);
    \draw[goodred,transform canvas={yshift=2pt}] (7)--(8,0);
    \draw[goodred,transform canvas={yshift=1pt}] (7)--(8,0);
    \draw[goodred] (8a)--(8b);
    \node at (0.5,0.3) {\color{blue}\small$1$};
    \node at (1.5,0.4) {\color{goodgreen}\small$2$};
    \node[rotate=45] at (4.6,0.6) {\color{goodgreen}\small$2k-2$};
    \node at (5.6,0.4) {\color{goodmagenta}\small$k$};
    \node[rotate=45] at (5.6,-0.6) {\color{cyan}\small$k-1$};
    \node[rotate=45] at (6.5,-0.6) {\color{cyan}\small$2k-2$};
    \node at (6.7,0.7) {\color{goodred}\small$1$};
    \node at (7.5,0.4) {\color{goodred}\small$2$};
    \node at (8.3,0.7) {\color{goodred}\small$1$};
    \node at (8.7,-0.3) {\color{goodred}\small$1$};
    \node at (8.9,0.3) {\color{cyan}\small$k-1$};
    \node at (8.1,-0.7) {\color{cyan}\small$k-1$};
    \end{tikzpicture}}
    \label{branes:twocones5d2}
\end{equation}
\end{subequations}
The corresponding magnetic quivers read
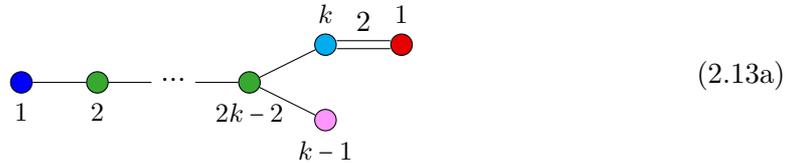
\begin{subequations}
\begin{equation}
    \raisebox{-.5\height}{\begin{tikzpicture}
    \node[gaugeb,label=below:{\small$1$}] (1) at (0,0) {};
    \node[gaugeg,label=below:{\small$2$}] (2) at (1,0) {};
    \node (3) at (2,0) {$\cdots$};
    \node[gaugeg,label=below:{\small$2k-2$}] (4) at (3,0) {};
    \node[gaugec,label=above:{\small$k$}] (5a) at (4,0.5) {};
    \node[gaugem,label=below:{\small$k-1$}] (5b) at (4,-0.5) {};
    \node[gauger,label=above:{\small$1$}] (6a) at (5,0.5) {};
    \draw (1)--(2)--(3)--(4)--(5a) (5b)--(4);
    \draw[transform canvas={yshift=-1.5pt}] (5a)--(6a);
    \draw[transform canvas={yshift=1.5pt}] (5a)--(6a);
    \node at (4.5,0.8) {$2$};
    \end{tikzpicture}}
    \label{magnetic:twocones5d1}
\end{equation}
and
\begin{equation}
    \raisebox{-.5\height}
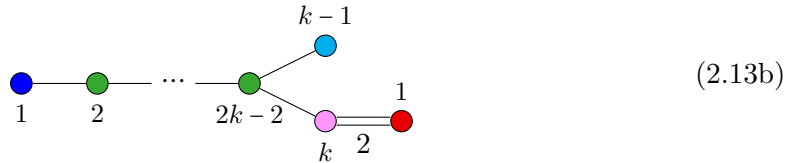
{\begin{tikzpicture}
    \node[gaugeb,label=below:{\small$1$}] (1) at (0,0) {};
    \node[gaugeg,label=below:{\small$2$}] (2) at (1,0) {};
    \node (3) at (2,0) {$\cdots$};
    \node[gaugeg,label=below:{\small$2k-2$}] (4) at (3,0) {};
    \node[gaugec,label=above:{\small$k-1$}] (5a) at (4,0.5) {};
    \node[gaugem,label=below:{\small$k$}] (5b) at (4,-0.5) {};
    \node[gauger,label=above:{\small$1$}] (6b) at (5,-0.5) {};
    \draw (1)--(2)--(3)--(4)--(5a) (5b)--(4);
    \draw[transform canvas={yshift=-1.5pt}] (5b)--(6b);
    \draw[transform canvas={yshift=1.5pt}] (5b)--(6b);
    \node at (4.5,-0.8) {$2$};
    \end{tikzpicture}}
    \label{magnetic:twocones5d2}
\end{equation}
\end{subequations}
Upon ungauging the red U$(1)$ gauge nodes in the two quivers we obtain the expected quivers. Crucial in obtaining the second magnetic quiver is the involvement of the NS5 brane. Let us use this knowledge to identify a different Higgs phase in the 3d system.

In presence of an $O5^-$ plane, a D3 brane cannot span between a D5 and its image in a supersymmetric way. However, we can use the presence of the NS5 brane. A D3 brane can span between a D5 and the NS5 and between the NS5 and the mirror image of the D5, see Figure \ref{fig:S-ruleConfig}. This allows us to go to a second maximal Higgs phase in the brane construction:
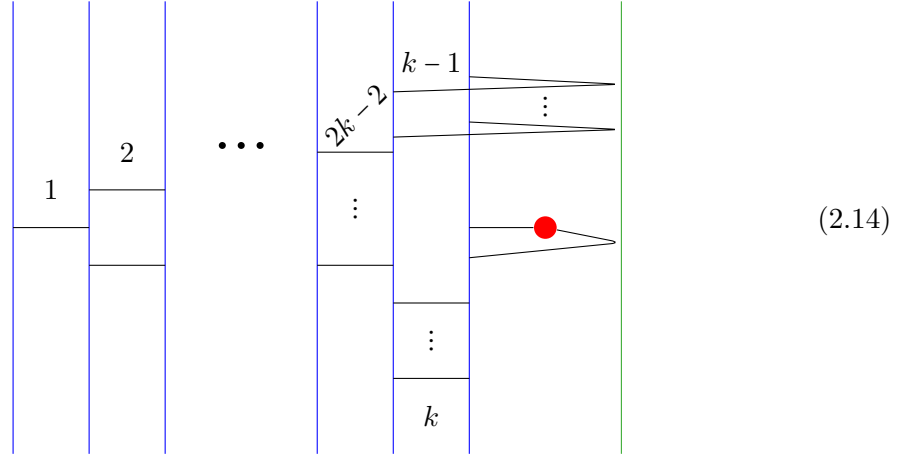
\begin{equation}
    \raisebox{-.5\height}{\begin{tikzpicture}
    \draw[blue] (0,-3)--(0,3) (1,-3)--(1,3) (2,-3)--(2,3) (4,-3)--(4,3) (5,-3)--(5,3) (6,-3)--(6,3);
    \draw[goodgreen] (8,-3)--(8,3);
    \node at (3,1) {\Huge$\cdots$};
    \node[NS] (1) at (7,0) {};
    \draw (0,0)--(1,0);
    \node at (0.5,0.5) {$1$};
    \node at (1.5,1) {$2$};
    \draw (1,0.5)--(2,0.5) (1,-0.5)--(2,-0.5);
    \node[rotate=45] at (4.5,1.5) {$2k-2$};
    \draw (4,1)--(5,1) (4,-0.5)--(5,-0.5);
    \node at (4.5,0.25) {$\vdots$};
    \node at (5.5,-2.5) {$k$};
    \draw (5,-1)--(6,-1) (5,-2)--(6,-2);
    \node at (5.5,-1.5) {$\vdots$};
    \node at (5.5,2.2) {$k-1$};
    \draw[rounded corners] (6,2)--(8,1.9)--(5,1.8);
    \node at (7,1.6) {$\vdots$};
    \draw[rounded corners] (6,1.4)--(8,1.3)--(5,1.2);
    \draw[rounded corners] (6,0)--(1)--(8,-0.2)--(6,-0.4);
    \end{tikzpicture}}
    \label{branes:Sp(k)-[D2k]C2}
\end{equation}
We have therefore identified the second cone. From \eqref{branes:Sp(k)-[D2k]C2} we can read the magnetic quiver:
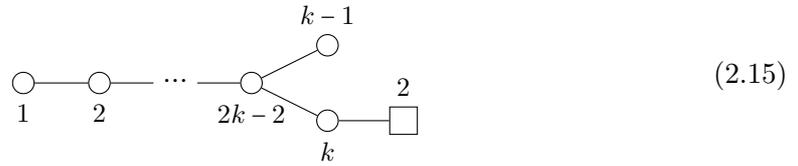
\begin{equation}
    \raisebox{-.5\height}{\begin{tikzpicture}
    \node[gauge,label=below:{\small$1$}] (1) at (0,0) {};
    \node[gauge,label=below:{\small$2$}] (2) at (1,0) {};
    \node (3) at (2,0) {$\cdots$};
    \node[gauge,label=below:{\small$2k-2$}] (4) at (3,0) {};
    \node[gauge,label=above:{\small$k-1$}] (5a) at (4,0.5) {};
    \node[gauge,label=below:{\small$k$}] (5b) at (4,-0.5) {};
    \node[flavour,label=above:{\small$2$}] (6b) at (5,-0.5) {};
    \draw (1)--(2)--(3)--(4)--(5a) (6b)--(5b)--(4);
    \end{tikzpicture}}
    \label{magnetic:Sp(k)-[D2k]C2}
\end{equation}
which is what we already expected to find, see \eqref{eq:KPC2}. The intersection of the two cones is readily found from the brane set up by aligning D3 branes, thus forming a non maximally decomposed brane configuration:
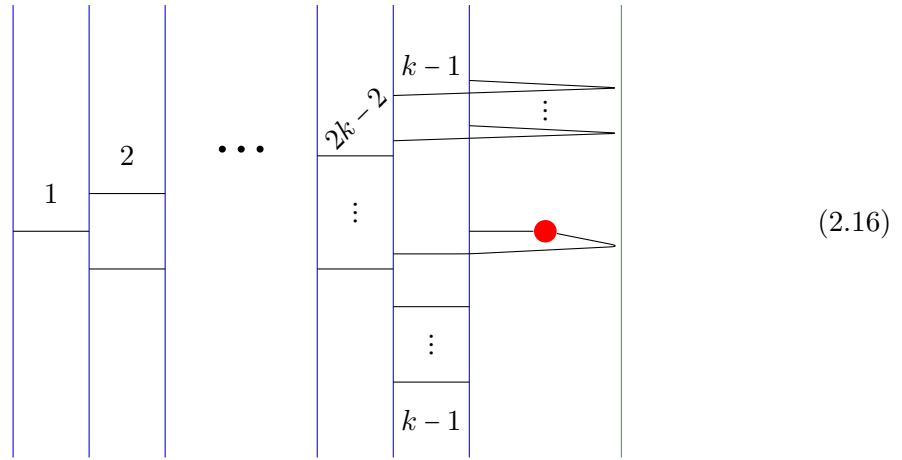
\begin{equation}
    \raisebox{-.5\height}{\begin{tikzpicture}
    \draw[blue] (0,-3)--(0,3) (1,-3)--(1,3) (2,-3)--(2,3) (4,-3)--(4,3) (5,-3)--(5,3) (6,-3)--(6,3);
    \draw[goodgreen] (8,-3)--(8,3);
    \node at (3,1) {\Huge$\cdots$};
    \node[NS] (1) at (7,0) {};
    \draw (0,0)--(1,0);
    \node at (0.5,0.5) {$1$};
    \node at (1.5,1) {$2$};
    \draw (1,0.5)--(2,0.5) (1,-0.5)--(2,-0.5);
    \node[rotate=45] at (4.5,1.5) {$2k-2$};
    \draw (4,1)--(5,1) (4,-0.5)--(5,-0.5);
    \node at (4.5,0.25) {$\vdots$};
    \node at (5.5,-2.5) {$k-1$};
    \draw (5,-1)--(6,-1) (5,-2)--(6,-2);
    \node at (5.5,-1.5) {$\vdots$};
    \node at (5.5,2.2) {$k-1$};
    \draw[rounded corners] (6,2)--(8,1.9)--(5,1.8);
    \node at (7,1.6) {$\vdots$};
    \draw[rounded corners] (6,1.4)--(8,1.3)--(5,1.2);
    \draw[rounded corners] (6,0)--(1)--(8,-0.2)--(6,-0.3)--(5,-0.3);
    \end{tikzpicture}}
    \label{branes:Sp(k)-[D2k]Int}
\end{equation}
The magnetic quiver for the intersection reads
\begin{equation}
    \raisebox{-.5\height}{\begin{tikzpicture}
    \node[gauge,label=below:{\small$1$}] (1) at (0,0) {};
    \node[gauge,label=below:{\small$2$}] (2) at (1,0) {};
    \node (3) at (2,0) {$\cdots$};
    \node[gauge,label=below:{\small$2k-2$}] (4) at (3,0) {};
    \node[gauge,label=above:{\small$k-1$}] (5a) at (4,0.5) {};
    \node[gauge,label=below:{\small$k-1$}] (5b) at (4,-0.5) {};
    \node[flavour,label=left:{\small$1$}] (4f) at (3,1) {};
    \draw (1)--(2)--(3)--(4)--(5a) (4f)--(4)--(5b);
    \end{tikzpicture}}\;, 
    \label{magnetic:Sp(k)-[D2k]Int}
\end{equation}
whose Coulomb branch is indeed $\overline{\mathcal{O}}^{[2^{2k-2},1^4]}_{\mathrm{O}}$, as expected from \eqref{eq:KPInt}. Note that using the quiver subtraction algorithm \cite{Bourget:2019aer} on the magnetic quivers to obtain the Hasse diagram of the Higgs branch of \eqref{eq:TwoConeQuiver} yields \eqref{hasse:Sp(k)-[D2k]}. 

We have seen how the theory \eqref{eq:TwoConeQuiver} has a Higgs branch made up of two cones, which correspond to phases in a brane system. Armed with this new understanding, we now turn to generalizations to theories where an arbitrary number of cones are involved. 

\subsection{A Tale of N Cones}
\label{sec:3d}
\label{sec:Umagquiv}

We start by expanding the brane construction \eqref{branes:Sp(k)-[D2k]full} to
\begin{equation}
    \makebox[\textwidth][c]{
    \raisebox{-.5\height}{\begin{tikzpicture}[scale=0.7]
    \draw[goodred] (0,-3)--(0,3) (1,-3)--(1,3) (2,-3)--(2,3) (4,-3)--(4,3) (5,-3)--(5,3);
    \draw[blue] (6,-2)--(8,2) (8,-2)--(10,2);
    \draw[goodgreen] (9,-2)--(11,2);
    \draw[blue] (10,-2)--(12,2) (12,-2)--(14,2);
    \draw[goodred] (15,-3)--(15,3) (16,-3)--(16,3) (18,-3)--(18,3) (19,-3)--(19,3) (20,-3)--(20,3);
    \draw[dashed] (-2,0)--(22,0);
    \draw[double] (0,0)--(2.5,0) (3.5,0)--(16.5,0) (17.5,0)--(20,0);
    \draw [decorate,decoration={brace,amplitude=5pt},xshift=0pt,yshift=-0.2cm]
    (5.2,-3.2)--(-0.2,-3.2) node [black,midway,yshift=-0.4cm] {$p$ NS5};
    \draw [decorate,decoration={brace,amplitude=5pt},xshift=0pt,yshift=-0.2cm]
    (7.8,2.2)--(10.2,2.2) node [black,midway,yshift=0.4cm] {$n$ D5};
    \node at (0.5,0.5) {$k_1$};
    \node at (1.5,0.5) {$k_2$};
    \node at (4.5,0.5) {$k_{p-1}$};
    \node at (6,0.5) {$k_p$ D3};
    \node at (8.8,-2.5) {$O5^-$};
    \draw [decorate,decoration={brace,amplitude=5pt},xshift=0pt,yshift=-0.2cm]
    (20.5,-3.2)--(9.5,-3.2) node [black,midway,yshift=-0.4cm] {Mirror images};
    \node at (3,2) {\Huge$\cdots$};
    \node at (8.5,1) {\Huge$\cdots$};
    \node at (17,2) {\Huge$\cdots$};
    \node at (12.5,1) {\Huge$\cdots$};
    \end{tikzpicture}}
    }
    \label{branes:Set-Up-USPfull}
\end{equation}
with $k_1 \leq k_2 \leq \dots \leq k_p$. The brane system is only consistent for $k_p$ even. We read the following electric quiver:
\begin{equation}
    \raisebox{-.5\height}{\begin{tikzpicture}
    \node[gauge,label=below:{$\mathrm{U}(k_1)$}] (1) at (0,0) {};
    \node[gauge,label=below:{$\mathrm{U}(k_2)$}] (2) at (2,0) {};
    \node (d) at (3,0) {$\cdots$};
    \node[gauge,label=below:{$\mathrm{U}(k_{p-1})$}] (3) at (4,0) {};
    \node[gauge,label=below:{$\mathrm{USp}(k_p)$}] (4) at (6,0) {};
    \node[flavour,label=below:{$D_n$}] (5) at (8,0) {};
    \draw (1)--(2)--(d)--(3)--(4)--(5);
    \end{tikzpicture}} \qquad,\quad k_p\textnormal{ even}\;.
    \label{electric:uspstart}
\end{equation}
Again, we resort to 2-dimensional drawings, looking at \eqref{branes:Set-Up-USPfull} ``from above'', and depicting only the physical space:
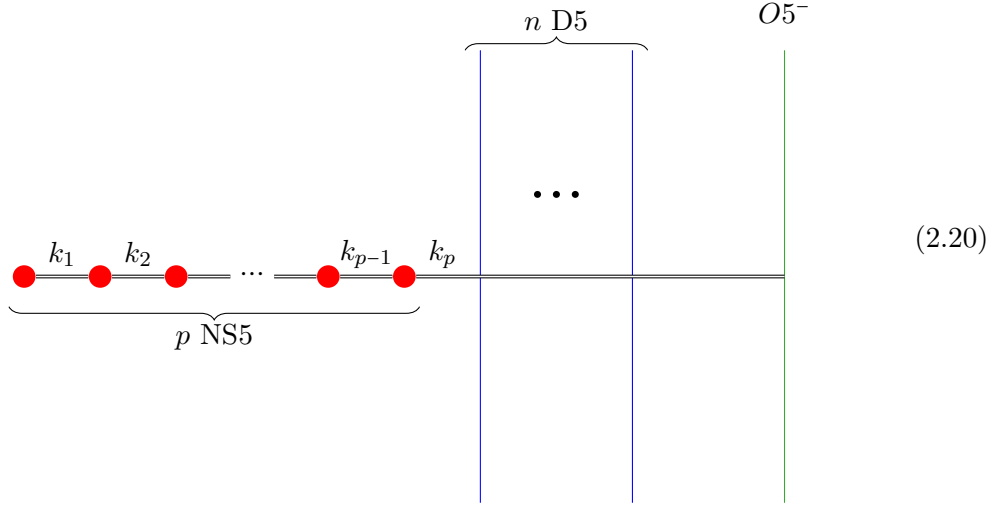
\begin{equation}
    \raisebox{-.5\height}{\begin{tikzpicture}
    \node[NS] (1) at (0,0) {};
    \node[NS] (2) at (1,0) {};
    \node[NS] (3) at (2,0) {};
    \node (4) at (3,0) {$\cdots$};
    \node[NS] (5) at (4,0) {};
    \node[NS] (6) at (5,0) {};
    \draw[blue] (6,-3)--(6,3) (8,-3)--(8,3);
    \draw[goodgreen] (10,-3)--(10,3);
    \draw[double] (1)--(2)--(3)--(4)--(5)--(6)--(10,0);
    \node at (7,1) {\Huge$\cdots$};
    \node at (0.5,0.3) {$k_1$};
    \node at (1.5,0.3) {$k_2$};
    \node at (4.5,0.3) {$k_{p-1}$};
    \node at (5.5,0.3) {$k_p$};
    \draw [decorate,decoration={brace,amplitude=5pt},xshift=0pt,yshift=-0.2cm]
    (5.2,-0.2)--(-0.2,-0.2) node [black,midway,yshift=-0.4cm] {$p$ NS5};
    \draw [decorate,decoration={brace,amplitude=5pt},xshift=0pt,yshift=-0.2cm]
    (5.8,3.2)--(8.2,3.2) node [black,midway,yshift=0.4cm] {$n$ D5};
    \node at (10,3.5) {$O5^-$};
    \end{tikzpicture}}
\end{equation}
We want to go to the Higgs phase of this brane set up and read off magnetic quivers. This is best done after moving the $p$ NS5 branes past $D5$ branes, as in Section \ref{sec:2cones}. In order to keep track of the annihilation of D3 branes, let us use a reparameterization:
\begin{equation}
    k_j=\sum_{i=1}^{j}l_i\;
\end{equation}
for $1 \leq j \leq p$, and separate the NS5 vertically (in our depiction, not physically)
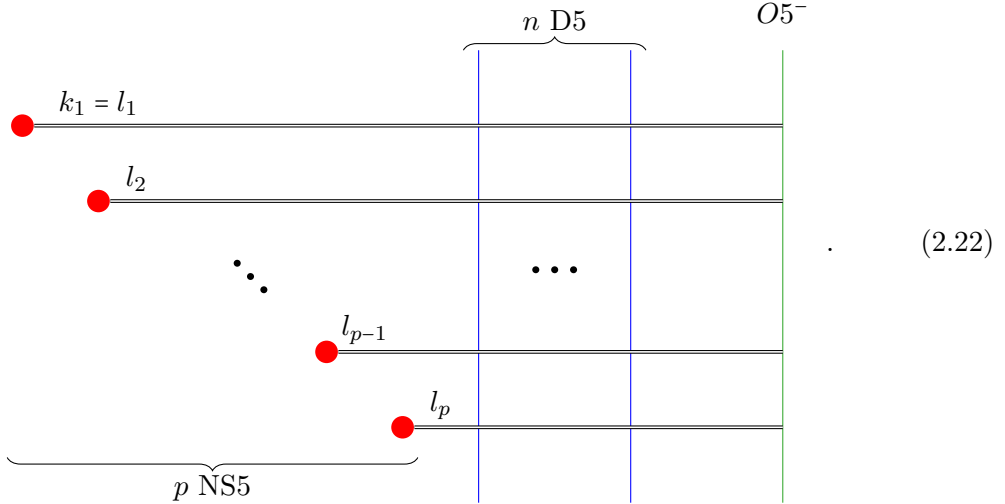
\begin{equation}
    \raisebox{-.5\height}{\begin{tikzpicture}
    \node[NS] (1) at (0,2) {};
    \node[NS] (2) at (1,1) {};
    \node (4) at (3,0) {\Huge$\ddots$};
    \node[NS] (5) at (4,-1) {};
    \node[NS] (6) at (5,-2) {};
    \draw[blue] (6,-3)--(6,3) (8,-3)--(8,3);
    \draw[goodgreen] (10,-3)--(10,3);
    \draw[double] (1)--(10,2) (2)--(10,1) (5)--(10,-1) (6)--(10,-2);
    \node at (1,2.3) {$k_1=l_1$};
    \node at (1.5,1.3) {$l_2$};
    \node at (4.5,-0.7) {$l_{p-1}$};
    \node at (5.5,-1.7) {$l_p$};
    \node at (7,0) {\Huge$\cdots$};
    \draw [decorate,decoration={brace,amplitude=5pt},xshift=0pt,yshift=-0.2cm]
    (5.2,-2.2)--(-0.2,-2.2) node [black,midway,yshift=-0.4cm] {$p$ NS5};
    \draw [decorate,decoration={brace,amplitude=5pt},xshift=0pt,yshift=-0.2cm]
    (5.8,3.2)--(8.2,3.2) node [black,midway,yshift=0.4cm] {$n$ D5};
    \node at (10,3.5) {$O5^-$};
    \end{tikzpicture}}\;.
\end{equation}
We obtain the bounds for complete Higgsing
\begin{equation}
    l_1\leq l_2\leq\dots\leq l_{p-1}\leq l_p\leq n\;.
\end{equation}
If this bound is violated, then even in the full Higgs phase of the brane set up some D3 are free to move along the NS5 branes indicating the Coulomb moduli of the effective theory at a general point on the Higgs branch.\\

If some of the $l_i=n$ then we have a situation similar to the one in Section \ref{sec:2cones}: After performing suitable Hanany-Witten transitions there are NS5 between the rightmost D5 and the $O5^-$ plane and the Higgs branch consists of multiple cones. To be precise, if
\begin{equation}
    \begin{split}
        l_i&<n \qquad \textrm{for} \quad 1\leq i\leq x\\
        l_j&=n \qquad \textrm{for} \quad x+1\leq j\leq p\;,
    \end{split}
\end{equation}
then there are $p-x$ NS5 branes in the last interval, and therefore $p-x+1$ cones in the Higgs branch of \eqref{electric:uspstart}. In this work we focus for simplicity on a special 2-parameter $(p,n)$ family of electric quivers {$\mathsf{Q}_e$} with $l_i=n$ for all $1\leq i\leq p$: 
\begin{equation}
    \raisebox{-.5\height}{\begin{tikzpicture}
    \node at (-2,0) {$\mathsf{Q}_e (p,n)$};
    \node at (-1,0) {$=$};
    \node[gauge,label=below:{$\mathrm{U}(n)$}] (1) at (0,0) {};
    \node[gauge,label=below:{$\mathrm{U}(2n)$}] (2) at (2,0) {};
    \node (d) at (3,0) {$\cdots$};
    \node[gauge,label=below:{$\mathrm{U}((p-1)n)$}] (3) at (4,0) {};
    \node[gauge,label=below:{$\mathrm{USp}(pn)$}] (4) at (6,0) {};
    \node[flavour,label=below:{$D_n$}] (5) at (8,0) {};
    \draw (1)--(2)--(d)--(3)--(4)--(5);
    \end{tikzpicture}}\qquad,\quad pn\textnormal{ even}\;.
    \label{electric:usp}
\end{equation}
We now demonstrate, by using the brane configurations discussed in Section \ref{sec:NewBraneSys}, that a member of this family has a Higgs branch consisting of exactly $p+1$ cones, and we obtain the magnetic quivers for each of them. 

\paragraph{Magnetic Quivers and the Higgs Variety.}
The brane set up for \eqref{electric:usp}, after suitable Hanany-Witten transitions, is given by:
\begin{equation}
   \raisebox{-.5\height}{ \begin{tikzpicture}
    \draw[blue] (0,-3)--(0,3) (1,-3)--(1,3) (2,-3)--(2,3) (4,-3)--(4,3) (5,-3)--(5,3) (6,-3)--(6,3);
    \draw[goodgreen] (9,-3)--(9,3);
    \node at (3,1) {\Huge$\cdots$};
    \node at (7.5,-0.2) {$\cdots$};
    \node[NS] (1) at (7,0) {};
    \node[NS] (2) at (8,0) {};
    \draw[double] (0,0)--(2.2,0) (3.8,0)--(9,0);
    \node at (0.5,0.3) {$p$};
    \node at (1.5,0.3) {$2p$};
    \node[rotate=70] at (4.5,0.7) {$(n-2)p$};
    \node[rotate=70] at (5.5,0.7) {$(n-1)p$};
    \node at (6.5,0.3) {$pn$};
    \draw [decorate,decoration={brace,amplitude=5pt},xshift=0pt,yshift=-0.2cm]
    (8.2,-0.2)--(6.8,-0.2) node [black,midway,yshift=-0.4cm] {$p$};
    \draw [decorate,decoration={brace,amplitude=5pt},xshift=0pt,yshift=-0.2cm]
    (6.2,-3.2)--(-0.2,-3.2) node [black,midway,yshift=-0.4cm] {$n$};
    \end{tikzpicture}}\;.
    \label{branes:uspH}
\end{equation}
From which we can obtain all $p+1$ maximal Higgs phases $P_l$, labeled by $l=0,\dots,p$:
\begin{equation}
\raisebox{-.5\height}{
    \begin{tikzpicture}
    \node at (-2,0) {$P_l=$};
    \draw[blue] (0,-3)--(0,3) (1,-3)--(1,3) (2,-3)--(2,3) (4,-3)--(4,3) (5,-3)--(5,3) (6,-3)--(6,3);
    \draw[goodgreen] (8,-3)--(8,3);
    \node at (3,1) {\Huge$\cdots$};
    \node[NS] (1) at (7,1) {};
    \node at (7,0.5) {$\vdots$};
    \node[NS] (2) at (7,0) {};
    \node[NS] (3) at (7,-1) {};
    \node[NS] (4) at (7,-2) {};
    \draw (0,0)--(1,0) (0,-1)--(1,-1);
    \node at (0.5,-0.5) {$\vdots$};
    \node at (0.5,0.5) {$p$};
    \node at (1.5,1) {$2p$};
    \draw (1,0.5)--(2,0.5) (1,-0.5)--(2,-0.5);
    \node at (1.5,0) {$\vdots$};
    \node[rotate=70] at (4.5,1.7) {$(n-2)p$};
    \draw (4,1)--(5,1) (4,-0.5)--(5,-0.5);
    \node at (4.5,0.25) {$\vdots$};
    \node[rotate=90] at (5.5,-0.5) {$\frac{np}{2}-(p-l)$};
    \draw (5,-1.7)--(6,-1.7) (5,-2.7)--(6,-2.7);
    \node at (5.5,-2.2) {$\vdots$};
    \node at (5.5,1.4) {$\frac{np}{2}-l$};
    \draw[rounded corners] (6,2.5)--(8,2.4)--(5,2.3);
    \node at (7,2.1) {$\vdots$};
    \draw[rounded corners] (6,1.9)--(8,1.8)--(5,1.7);
    \draw[rounded corners] (6,-1)--(3)--(8,-1.2)--(6,-1.4);
    \node at (7,-1.6) {$\vdots$};
    \draw[rounded corners] (6,-2)--(4)--(8,-2.2)--(6,-2.4);
    \node at (7.5,-1.6) {$l$};
    \draw [decorate,decoration={brace,amplitude=5pt},xshift=0pt,yshift=0cm]
    (7.2,1.2)--(7.2,-0.2) node [black,midway,xshift=0.4cm] {};
    \node[rotate=90] at (7.5,0.5) {$p-l$};
    \draw [decorate,decoration={brace,amplitude=5pt},xshift=0pt,yshift=-0.2cm]
    (6.2,-3.2)--(-0.2,-3.2) node [black,midway,yshift=-0.4cm] {$n$};
    \end{tikzpicture}
    }
    \label{branes:uspHl}
\end{equation}
where the NS5 are separated vertically for clarity.
From \eqref{branes:uspHl} we can read the magnetic quivers {$\mathsf{Q}_m^l$} for each individual cone:
\begin{equation}
\raisebox{-.5\height}{
    \begin{tikzpicture}
    \node at (-2,0) {$\mathsf{Q}_m^l (p,n)$};
    \node at (-1,0) {$=$};
    \node[gauge,label=below:{\small$p$}] (1) at (0,0) {};
    \node[gauge,label=below:{\small$2p$}] (2) at (1,0) {};
    \node (3) at (2,0) {$\cdots$};
    \node[gauge,label=above:{\small$(n-2)p$}] (4) at (3,0) {};
    \node[gauge,label=above:{\small$\frac{np}{2}-l$}] (5a) at (4,0.5) {};
    \node[gauge,label=below:{\small$\frac{np}{2}-(p-l)$}] (5b) at (4,-0.5) {};
    \node[flavour,label=right:{\small$2(p-l)$}] (6a) at (5,0.5) {};
    \node[flavour,label=right:{\small$2l$}] (6b) at (5,-0.5) {};
    \draw (1)--(2)--(3)--(4)--(5a)--(6a) (6b)--(5b)--(4);
    \node at (8,0) {$l=0,\dots,p.$};
    \end{tikzpicture}
    }
    \label{magnetic:uspHl}
\end{equation}
which each have Coulomb branch global symmetry $so(2n)$ as expected.

\paragraph{The Hasse Diagram.} Using quiver subtraction on the magnetic quivers (\ref{magnetic:uspHl}), one can compute the Hasse diagrams for each Higgs phase. Identifying the intersections is also possible from the brane system. As an example, for $n=2$, i.e.\ for the theory \eqref{eq:quivUneq2}, equation (\ref{magnetic:uspHl}) reduces to 
\begin{equation}
\raisebox{-.5\height}{
    \begin{tikzpicture}
    \node[gauge,label=above:{\small$p-l$}] (5a) at (4,0.5) {};
    \node[gauge,label=below:{\small$l$}] (5b) at (4,-0.5) {};
    \node[flavour,label=right:{\small$2(p-l)$}] (6a) at (5,0.5) {};
    \node[flavour,label=right:{\small$2l$}] (6b) at (5,-0.5) {};
    \draw (5a)--(6a) (6b)--(5b);
    \node at (8,0) {$l=0,\dots,p,$};
    \end{tikzpicture}
    }
\end{equation}
which shows that every cone is a direct product, giving the Hasse diagram
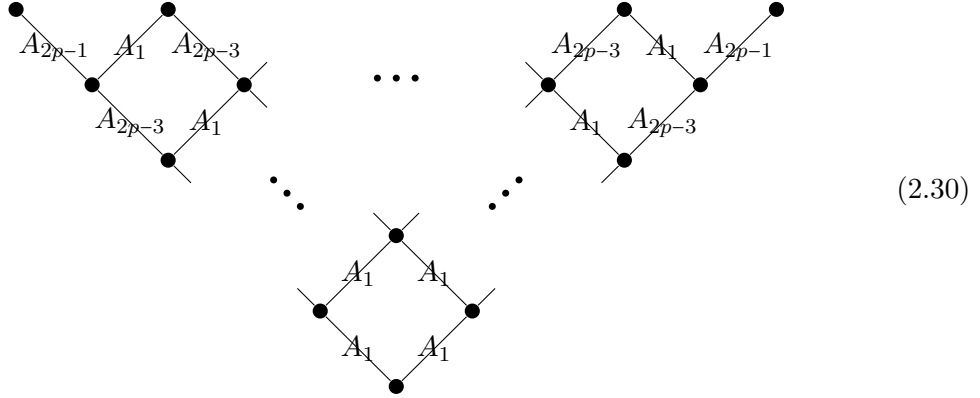
\begin{equation}
   \raisebox{-.5\height}{\begin{tikzpicture}
    \node[hasse] (a1) at (0,0) {};
    \node[hasse] (a2) at (1,1) {};
    \node[hasse] (a3) at (3,3) {};
    \node[hasse] (a4) at (4,4) {};
    \node[hasse] (a5) at (5,5) {};
    \node[hasse] (b1) at (-1,1) {};
    \node[hasse] (b2) at (0,2) {};
    \node[hasse] (b3) at (2,4) {};
    \node[hasse] (b4) at (3,5) {};
    \node[hasse] (c1) at (-3,3) {};
    \node[hasse] (c2) at (-2,4) {};
    \node[hasse] (d1) at (-4,4) {};
    \node[hasse] (d2) at (-3,5) {};
    \node[hasse] (e) at (-5,5) {};
    \draw (a1)--(a2)--(1.3,1.3) (2.7,2.7)--(a3)--(a4)--(a5) (b1)--(b2)--(0.3,2.3) (1.7,3.7)--(b3)--(b4) (c1)--(c2)--(-1.7,4.3) (d1)--(d2);
    \draw (a1)--(b1)--(-1.3,1.3) (-2.7,2.7)--(c1)--(d1)--(e) (a2)--(b2)--(-0.3,2.3) (-1.7,3.7)--(c2)--(d2) (a3)--(b3)--(1.7,4.3) (a4)--(b4);
    \node at (-1.5,2.5) {\rotatebox{-45}{\Huge$\cdots$}};
    \node at (0,4) {\Huge$\cdots$};
    \node at (1.5,2.5) {\rotatebox{45}{\Huge$\cdots$}};
    \node at ($(a1)!0.5!(a2)$) {$A_1$};
    \node at ($(b1)!0.5!(b2)$) {$A_1$};
    \node at ($(c1)!0.5!(c2)$) {$A_1$};
    \node at ($(d1)!0.5!(d2)$) {$A_1$};
    \node at ($(a1)!0.5!(b1)$) {$A_1$};
    \node at ($(a2)!0.5!(b2)$) {$A_1$};
    \node at ($(a3)!0.5!(b3)$) {$A_1$};
    \node at ($(a4)!0.5!(b4)$) {$A_1$};
    \node at ($(a3)!0.5!(a4)$) {$A_{2p-3}$};
    \node at ($(b3)!0.5!(b4)$) {$A_{2p-3}$};
    \node at ($(c1)!0.5!(d1)$) {$A_{2p-3}$};
    \node at ($(c2)!0.5!(d2)$) {$A_{2p-3}$};
    \node at ($(a4)!0.5!(a5)$) {$A_{2p-1}$};
    \node at ($(d1)!0.5!(e)$) {$A_{2p-1}$};
    \end{tikzpicture}}
    \label{Hasse:3d-nis2}
\end{equation}
For $n>2$, \eqref{Hasse:3d-nis2} is contained as a subdiagram, but the full diagram is very intricate, although straightforward to compute. Instead of going in this direction, in the next subsection we keep $n=2$, but focus on the comparison with a direct computation of the classical Higgs branch, which may involve nilpotent operators. 

\subsection{Hyper-K\"ahler Quotient and the Higgs Scheme}
\label{sec:scheme}

In the following we compare the Higgs varieties obtained through magnetic quivers in Section \ref{sec:Umagquiv} with the hyper-K\"ahler quotient.

\subsubsection{Illustrative Example} 
Let us first focus on the case $p=n=2$, for which the Hasse diagram reads
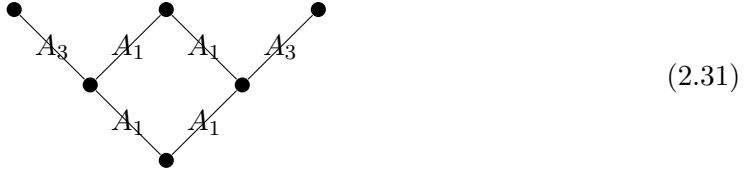
\begin{equation}
\raisebox{-.5\height}{
    \begin{tikzpicture}
    \node[hasse] (a1) at (0,0) {};
    \node[hasse] (a2) at (1,1) {};
    \node[hasse] (a3) at (2,2) {};
    \node[hasse] (b1) at (-1,1) {};
    \node[hasse] (b2) at (0,2) {};
    \node[hasse] (c1) at (-2,2) {};
    \draw (a1)--(a2)--(a3) (b1)--(b2) (a1)--(b1)--(c1) (a2)--(b2);
    \node at ($(a1)!0.5!(a2)$) {$A_1$};
    \node at ($(b1)!0.5!(b2)$) {$A_1$};
    \node at ($(a1)!0.5!(b1)$) {$A_1$};
    \node at ($(a2)!0.5!(b2)$) {$A_1$};
    \node at ($(a2)!0.5!(a3)$) {$A_3$};
    \node at ($(b1)!0.5!(c1)$) {$A_3$};
    \end{tikzpicture}
    }
\end{equation}
In the following, we denote a geometric space with its schematic Hasse diagram, as a slight abuse of notation. The surgery formula states, that the Hilbert series of a space consisting of several cones $\mathcal{C}_i$ ($i=1, \dots , \alpha)$ is given by
\begin{equation}
    \HS=\sum_{i=1}^{\alpha}\HS(\mathcal{C}_i)-\sum_{1\leq i<j\leq\alpha}\HS(\mathcal{C}_i\cap\mathcal{C}_j)+\dots+(-1)^{\alpha+1}\HS\left(\bigcap_i\mathcal{C}_i\right)\,.
    \label{eq:sieve}
\end{equation}
Hence the Hilbert series of the Higgs variety of $\mathrm{U}(2)-\mathrm{USp}(4)-[D_2]$ is given by
\begin{equation}
\begin{split}
    \HS(\HV(\mathrm{U}(2)-\mathrm{USp}(4)-[D_2]))=&\HS  \left(\raisebox{-.3\height}{\begin{tikzpicture}[scale=0.3]
    \node[hasse] (a1) at (0,0) {};
    \node[hasse] (b1) at (-1,1) {};
    \node[hasse] (c1) at (-2,2) {};
    \draw (a1)--(b1)--(c1);
    \end{tikzpicture}} \right)+\HS \left(\raisebox{-.3\height}{\begin{tikzpicture}[scale=0.3]
    \node[hasse] (a1) at (0,0) {};
    \node[hasse] (a2) at (1,1) {};
    \node[hasse] (b1) at (-1,1) {};
    \node[hasse] (b2) at (0,2) {};
    \draw (a1)--(a2) (b1)--(b2) (a1)--(b1) (a2)--(b2);
    \end{tikzpicture}} \right)+\HS \left(\raisebox{-.3\height}{\begin{tikzpicture}[scale=0.3]
    \node[hasse] (a1) at (0,0) {};
    \node[hasse] (a2) at (1,1) {};
    \node[hasse] (a3) at (2,2) {};
    \draw (a1)--(a2)--(a3);
    \end{tikzpicture}} \right)\\
    &-\HS \left(\raisebox{-.3\height}{\begin{tikzpicture}[scale=0.3]
    \node[hasse] (a1) at (0,0) {};
    \node[hasse] (b1) at (-1,1) {};
    \draw (a1)--(b1);
    \end{tikzpicture}} \right)-\HS \left(\raisebox{-.3\height}{\begin{tikzpicture}[scale=0.3]
    \node[hasse] (a1) at (0,0) {};
    \end{tikzpicture}} \right)-\HS \left(\raisebox{-.3\height}{\begin{tikzpicture}[scale=0.3]
    \node[hasse] (a1) at (0,0) {};
    \node[hasse] (a2) at (1,1) {};
    \draw (a1)--(a2);
    \end{tikzpicture}} \right)\\
    &+\HS \left(\raisebox{-.3\height}{\begin{tikzpicture}[scale=0.3]
    \node[hasse] (a1) at (0,0) {};
    \end{tikzpicture}} \right)\\
    &=\frac{(1-t^6)(1-t^8)}{(1-t^2)^3(1-t^4)^3}+\frac{(1-t^4)^2}{(1-t^2)^6}+\frac{(1-t^6)(1-t^8)}{(1-t^2)^3(1-t^4)^3}\\
    &-\frac{(1-t^4)}{(1-t^2)^3}-1-\frac{(1-t^4)}{(1-t^2)^3}\\
    &+1\,.
\end{split}
\label{HS:3d-3cones-Higgs-variety}
\end{equation}
Where the last expression is the explicit unrefined Hilbert series.

Comparing this to the Hilbert series obtained through the hyper-K\"ahler quotient we find a mismatch. This is an indication that there are nilpotent operators in the Higgs ring \cite{Bourget:2019rtl}, and the Higgs branch is a non-reduced scheme, i.e.\ a Higgs scheme, and we hence need to modify \eqref{HS:3d-3cones-Higgs-variety} accordingly.\footnote{As is discussed in detail in \cite{Bourget:2019rtl} magnetic quivers only provide access to the Higgs variety. If there are nilpotent operators in the full Higgs branch chiral ring, then the Higgs branch is a non-reduced scheme.} The Hilbert series of the Higgs scheme can be computed directly via the hyper-K\"ahler quotient, however the decomposition into various cones with multiplicity is not straight forward to obtain in general. After an educated guess for the pre-factors, we find that the Hilbert series of the Higgs scheme reads
\begin{equation}
\begin{split}
    \HS(\HSCH(\mathrm{U}(2)-\mathrm{USp}(4)-[D_2]))=&\HS\left(\raisebox{-.3\height}{\begin{tikzpicture}[scale=0.3]
    \node[hasse] (a1) at (0,0) {};
    \node[hasse] (b1) at (-1,1) {};
    \node[hasse] (c1) at (-2,2) {};
    \draw (a1)--(b1)--(c1);
    \end{tikzpicture}}\right)
    +(1+t^2)\HS\left(\raisebox{-.3\height}{\begin{tikzpicture}[scale=0.3]
    \node[hasse] (a1) at (0,0) {};
    \node[hasse] (a2) at (1,1) {};
    \node[hasse] (b1) at (-1,1) {};
    \node[hasse] (b2) at (0,2) {};
    \draw (a1)--(a2) (b1)--(b2) (a1)--(b1) (a2)--(b2);
    \end{tikzpicture}}\right)
    +\HS\left(\raisebox{-.3\height}{\begin{tikzpicture}[scale=0.3]
    \node[hasse] (a1) at (0,0) {};
    \node[hasse] (a2) at (1,1) {};
    \node[hasse] (a3) at (2,2) {};
    \draw (a1)--(a2)--(a3);
    \end{tikzpicture}}\right)\\
    &-(1+t^2)\HS\left(\raisebox{-.3\height}{\begin{tikzpicture}[scale=0.3]
    \node[hasse] (a1) at (0,0) {};
    \node[hasse] (b1) at (-1,1) {};
    \draw (a1)--(b1);
    \end{tikzpicture}}\right)
    -\HS\left(\begin{tikzpicture}[scale=0.3]
    \node[hasse] (a1) at (0,0) {};
    \end{tikzpicture}\right)
    -(1+t^2)\HS\left(\raisebox{-.3\height}{\begin{tikzpicture}[scale=0.3]
    \node[hasse] (a1) at (0,0) {};
    \node[hasse] (a2) at (1,1) {};
    \draw (a1)--(a2);
    \end{tikzpicture}}\right)\\
    &+(1+t^2)\HS(\begin{tikzpicture}[scale=0.3]
    \node[hasse] (a1) at (0,0) {};
    \end{tikzpicture})\\
    &=\frac{(1-t^6)(1-t^8)}{(1-t^2)^3(1-t^4)^3}+(1+t^2)\frac{(1-t^4)^2}{(1-t^2)^6}+\frac{(1-t^6)(1-t^8)}{(1-t^2)^3(1-t^4)^3}\\
    &-(1+t^2)\frac{(1-t^4)}{(1-t^2)^3}-1-(1+t^2)\frac{(1-t^4)}{(1-t^2)^3}\\
    &+(1+t^2)1  \,,
\end{split}
\label{HS:3d-3cones-Higgs-scheme}
\end{equation}
which may be rewriten as
\begin{equation}
\begin{split}
    \HS(\HSCH(\mathrm{U}(2)-\mathrm{USp}(4)-[D_2]))=&\HS\left(\raisebox{-.3\height}{\begin{tikzpicture}[scale=0.3]
    \node[hasse] (a1) at (0,0) {};
    \node[hasse] (b1) at (-1,1) {};
    \node[hasse] (c1) at (-2,2) {};
    \draw (a1)--(b1)--(c1);
    \end{tikzpicture}}\right)
    +(1+t^2)\HS\left(\raisebox{-.3\height}{\begin{tikzpicture}[scale=0.3]
    \node[hasse] (a1) at (0,0) {};
    \node[hasse] (a2) at (1,1) {};
    \node[hasse] (b1) at (-1,1) {};
    \node[hasse] (b2) at (0,2) {};
    \draw (a1)--(a2) (b1)--(b2) (a1)--(b1) (a2)--(b2);
    \end{tikzpicture}}\right)
    +\HS\left(\raisebox{-.3\height}{\begin{tikzpicture}[scale=0.3]
    \node[hasse] (a1) at (0,0) {};
    \node[hasse] (a2) at (1,1) {};
    \node[hasse] (a3) at (2,2) {};
    \draw (a1)--(a2)--(a3);
    \end{tikzpicture}}\right)\\
    &-(1+t^2)\HS\left(\raisebox{-.3\height}{\begin{tikzpicture}[scale=0.3]
    \node[hasse] (a1) at (0,0) {};
    \node[hasse] (b1) at (-1,1) {};
    \draw (a1)--(b1);
    \end{tikzpicture}}\right)
    -(1+t^2)\HS\left(\raisebox{-.3\height}{\begin{tikzpicture}[scale=0.3]
    \node[hasse] (a1) at (0,0) {};
    \node[hasse] (a2) at (1,1) {};
    \draw (a1)--(a2);
    \end{tikzpicture}}\right)\\
    &+(t^2)\HS(\begin{tikzpicture}[scale=0.3]
    \node[hasse] (a1) at (0,0) {};
    \end{tikzpicture})\\
    &=\frac{(1-t^6)(1-t^8)}{(1-t^2)^3(1-t^4)^3}+(1+t^2)\frac{(1-t^4)^2}{(1-t^2)^6}+\frac{(1-t^6)(1-t^8)}{(1-t^2)^3(1-t^4)^3}\\
    &-(1+t^2)\frac{(1-t^4)}{(1-t^2)^3}-(1+t^2)\frac{(1-t^4)}{(1-t^2)^3}\\
    &+(t^2)1 \,.
\end{split}
\label{HS:3d-3cones-Higgs-scheme-other}
\end{equation}
We can read from the Hilbert series \eqref{HS:3d-3cones-Higgs-scheme-other} that the middle cone has a multiplicity of two, while the outer cones have a multiplicity of one. To be more precise, only the maximal leaf of the middle cone carries multiplicity two, while all other leaves carry multiplicity one. In order to simplify notation, we write equations like \eqref{HS:3d-3cones-Higgs-scheme-other} as
\begin{equation}
\raisebox{-.5\height}{
    \begin{tikzpicture}[scale=1.3]
    \node at (-5,1) {$\HS(\HSCH(\mathrm{U}(2)-\mathrm{USp}(4)-[D_2]))=$};
    \node[rec] (a1) at (0,0) {$t^2$};
    \node[rec] (a2) at (1,1) {$-(1+t^2)$};
    \node[rec] (a3) at (2,2) {$1$};
    \node[rec] (b1) at (-1,1) {$-(1+t^2)$};
    \node[rec] (b2) at (0,2) {$(1+t^2)$};
    \node[rec] (c1) at (-2,2) {$1$};
    \draw (a1)--(a2)--(a3) (b1)--(b2) (a1)--(b1)--(c1) (a2)--(b2);
    \node at ($(a1)!0.5!(a2)$) {$A_1$};
    \node at ($(b1)!0.5!(b2)$) {$A_1$};
    \node at ($(a1)!0.5!(b1)$) {$A_1$};
    \node at ($(a2)!0.5!(b2)$) {$A_1$};
    \node at ($(a2)!0.5!(a3)$) {$A_3$};
    \node at ($(b1)!0.5!(c1)$) {$A_3$};
    \end{tikzpicture}
    }
    \label{HiggsScheme:3cones}
\end{equation}
We can compare this to the Higgs variety 
\begin{equation}
\raisebox{-.5\height}{
    \begin{tikzpicture}[scale=1.3]
    \node at (-5,1) {$\HS(\HV(\mathrm{U}(2)-\mathrm{USp}(4)-[D_2]))=$};
    \node[rec] (a1) at (0,0) {$0$};
    \node[rec] (a2) at (1,1) {$-1$};
    \node[rec] (a3) at (2,2) {$1$};
    \node[rec] (b1) at (-1,1) {$-1$};
    \node[rec] (b2) at (0,2) {$1$};
    \node[rec] (c1) at (-2,2) {$1$};
    \draw (a1)--(a2)--(a3) (b1)--(b2) (a1)--(b1)--(c1) (a2)--(b2);
    \node at ($(a1)!0.5!(a2)$) {$A_1$};
    \node at ($(b1)!0.5!(b2)$) {$A_1$};
    \node at ($(a1)!0.5!(b1)$) {$A_1$};
    \node at ($(a2)!0.5!(b2)$) {$A_1$};
    \node at ($(a2)!0.5!(a3)$) {$A_3$};
    \node at ($(b1)!0.5!(c1)$) {$A_3$};
    \end{tikzpicture}
    }
    \label{HiggsVar:3cones}
\end{equation}
and see that the reduction is accomplished by setting all higher order prefactors in $t$ to $0$, as one would expect.\\

\paragraph{Identification of nilpotent operators.} The identification of nilpotent operators is non-trivial. Therefore we only discuss the simplest example. For $n=2$ and $p=2$ the electric quiver in 4 supercharges notation becomes
\begin{equation}
    \raisebox{-.5\height}{\begin{tikzpicture}
    \node[gauge,label=below:{$\mathrm{U}(2)$}] (3) at (4,0) {};
    \node[gauge,label=below:{$\mathrm{USp}(4)$}] (4) at (6,0) {};
    \node[flavour,label=below:{$D_2$}] (5) at (8,0) {};
    \draw[->,thick] (3) .. controls (5,-0.5) .. (4);
    \draw[->,thick] (4) .. controls (5,0.5) .. (3);
    \draw[->,thick] (4)--(5);
    \draw[->,thick] (3) to [out=45,in=135,looseness=10] (3);
    \draw[->,thick] (4) to [out=45,in=135,looseness=10] (4);
    \node at (4,0.7) {$\Phi_U$};
    \node at (6,0.7) {$\Phi_S$};
    \node at (5,0.6) {$A$};
    \node at (5,-0.6) {$B$};
    \node at (7,0.3) {$C$};
    \end{tikzpicture}}\;.
\end{equation}
It is instructive to compare character expansion of the Hilbert Series and its PL for both the Higgs scheme
\begin{equation}
    \begin{split}
        \mathrm{HS}(\mathcal{HV})=& \, 1+([2,0]+[0,2])t^2+([2,0]+[0,2]+\underbrace{[4,0]+[2,2]+[0,4]+2[0,0]}_{\mathrm{Sym}^2([2,0]+[0,2])})t^4\\
        &+([6,0]+[4,2]+[2,4]+[0,6]+[4,0]+[0,4]+2[2,0]+2[0,2])t^6\\
        &+O(t^{8}) \\    
        \mathrm{PL}(\mathcal{HV})=& , ([2,0]+[0,2])t^2+([2,0]+[0,2])t^4\\
        &+(2[2,2]+[2,0]+[0,2]+2[0,0])t^6\\
        &+O(t^{8})
    \end{split}
\end{equation}
and the Higgs variety
\begin{equation}
    \begin{split}
        \mathrm{HS}(\mathcal{HS})=& \, 1+([2,0]+[0,2])t^2+([2,0]+[0,2]+[4,0]+[2,2]+[0,4]+2[0,0])t^4\\
        &+([6,0]+[4,2]+[2,4]+[0,6]+[4,0]+[0,4]+2[2,0]+2[0,2]{\color{red}+[2,2]})t^6\\
        &+([8,0]+[6,2]+[4,4]+[2,6]+[0,8]+[6,0]+[0,6]+3[4,0]+3[0,4]\\
        &+O(t^{8}) \\ 
                \mathrm{PL}(\mathcal{HS})=& \, ([2,0]+[0,2])t^2+([2,0]+[0,2])t^4\\
        &+({\color{red}1}[2,2]+[2,0]+[0,2]+2[0,0])t^6\\
        &+O(t^{8}) \, . 
    \end{split}
\end{equation}
We can see that both the Higgs scheme and the Higgs variety have a generator in the adjoint of $\mathfrak{so}(4)$ at degree 2 and 4. However for the Higgs variety there is a relation in the $[2,2]$ at degree 6 which is not present in the Higgs scheme. The components of this must be our nilpotent operators, and adding those as relation to the ideal of the Higgs scheme will make it the radical ideal of the Higgs variety. This is a difficult thing to check, since we would need to compute a Gr\"obner Basis which for our rings in question is computationally expensive and not doable in reasonable time. We can still identify what the predicted nilpotent operator looks like. 

Let $\Omega$ be the USp$(4)$ invariant tensor. The superpotential is
\begin{equation}
    W=Tr(C\Omega\Phi_S\Omega C^T)-Tr(A\Omega\Phi_S\Omega B)+Tr(\Omega B\Phi_U A)\; , 
\end{equation}
from which we obtain the F-term relations
\begin{equation}
    \begin{split}
        F_S&=\frac{\partial W}{\partial \Phi_S}=\Omega C^T C\Omega-\frac{1}{2}(\Omega BA\Omega+\Omega A^T B^T\Omega) \overset{!}{=} 0\\
        &\qquad \Leftrightarrow C^T C=\frac{1}{2}(BA+A^TB^T)\\
        F_U&=\frac{\partial W}{\partial \Phi_U}=A\Omega B \overset{!}{=} 0
    \end{split}
\end{equation}

The gauge invariant generators of the Higgs scheme are
\begin{equation}
    \begin{split}
        M&=C\Omega C^T\\
        N&=C\Omega(BA-A^TB^T)\Omega C^T
    \end{split}
\end{equation}
transforming in the adjoint of $\mathfrak{so}(4)$, i.e. the $[2,0]+[0,2]$ representation, at degree 2 and 4 respectively. 
Let us call $M_L$ the part of $M$ which transforms in the $[2,0]$, $M_R$ the part of $M$ which transforms in the $[0,2]$, and likewise $N_L$ and $N_R$ the parts that make $N$. The two independent $[2,2]$ operators at degree 6 are built from $M_L\otimes N_R$ and $M_R\otimes N_L$.

Using F-term relations we find that  
\begin{equation}
    MN+NM=0\;.
\end{equation}
This relation transforms in $[2,2]+[0,0]$ at degree 6, and the $[2,2]$ part is made from a linear combination of $M_L\otimes N_R$ and $M_R\otimes N_L$. The gauge invariant operator at degree 6 which transforms in $[2,2]$, which from the Hilbert series computations above is nilpotent, is an independent linear combination of $M_L\otimes N_R$ and $M_R\otimes N_L$.

\paragraph{Generic 3 Cone case.}
We note that \eqref{HiggsScheme:3cones} turns out to be true for all values of $n\geq2$ as well, if we set all other nodes in the Hasse diagram to $0$. As an example, consider the case $n=3$. The Hasse diagram is given by
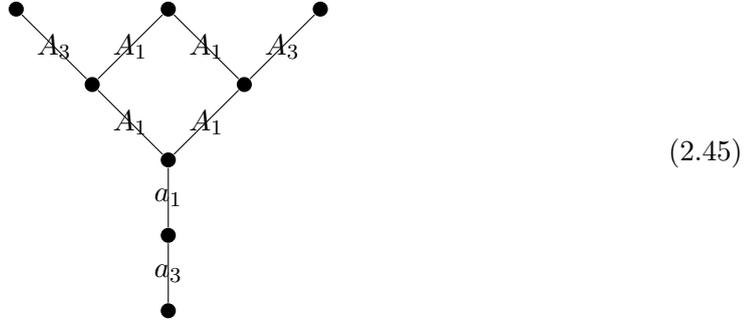
\begin{equation}
\raisebox{-.5\height}{
    \begin{tikzpicture}
    \node[hasse] (oo) at (0,-2) {};
    \node[hasse] (o) at (0,-1) {};
    \node[hasse] (a1) at (0,0) {};
    \node[hasse] (a2) at (1,1) {};
    \node[hasse] (a3) at (2,2) {};
    \node[hasse] (b1) at (-1,1) {};
    \node[hasse] (b2) at (0,2) {};
    \node[hasse] (c1) at (-2,2) {};
    \draw (a1)--(a2)--(a3) (b1)--(b2) (a1)--(b1)--(c1) (a2)--(b2) (oo)--(o)--(a1);
    \node at ($(a1)!0.5!(a2)$) {$A_1$};
    \node at ($(b1)!0.5!(b2)$) {$A_1$};
    \node at ($(a1)!0.5!(b1)$) {$A_1$};
    \node at ($(a2)!0.5!(b2)$) {$A_1$};
    \node at ($(a2)!0.5!(a3)$) {$A_3$};
    \node at ($(b1)!0.5!(c1)$) {$A_3$};
    \node at ($(o)!0.5!(a1)$) {$a_1$};
    \node at ($(oo)!0.5!(o)$) {$a_3$};
    \end{tikzpicture}
    }
\end{equation}
and the Higgs scheme reads as follows:
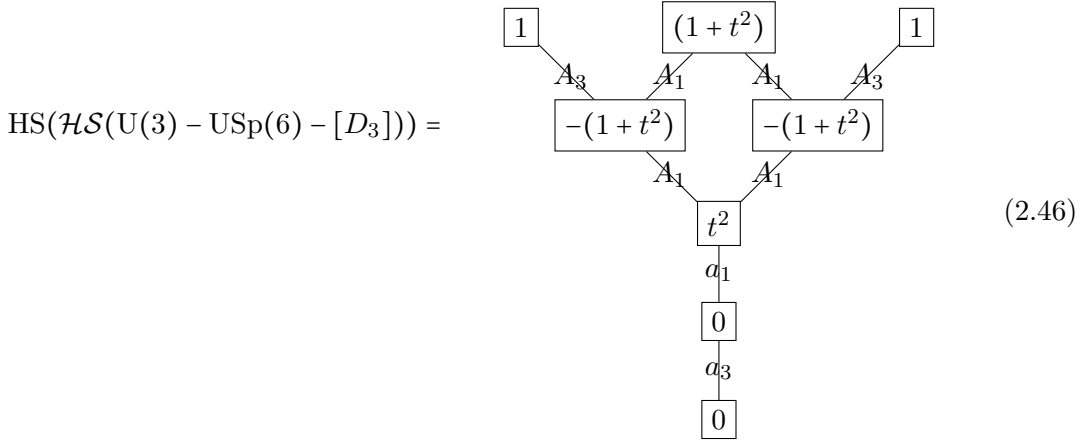
\begin{equation}
\raisebox{-.5\height}{
    \begin{tikzpicture}[scale=1.3]
    \node at (-5,1) {$\HS(\HSCH(\mathrm{U}(3)-\mathrm{USp}(6)-[D_3]))=$};
    \node[rec] (oo) at (0,-2) {$0$};
    \node[rec] (o) at (0,-1) {$0$};
    \node[rec] (a1) at (0,0) {$t^2$};
    \node[rec] (a2) at (1,1) {$-(1+t^2)$};
    \node[rec] (a3) at (2,2) {$1$};
    \node[rec] (b1) at (-1,1) {$-(1+t^2)$};
    \node[rec] (b2) at (0,2) {$(1+t^2)$};
    \node[rec] (c1) at (-2,2) {$1$};
    \draw (a1)--(a2)--(a3) (b1)--(b2) (a1)--(b1)--(c1) (a2)--(b2) (oo)--(o)--(a1);
    \node at ($(a1)!0.5!(a2)$) {$A_1$};
    \node at ($(b1)!0.5!(b2)$) {$A_1$};
    \node at ($(a1)!0.5!(b1)$) {$A_1$};
    \node at ($(a2)!0.5!(b2)$) {$A_1$};
    \node at ($(a2)!0.5!(a3)$) {$A_3$};
    \node at ($(b1)!0.5!(c1)$) {$A_3$};
    \node at ($(o)!0.5!(a1)$) {$a_1$};
    \node at ($(oo)!0.5!(o)$) {$a_3$};
    \end{tikzpicture}
    }
    \label{eq:diagramWithTail}
\end{equation}

\subsubsection{General case}

To derive the weighted sum over the Coulomb branches of the magnetic quivers ${\mathsf{Q}_m^l}$ and their intersections, for the general case, we need  
\begin{enumerate}
    \item[$(i)$] Expressions for the weighting factors as polynomials in $t$ for each relevant node in the Hasse diagram; 
    \item[$(ii)$] The quiver diagrams for each node. 
\end{enumerate} 
Let us start with $(i)$. It is useful to recall the notion of $q$-analog. The $q$-analog of the positive integer $n$ is 
\begin{equation}
    [n]_q \coloneqq \frac{1-q^{n}}{1-q} = 1 + q + \dots + q^{n-1} \, . 
\end{equation}
The integer $n$ is recovered when taking the limit $q \rightarrow 1$. This notation is standard in combinatorics and special function literature, but in this paper we trade $q$ for $t^2$ for consistency with the usual Hilbert series notations. Similarly, there is a $t^2$-analog for the $r$-multinomial coefficient, 
\begin{equation}
\label{eq:binomialAnalog}
    \binom{n_1 + \dots + n_r}{n_1 , \dots , n_r}_{t^2} \coloneqq \frac{(t^2;t^2)_{n_1 + \dots + n_r}}{(t^2;t^2)_{n_1} \cdots (t^2;t^2)_{n_r} } = \frac{\prod_{k=1}^{n_1 + \dots + n_r} (1-t^{2k})}{ \prod_{i=1}^{r} \prod_{k=1}^{n_i} (1-t^{2k})}
\end{equation}
where $(t^2;t^2)_{n} = \prod_{k=1}^{n} (1-t^{2k})$ is the $t^2$-Pochhammer symbol.

Moving on to $(ii)$, the magnetic quivers ${\mathsf{Q}_m^l}$, for $l=0,\dots,p$, correspond to the top row of the Hasse diagram. There are $p+1$ rows in the diagram, that we label from top to bottom by $k=0,\dots ,p$. Introducing the notation ${\mathsf{Q}_m^{k,l}}$ to label the quivers according to row, such that ${\mathsf{Q}_m^{l}}\equiv {\mathsf{Q}_m^{0,l}}$, each magnetic quiver ${\mathsf{Q}_m^{k,l}}$ can be seen to be located at the intersection of its dominant quivers ${\mathsf{Q}_m^{k-1,l}}$ and ${\mathsf{Q}_m^{k-1,l+1}}$.
The magnetic quiver at each intersection of the Hasse diagram is simply the maximal quiver that can be subtracted from its dominant quivers. Thus, the diagram for each magnetic quiver $\mathsf{Q}_m^{k,l}$ follows from \eqref{magnetic:uspHl} as:
\begin{equation}
\raisebox{-.5\height}{
    \begin{tikzpicture}
    \node at (-2,0) {$\mathsf{Q}_m^{k,l}$};
    \node at (-1,0) {$=$};
    \node[gauge,label=below:{\small$p$}] (1) at (0,0) {};
    \node[gauge,label=below:{\small$2p$}] (2) at (1,0) {};
    \node (3) at (2,0) {$\cdots$};
    \node[gauge,label=above:{\small$(n-2)p$}] (4) at (3,0) {};
    \node[gauge,label=above:{\small$\frac{np}{2}-k-l$}] (5a) at (4,0.5) {};
    \node[gauge,label=below:{\small$\frac{np}{2}-(p-l)$}] (5b) at (4,-0.5) {};
    \node[flavour,label=right:{\small$2(p-k-l)$}] (6a) at (5,0.5) {};
    \node[flavour,label=right:{\small$2l$}] (6b) at (5,-0.5) {};
    \node[flavour,label=left:{\small$k$}] (6c) at (3,-0.5) {};
    \draw (1)--(2)--(3)--(4)--(5a)--(6a) (6b)--(5b)--(4)(4)--(6c);
    \end{tikzpicture}
    }
    \label{magnetic:uspHkl}
\end{equation}
for $k=0,\dots,p$ and $l=0,\dots,p-k$.

The relevant Hasse diagram is always of a triangular form, since the nodes associated with the tail carry zero weight, as illustrated in \eqref{eq:diagramWithTail}. We locate each node within this triangle using the integer coordinates $(j,k,l)$, where $j + k + l = p$. The bottom node is assigned the coordinates $(0,p,0)$, and the top left/right nodes are assigned $(p,0,0)$ and $(0,0,p)$ respectively.

We can summarize our conjecture for the construction of the general case Hilbert series of the Higgs Scheme comprising the magnetic quivers as follows:
\begin{equation}
   \mathrm{HS} ( \mathcal{HS} ) = \sum\limits_{k=0}^{p}  {\left( { - 1} \right)^k}{t^{k\left( {k - 1} \right)}} \sum\limits_{l=0}^{k}  \binom{p}{p-k-l,k,l}_{t^2} \mathrm{HS}\left(\mathcal{C}\left(\mathsf{Q}_m^{k,l}\right)\right)
   \label{eq:conjectureMultiplicities} \,.
\end{equation}
This construction has been tested for various low rank quivers. In all the cases studied, it yields exactly the Higgs branch Hilbert series (both refined and unrefined) of the corresponding electric quivers $\mathsf{Q}_e$. The coefficients that appear as multiplicities are up to a prefactor the multinomial $t^2$-analogs $\binom{j+k+l}{j,k,l}_{t^2}$. When evaluated at $t=1$ they reduce to the usual multinomial coefficients that can be constructed using the Pascal pyramid, see Figure \ref{fig:pyramid}. The conjecture \eqref{eq:conjectureMultiplicities} is based on observations, and one can suspect a deeper combinatorial explanation should exist. We leave this as an open question.

\paragraph{Example.}
As an illustration, consider $p=3$, $n=2$. For the Higgs Variety we have the usual multiplicities as given by the sieve formula: 
\begin{equation}
    \begin{tikzpicture}[scale=1.5]
    \node at (-5,1.5) {$\HS(\HV(\mathrm{U}(2)-\mathrm{U}(4)-\mathrm{USp}(6)-[D_2]))=$};
    \node[rec] (a1) at (0,0) {$0$};
    \node[rec] (a2) at (1,1) {$0$};
    \node[rec] (a3) at (2,2) {$-1$};
    \node[rec] (a4) at (3,3) {$1$};
    \node[rec] (b1) at (-1,1) {$0$};
    \node[rec] (b2) at (0,2) {$-1$};
    \node[rec] (b3) at (1,3) {$1$};
    \node[rec] (c1) at (-2,2) {$-1$};
    \node[rec] (c2) at (-1,3) {$1$};
    \node[rec] (d1) at (-3,3) {$1$};
    \draw (a1)--(a2)--(a3)--(a4) (b1)--(b2)--(b3) (c1)--(c2) (a1)--(b1)--(c1)--(d1) (a2)--(b2)--(c2) (a3)--(b3);
    \node at ($(a1)!0.5!(a2)$) {$A_1$};
    \node at ($(b1)!0.5!(b2)$) {$A_1$};
    \node at ($(c1)!0.5!(c2)$) {$A_1$};
    \node at ($(a1)!0.5!(b1)$) {$A_1$};
    \node at ($(a2)!0.5!(b2)$) {$A_1$};
    \node at ($(a3)!0.5!(b3)$) {$A_1$};
    \node at ($(a2)!0.5!(a3)$) {$A_3$};
    \node at ($(b2)!0.5!(b3)$) {$A_3$};
    \node at ($(b1)!0.5!(c1)$) {$A_3$};
    \node at ($(b2)!0.5!(c2)$) {$A_3$};
    \node at ($(c1)!0.5!(d1)$) {$A_5$};
    \node at ($(a3)!0.5!(a4)$) {$A_5$};
    \end{tikzpicture}
\end{equation}
For the Higgs Scheme, we compute instead the multinomial $t^2$-analogs, which evaluate to 
\begin{equation}
\HS(\HSCH(\mathrm{U}(2)-\mathrm{U}(4)-\mathrm{USp}(6)-[D_2]))= 
     \raisebox{-.5\height}{\begin{tikzpicture}[scale=1.5]
    \node[draw,fill=white,inner sep=3pt] (a1) at (0,0) {\scalebox{.8}{$-t^6$}};
    \node[draw,fill=white,inner sep=3pt] (a2) at (1,1) {\scalebox{.8}{$(t^2+t^4+t^6)$}};
    \node[draw,fill=white,inner sep=3pt] (a3) at (2,2) {\scalebox{.8}{$-(1+t^2+t^4)$}};
    \node[draw,fill=white,inner sep=3pt] (a4) at (3,3) {\scalebox{.8}{$1$}};
    \node[draw,fill=white,inner sep=3pt] (b1) at (-1,1) {\scalebox{.8}{$(t^2+t^4+t^6)$}};
    \node[draw,fill=white,inner sep=3pt] (b2) at (0,2) {\scalebox{.8}{$-(1+2t^2+2t^4+t^6)$}};
    \node[draw,fill=white,inner sep=3pt] (b3) at (1,3) {\scalebox{.8}{$(1+t^2+t^4)$}};
    \node[draw,fill=white,inner sep=3pt] (c1) at (-2,2) {\scalebox{.8}{$-(1+t^2+t^4)$}};
    \node[draw,fill=white,inner sep=3pt] (c2) at (-1,3) {\scalebox{.8}{$(1+t^2+t^4)$}};
    \node[draw,fill=white,inner sep=3pt] (d1) at (-3,3) {\scalebox{.8}{$1$}};
    \draw (a1)--(a2)--(a3)--(a4) (b1)--(b2)--(b3) (c1)--(c2) (a1)--(b1)--(c1)--(d1) (a2)--(b2)--(c2) (a3)--(b3);
    \node at ($(a1)!0.5!(a2)$) {$A_1$};
    \node at ($(b1)!0.5!(b2)$) {$A_1$};
    \node at ($(c1)!0.5!(c2)$) {$A_1$};
    \node at ($(a1)!0.5!(b1)$) {$A_1$};
    \node at ($(a2)!0.5!(b2)$) {$A_1$};
    \node at ($(a3)!0.5!(b3)$) {$A_1$};
    \node at ($(a2)!0.5!(a3)$) {$A_3$};
    \node at ($(b2)!0.5!(b3)$) {$A_3$};
    \node at ($(b1)!0.5!(c1)$) {$A_3$};
    \node at ($(b2)!0.5!(c2)$) {$A_3$};
    \node at ($(c1)!0.5!(d1)$) {$A_5$};
    \node at ($(a3)!0.5!(a4)$) {$A_5$};
    \end{tikzpicture}}
    \label{eq:diagU2U4Usp6D2mult}
\end{equation}
This can be checked against a direct computation using the Hilbert series for each cone, as given by the following diagram: 
\begin{equation}
         \raisebox{-.5\height}{\begin{tikzpicture}[scale=1.5]
    \node[draw,fill=white,inner sep=3pt] (a1) at (0,0) {\scalebox{.7}{$1$}};
    \node[draw,fill=white,inner sep=3pt] (a2) at (1,1) {\scalebox{.7}{$\frac{1-t^4}{\left(1-t^2\right)^3}$}};
    \node[draw,fill=white,inner sep=3pt] (a3) at (2,2) {\scalebox{.7}{$\frac{\left(1-t^6\right) \left(1-t^8\right)}{\left(1-t^2\right)^3 \left(1-t^4\right)^3}$}};
    \node[draw,fill=white,inner sep=3pt] (a4) at (3,3) {\scalebox{.7}{$\frac{\left(1-t^8\right) \left(1-t^{10}\right) \left(1-t^{12}\right)}{\left(1-t^2\right)^3 \left(1-t^4\right)^3 \left(1-t^6\right)^3}$}};
    \node[draw,fill=white,inner sep=3pt] (b1) at (-1,1) {\scalebox{.7}{$\frac{1-t^4}{\left(1-t^2\right)^3}$}};
    \node[draw,fill=white,inner sep=3pt] (b2) at (0,2) {\scalebox{.7}{$\frac{\left(1-t^4\right)^2}{\left(1-t^2\right)^6}$}};
    \node[draw,fill=white,inner sep=3pt] (b3) at (1,3) {\scalebox{.7}{$\frac{\left(1-t^6\right) \left(1-t^8\right)}{\left(1-t^2\right)^6 \left(1-t^4\right)^2}$}};
    \node[draw,fill=white,inner sep=3pt] (c1) at (-2,2) {\scalebox{.7}{$\frac{\left(1-t^6\right) \left(1-t^8\right)}{\left(1-t^2\right)^3 \left(1-t^4\right)^3}$}};
    \node[draw,fill=white,inner sep=3pt] (c2) at (-1,3) {\scalebox{.7}{$\frac{\left(1-t^6\right) \left(1-t^8\right)}{\left(1-t^2\right)^6 \left(1-t^4\right)^2}$}};
    \node[draw,fill=white,inner sep=3pt] (d1) at (-3,3) {\scalebox{.7}{$\frac{\left(1-t^8\right) \left(1-t^{10}\right) \left(1-t^{12}\right)}{\left(1-t^2\right)^3 \left(1-t^4\right)^3 \left(1-t^6\right)^3}$}};
    \draw (a1)--(a2)--(a3)--(a4) (b1)--(b2)--(b3) (c1)--(c2) (a1)--(b1)--(c1)--(d1) (a2)--(b2)--(c2) (a3)--(b3);
    \node at ($(a1)!0.5!(a2)$) {$A_1$};
    \node at ($(b1)!0.5!(b2)$) {$A_1$};
    \node at ($(c1)!0.5!(c2)$) {$A_1$};
    \node at ($(a1)!0.5!(b1)$) {$A_1$};
    \node at ($(a2)!0.5!(b2)$) {$A_1$};
    \node at ($(a3)!0.5!(b3)$) {$A_1$};
    \node at ($(a2)!0.5!(a3)$) {$A_3$};
    \node at ($(b2)!0.5!(b3)$) {$A_3$};
    \node at ($(b1)!0.5!(c1)$) {$A_3$};
    \node at ($(b2)!0.5!(c2)$) {$A_3$};
    \node at ($(c1)!0.5!(d1)$) {$A_5$};
    \node at ($(a3)!0.5!(a4)$) {$A_5$};
    \end{tikzpicture}}
    \label{eq:diagU2U4Usp6D2HS}
\end{equation}
Combining \eqref{eq:diagU2U4Usp6D2mult} and \eqref{eq:diagU2U4Usp6D2HS} one finds the Hilbert series 
\begin{equation}
\frac{\left( \begin{array}{c}
1 +4   t^2+14 t^4+40 t^6+83 t^8+143 t^{10}+181 t^{12}+172 t^{14} \\ +98 t^{16}+21 t^{18}-35 t^{20}-31 t^{22}-10 t^{24}+4 t^{26}+4 t^{28}-t^{30}
\end{array} \right) }{\left(1-t^2\right)^2 \left(1-t^4\right)^2 \left(1-t^6\right)^2} \, , 
\end{equation}
which precisely agrees with the result from the hyper-K\"ahler quotient. This was also checked on the level of refined Hilbert series.

\subsection{Global Form of Flavour Symmetry}
\label{sec:globalSymm}
The global symmetry of the classical Higgs branch of \eqref{electric:usp} is
\begin{equation}
    \label{eq:flavourSymm}
    \mathrm{PSO}(2n)\rtimes\mathbb{Z}_2 \, . 
\end{equation}
The PSO$(2n)$ factor is due to the fact that only representations in the adjoint lattice of $D_n$ show up in the Higgs branch chiral ring. This can also be seen from the magnetic quivers \eqref{magnetic:uspHl} which are all balanced. The $\mathbb{Z}_2$ parity is a symmetry exchanging isomorphic cones in the Higgs branch. This can be seen from magnetic quivers as exchanging $l\leftrightarrow(p-l)$ in \eqref{magnetic:uspHl}, i.e.\ exchanging the two spinor Dynkin labels of $D_n$ (outer automorphism of the $D_n$ algebra).

For $p=1$ the $\mathbb{Z}_2$ parity symmetry has been described in \cite{Gaiotto:2008ak} and \cite{Assel:2018exy}, where it has been pointed out that in addition to acting on the classical Higgs branch, it also acts on the quantum Coulomb branch. We see that this is the case for all $p$ in the next subsection.

\subsection{3d Coulomb Branch and Full Moduli Space}
\label{sec:3dFullMod}
The appearance of many cones in the classical Higgs branch is an indication for many singularities in the Coulomb branch. To see this one can use for example the \emph{inversion} argument of \cite{Grimminger:2020dmg} which is reviewed in Appendix \ref{app:badCB}. We can use inversion to conjecture the Hasse diagram of the quantum Coulomb branch based on the Hasse diagram of the classical Higgs branch. Since Hasse diagrams become very complicated we focus at the simplest examples.

To check our claims one should in principle do a detailed study of bad Coulomb branches with several singular points along the lines of \cite{Assel:2017jgo,Assel:2018exy}, which is challenging. We hope to report on this in future work.

\paragraph{2 Cones.} Inversion of the classical Higgs branch Hasse diagram of USp$(2k)-[D_{2k}]$ was already successfully used in \cite{Grimminger:2020dmg} to produce the Hasse diagram of the quantum Coulomb branch, matching the results of \cite{Assel:2018exy}. For the simplest case of $k=1$ we get
\begin{equation}
    \raisebox{-.5\height}{
    \begin{tikzpicture}
    	\node (cH) at (0,0) {$\begin{tikzpicture}
        	\node[hasse] (0) at (0,0) {};
        	\node[hasse] (11) at (-1,1) {};
        	\node[hasse] (12) at (1,1) {};
        	\draw (0)--(11) (12)--(0);
        	\node at (-0.5,0.5) {$A_1$};
        	\node at (0.5,0.5) {$A_1$};
    	\end{tikzpicture}$};
    	\node (qC) at (4,0) {$\begin{tikzpicture}
        	\node[hasse] (0) at (0,0) {};
        	\node[hasse] (11) at (-1,-1) {};
        	\node[hasse] (12) at (1,-1) {};
        	\draw[red] (0)--(11) (12)--(0);
        	\node at (-0.5,-0.5) {$a_1$};
        	\node at (0.5,-0.5) {$a_1$};
    \end{tikzpicture}$};
    \draw[->] (cH)--(qC);
    \node at (2,0.5) {$\mathcal{I}$};
    \end{tikzpicture}}
\end{equation}
and the Hasse diagram for the entire moduli space is given by
\begin{equation}
     \raisebox{-.5\height}{\begin{tikzpicture}
        	\node[hasse] (0) at (0,0) {};
        	\node[hasse] (11) at (-1,-1) {};
        	\node[hasse] (12) at (1,-1) {};
        	\draw[red] (0)--(11) (12)--(0);
        	\node at (-0.5,-0.5) {$a_1$};
        	\node at (0.5,-0.5) {$a_1$};

            \node[hasse] (11h) at (-2,0) {};
            \node[hasse] (12h) at (2,0) {};
            \draw[blue] (11)--(11h) (12)--(12h);
            \node at (-1.5,-0.5) {$A_1$};
            \node at (1.5,-0.5) {$A_1$};
    \end{tikzpicture}}
\end{equation}
The physics at the two most singular points in the Coulomb branch is equivalent, and the local moduli space geometry is that of U$(1)-[2]$. The two cones, which intersect in the classical Higgs branch, emanate from different points in the quantum Coulomb branch and hence are separated in the full moduli space \cite{Seiberg:1996nz}. This is an effect also observed in 4d $\mathcal{N}=2$ theories \cite{Seiberg:1994aj,Argyres:1996eh}. The $\mathbb{Z}_2$ symmetry of \eqref{eq:flavourSymm} is visible as a vertical reflection of the Hasse diagram of the full moduli space.

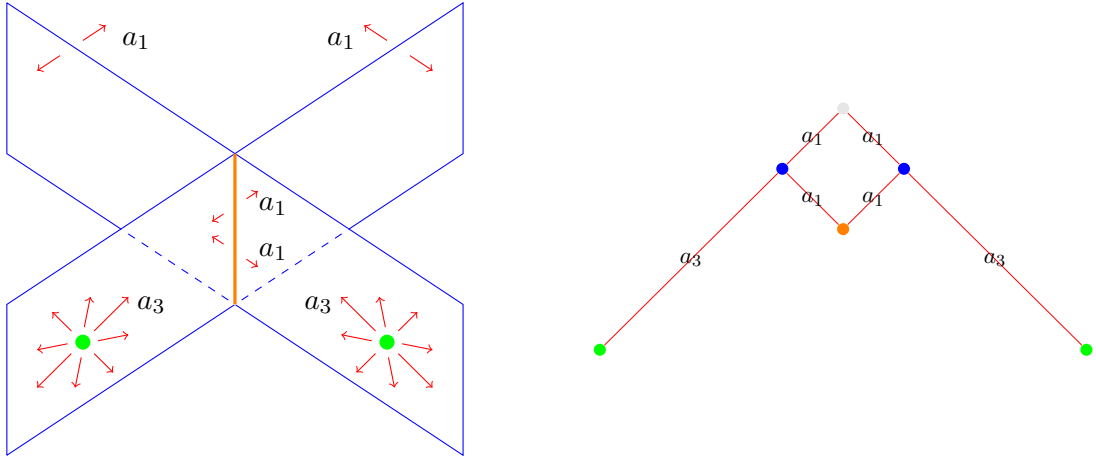
\begin{figure}
\centering
\begin{tikzpicture}
\newcommand{\stararound}{\begin{tikzpicture}
        	\draw[->,red] (.2,.02)--(.6,.1);
        	\draw[->,red] (-.2,-.02)--(-.6,-.1);
        	\draw[->,red] (-.02,-.2)--(-.1,-.6);
        	\draw[->,red] (.02,.2)--(.1,.6);
        	\draw[->,red] (.15,.15)--(.6,.6);
        	\draw[->,red] (-.15,.15)--(-.4,.4);
        	\draw[->,red] (.15,-.15)--(.4,-.4);
        	\draw[->,red] (-.15,-.15)--(-.6,-.6);
    \end{tikzpicture}}
\newcommand{\stararoundd}{\begin{tikzpicture}
        	\draw[->,red] (-.2,.02)--(-.6,.1);
        	\draw[->,red] (.2,-.02)--(.6,-.1);
        	\draw[->,red] (.02,-.2)--(.1,-.6);
        	\draw[->,red] (-.02,.2)--(-.1,.6);
        	\draw[->,red] (-.15,.15)--(-.6,.6);
        	\draw[->,red] (.15,.15)--(.4,.4);
        	\draw[->,red] (-.15,-.15)--(-.4,-.4);
        	\draw[->,red] (.15,-.15)--(.6,-.6);
    \end{tikzpicture}}
\draw[blue] (-3,-3)--(3,1)--(3,3)--(-3,-1)--(-3,-3);
\draw[blue] (-3,3)--(3,-1)--(3,-3)--(-3,1)--(-3,3);
\draw[white,thick,dashed] (0,-1)--(1.5,0);
\draw[white,thick,dashed] (0,-1)--(-1.5,0);
\draw[orange,very thick] (0,-1)--(0,1);
\node[hasse,green] (1) at (-2,-1.5) {};
\node[hasse,green] at (2,-1.5) {};
\node at (-2,-1.5) {\stararound};
\node at (2,-1.5) {\stararoundd};
\draw[->,red] (.15,.1+.3)--(.3,.2+.3);
\draw[->,red] (-.15,-.1+.3)--(-.3,-.2+.3);
\draw[->,red] (-.15,.1-.3)--(-.3,.2-.3);
\draw[->,red] (.15,-.1-.3)--(.3,-.2-.3);
\draw[->,red] (-2,2.5)--(-2+.3,2.5+.2);
\draw[->,red] (-2-.3,2.5-.2)--(-2-.3-.3,2.5-.2-.2);
\draw[->,red] (+2,2.5)--(+2-.3,2.5+.2);
\draw[->,red] (+2+.3,2.5-.2)--(+2+.3+.3,2.5-.2-.2);
\node at (-1.1,-1) {$a_3$};
\node at (1.1,-1) {$a_3$};
\node at (.5,-.3) {$a_1$};
\node at (.5,.3) {$a_1$};
\node at (-1.3,2.5) {$a_1$};
\node at (1.4,2.5) {$a_1$};
\node at (8,0) {\scalebox{.8}{\begin{tikzpicture}
        	\node[hasse,black!10] (0) at (0,0) {};
        	\node[hasse,blue] (11) at (-1,-1) {};
        	\node[hasse,blue] (12) at (1,-1) {};
        	\node[hasse,green] (21) at (-4,-4) {};
        	\node[hasse,orange] (22) at (0,-2) {};
        	\node[hasse,green] (23) at (4,-4) {};
        	\draw[red] (0)--(11)--(21) (11)--(22)--(12)--(23) (12)--(0);
        	\node at (-0.5,-0.5) {$a_1$};
        	\node at (0.5,-0.5) {$a_1$};
        	\node at (-0.5,-1.5) {$a_1$};
        	\node at (0.5,-1.5) {$a_1$};
        	\node at (-2.5,-2.5) {$a_3$};
        	\node at (2.5,-2.5) {$a_3$};
    \end{tikzpicture}}};
\end{tikzpicture}
\caption{Artist's impression of the Coulomb branch of \eqref{electric:usp} with $p=n=2$ (left). The nodes of the Hasse diagram (right) are colored in accordance with the depiction on the left. Note that on the drawing every elementary transverse slice is shown with real dimension 1, instead of quaternionic dimension 1 and 3 for $a_1$ and $a_3$. }
\label{fig:artitView}
\end{figure}

\paragraph{3 Cones.} Inversion of the classical Higgs branch Hasse diagram of U$(2)-$USp$(4)-[D_2]$ gives the conjectured quantum Coulomb branch Hasse diagram. See the Appendix \ref{app:badCB} for details on inversion, and a discussion on lowest leaf with non-zero dimension.
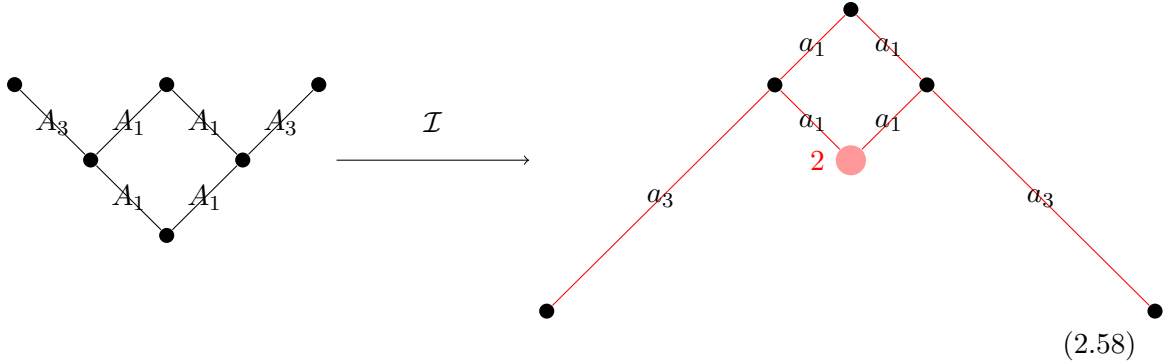
\begin{equation}
    \raisebox{-.5\height}{ \begin{tikzpicture}
    	\node (cH) at (0,0) {$\begin{tikzpicture}
        	\node[hasse] (0) at (0,0) {};
        	\node[hasse] (11) at (-1,1) {};
        	\node[hasse] (12) at (1,1) {};
        	\node[hasse] (21) at (-2,2) {};
        	\node[hasse] (22) at (0,2) {};
        	\node[hasse] (23) at (2,2) {};
        	\draw (0)--(11)--(21) (11)--(22)--(12)--(23) (12)--(0);
        	\node at (-0.5,0.5) {$A_1$};
        	\node at (0.5,0.5) {$A_1$};
        	\node at (-0.5,1.5) {$A_1$};
        	\node at (0.5,1.5) {$A_1$};
        	\node at (-1.5,1.5) {$A_3$};
        	\node at (1.5,1.5) {$A_3$};
    	\end{tikzpicture}$};
    	\node (qC) at (9,0) {$\begin{tikzpicture}
        	\node[hasse] (0) at (0,0) {};
        	\node[hasse] (11) at (-1,-1) {};
        	\node[hasse] (12) at (1,-1) {};
        	\node[hasse] (21) at (-4,-4) {};
        	\node[hassebr,label=left:{\color{red}$2$}] (22) at (0,-2) {};
        	\node[hasse] (23) at (4,-4) {};
        	\draw[red] (0)--(11)--(21) (11)--(22)--(12)--(23) (12)--(0);
        	\node at (-0.5,-0.5) {$a_1$};
        	\node at (0.5,-0.5) {$a_1$};
        	\node at (-0.5,-1.5) {$a_1$};
        	\node at (0.5,-1.5) {$a_1$};
        	\node at (-2.5,-2.5) {$a_3$};
        	\node at (2.5,-2.5) {$a_3$};
    \end{tikzpicture}$};
    \draw[->] (cH)--(qC);
    \node at (3.5,0.5) {$\mathcal{I}$};
    \end{tikzpicture}}
\end{equation}
We observe that the dimension of the Coulomb branch read from the Hasse diagram is 4, which fits the rank of the theory. The Coulomb branch consists of 6 leaves. The two most singular points (the bottom left and right leaves) are equivalent, and exchanged by the $\mathbb{Z}_2$ symmetry discussed in Section \ref{sec:globalSymm}. A further lowest leaf $\begin{tikzpicture}
    \node[hassebr] at (0,0) {};
\end{tikzpicture}$ is of dimension two (See Appendix \ref{app:badCB}). An artistic depiction of the Coulomb branch is given in Figure \ref{fig:artitView}. The massless degrees of freedom present at each leaf may be indicated by transverse slices as follows:
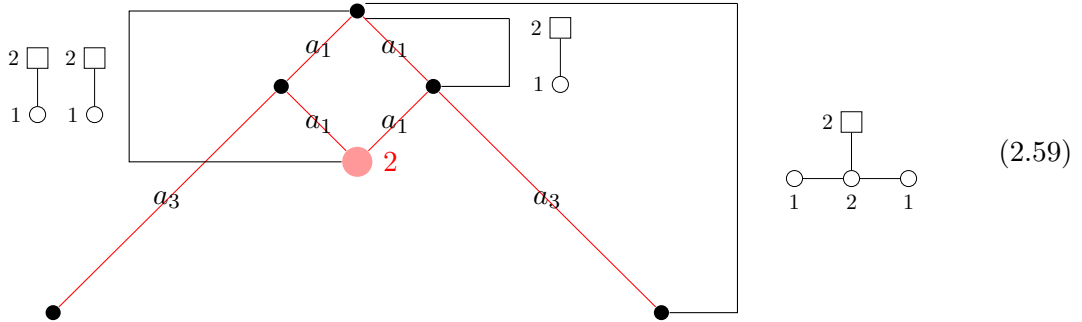
\begin{equation}
	 \raisebox{-.5\height}{\begin{tikzpicture}
        	\node[hasse] (0) at (0,0) {};
        	\node[hasse] (11) at (-1,-1) {};
        	\node[hasse] (12) at (1,-1) {};
        	\node[hasse] (21) at (-4,-4) {};
        	\node[hassebr,label=right:{\color{red}$2$}] (22) at (0,-2) {};
        	\node[hasse] (23) at (4,-4) {};
        	\draw[red] (0)--(11)--(21) (11)--(22)--(12)--(23) (12)--(0);
        	\node at (-0.5,-0.5) {$a_1$};
        	\node at (0.5,-0.5) {$a_1$};
        	\node at (-0.5,-1.5) {$a_1$};
        	\node at (0.5,-1.5) {$a_1$};
        	\node at (-2.5,-2.5) {$a_3$};
        	\node at (2.5,-2.5) {$a_3$};
        	
        	\draw (23)--(5,-4)--(5,0.1);
        	\draw[transform canvas={yshift=0.1cm}] (5,0)--(0);
        	\node at (6.5,-2) {$\scalebox{0.75}{\begin{tikzpicture}
        		\node[gauge,label=below:{$1$}] (ms1) at (0,0) {};
        		\node[gauge,label=below:{$2$}] (ms2) at (1,0) {};
        		\node[gauge,label=below:{$1$}] (ms3) at (2,0) {};
        		\node[flavour,label=left:{$2$}] (ms2f) at (1,1) {};
        		\draw (ms1)--(ms2)--(ms3) (ms2)--(ms2f);
        	\end{tikzpicture}}$};
        	
        	\draw (12)--(2,-1)--(2,-0.1);
        	\draw[transform canvas={yshift=-0.1cm}] (2,0)--(0);
        	\node at (2.5,-0.6) {$\scalebox{0.75}{\begin{tikzpicture}
        		\node[gauge,label=left:{$1$}] (ms2) at (1,0) {};
        		\node[flavour,label=left:{$2$}] (ms2f) at (1,1) {};
        		\draw (ms2)--(ms2f);
        	\end{tikzpicture}}$};
			
			\draw (22)--(-3,-2)--(-3,0)--(0);
        	\node at (-4,-1) {$\scalebox{0.75}{\begin{tikzpicture}
        		\node[gauge,label=left:{$1$}] (ms2x) at (0,0) {};
        		\node[flavour,label=left:{$2$}] (ms2fx) at (0,1) {};
        		\draw (ms2x)--(ms2fx);
        		\node[gauge,label=left:{$1$}] (ms2) at (1,0) {};
        		\node[flavour,label=left:{$2$}] (ms2f) at (1,1) {};
        		\draw (ms2)--(ms2f);
        	\end{tikzpicture}}$};
    \end{tikzpicture}}
    \label{eq:InversionSlices}
\end{equation}
We notice that nowhere on the Coulomb branch are all the fields in U$(2)-$USp$(4)-[D_2]$ massless. For example, at one of the two most singular points the massless states are
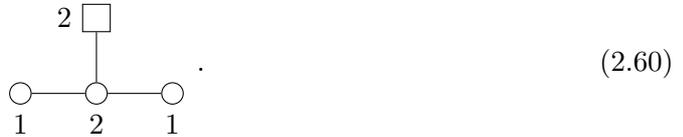
\begin{equation}
   \raisebox{-.5\height}{ \begin{tikzpicture}
        		\node[gauge,label=below:{$1$}] (ms1) at (0,0) {};
        		\node[gauge,label=below:{$2$}] (ms2) at (1,0) {};
        		\node[gauge,label=below:{$1$}] (ms3) at (2,0) {};
        		\node[flavour,label=left:{$2$}] (ms2f) at (1,1) {};
        		\draw (ms1)--(ms2)--(ms3) (ms2)--(ms2f);
        	\end{tikzpicture}}\;.
\end{equation}
The effect is similar to the case of SU$(2)-[2]$, where the massless states at one of the most singular points in the Coulomb branch are U$(1)-[2]$.\\

The Hasse diagram of the full moduli space can be obtained from further local inversions \cite{Grimminger:2020dmg}:

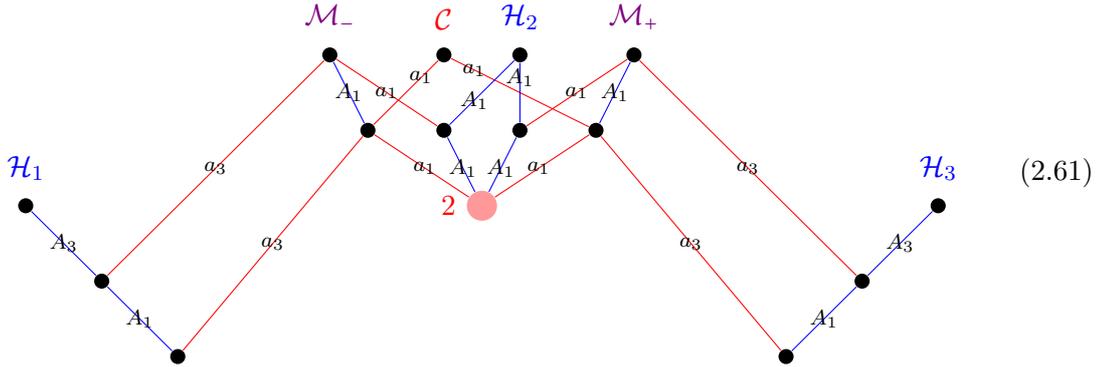
\begin{equation}
	 \raisebox{-.5\height}{\begin{tikzpicture}
        \node[hasse] (0) at (-0.5,0) {};
        \node[hasse] (11) at (-1.5,-1) {};
        \node[hasse] (12) at (1.5,-1) {};
        \node[hasse] (21) at (-4,-4) {};
        \node[hassebr,label=left:{\color{red}$2$}] (22) at (0,-2) {};
        \node[hasse] (23) at (4,-4) {};
        \draw[red] (0)--(11)--(21) (11)--(22)--(12)--(23) (12)--(0);
        
        \node[hasse] (21h1) at (-5,-3) {};
        \node[hasse] (21h2) at (-6,-2) {};
        \draw[blue] (21)--(21h1)--(21h2);
        
        \node[hasse] (11h) at (-2,0) {};
        \draw[blue] (11)--(11h);
        
        \node[hasse] (22h1) at (-0.5,-1) {};
        \node[hasse] (22h2) at (0.5,-1) {};
        \node[hasse] (22h3) at (0.5,0) {};
        \draw[blue] (22)--(22h1)--(22h3)--(22h2)--(22);
        
        \node[hasse] (12h) at (2,0) {};
        \draw[blue] (12)--(12h);
        
        \node[hasse] (23h1) at (5,-3) {};
        \node[hasse] (23h2) at (6,-2) {};
        \draw[blue] (23)--(23h1)--(23h2);
        
        \draw[red] (21h1)--(11h)--(22h1) (22h2)--(12h)--(23h1);
        
        \node at ($(0)!0.3!(11)$) {\scriptsize$a_1$};
        \node at ($(0)!0.2!(12)$) {\scriptsize$a_1$};
        \node at ($(11)!0.5!(22)$) {\scriptsize$a_1$};
        \node at ($(12)!0.5!(22)$) {\scriptsize$a_1$};
        \node at ($(22h1)!0.5!(11h)$) {\scriptsize$a_1$};
        \node at ($(22h2)!0.5!(12h)$) {\scriptsize$a_1$};
        \node at ($(11)!0.5!(21)$) {\scriptsize$a_3$};
        \node at ($(12)!0.5!(23)$) {\scriptsize$a_3$};
        \node at ($(23h1)!0.5!(12h)$) {\scriptsize$a_3$};
        \node at ($(21h1)!0.5!(11h)$) {\scriptsize$a_3$};
        
        \node at ($(21)!0.5!(21h1)$) {\scriptsize$A_1$};
        \node at ($(21h1)!0.5!(21h2)$) {\scriptsize$A_3$};
        \node at ($(23)!0.5!(23h1)$) {\scriptsize$A_1$};
        \node at ($(23h1)!0.5!(23h2)$) {\scriptsize$A_3$};
        \node at ($(11)!0.5!(11h)$) {\scriptsize$A_1$};
        \node at ($(12)!0.5!(12h)$) {\scriptsize$A_1$};
        \node at ($(22)!0.5!(22h1)$) {\scriptsize$A_1$};
        \node at ($(22)!0.5!(22h2)$) {\scriptsize$A_1$};
        \node at ($(22h1)!0.4!(22h3)$) {\scriptsize$A_1$};
        \node at ($(22h2)!0.7!(22h3)$) {\scriptsize$A_1$};

        \node[above = 0.1cm of 0] {\color{red}$\mathcal{C}$};
        
        \node[above = 0.1cm of 21h2] {\color{blue}$\mathcal{H}_1$};
        \node[above = 0.1cm of 22h3] {\color{blue}$\mathcal{H}_2$};
        \node[above = 0.1cm of 23h2] {\color{blue}$\mathcal{H}_3$};
        
        \node[above = 0.1cm of 11h] {\color{violet}$\mathcal{M}_{-}$};
        \node[above = 0.1cm of 12h] {\color{violet}$\mathcal{M}_{+}$};
	\end{tikzpicture}}
\end{equation}
We can see that the three cones, $\mathcal{H}_1$, $\mathcal{H}_2$ and $\mathcal{H}_3$, which make up the classical Higgs branch are still present in the quantum moduli space. However, the three cones do not intersect as they do in the classical Higgs branch. The classical intersections appear as Higgs directions on less singular Coulomb branch loci, leading to mixed branches $\mathcal{M}_{-}$ and $\mathcal{M}_{+}$. The fate of nilpotent operators in the quantum moduli space is unclear.

\paragraph{General Case, physics at the most singular point.} For the general case, with $p+1$ cones, inversion indicates that there are exactly two most singular points in the Coulomb branch, and that the local moduli space geometry at these two points is the same.\footnote{This is not the case for the linear quiver wherein all $\mathrm{U}(n_i)\to \mathrm{SU}(n_i)$, see Section \ref{sec:5d}. Another example is SU$(k)-[2k-2]$ for $k>2$: there are two most singular points which have non-isomorphic Higgs branches (mesonic \& baryonic). See Appendix \ref{app:badCB} and in particular \eqref{eq:fullMSSU3with4} for an explicit example.} The set of massless states at either of the two most singular points in the Coulomb branch of
\begin{equation}
    \raisebox{-.5\height}{ \begin{tikzpicture}
        \node[gauge,label=below:{$\mathrm{U}(n)$}] (1) at (0,0) {};
    \node[gauge,label=below:{$\mathrm{U}(2n)$}] (2) at (2,0) {};
    \node (d) at (3,0) {$\cdots$};
    \node[gauge,label=below:{$\mathrm{U}((p-1)n)$}] (3) at (4,0) {};
    \node[gauge,label=below:{$\mathrm{USp}(pn)$}] (4) at (6,0) {};
    \node[flavour,label=below:{$D_n$}] (5) at (8,0) {};
    \draw (1)--(2)--(d)--(3)--(4)--(5);
    \end{tikzpicture}}
\end{equation}
is described by the 3d mirror of the magnetic quiver ($l=0$ in \eqref{magnetic:uspHl}):
\begin{equation}
    \raisebox{-.5\height}{ \begin{tikzpicture}
        \node[gauge,label=below:{\small$p$}] (1) at (0,0) {};
        \node[gauge,label=below:{\small$2p$}] (2) at (1,0) {};
        \node (3) at (2,0) {$\cdots$};
        \node[gauge,label=above:{\small$(n-2)p$}] (4) at (3,0) {};
        \node[gauge,label=above:{\small$\frac{np}{2}$}] (5a) at (4,0.5) {};
        \node[gauge,label=below:{\small$\frac{np}{2}-p$}] (5b) at (4,-0.5) {};
        \node[flavour,label=right:{\small$2p$}] (6a) at (5,0.5) {};
        \draw (1)--(2)--(3)--(4)--(5a)--(6a) (5b)--(4);
    \end{tikzpicture}}\;.
\end{equation}
We only know such a 3d mirror for $2\leq n\leq4$, since we can exploit $D_2=A_1A_1$ and $D_3=A_3$, as well as triality of $D_4$ to compute the 3d mirror of the magnetic quivers in question.

The massless states at the most singular Coulomb branch points are:\\
For $n=2$:
\begin{equation}
    \raisebox{-.5\height}{\scalebox{.7}{\begin{tikzpicture}
        \node[gauge,label=below:{$\mathrm{U}(2)$}] (1) at (0,0) {};
        \node[gauge,label=below:{$\mathrm{U}(4)$}] (2) at (2,0) {};
        \node (d) at (3,0) {$\cdots$};
        \node[gauge,label=below:{$\mathrm{U}((p-1)2)$}] (3) at (4,0) {};
        \node[gauge,label=below:{$\mathrm{USp}(2p)$}] (4) at (6,0) {};
        \node[flavour,label=below:{$D_2$}] (5) at (8,0) {};
        \draw (1)--(2)--(d)--(3)--(4)--(5);
    \end{tikzpicture}}}\qquad\rightarrow\qquad\raisebox{-.5\height}{\scalebox{.7}{\begin{tikzpicture}
        \node[gauge,label=below:{$1$}] (1) at (1,0) {};
        \node (2) at (2,0) {$\cdots$};
        \node[gauge,label=below:{$p-1$}] (3) at (3,0) {};
        \node[gauge,label=below:{$p$}] (4) at (4,0) {};
        \node[flavour,label=left:{$2$}] (4f) at (4,1) {};
        \node[gauge,label=below:{$p-1$}] (5) at (5,0) {};
        \node (6) at (6,0) {$\cdots$};
        \node[gauge,label=below:{$1$}] (7) at (7,0) {};
        \draw (1)--(2)--(3)--(4)--(5)--(6)--(7) (4)--(4f);
    \end{tikzpicture}}}\;.
\end{equation}
For $n=3$, $p=2x$:
\begin{equation}
    \raisebox{-.5\height}{\scalebox{.7}{\begin{tikzpicture}
        \node[gauge,label=below:{$\mathrm{U}(3)$}] (1) at (0,0) {};
        \node[gauge,label=below:{$\mathrm{U}(6)$}] (2) at (2,0) {};
        \node (d) at (3,0) {$\cdots$};
        \node[gauge,label=below:{$\mathrm{U}((2x-1)3)$}] (3) at (4,0) {};
        \node[gauge,label=below:{$\mathrm{USp}(6x)$}] (4) at (6,0) {};
        \node[flavour,label=below:{$D_3$}] (5) at (8,0) {};
        \draw (1)--(2)--(d)--(3)--(4)--(5);
    \end{tikzpicture}}}\qquad\rightarrow\qquad\raisebox{-.5\height}{\scalebox{.7}{\begin{tikzpicture}
        \node[gauge,label=below:{$1$}] (1) at (1,0) {};
        \node (2) at (2,0) {$\cdots$};
        \node[gauge,label=below:{$3x-1$}] (3) at (3,0) {};
        \node[gauge,label=below:{$3x$}] (4) at (4,0) {};
        \node[flavour,label=left:{$4$}] (4f) at (4,1) {};
        \node[gauge,label=below:{$3x-3$}] (5) at (5,0) {};
        \node (6) at (6,0) {$\cdots$};
        \node[gauge,label=below:{$3$}] (7) at (7,0) {};
        \draw (1)--(2)--(3)--(4)--(5)--(6)--(7) (4)--(4f);
    \end{tikzpicture}}}\;.
\end{equation}
For $n=4$:
\begin{equation}
    \raisebox{-.5\height}{\scalebox{.7}{\begin{tikzpicture}
        \node[gauge,label=below:{$\mathrm{U}(4)$}] (1) at (0,0) {};
        \node[gauge,label=below:{$\mathrm{U}(8)$}] (2) at (2,0) {};
        \node (d) at (3,0) {$\cdots$};
        \node[gauge,label=below:{$\mathrm{U}((p-1)4)$}] (3) at (4,0) {};
        \node[gauge,label=below:{$\mathrm{USp}(4p)$}] (4) at (6,0) {};
        \node[flavour,label=below:{$D_4$}] (5) at (8,0) {};
        \draw (1)--(2)--(d)--(3)--(4)--(5);
    \end{tikzpicture}}}\qquad\rightarrow\qquad\raisebox{-.5\height}{\scalebox{.7}{\begin{tikzpicture}
        \node[gauge,label=below:{$\mathrm{U}(1)$}] (1) at (1,0) {};
        \node (2) at (2,0) {$\cdots$};
        \node[gauge,label=below:{$\mathrm{U}(2p-2)$}] (3) at (3,0) {};
        \node[gauge,label=below:{$\mathrm{U}(2p-1)$}] (4) at (5,0) {};
        \node[gauge,label=below:{$\mathrm{USp}(2p)$}] (5) at (7,0) {};
        \node[flavour,label=below:{$D_4$}] (6) at (9,0) {};
        \draw (1)--(2)--(3)--(4)--(5)--(6);
    \end{tikzpicture}}}\;.
\end{equation}

\section{$\mathrm{SU}(n)-\mathrm{SU}(2n)-\cdots-\mathrm{SU}((p-1)n)-\mathrm{USp}(pn)-[D_n]$ on Brane Webs}
\label{sec:5d}
We now turn to constructions of 5-brane webs with 7-branes and an $O7^-$ orientifold plane. The $O7^-$ is quantum mechanically resolved into a pair of mutually non-local $7$ branes, such that their combined monodromy equals that of the $O7^-$ \cite{Sen:1996vd,Zafrir:2015rga}. We pick the resolution into a $[1,1]7$ and a $[1,-1]7$ brane.

\subsection{General Family}
We study the classical Higgs branch of the following theory, which is obtained by changing all U nodes in \eqref{electric:usp} to SU nodes:
\begin{equation}
   \raisebox{-.5\height}{  \begin{tikzpicture}
    \node[gauge,label=below:{$\mathrm{SU}(n)$}] (1) at (0,0) {};
    \node[gauge,label=below:{$\mathrm{SU}(2n)$}] (2) at (2,0) {};
    \node (d) at (3,0) {$\cdots$};
    \node[gauge,label=below:{$\mathrm{SU}((p-1)n)$}] (3) at (4,0) {};
    \node[gauge,label=below:{$\mathrm{USp}(pn)$}] (4) at (6,0) {};
    \node[flavour,label=below:{$D_n$}] (5) at (8,0) {};
    \draw (1)--(2)--(d)--(3)--(4)--(5);
    \end{tikzpicture}} \qquad,\quad pn\textnormal{ even}\;.
    \label{electric:suspstart}
\end{equation}
After resolving the $O7^-$ and suitable Hanany-Witten transitions, we can read the magnetic quivers from the brane web
\begin{equation}
   \raisebox{-.5\height}{  \begin{tikzpicture}
    \node[sev] (1) at (0,0) {};
    \node[sev] (2) at (1,0) {};
    \node[sev] (3) at (2,0) {};
    \node (4) at (3,0) {$\cdots$};
    \node[sev] (5) at (4,0) {};
    \node[sev] (6) at (5,0) {};
    \node[sev] (7) at (6,0) {};
    \node[sev] (8a) at (7,1) {};
    \node[sev] (8b) at (7,-1) {};
    \node[sev] (9a) at (9,1) {};
    \node[sev] (9b) at (9,-1) {};
    \node[sev] (10a) at (11,1) {};
    \node[sev] (10b) at (11,-1) {};
    \draw[double] (1)--(2)--(3)--(4)--(5)--(6)--(7)--(10,0)--(10a) (10b)--(10,0);
    \draw (8a)--(8b) (9a)--(9b);
    \node at (8,0.5) {\Huge$\cdots$};
    \draw [decorate,decoration={brace,amplitude=5pt},xshift=0pt,yshift=-0.2cm]
    (6.2,-0.2)--(-0.2,-0.2) node [black,midway,yshift=-0.4cm] {$n$};
    \draw [decorate,decoration={brace,amplitude=5pt},xshift=0pt,yshift=0.2cm]
    (6.8,1.2)--(9.2,1.2) node [black,midway,yshift=0.4cm] {$p$};
    \node at (0.5,0.3) {\small$p$};
    \node at (1.5,0.3) {\small$2p$};
    \node[rotate=45] at (4.6,0.6) {\small$(n-2)p$};
    \node[rotate=45] at (5.6,0.6) {\small$(n-1)p$};
    \node at (6.5,0.3) {\small$np$};
    \node at (10.3,0.7) {\small$\frac{np}{2}$};
    \node at (10.3,-0.7) {\small$\frac{np}{2}$};
    \end{tikzpicture}}\;.
    \label{eq:braneWebFiniteCouplingGeneric}
\end{equation}
While this essentially looks like a 5d version of the brane system \eqref{branes:uspH} (they are T-dual after all) there is a crucial difference in reading the magnetic quiver: The NS5 branes in \eqref{eq:braneWebFiniteCouplingGeneric} contribute to the magnetic quiver as gauge nodes rather than flavor nodes (as was the case in \eqref{branes:uspH}). In \eqref{branes:uspHl} and \eqref{magnetic:uspHl} the number $l$ of NS5 branes attached to D3 branes is used to label the inequivalent maximal Higgs phases and their magnetic quiver representations. We can use the same label $l$ in the present construction; however, since the $p$ NS5 branes contribute as gauge nodes to the magnetic quiver, it is of crucial importance which one of them combines with D5 branes. Hence for every value of $l\in\{0,\dots,p\}$ there are ${p\choose l}$ different cones. The total number of cones in the Higgs branch is:
\begin{equation}
    \#(\mathrm{cones})=\sum_{l=0}^{p}{p\choose l}=2^p \, . 
\end{equation}
Each one of the $p$ NS5 branes can either connect with a D5 brane or not, hence the $2^p$ options.\\

In the next subsection, we study the case $p=2$ where four cones show up. When $p>2$, the combinatorial complexity grows very fast, as illustrated with a quick survey of the $p=3$ case in Appendix \ref{app:eightcones}.

\subsection{Four Cone Example}
Let us study the simplest case involving more than 2 cones, $p=2$ in \eqref{electric:suspstart}:
\begin{equation}
\label{eq:quiverSUn}
    \raisebox{-.5\height}{ \begin{tikzpicture}
    \node[gauge,label=below:{$\mathrm{SU}(n)$}] (1) at (0,0) {};
    \node[gauge,label=below:{$\mathrm{USp}(2n)$}] (2) at (2,0) {};
    \node[flavour,label=below:{$D_n$}] (3) at (4,0) {};
    \draw (1)--(2)--(3);
    \end{tikzpicture}}
\end{equation}
The relevant brane system for finite coupling is \eqref{eq:braneWebFiniteCouplingGeneric} with $p=2$. 
There are 4 different maximal subdivisions, and 4 corresponding magnetic quivers.
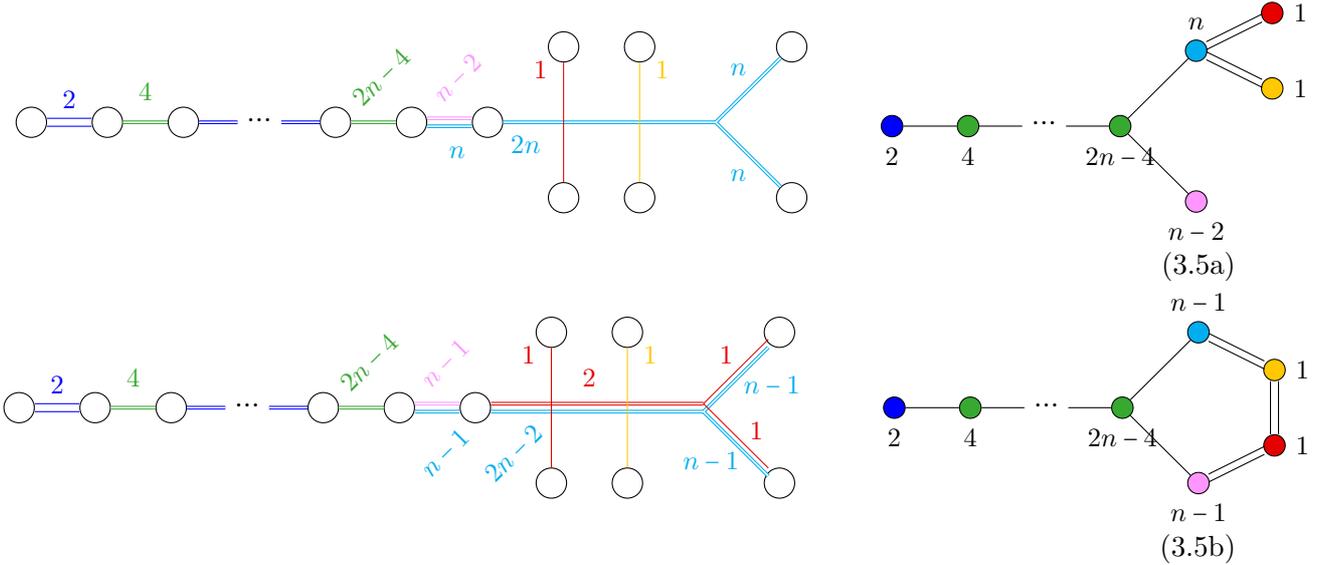
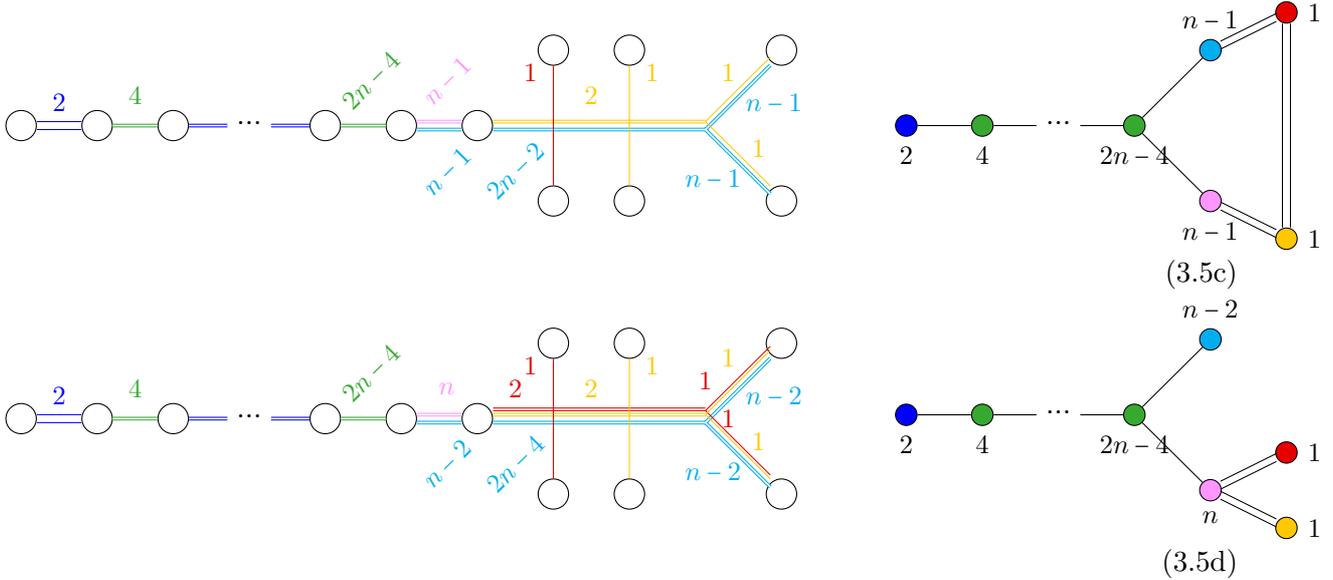
\begin{subequations}
    \begin{equation}
    \label{magnetic:fourcones5d1}
        \makebox[\textwidth][c]{\raisebox{-.5\height}{ \begin{tikzpicture}
    \node[sev] (1) at (0,0) {};
    \node[sev] (2) at (1,0) {};
    \node[sev] (3) at (2,0) {};
    \node (4) at (3,0) {$\cdots$};
    \node[sev] (5) at (4,0) {};
    \node[sev] (6) at (5,0) {};
    \node[sev] (7) at (6,0) {};
    \node[sev] (8a) at (7,1) {};
    \node[sev] (8b) at (7,-1) {};
    \node[sev] (9a) at (8,1) {};
    \node[sev] (9b) at (8,-1) {};
    \node[sev] (10a) at (10,1) {};
    \node[sev] (10b) at (10,-1) {};
    \draw[blue,transform canvas={yshift=-1.5pt}] (1)--(2);
    \draw[blue,transform canvas={yshift=1.5pt}] (1)--(2);
    \draw[goodgreen,double] (2)--(3);
    \draw[blue,double] (3)--(4)--(5);
    \draw[goodgreen,double] (5)--(6);
    \draw[goodmagenta,double,transform canvas={yshift=1.5pt}] (6)--(7);
    \draw[cyan,double,transform canvas={yshift=-1.5pt}] (6)--(7);
    \draw[cyan,double] (7)--(9,0)--(10a) (10b)--(9,0);
    \draw[goodred] (8a)--(8b);
    \draw[goodorange] (9a)--(9b);
    \node at (0.5,0.3) {\color{blue}\small$2$};
    \node at (1.5,0.4) {\color{goodgreen}\small$4$};
    \node[rotate=45] at (4.6,0.6) {\color{goodgreen}\small$2n-4$};
    \node[rotate=45] at (5.6,0.6) {\color{goodmagenta}\small$n-2$};
    \node at (6.7,0.7) {\color{goodred}\small$1$};
    \node at (8.3,0.7) {\color{goodorange}\small$1$};
    \node at (5.6,-0.4) {\color{cyan}\small$n$};
    \node at (6.5,-0.3) {\color{cyan}\small$2n$};
    \node at (9.3,0.7) {\color{cyan}\small$n$};
    \node at (9.3,-0.7) {\color{cyan}\small$n$};
    \end{tikzpicture}}  
    \qquad \raisebox{-.5\height}{\begin{tikzpicture}
    \node[gaugeb,label=below:{\small$2$}] (1) at (0,0) {};
    \node[gaugeg,label=below:{\small$4$}] (2) at (1,0) {};
    \node (3) at (2,0) {$\cdots$};
    \node[gaugeg,label=below:{\small$2n-4$}] (4) at (3,0) {};
    \node[gaugec,label=above:{\small$n$}] (5a) at (4,1) {};
    \node[gaugem,label=below:{\small$n-2$}] (5b) at (4,-1) {};
    \node[gauger,label=right:{\small$1$}] (6ar) at (5,1.5) {};
    \node[gaugeo,label=right:{\small$1$}] (6ay) at (5,0.5) {};
    \draw (1)--(2)--(3)--(4)--(5a) (5b)--(4);
    \draw[transform canvas={yshift=-1.5pt}] (5a)--(6ar);
    \draw[transform canvas={yshift=1.5pt}] (5a)--(6ar);
    \draw[transform canvas={yshift=-1.5pt}] (5a)--(6ay);
    \draw[transform canvas={yshift=1.5pt}] (5a)--(6ay);
    \end{tikzpicture}}}
    \end{equation}
    \begin{equation}
    \label{magnetic:fourcones5d2}
        \makebox[\textwidth][c]{\raisebox{-.5\height}{\begin{tikzpicture}
    \node[sev] (1) at (0,0) {};
    \node[sev] (2) at (1,0) {};
    \node[sev] (3) at (2,0) {};
    \node (4) at (3,0) {$\cdots$};
    \node[sev] (5) at (4,0) {};
    \node[sev] (6) at (5,0) {};
    \node[sev] (7) at (6,0) {};
    \node[sev] (8a) at (7,1) {};
    \node[sev] (8b) at (7,-1) {};
    \node[sev] (9a) at (8,1) {};
    \node[sev] (9b) at (8,-1) {};
    \node[sev] (10a) at (10,1) {};
    \node[sev] (10b) at (10,-1) {};
    \draw[blue,transform canvas={yshift=-1.5pt}] (1)--(2);
    \draw[blue,transform canvas={yshift=1.5pt}] (1)--(2);
    \draw[goodgreen,double] (2)--(3);
    \draw[blue,double] (3)--(4)--(5);
    \draw[goodgreen,double] (5)--(6);
    \draw[goodmagenta,double,transform canvas={yshift=1.5pt}] (6)--(7);
    \draw[cyan,double,transform canvas={yshift=-1.5pt}] (6)--(7);
    \draw[cyan,double,transform canvas={yshift=-1.5pt}] (7)--(9,0)--(10a) (10b)--(9,0);
    \draw[goodred,transform canvas={yshift=1.5pt}] (9,0)--(10a) (10b)--(9,0);
    \draw[goodred,transform canvas={yshift=2pt}] (7)--(9,0);
    \draw[goodred,transform canvas={yshift=1pt}] (7)--(9,0);
    \draw[goodred] (8a)--(8b);
    \draw[goodorange] (9a)--(9b);
    \node at (0.5,0.3) {\color{blue}\small$2$};
    \node at (1.5,0.4) {\color{goodgreen}\small$4$};
    \node[rotate=45] at (4.6,0.6) {\color{goodgreen}\small$2n-4$};
    \node[rotate=45] at (5.6,0.6) {\color{goodmagenta}\small$n-1$};
    \node[rotate=45] at (5.6,-0.6) {\color{cyan}\small$n-1$};
    \node[rotate=45] at (6.5,-0.6) {\color{cyan}\small$2n-2$};
    \node at (6.7,0.7) {\color{goodred}\small$1$};
    \node at (7.5,0.4) {\color{goodred}\small$2$};
    \node at (9.3,0.7) {\color{goodred}\small$1$};
    \node at (9.7,-0.3) {\color{goodred}\small$1$};
    \node at (8.3,0.7) {\color{goodorange}\small$1$};
    \node at (9.9,0.3) {\color{cyan}\small$n-1$};
    \node at (9.1,-0.7) {\color{cyan}\small$n-1$};
    \end{tikzpicture}}
    \qquad \raisebox{-.5\height}{\begin{tikzpicture}
    \node[gaugeb,label=below:{\small$2$}] (1) at (0,0) {};
    \node[gaugeg,label=below:{\small$4$}] (2) at (1,0) {};
    \node (3) at (2,0) {$\cdots$};
    \node[gaugeg,label=below:{\small$2n-4$}] (4) at (3,0) {};
    \node[gaugec,label=above:{\small$n-1$}] (5a) at (4,1) {};
    \node[gaugem,label=below:{\small$n-1$}] (5b) at (4,-1) {};
    \node[gaugeo,label=right:{\small$1$}] (6ay) at (5,0.5) {};
    \node[gauger,label=right:{\small$1$}] (6br) at (5,-0.5) {};
    \draw (1)--(2)--(3)--(4)--(5a) (5b)--(4);
    \draw[transform canvas={yshift=-1.5pt}] (5b)--(6br);
    \draw[transform canvas={yshift=1.5pt}] (5b)--(6br);
    \draw[transform canvas={yshift=-1.5pt}] (5a)--(6ay);
    \draw[transform canvas={yshift=1.5pt}] (5a)--(6ay);
    \draw[transform canvas={xshift=-1.5pt}] (6ay)--(6br);
    \draw[transform canvas={xshift=1.5pt}] (6ay)--(6br);
    \end{tikzpicture}}}
    \end{equation}
    \begin{equation}
    \label{magnetic:fourcones5d3}
        \makebox[\textwidth][c]{\raisebox{-.5\height}{  \begin{tikzpicture}
    \node[sev] (1) at (0,0) {};
    \node[sev] (2) at (1,0) {};
    \node[sev] (3) at (2,0) {};
    \node (4) at (3,0) {$\cdots$};
    \node[sev] (5) at (4,0) {};
    \node[sev] (6) at (5,0) {};
    \node[sev] (7) at (6,0) {};
    \node[sev] (8a) at (7,1) {};
    \node[sev] (8b) at (7,-1) {};
    \node[sev] (9a) at (8,1) {};
    \node[sev] (9b) at (8,-1) {};
    \node[sev] (10a) at (10,1) {};
    \node[sev] (10b) at (10,-1) {};
    \draw[blue,transform canvas={yshift=-1.5pt}] (1)--(2);
    \draw[blue,transform canvas={yshift=1.5pt}] (1)--(2);
    \draw[goodgreen,double] (2)--(3);
    \draw[blue,double] (3)--(4)--(5);
    \draw[goodgreen,double] (5)--(6);
    \draw[goodmagenta,double,transform canvas={yshift=1.5pt}] (6)--(7);
    \draw[cyan,double,transform canvas={yshift=-1.5pt}] (6)--(7);
    \draw[cyan,double,transform canvas={yshift=-1.5pt}] (7)--(9,0)--(10a) (10b)--(9,0);
    \draw[goodorange,transform canvas={yshift=1.5pt}] (9,0)--(10a) (10b)--(9,0);
    \draw[goodorange,transform canvas={yshift=2pt}] (7)--(9,0);
    \draw[goodorange,transform canvas={yshift=1pt}] (7)--(9,0);
    \draw[goodred] (8a)--(8b);
    \draw[goodorange] (9a)--(9b);
    \node at (0.5,0.3) {\color{blue}\small$2$};
    \node at (1.5,0.4) {\color{goodgreen}\small$4$};
    \node[rotate=45] at (4.6,0.6) {\color{goodgreen}\small$2n-4$};
    \node[rotate=45] at (5.6,0.6) {\color{goodmagenta}\small$n-1$};
    \node[rotate=45] at (5.6,-0.6) {\color{cyan}\small$n-1$};
    \node[rotate=45] at (6.5,-0.6) {\color{cyan}\small$2n-2$};
    \node at (6.7,0.7) {\color{goodred}\small$1$};
    \node at (7.5,0.4) {\color{goodorange}\small$2$};
    \node at (9.3,0.7) {\color{goodorange}\small$1$};
    \node at (9.7,-0.3) {\color{goodorange}\small$1$};
    \node at (8.3,0.7) {\color{goodorange}\small$1$};
    \node at (9.9,0.3) {\color{cyan}\small$n-1$};
    \node at (9.1,-0.7) {\color{cyan}\small$n-1$};
    \end{tikzpicture}}
    \qquad \raisebox{-.5\height}{ \begin{tikzpicture}
    \node[gaugeb,label=below:{\small$2$}] (1) at (0,0) {};
    \node[gaugeg,label=below:{\small$4$}] (2) at (1,0) {};
    \node (3) at (2,0) {$\cdots$};
    \node[gaugeg,label=below:{\small$2n-4$}] (4) at (3,0) {};
    \node[gaugec,label=above:{\small$n-1$}] (5a) at (4,1) {};
    \node[gaugem,label=below:{\small$n-1$}] (5b) at (4,-1) {};
    \node[gauger,label=right:{\small$1$}] (6ar) at (5,1.5) {};
    \node[gaugeo,label=right:{\small$1$}] (6by) at (5,-1.5) {};
    \draw (1)--(2)--(3)--(4)--(5a) (5b)--(4);
    \draw[transform canvas={yshift=-1.5pt}] (5a)--(6ar);
    \draw[transform canvas={yshift=1.5pt}] (5a)--(6ar);
    \draw[transform canvas={yshift=-1.5pt}] (5b)--(6by);
    \draw[transform canvas={yshift=1.5pt}] (5b)--(6by);
    \draw[transform canvas={xshift=-1.5pt}] (6ar)--(6by);
    \draw[transform canvas={xshift=1.5pt}] (6ar)--(6by);
    \end{tikzpicture}}}
    \end{equation}
    \begin{equation}
    \label{magnetic:fourcones5d4}
        \makebox[\textwidth][c]{\raisebox{-.5\height}{ \begin{tikzpicture}
    \node[sev] (1) at (0,0) {};
    \node[sev] (2) at (1,0) {};
    \node[sev] (3) at (2,0) {};
    \node (4) at (3,0) {$\cdots$};
    \node[sev] (5) at (4,0) {};
    \node[sev] (6) at (5,0) {};
    \node[sev] (7) at (6,0) {};
    \node[sev] (8a) at (7,1) {};
    \node[sev] (8b) at (7,-1) {};
    \node[sev] (9a) at (8,1) {};
    \node[sev] (9b) at (8,-1) {};
    \node[sev] (10a) at (10,1) {};
    \node[sev] (10b) at (10,-1) {};
    \draw[blue,transform canvas={yshift=-1.5pt}] (1)--(2);
    \draw[blue,transform canvas={yshift=1.5pt}] (1)--(2);
    \draw[goodgreen,double] (2)--(3);
    \draw[blue,double] (3)--(4)--(5);
    \draw[goodgreen,double] (5)--(6);
    \draw[goodmagenta,double,transform canvas={yshift=1.5pt}] (6)--(7);
    \draw[cyan,double,transform canvas={yshift=-1.5pt}] (6)--(7);
    \draw[cyan,double,transform canvas={yshift=-1.5pt}] (7)--(9,0)--(10a) (10b)--(9,0);
    \draw[goodred,transform canvas={yshift=3pt}] (9,0)--(10a) (10b)--(9,0);
    \draw[goodred,transform canvas={yshift=4pt}] (7)--(9,0);
    \draw[goodred,transform canvas={yshift=3pt}] (7)--(9,0);
    \draw[goodorange,transform canvas={yshift=1.5pt}] (9,0)--(10a) (10b)--(9,0);
    \draw[goodorange,transform canvas={yshift=2pt}] (7)--(9,0);
    \draw[goodorange,transform canvas={yshift=1pt}] (7)--(9,0);
    \draw[goodred] (8a)--(8b);
    \draw[goodorange] (9a)--(9b);
    \node at (0.5,0.3) {\color{blue}\small$2$};
    \node at (1.5,0.4) {\color{goodgreen}\small$4$};
    \node[rotate=45] at (4.6,0.6) {\color{goodgreen}\small$2n-4$};
    \node at (5.6,0.4) {\color{goodmagenta}\small$n$};
    \node[rotate=45] at (5.6,-0.6) {\color{cyan}\small$n-2$};
    \node[rotate=45] at (6.5,-0.6) {\color{cyan}\small$2n-4$};
    \node at (6.7,0.7) {\color{goodred}\small$1$};
    \node at (6.5,0.4) {\color{goodred}\small$2$};
    \node at (9,0.5) {\color{goodred}\small$1$};
    \node at (9.3,0) {\color{goodred}\small$1$};
    \node at (7.5,0.4) {\color{goodorange}\small$2$};
    \node at (9.3,0.8) {\color{goodorange}\small$1$};
    \node at (9.7,-0.3) {\color{goodorange}\small$1$};
    \node at (8.3,0.7) {\color{goodorange}\small$1$};
    \node at (9.9,0.3) {\color{cyan}\small$n-2$};
    \node at (9.1,-0.7) {\color{cyan}\small$n-2$};
    \end{tikzpicture}}
    \qquad \raisebox{-.5\height}{ \begin{tikzpicture}
    \node[gaugeb,label=below:{\small$2$}] (1) at (0,0) {};
    \node[gaugeg,label=below:{\small$4$}] (2) at (1,0) {};
    \node (3) at (2,0) {$\cdots$};
    \node[gaugeg,label=below:{\small$2n-4$}] (4) at (3,0) {};
    \node[gaugec,label=above:{\small$n-2$}] (5a) at (4,1) {};
    \node[gaugem,label=below:{\small$n$}] (5b) at (4,-1) {};
    \node[gauger,label=right:{\small$1$}] (6br) at (5,-0.5) {};
    \node[gaugeo,label=right:{\small$1$}] (6by) at (5,-1.5) {};
    \draw (1)--(2)--(3)--(4)--(5a) (5b)--(4);
    \draw[transform canvas={yshift=-1.5pt}] (5b)--(6br);
    \draw[transform canvas={yshift=1.5pt}] (5b)--(6br);
    \draw[transform canvas={yshift=-1.5pt}] (5b)--(6by);
    \draw[transform canvas={yshift=1.5pt}] (5b)--(6by);
    \end{tikzpicture}}}
    \end{equation}
\end{subequations}

The four cones intersect in non-trivial ways, as indicated on the following  Hasse diagram: 
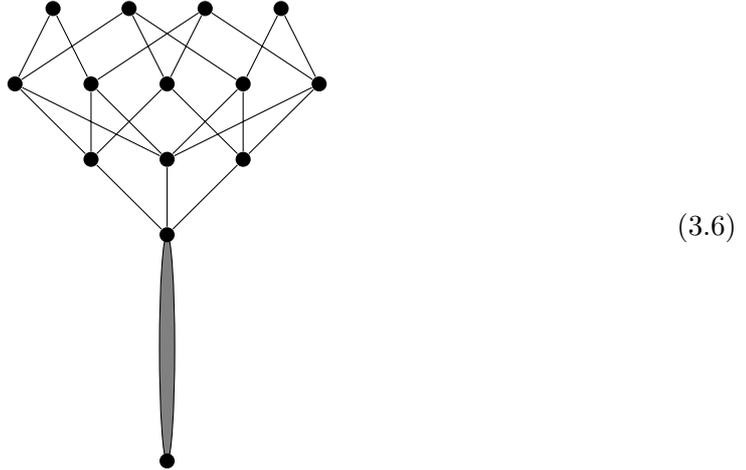
\begin{equation}
    \raisebox{-.5\height}{ \begin{tikzpicture}
    \draw[fill=gray] (0,-1.5) ellipse (.1 and 1.5);
    \node[hasse] at (0,-3) {};
    \node[hasse] (1) at (0,0) {};
    \node[hasse] (a1) at (-1,1) {};
    \node[hasse] (a2) at (0,1) {};
    \node[hasse] (a3) at (1,1) {};
    \node[hasse] (b1) at (-2,2) {};
    \node[hasse] (b2) at (-1,2) {};
    \node[hasse] (b3) at (0,2) {};
    \node[hasse] (b4) at (1,2) {};
    \node[hasse] (b5) at (2,2) {};
    \node[hasse] (c1) at (-1.5,3) {};
    \node[hasse] (c2) at (-0.5,3) {};
    \node[hasse] (c3) at (0.5,3) {};
    \node[hasse] (c4) at (1.5,3) {};
    \draw (1)--(a1)--(b1)--(c1)--(b2)--(a1)--(b3)--(c2)--(b1)--(a2)--(b2)--(c3)--(b3)--(a3)--(1)--(a2)--(b4)--(c4)--(b5)--(a3)--(b4)--(c2) (c3)--(b5)--(a2);
    \end{tikzpicture}}
    \label{eq:diagramCrown}
\end{equation}
where all lines in the top part are $A_1$ transitions. The magnetic quivers for the various leaf closures can be obtained by quiver subtraction. The bottom gray blob represents the complicated Coulomb branch Hasse diagram of the quiver 
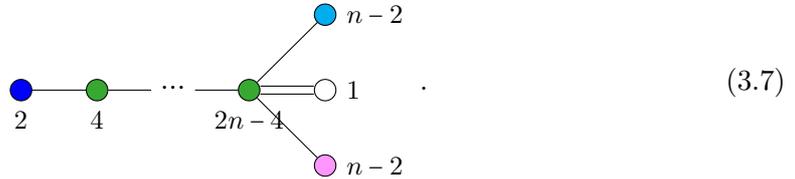
\begin{equation}
    \raisebox{-.5\height}{ \begin{tikzpicture}
    \node[gaugeb,label=below:{\small$2$}] (1) at (0,0) {};
    \node[gaugeg,label=below:{\small$4$}] (2) at (1,0) {};
    \node (3) at (2,0) {$\cdots$};
    \node[gaugeg,label=below:{\small$2n-4$}] (4) at (3,0) {};
    \node[gaugec,label=right:{\small$n-2$}] (5a) at (4,1) {};
    \node[gaugem,label=right:{\small$n-2$}] (5b) at (4,-1) {};
    \node[gauge,label=right:{\small$1$}] (6) at (4,0) {};
    \draw (1)--(2)--(3)--(4)--(5a) (5b)--(4);
    \draw[transform canvas={yshift=-1.5pt}] (4)--(6);
    \draw[transform canvas={yshift=1.5pt}] (4)--(6);
    \end{tikzpicture}}\;.
    \label{eq:stemQuiver}
\end{equation}

In order to check that this is correct, one can extract the Hilbert series for the Higgs variety $\mathcal{HV}$ of \eqref{eq:quiverSUn} as before. We apply the sieve formula \eqref{eq:sieve}, taking into account all possible intersections, giving (in the following equations, we represent only the top part of \eqref{eq:diagramCrown}, as the bottom part is identical in all cases). We denote
\begin{equation}
    \mathcal{C}_a=\mathcal{C}\eqref{magnetic:fourcones5d1}\textnormal{, }
    \mathcal{C}_b=\mathcal{C}\eqref{magnetic:fourcones5d2} \textnormal{, }
    \mathcal{C}_c=\mathcal{C}\eqref{magnetic:fourcones5d3} \textnormal{, and  }
    \mathcal{C}_d=\mathcal{C}\eqref{magnetic:fourcones5d4}\;.
\end{equation}
\begin{equation}
    \begin{split}
        \mathrm{HS}(\mathcal{HV}\eqref{eq:quiverSUn})  =&\mathrm{HS}(\mathcal{C}_a)
    +\mathrm{HS}(\mathcal{C}_b)
    +\mathrm{HS}(\mathcal{C}_c)
    +\mathrm{HS}(\mathcal{C}_d)\\
     & -\mathrm{HS}(\mathcal{C}_a\cap\mathcal{C}_b)
    -\mathrm{HS}(\mathcal{C}_a\cap\mathcal{C}_c)
    -\mathrm{HS}(\mathcal{C}_a\cap\mathcal{C}_d)\\
     & -\mathrm{HS}(\mathcal{C}_b\cap\mathcal{C}_c)
    -\mathrm{HS}(\mathcal{C}_b\cap\mathcal{C}_d)
    -\mathrm{HS}(\mathcal{C}_c\cap\mathcal{C}_d)\\
     & +\mathrm{HS}(\mathcal{C}_a\cap\mathcal{C}_b\cap\mathcal{C}_c)
    +\mathrm{HS}(\mathcal{C}_a\cap\mathcal{C}_b\cap\mathcal{C}_d)
    +\mathrm{HS}(\mathcal{C}_a\cap\mathcal{C}_c\cap\mathcal{C}_d)
    +\mathrm{HS}(\mathcal{C}_b\cap\mathcal{C}_c\cap\mathcal{C}_d)\\
     &-\mathrm{HS}(\mathcal{C}_a\cap\mathcal{C}_b\cap\mathcal{C}_c\cap\mathcal{C}_d)\\
    =&  \mathrm{HS} \left( \smallhasse{
    \node[hasse] (1) at (0,0) {};
    \node[hasse] (a1) at (-1,1) {};
    \node[hasse] (a2) at (0,1) {};
    \node[hassegray] (a3) at (1,1) {};
    \node[hasse] (b1) at (-2,2) {};
    \node[hasse] (b2) at (-1,2) {};
    \node[hassegray] (b3) at (0,2) {};
    \node[hassegray] (b4) at (1,2) {};
    \node[hassegray] (b5) at (2,2) {};
    \node[hasse] (c1) at (-1.5,3) {};
    \node[hassegray] (c2) at (-0.5,3) {};
    \node[hassegray] (c3) at (0.5,3) {};
    \node[hassegray] (c4) at (1.5,3) {};
    \draw  (1)--(a1) (1)--(a2) (a1)--(b1) (a1)--(b2) (a2)--(b1) (a2)--(b2) (b1)--(c1) (b2)--(c1);
    \draw[linegray] (1)--(a3) (a1)--(b3) (a2)--(b4) (a2)--(b5) (a3)--(b3) (a3)--(b4) (a3)--(b5) (b1)--(c2) (b2)--(c3) (b3)--(c2) (b3)--(c3) (b4)--(c2) (b4)--(c4) (b5)--(c3) (b5)--(c4);}  \right) 
    +  \mathrm{HS} \left( \smallhasse{
    \node[hasse] (1) at (0,0) {};
    \node[hasse] (a1) at (-1,1) {};
    \node[hasse] (a2) at (0,1) {};
    \node[hasse] (a3) at (1,1) {};
    \node[hasse] (b1) at (-2,2) {};
    \node[hassegray] (b2) at (-1,2) {};
    \node[hasse] (b3) at (0,2) {};
    \node[hasse] (b4) at (1,2) {};
    \node[hassegray] (b5) at (2,2) {};
    \node[hassegray] (c1) at (-1.5,3) {};
    \node[hasse] (c2) at (-0.5,3) {};
    \node[hassegray] (c3) at (0.5,3) {};
    \node[hassegray] (c4) at (1.5,3) {};
    \draw  (1)--(a1) (1)--(a2) (1)--(a3) (a1)--(b1) (a1)--(b3) (a2)--(b1) (a2)--(b4) (a3)--(b3) (a3)--(b4) (b1)--(c2) (b3)--(c2) (b4)--(c2);
    \draw[linegray]  (a2)--(b2) (a1)--(b2) (a2)--(b5) (a3)--(b5) (b1)--(c1) (b2)--(c1) (b2)--(c3) (b3)--(c3) (b5)--(c3) (b5)--(c4) (b4)--(c4);}  \right)   
    +  \mathrm{HS} \left( \smallhasse{
    \node[hasse] (1) at (0,0) {};
    \node[hasse] (a1) at (-1,1) {};
    \node[hasse] (a2) at (0,1) {};
    \node[hasse] (a3) at (1,1) {};
    \node[hassegray] (b1) at (-2,2) {};
    \node[hasse] (b2) at (-1,2) {};
    \node[hasse] (b3) at (0,2) {};
    \node[hassegray] (b4) at (1,2) {};
    \node[hasse] (b5) at (2,2) {};
    \node[hassegray] (c1) at (-1.5,3) {};
    \node[hassegray] (c2) at (-0.5,3) {};
    \node[hasse] (c3) at (0.5,3) {};
    \node[hassegray] (c4) at (1.5,3) {};
    \draw  (1)--(a1) (1)--(a2) (1)--(a3) (a1)--(b2) (a1)--(b3) (a2)--(b2) (a2)--(b5) (a3)--(b3) (a3)--(b5) (b2)--(c3) (b3)--(c3) (b5)--(c3);
    \draw[linegray] (a2)--(b1) (a1)--(b1) (a2)--(b4) (a3)--(b4) (b1)--(c1) (b1)--(c2) (b2)--(c1) (b3)--(c2) (b4)--(c2) (b4)--(c4) (b5)--(c4);}  \right) 
    +   \mathrm{HS} \left( \smallhasse{
    \node[hasse] (1) at (0,0) {};
    \node[hassegray] (a1) at (-1,1) {};
    \node[hasse] (a2) at (0,1) {};
    \node[hasse] (a3) at (1,1) {};
    \node[hassegray] (b1) at (-2,2) {};
    \node[hassegray] (b2) at (-1,2) {};
    \node[hassegray] (b3) at (0,2) {};
    \node[hasse] (b4) at (1,2) {};
    \node[hasse] (b5) at (2,2) {};
    \node[hassegray] (c1) at (-1.5,3) {};
    \node[hassegray] (c2) at (-0.5,3) {};
    \node[hassegray] (c3) at (0.5,3) {};
    \node[hasse] (c4) at (1.5,3) {};
    \draw  (1)--(a3) (1)--(a2) (a3)--(b4) (a3)--(b5) (a2)--(b4) (a2)--(b5) (b4)--(c4) (b5)--(c4);
    \draw[linegray] (1)--(a1) (a1)--(b3) (a2)--(b1) (a2)--(b2) (a1)--(b3) (a1)--(b1) (a1)--(b2) (b1)--(c1) (b1)--(c2) (b2)--(c3) (b3)--(c2) (b3)--(c3) (b1)--(c2) (b2)--(c4) (b2)--(c1) (b2)--(c1);}  \right)   \\ 
    & -  \mathrm{HS} \left( \smallhasse{
    \node[hasse] (1) at (0,0) {};
    \node[hasse] (a1) at (-1,1) {};
    \node[hasse] (a2) at (0,1) {};
    \node[hassegray] (a3) at (1,1) {};
    \node[hasse] (b1) at (-2,2) {};
    \node[hassegray] (b2) at (-1,2) {};
    \node[hassegray] (b3) at (0,2) {};
    \node[hassegray] (b4) at (1,2) {};
    \node[hassegray] (b5) at (2,2) {};
    \node[hassegray] (c1) at (-1.5,3) {};
    \node[hassegray] (c2) at (-0.5,3) {};
    \node[hassegray] (c3) at (0.5,3) {};
    \node[hassegray] (c4) at (1.5,3) {};
    \draw[black] (1)--(a1)--(b1)--(a2)--(1); 
    \draw[linegray] (1)--(a3) (a1)--(b3) (a2)--(b4) (a2)--(b5) (a3)--(b3) (a3)--(b4) (a3)--(b5) (b1)--(c2) (b2)--(c3) (b3)--(c2) (b3)--(c3) (b4)--(c2) (b4)--(c4) (b5)--(c3) (b5)--(c4) (a1)--(b2)  (a2)--(b2) (b1)--(c1) (b2)--(c1);} \right)
    -  \mathrm{HS} \left( \smallhasse{
    \node[hasse] (1) at (0,0) {};
    \node[hasse] (a1) at (-1,1) {};
    \node[hasse] (a2) at (0,1) {};
    \node[hassegray] (a3) at (1,1) {};
    \node[hassegray] (b1) at (-2,2) {};
    \node[hasse] (b2) at (-1,2) {};
    \node[hassegray] (b3) at (0,2) {};
    \node[hassegray] (b4) at (1,2) {};
    \node[hassegray] (b5) at (2,2) {};
    \node[hassegray] (c1) at (-1.5,3) {};
    \node[hassegray] (c2) at (-0.5,3) {};
    \node[hassegray] (c3) at (0.5,3) {};
    \node[hassegray] (c4) at (1.5,3) {};
    \draw (1)--(a1)--(b2)--(a2)--(1);
    \draw[linegray] (1)--(a3) (a1)--(b3) (a2)--(b4) (a2)--(b5) (a3)--(b3) (a3)--(b4) (a3)--(b5) (b1)--(c2) (b2)--(c3) (b3)--(c2) (b3)--(c3) (b4)--(c2) (b4)--(c4) (b5)--(c3) (b5)--(c4) (a1)--(b1)  (a2)--(b1) (b1)--(c1) (b2)--(c1);}\right)   
    -  \mathrm{HS} \left( \smallhasse{
    \node[hasse] (1) at (0,0) {};
    \node[hassegray] (a1) at (-1,1) {};
    \node[hasse] (a2) at (0,1) {};
    \node[hassegray] (a3) at (1,1) {};
    \node[hassegray] (b1) at (-2,2) {};
    \node[hassegray] (b2) at (-1,2) {};
    \node[hassegray] (b3) at (0,2) {};
    \node[hassegray] (b4) at (1,2) {};
    \node[hassegray] (b5) at (2,2) {};
    \node[hassegray] (c1) at (-1.5,3) {};
    \node[hassegray] (c2) at (-0.5,3) {};
    \node[hassegray] (c3) at (0.5,3) {};
    \node[hassegray] (c4) at (1.5,3) {};
    \draw (1)--(a2);
    \draw[linegray] (1)--(a3) (a1)--(b3) (a2)--(b4) (a2)--(b5) (a3)--(b3) (a3)--(b4) (a3)--(b5) (b1)--(c2) (b2)--(c3) (b3)--(c2) (b3)--(c3) (b4)--(c2) (b4)--(c4) (b5)--(c3) (b5)--(c4) (1)--(a1) (a1)--(b1) (a1)--(b2) (a2)--(b1) (a2)--(b2) (b1)--(c1) (b2)--(c1);} \right)\\
    & -  \mathrm{HS} \left( \smallhasse{
    \node[hasse] (1) at (0,0) {};
    \node[hasse] (a1) at (-1,1) {};
    \node[hasse] (a2) at (0,1) {};
    \node[hasse] (a3) at (1,1) {};
    \node[hassegray] (b1) at (-2,2) {};
    \node[hassegray] (b2) at (-1,2) {};
    \node[hasse] (b3) at (0,2) {};
    \node[hassegray] (b4) at (1,2) {};
    \node[hassegray] (b5) at (2,2) {};
    \node[hassegray] (c1) at (-1.5,3) {};
    \node[hassegray] (c2) at (-0.5,3) {};
    \node[hassegray] (c3) at (0.5,3) {};
    \node[hassegray] (c4) at (1.5,3) {};
    \draw (1)--(a1)--(b3)--(a3)--(1) (1)--(a2);
    \draw[linegray]  (a2)--(b4) (a2)--(b5) (a3)--(b4) (a3)--(b5) (b1)--(c2) (b2)--(c3) (b3)--(c2) (b3)--(c3) (b4)--(c2) (b4)--(c4) (b5)--(c3) (b5)--(c4) (a1)--(b1) (a1)--(b2) (a2)--(b1) (a2)--(b2) (b1)--(c1) (b2)--(c1);} \right)
    -  \mathrm{HS} \left( \smallhasse{
    \node[hasse] (1) at (0,0) {};
    \node[hassegray] (a1) at (-1,1) {};
    \node[hasse] (a2) at (0,1) {};
    \node[hasse] (a3) at (1,1) {};
    \node[hassegray] (b1) at (-2,2) {};
    \node[hassegray] (b2) at (-1,2) {};
    \node[hassegray] (b3) at (0,2) {};
    \node[hasse] (b4) at (1,2) {};
    \node[hassegray] (b5) at (2,2) {};
    \node[hassegray] (c1) at (-1.5,3) {};
    \node[hassegray] (c2) at (-0.5,3) {};
    \node[hassegray] (c3) at (0.5,3) {};
    \node[hassegray] (c4) at (1.5,3) {};
    \draw (1)--(a2)--(b4)--(a3)--(1);
    \draw[linegray] (a1)--(b3) (a2)--(b5) (a3)--(b3) (a3)--(b5) (b1)--(c2) (b2)--(c3) (b3)--(c2) (b3)--(c3) (b4)--(c2) (b4)--(c4) (b5)--(c3) (b5)--(c4) (1)--(a1) (a1)--(b1) (a1)--(b2) (a2)--(b1) (a2)--(b2) (b1)--(c1) (b2)--(c1);} \right)
    -  \mathrm{HS} \left( \smallhasse{
    \node[hasse] (1) at (0,0) {};
    \node[hassegray] (a1) at (-1,1) {};
    \node[hasse] (a2) at (0,1) {};
    \node[hasse] (a3) at (1,1) {};
    \node[hassegray] (b1) at (-2,2) {};
    \node[hassegray] (b2) at (-1,2) {};
    \node[hassegray] (b3) at (0,2) {};
    \node[hassegray] (b4) at (1,2) {};
    \node[hasse] (b5) at (2,2) {};
    \node[hassegray] (c1) at (-1.5,3) {};
    \node[hassegray] (c2) at (-0.5,3) {};
    \node[hassegray] (c3) at (0.5,3) {};
    \node[hassegray] (c4) at (1.5,3) {};
    \draw[black] (1)--(a2)--(b5)--(a3)--(1); 
    \draw[linegray] (a1)--(b3) (a2)--(b4) (a3)--(b3) (a3)--(b4) (b1)--(c2) (b2)--(c3) (b3)--(c2) (b3)--(c3) (b4)--(c2) (b4)--(c4) (b5)--(c3) (b5)--(c4) (1)--(a1) (a1)--(b1) (a1)--(b2) (a2)--(b1) (a2)--(b2) (b1)--(c1) (b2)--(c1);} \right) \\
    & +   \mathrm{HS} \left( \smallhasse{
    \node[hasse] (1) at (0,0) {};
    \node[hasse] (a1) at (-1,1) {};
    \node[hasse] (a2) at (0,1) {};
    \node[hassegray] (a3) at (1,1) {};
    \node[hassegray] (b1) at (-2,2) {};
    \node[hassegray] (b2) at (-1,2) {};
    \node[hassegray] (b3) at (0,2) {};
    \node[hassegray] (b4) at (1,2) {};
    \node[hassegray] (b5) at (2,2) {};
    \node[hassegray] (c1) at (-1.5,3) {};
    \node[hassegray] (c2) at (-0.5,3) {};
    \node[hassegray] (c3) at (0.5,3) {};
    \node[hassegray] (c4) at (1.5,3) {};
    \draw (a1)--(1)--(a2);
    \draw[linegray] (1)--(a3) (a1)--(b3) (a2)--(b4) (a2)--(b5) (a3)--(b3) (a3)--(b4) (a3)--(b5) (b1)--(c2) (b2)--(c3) (b3)--(c2) (b3)--(c3) (b4)--(c2) (b4)--(c4) (b5)--(c3) (b5)--(c4) (a1)--(b1) (a1)--(b2) (a2)--(b1) (a2)--(b2) (b1)--(c1) (b2)--(c1);} \right)  
    + \mathrm{HS} \left( \smallhasse{
    \node[hasse] (1) at (0,0) {};
    \node[hassegray] (a1) at (-1,1) {};
    \node[hasse] (a2) at (0,1) {};
    \node[hassegray] (a3) at (1,1) {};
    \node[hassegray] (b1) at (-2,2) {};
    \node[hassegray] (b2) at (-1,2) {};
    \node[hassegray] (b3) at (0,2) {};
    \node[hassegray] (b4) at (1,2) {};
    \node[hassegray] (b5) at (2,2) {};
    \node[hassegray] (c1) at (-1.5,3) {};
    \node[hassegray] (c2) at (-0.5,3) {};
    \node[hassegray] (c3) at (0.5,3) {};
    \node[hassegray] (c4) at (1.5,3) {};
    \draw (1)--(a2);
    \draw[linegray] (1)--(a3) (a1)--(b3) (a2)--(b4) (a2)--(b5) (a3)--(b3) (a3)--(b4) (a3)--(b5) (b1)--(c2) (b2)--(c3) (b3)--(c2) (b3)--(c3) (b4)--(c2) (b4)--(c4) (b5)--(c3) (b5)--(c4) (1)--(a1) (a1)--(b1) (a1)--(b2) (a2)--(b1) (a2)--(b2) (b1)--(c1) (b2)--(c1);} \right) 
    +  \mathrm{HS} \left( \smallhasse{
    \node[hasse] (1) at (0,0) {};
    \node[hassegray] (a1) at (-1,1) {};
    \node[hasse] (a2) at (0,1) {};
    \node[hassegray] (a3) at (1,1) {};
    \node[hassegray] (b1) at (-2,2) {};
    \node[hassegray] (b2) at (-1,2) {};
    \node[hassegray] (b3) at (0,2) {};
    \node[hassegray] (b4) at (1,2) {};
    \node[hassegray] (b5) at (2,2) {};
    \node[hassegray] (c1) at (-1.5,3) {};
    \node[hassegray] (c2) at (-0.5,3) {};
    \node[hassegray] (c3) at (0.5,3) {};
    \node[hassegray] (c4) at (1.5,3) {};
    \draw (1)--(a2);
    \draw[linegray] (1)--(a3) (a1)--(b3) (a2)--(b4) (a2)--(b5) (a3)--(b3) (a3)--(b4) (a3)--(b5) (b1)--(c2) (b2)--(c3) (b3)--(c2) (b3)--(c3) (b4)--(c2) (b4)--(c4) (b5)--(c3) (b5)--(c4) (1)--(a1) (a1)--(b1) (a1)--(b2) (a2)--(b1) (a2)--(b2) (b1)--(c1) (b2)--(c1);} \right)
    +   \mathrm{HS} \left( \smallhasse{
    \node[hasse] (1) at (0,0) {};
    \node[hassegray] (a1) at (-1,1) {};
    \node[hasse] (a2) at (0,1) {};
    \node[hasse] (a3) at (1,1) {};
    \node[hassegray] (b1) at (-2,2) {};
    \node[hassegray] (b2) at (-1,2) {};
    \node[hassegray] (b3) at (0,2) {};
    \node[hassegray] (b4) at (1,2) {};
    \node[hassegray] (b5) at (2,2) {};
    \node[hassegray] (c1) at (-1.5,3) {};
    \node[hassegray] (c2) at (-0.5,3) {};
    \node[hassegray] (c3) at (0.5,3) {};
    \node[hassegray] (c4) at (1.5,3) {};
    \draw (a3)--(1)--(a2);
    \draw[linegray] (1)--(a1) (a1)--(b3) (a2)--(b4) (a2)--(b5) (a3)--(b3) (a3)--(b4) (a3)--(b5) (b1)--(c2) (b2)--(c3) (b3)--(c2) (b3)--(c3) (b4)--(c2) (b4)--(c4) (b5)--(c3) (b5)--(c4) (a1)--(b1) (a1)--(b2) (a2)--(b1) (a2)--(b2) (b1)--(c1) (b2)--(c1);} \right)  \\ 
    & -  \mathrm{HS} \left( \smallhasse{
    \node[hasse] (1) at (0,0) {};
    \node[hassegray] (a1) at (-1,1) {};
    \node[hasse] (a2) at (0,1) {};
    \node[hassegray] (a3) at (1,1) {};
    \node[hassegray] (b1) at (-2,2) {};
    \node[hassegray] (b2) at (-1,2) {};
    \node[hassegray] (b3) at (0,2) {};
    \node[hassegray] (b4) at (1,2) {};
    \node[hassegray] (b5) at (2,2) {};
    \node[hassegray] (c1) at (-1.5,3) {};
    \node[hassegray] (c2) at (-0.5,3) {};
    \node[hassegray] (c3) at (0.5,3) {};
    \node[hassegray] (c4) at (1.5,3) {};
    \draw (1)--(a2);
    \draw[linegray] (1)--(a3) (a1)--(b3) (a2)--(b4) (a2)--(b5) (a3)--(b3) (a3)--(b4) (a3)--(b5) (b1)--(c2) (b2)--(c3) (b3)--(c2) (b3)--(c3) (b4)--(c2) (b4)--(c4) (b5)--(c3) (b5)--(c4) (1)--(a1) (a1)--(b1) (a1)--(b2) (a2)--(b1) (a2)--(b2) (b1)--(c1) (b2)--(c1);} \right)
    \end{split}
\end{equation} 
In this equation, the first line contains the four cones, the next two lines contain the $\binom{4}{2} = 6$ pairwise intersections, the fourth line contains the $\binom{4}{3} = 4$ threefold intersections, and the last line is the intersection of all four cones -- which is not a point. In each case, the Hilbert series (HS) can be obtained as the Coulomb branch HS for the magnetic quiver corresponding to the highest painted vertex, if there is one, see Table \ref{tab:HS}. Interestingly, the intersections of certain cones are themselves unions of several cones, and we can write for instance 
\begin{equation}
  \mathrm{HS}(\mathcal{C}_b\cap\mathcal{C}_c)=\mathrm{HS} \left( \smallhasse{
    \node[hasse] (1) at (0,0) {};
    \node[hasse] (a1) at (-1,1) {};
    \node[hasse] (a2) at (0,1) {};
    \node[hasse] (a3) at (1,1) {};
    \node[hassegray] (b1) at (-2,2) {};
    \node[hassegray] (b2) at (-1,2) {};
    \node[hasse] (b3) at (0,2) {};
    \node[hassegray] (b4) at (1,2) {};
    \node[hassegray] (b5) at (2,2) {};
    \node[hassegray] (c1) at (-1.5,3) {};
    \node[hassegray] (c2) at (-0.5,3) {};
    \node[hassegray] (c3) at (0.5,3) {};
    \node[hassegray] (c4) at (1.5,3) {};
    \draw (1)--(a1)--(b3)--(a3)--(1) (1)--(a2);
    \draw[linegray]  (a2)--(b4) (a2)--(b5) (a3)--(b4) (a3)--(b5) (b1)--(c2) (b2)--(c3) (b3)--(c2) (b3)--(c3) (b4)--(c2) (b4)--(c4) (b5)--(c3) (b5)--(c4) (a1)--(b1) (a1)--(b2) (a2)--(b1) (a2)--(b2) (b1)--(c1) (b2)--(c1);} \right)  =   \mathrm{HS} \left( \smallhasse{
    \node[hasse] (1) at (0,0) {};
    \node[hasse] (a1) at (-1,1) {};
    \node[hassegray] (a2) at (0,1) {};
    \node[hasse] (a3) at (1,1) {};
    \node[hassegray] (b1) at (-2,2) {};
    \node[hassegray] (b2) at (-1,2) {};
    \node[hasse] (b3) at (0,2) {};
    \node[hassegray] (b4) at (1,2) {};
    \node[hassegray] (b5) at (2,2) {};
    \node[hassegray] (c1) at (-1.5,3) {};
    \node[hassegray] (c2) at (-0.5,3) {};
    \node[hassegray] (c3) at (0.5,3) {};
    \node[hassegray] (c4) at (1.5,3) {};
    \draw (1)--(a1)--(b3)--(a3)--(1);
    \draw[linegray]  (1)--(a2) (a2)--(b4) (a2)--(b5) (a3)--(b4) (a3)--(b5) (b1)--(c2) (b2)--(c3) (b3)--(c2) (b3)--(c3) (b4)--(c2) (b4)--(c4) (b5)--(c3) (b5)--(c4) (a1)--(b1) (a1)--(b2) (a2)--(b1) (a2)--(b2) (b1)--(c1) (b2)--(c1);} \right) + \mathrm{HS} \left( \smallhasse{
    \node[hasse] (1) at (0,0) {};
    \node[hassegray] (a1) at (-1,1) {};
    \node[hasse] (a2) at (0,1) {};
    \node[hassegray] (a3) at (1,1) {};
    \node[hassegray] (b1) at (-2,2) {};
    \node[hassegray] (b2) at (-1,2) {};
    \node[hassegray] (b3) at (0,2) {};
    \node[hassegray] (b4) at (1,2) {};
    \node[hassegray] (b5) at (2,2) {};
    \node[hassegray] (c1) at (-1.5,3) {};
    \node[hassegray] (c2) at (-0.5,3) {};
    \node[hassegray] (c3) at (0.5,3) {};
    \node[hassegray] (c4) at (1.5,3) {};
    \draw (1)--(a2);
    \draw[linegray] (1)--(a3) (a1)--(b3) (a2)--(b4) (a2)--(b5) (a3)--(b3) (a3)--(b4) (a3)--(b5) (b1)--(c2) (b2)--(c3) (b3)--(c2) (b3)--(c3) (b4)--(c2) (b4)--(c4) (b5)--(c3) (b5)--(c4) (1)--(a1) (a1)--(b1) (a1)--(b2) (a2)--(b1) (a2)--(b2) (b1)--(c1) (b2)--(c1);} \right) -  \mathrm{HS} \left( \smallhasse{
    \node[hasse] (1) at (0,0) {};
    \node[hassegray] (a1) at (-1,1) {};
    \node[hassegray] (a2) at (0,1) {};
    \node[hassegray] (a3) at (1,1) {};
    \node[hassegray] (b1) at (-2,2) {};
    \node[hassegray] (b2) at (-1,2) {};
    \node[hassegray] (b3) at (0,2) {};
    \node[hassegray] (b4) at (1,2) {};
    \node[hassegray] (b5) at (2,2) {};
    \node[hassegray] (c1) at (-1.5,3) {};
    \node[hassegray] (c2) at (-0.5,3) {};
    \node[hassegray] (c3) at (0.5,3) {};
    \node[hassegray] (c4) at (1.5,3) {};
    \draw[linegray] (1)--(a2) (1)--(a3) (a1)--(b3) (a2)--(b4) (a2)--(b5) (a3)--(b3) (a3)--(b4) (a3)--(b5) (b1)--(c2) (b2)--(c3) (b3)--(c2) (b3)--(c3) (b4)--(c2) (b4)--(c4) (b5)--(c3) (b5)--(c4) (1)--(a1) (a1)--(b1) (a1)--(b2) (a2)--(b1) (a2)--(b2) (b1)--(c1) (b2)--(c1);} \right)  
\end{equation}
for the intersection of \eqref{magnetic:fourcones5d2} and \eqref{magnetic:fourcones5d3}. Expanding similarly all the cones, we finally get 
\begin{eqnarray}
\label{eq:resultCrown}
    \mathrm{HS}(\mathcal{HV}\eqref{eq:quiverSUn}) &=&  \mathrm{HS} \left( \smallhasse{
    \node[hasse] (1) at (0,0) {};
    \node[hasse] (a1) at (-1,1) {};
    \node[hasse] (a2) at (0,1) {};
    \node[hassegray] (a3) at (1,1) {};
    \node[hasse] (b1) at (-2,2) {};
    \node[hasse] (b2) at (-1,2) {};
    \node[hassegray] (b3) at (0,2) {};
    \node[hassegray] (b4) at (1,2) {};
    \node[hassegray] (b5) at (2,2) {};
    \node[hasse] (c1) at (-1.5,3) {};
    \node[hassegray] (c2) at (-0.5,3) {};
    \node[hassegray] (c3) at (0.5,3) {};
    \node[hassegray] (c4) at (1.5,3) {};
    \draw  (1)--(a1) (1)--(a2) (a1)--(b1) (a1)--(b2) (a2)--(b1) (a2)--(b2) (b1)--(c1) (b2)--(c1);
    \draw[linegray] (1)--(a3) (a1)--(b3) (a2)--(b4) (a2)--(b5) (a3)--(b3) (a3)--(b4) (a3)--(b5) (b1)--(c2) (b2)--(c3) (b3)--(c2) (b3)--(c3) (b4)--(c2) (b4)--(c4) (b5)--(c3) (b5)--(c4);}  \right)  +  \mathrm{HS} \left( \smallhasse{
    \node[hasse] (1) at (0,0) {};
    \node[hasse] (a1) at (-1,1) {};
    \node[hasse] (a2) at (0,1) {};
    \node[hasse] (a3) at (1,1) {};
    \node[hasse] (b1) at (-2,2) {};
    \node[hassegray] (b2) at (-1,2) {};
    \node[hasse] (b3) at (0,2) {};
    \node[hasse] (b4) at (1,2) {};
    \node[hassegray] (b5) at (2,2) {};
    \node[hassegray] (c1) at (-1.5,3) {};
    \node[hasse] (c2) at (-0.5,3) {};
    \node[hassegray] (c3) at (0.5,3) {};
    \node[hassegray] (c4) at (1.5,3) {};
    \draw  (1)--(a1) (1)--(a2) (1)--(a3) (a1)--(b1) (a1)--(b3) (a2)--(b1) (a2)--(b4) (a3)--(b3) (a3)--(b4) (b1)--(c2) (b3)--(c2) (b4)--(c2);
    \draw[linegray]  (a2)--(b2) (a1)--(b2) (a2)--(b5) (a3)--(b5) (b1)--(c1) (b2)--(c1) (b2)--(c3) (b3)--(c3) (b5)--(c3) (b5)--(c4) (b4)--(c4);}  \right)   +  \mathrm{HS} \left( \smallhasse{
    \node[hasse] (1) at (0,0) {};
    \node[hasse] (a1) at (-1,1) {};
    \node[hasse] (a2) at (0,1) {};
    \node[hasse] (a3) at (1,1) {};
    \node[hassegray] (b1) at (-2,2) {};
    \node[hasse] (b2) at (-1,2) {};
    \node[hasse] (b3) at (0,2) {};
    \node[hassegray] (b4) at (1,2) {};
    \node[hasse] (b5) at (2,2) {};
    \node[hassegray] (c1) at (-1.5,3) {};
    \node[hassegray] (c2) at (-0.5,3) {};
    \node[hasse] (c3) at (0.5,3) {};
    \node[hassegray] (c4) at (1.5,3) {};
    \draw  (1)--(a1) (1)--(a2) (1)--(a3) (a1)--(b2) (a1)--(b3) (a2)--(b2) (a2)--(b5) (a3)--(b3) (a3)--(b5) (b2)--(c3) (b3)--(c3) (b5)--(c3);
    \draw[linegray] (a2)--(b1) (a1)--(b1) (a2)--(b4) (a3)--(b4) (b1)--(c1) (b1)--(c2) (b2)--(c1) (b3)--(c2) (b4)--(c2) (b4)--(c4) (b5)--(c4);}  \right) + \mathrm{HS} \left( \smallhasse{
    \node[hasse] (1) at (0,0) {};
    \node[hassegray] (a1) at (-1,1) {};
    \node[hasse] (a2) at (0,1) {};
    \node[hasse] (a3) at (1,1) {};
    \node[hassegray] (b1) at (-2,2) {};
    \node[hassegray] (b2) at (-1,2) {};
    \node[hassegray] (b3) at (0,2) {};
    \node[hasse] (b4) at (1,2) {};
    \node[hasse] (b5) at (2,2) {};
    \node[hassegray] (c1) at (-1.5,3) {};
    \node[hassegray] (c2) at (-0.5,3) {};
    \node[hassegray] (c3) at (0.5,3) {};
    \node[hasse] (c4) at (1.5,3) {};
    \draw  (1)--(a3) (1)--(a2) (a3)--(b4) (a3)--(b5) (a2)--(b4) (a2)--(b5) (b4)--(c4) (b5)--(c4);
    \draw[linegray] (1)--(a1) (a1)--(b3) (a2)--(b1) (a2)--(b2) (a1)--(b3) (a1)--(b1) (a1)--(b2) (b1)--(c1) (b1)--(c2) (b2)--(c3) (b3)--(c2) (b3)--(c3) (b1)--(c2) (b2)--(c4) (b2)--(c1) (b2)--(c1);}  \right) \nonumber \\ 
    & & -  \mathrm{HS} \left( \smallhasse{
    \node[hasse] (1) at (0,0) {};
    \node[hasse] (a1) at (-1,1) {};
    \node[hasse] (a2) at (0,1) {};
    \node[hassegray] (a3) at (1,1) {};
    \node[hasse] (b1) at (-2,2) {};
    \node[hassegray] (b2) at (-1,2) {};
    \node[hassegray] (b3) at (0,2) {};
    \node[hassegray] (b4) at (1,2) {};
    \node[hassegray] (b5) at (2,2) {};
    \node[hassegray] (c1) at (-1.5,3) {};
    \node[hassegray] (c2) at (-0.5,3) {};
    \node[hassegray] (c3) at (0.5,3) {};
    \node[hassegray] (c4) at (1.5,3) {};
    \draw[black] (1)--(a1)--(b1)--(a2)--(1); 
    \draw[linegray] (1)--(a3) (a1)--(b3) (a2)--(b4) (a2)--(b5) (a3)--(b3) (a3)--(b4) (a3)--(b5) (b1)--(c2) (b2)--(c3) (b3)--(c2) (b3)--(c3) (b4)--(c2) (b4)--(c4) (b5)--(c3) (b5)--(c4) (a1)--(b2)  (a2)--(b2) (b1)--(c1) (b2)--(c1);} \right)     -  \mathrm{HS} \left( \smallhasse{
    \node[hasse] (1) at (0,0) {};
    \node[hasse] (a1) at (-1,1) {};
    \node[hasse] (a2) at (0,1) {};
    \node[hassegray] (a3) at (1,1) {};
    \node[hassegray] (b1) at (-2,2) {};
    \node[hasse] (b2) at (-1,2) {};
    \node[hassegray] (b3) at (0,2) {};
    \node[hassegray] (b4) at (1,2) {};
    \node[hassegray] (b5) at (2,2) {};
    \node[hassegray] (c1) at (-1.5,3) {};
    \node[hassegray] (c2) at (-0.5,3) {};
    \node[hassegray] (c3) at (0.5,3) {};
    \node[hassegray] (c4) at (1.5,3) {};
    \draw (1)--(a1)--(b2)--(a2)--(1);
    \draw[linegray] (1)--(a3) (a1)--(b3) (a2)--(b4) (a2)--(b5) (a3)--(b3) (a3)--(b4) (a3)--(b5) (b1)--(c2) (b2)--(c3) (b3)--(c2) (b3)--(c3) (b4)--(c2) (b4)--(c4) (b5)--(c3) (b5)--(c4) (a1)--(b1)  (a2)--(b1) (b1)--(c1) (b2)--(c1);}
\right) -    \mathrm{HS} \left( \smallhasse{
    \node[hasse] (1) at (0,0) {};
    \node[hasse] (a1) at (-1,1) {};
    \node[hassegray] (a2) at (0,1) {};
    \node[hasse] (a3) at (1,1) {};
    \node[hassegray] (b1) at (-2,2) {};
    \node[hassegray] (b2) at (-1,2) {};
    \node[hasse] (b3) at (0,2) {};
    \node[hassegray] (b4) at (1,2) {};
    \node[hassegray] (b5) at (2,2) {};
    \node[hassegray] (c1) at (-1.5,3) {};
    \node[hassegray] (c2) at (-0.5,3) {};
    \node[hassegray] (c3) at (0.5,3) {};
    \node[hassegray] (c4) at (1.5,3) {};
    \draw (1)--(a1)--(b3)--(a3)--(1);
    \draw[linegray]  (1)--(a2) (a2)--(b4) (a2)--(b5) (a3)--(b4) (a3)--(b5) (b1)--(c2) (b2)--(c3) (b3)--(c2) (b3)--(c3) (b4)--(c2) (b4)--(c4) (b5)--(c3) (b5)--(c4) (a1)--(b1) (a1)--(b2) (a2)--(b1) (a2)--(b2) (b1)--(c1) (b2)--(c1);} \right)  -  \mathrm{HS} \left( \smallhasse{
    \node[hasse] (1) at (0,0) {};
    \node[hassegray] (a1) at (-1,1) {};
    \node[hasse] (a2) at (0,1) {};
    \node[hasse] (a3) at (1,1) {};
    \node[hassegray] (b1) at (-2,2) {};
    \node[hassegray] (b2) at (-1,2) {};
    \node[hassegray] (b3) at (0,2) {};
    \node[hasse] (b4) at (1,2) {};
    \node[hassegray] (b5) at (2,2) {};
    \node[hassegray] (c1) at (-1.5,3) {};
    \node[hassegray] (c2) at (-0.5,3) {};
    \node[hassegray] (c3) at (0.5,3) {};
    \node[hassegray] (c4) at (1.5,3) {};
    \draw (1)--(a2)--(b4)--(a3)--(1);
    \draw[linegray] (a1)--(b3) (a2)--(b5) (a3)--(b3) (a3)--(b5) (b1)--(c2) (b2)--(c3) (b3)--(c2) (b3)--(c3) (b4)--(c2) (b4)--(c4) (b5)--(c3) (b5)--(c4) (1)--(a1) (a1)--(b1) (a1)--(b2) (a2)--(b1) (a2)--(b2) (b1)--(c1) (b2)--(c1);} \right)  -  \mathrm{HS} \left( \smallhasse{
    \node[hasse] (1) at (0,0) {};
    \node[hassegray] (a1) at (-1,1) {};
    \node[hasse] (a2) at (0,1) {};
    \node[hasse] (a3) at (1,1) {};
    \node[hassegray] (b1) at (-2,2) {};
    \node[hassegray] (b2) at (-1,2) {};
    \node[hassegray] (b3) at (0,2) {};
    \node[hassegray] (b4) at (1,2) {};
    \node[hasse] (b5) at (2,2) {};
    \node[hassegray] (c1) at (-1.5,3) {};
    \node[hassegray] (c2) at (-0.5,3) {};
    \node[hassegray] (c3) at (0.5,3) {};
    \node[hassegray] (c4) at (1.5,3) {};
    \draw[black] (1)--(a2)--(b5)--(a3)--(1); 
    \draw[linegray] (a1)--(b3) (a2)--(b4) (a3)--(b3) (a3)--(b4) (b1)--(c2) (b2)--(c3) (b3)--(c2) (b3)--(c3) (b4)--(c2) (b4)--(c4) (b5)--(c3) (b5)--(c4) (1)--(a1) (a1)--(b1) (a1)--(b2) (a2)--(b1) (a2)--(b2) (b1)--(c1) (b2)--(c1);}
\right)  \nonumber  \\ & &  +   \mathrm{HS} \left( \smallhasse{
    \node[hasse] (1) at (0,0) {};
    \node[hasse] (a1) at (-1,1) {};
    \node[hassegray] (a2) at (0,1) {};
    \node[hassegray] (a3) at (1,1) {};
    \node[hassegray] (b1) at (-2,2) {};
    \node[hassegray] (b2) at (-1,2) {};
    \node[hassegray] (b3) at (0,2) {};
    \node[hassegray] (b4) at (1,2) {};
    \node[hassegray] (b5) at (2,2) {};
    \node[hassegray] (c1) at (-1.5,3) {};
    \node[hassegray] (c2) at (-0.5,3) {};
    \node[hassegray] (c3) at (0.5,3) {};
    \node[hassegray] (c4) at (1.5,3) {};
    \draw (a1)--(1);
    \draw[linegray] (1)--(a2) (1)--(a3) (a1)--(b3) (a2)--(b4) (a2)--(b5) (a3)--(b3) (a3)--(b4) (a3)--(b5) (b1)--(c2) (b2)--(c3) (b3)--(c2) (b3)--(c3) (b4)--(c2) (b4)--(c4) (b5)--(c3) (b5)--(c4) (a1)--(b1) (a1)--(b2) (a2)--(b1) (a2)--(b2) (b1)--(c1) (b2)--(c1);} \right) + \mathrm{HS} \left( \smallhasse{
    \node[hasse] (1) at (0,0) {};
    \node[hassegray] (a1) at (-1,1) {};
    \node[hasse] (a2) at (0,1) {};
    \node[hassegray] (a3) at (1,1) {};
    \node[hassegray] (b1) at (-2,2) {};
    \node[hassegray] (b2) at (-1,2) {};
    \node[hassegray] (b3) at (0,2) {};
    \node[hassegray] (b4) at (1,2) {};
    \node[hassegray] (b5) at (2,2) {};
    \node[hassegray] (c1) at (-1.5,3) {};
    \node[hassegray] (c2) at (-0.5,3) {};
    \node[hassegray] (c3) at (0.5,3) {};
    \node[hassegray] (c4) at (1.5,3) {};
    \draw (1)--(a2);
    \draw[linegray] (1)--(a3) (a1)--(b3) (a2)--(b4) (a2)--(b5) (a3)--(b3) (a3)--(b4) (a3)--(b5) (b1)--(c2) (b2)--(c3) (b3)--(c2) (b3)--(c3) (b4)--(c2) (b4)--(c4) (b5)--(c3) (b5)--(c4) (1)--(a1) (a1)--(b1) (a1)--(b2) (a2)--(b1) (a2)--(b2) (b1)--(c1) (b2)--(c1);} \right)   +   \mathrm{HS} \left( \smallhasse{
    \node[hasse] (1) at (0,0) {};
    \node[hassegray] (a1) at (-1,1) {};
    \node[hassegray] (a2) at (0,1) {};
    \node[hasse] (a3) at (1,1) {};
    \node[hassegray] (b1) at (-2,2) {};
    \node[hassegray] (b2) at (-1,2) {};
    \node[hassegray] (b3) at (0,2) {};
    \node[hassegray] (b4) at (1,2) {};
    \node[hassegray] (b5) at (2,2) {};
    \node[hassegray] (c1) at (-1.5,3) {};
    \node[hassegray] (c2) at (-0.5,3) {};
    \node[hassegray] (c3) at (0.5,3) {};
    \node[hassegray] (c4) at (1.5,3) {};
    \draw (a3)--(1);
    \draw[linegray] (1)--(a2) (1)--(a1) (a1)--(b3) (a2)--(b4) (a2)--(b5) (a3)--(b3) (a3)--(b4) (a3)--(b5) (b1)--(c2) (b2)--(c3) (b3)--(c2) (b3)--(c3) (b4)--(c2) (b4)--(c4) (b5)--(c3) (b5)--(c4) (a1)--(b1) (a1)--(b2) (a2)--(b1) (a2)--(b2) (b1)--(c1) (b2)--(c1);} \right)   \\ & & -  \mathrm{HS} \left( \smallhasse{
    \node[hasse] (1) at (0,0) {};
    \node[hassegray] (a1) at (-1,1) {};
    \node[hassegray] (a2) at (0,1) {};
    \node[hassegray] (a3) at (1,1) {};
    \node[hassegray] (b1) at (-2,2) {};
    \node[hassegray] (b2) at (-1,2) {};
    \node[hassegray] (b3) at (0,2) {};
    \node[hassegray] (b4) at (1,2) {};
    \node[hassegray] (b5) at (2,2) {};
    \node[hassegray] (c1) at (-1.5,3) {};
    \node[hassegray] (c2) at (-0.5,3) {};
    \node[hassegray] (c3) at (0.5,3) {};
    \node[hassegray] (c4) at (1.5,3) {};
    \draw[linegray] (1)--(a2) (1)--(a3) (a1)--(b3) (a2)--(b4) (a2)--(b5) (a3)--(b3) (a3)--(b4) (a3)--(b5) (b1)--(c2) (b2)--(c3) (b3)--(c2) (b3)--(c3) (b4)--(c2) (b4)--(c4) (b5)--(c3) (b5)--(c4) (1)--(a1) (a1)--(b1) (a1)--(b2) (a2)--(b1) (a2)--(b2) (b1)--(c1) (b2)--(c1);} \right)  \nonumber 
\end{eqnarray}

We check explicitly our prediction \eqref{eq:resultCrown} in the case $n=2$. In this case the quiver \eqref{eq:stemQuiver} is trivial, so the Hasse diagram reduces to the top part of \eqref{eq:diagramCrown}. The Hilbert series for the intersections are computed using the monopole formula, and the result is shown in Table \ref{tab:HS}. Plugging in the values into \eqref{eq:resultCrown}, one finds 
\begin{equation}
\frac{1 + 6 t^2 + 35 t^4 + 77 t^6 + 101 t^8 + 19 t^{10} - 31 t^{12} - 5 t^{14} + 
 6 t^{16} - t^{18}}{(1 - t^2)^3  (1 - t^4)^3}    \, . 
\end{equation}
This is the correct value for the Higgs branch Hilbert series of the electric quiver, computed via the hyper-K\"ahler quotient. 
Similarly, for $p=2$ and $n=3$ the values are also recorded in Table \ref{tab:HS}.  The resulting Hilbert series is 
  \begin{equation}
    \frac{ \left( \begin{array}{c}
1+12 t^2+99 t^4+619 t^6+2975 t^8+11260 t^{10}+33860 t^{12} \\ +81771 t^{14} +159921 t^{16}   +254762 t^{18}+331011 t^{20}+349336 t^{22}+295316 t^{24}\\ +193898
   t^{26}+91901 t^{28}+24715 t^{30}-2435 t^{32}   -5540 t^{34}-2080 t^{36}\\ -t^{38}+247 t^{40}+56 t^{42}-15 t^{44}-8 t^{46}-t^{48}   
\end{array}    \right) }{
(1-t^2)^4 (1-t^4)^6 (1-t^6)^4
   } \, . 
   \end{equation}
Again, this is in agreement with a direct computation from the quiver \eqref{eq:quiverSUn}. 

This implies in particular that for these Higgs branches, the Higgs variety and the Higgs scheme coincide, i.e. there are no nilpotent operators. One can conjecture this remains true for $p=2$ and higher values of $n$. However, for $p>2$, nilpotent operators do show up, as shown in the last paragraph of Appendix \ref{app:eightcones}. Our techniques, based on the knowledge of the Hilbert series (but not the geometry) of the Higgs scheme on the one hand, and of the geometry of the Higgs variety from magnetic quivers on the other hand, do not allow us to determine unambiguously the structure of the nilpotent operators.

\afterpage{
\begin{landscape}
 \begin{table}
 \centering
       \begin{tabular}{|c|c|c|}
     \hline   Higgs phase    & $n=2$ & $n=3$ \\ \hline 
         $\mathrm{HS} \left( \smallhasse{
    \node[hasse] (1) at (0,0) {};
    \node[hasse] (a1) at (-1,1) {};
    \node[hasse] (a2) at (0,1) {};
    \node[hassegray] (a3) at (1,1) {};
    \node[hasse] (b1) at (-2,2) {};
    \node[hasse] (b2) at (-1,2) {};
    \node[hassegray] (b3) at (0,2) {};
    \node[hassegray] (b4) at (1,2) {};
    \node[hassegray] (b5) at (2,2) {};
    \node[hasse] (c1) at (-1.5,3) {};
    \node[hassegray] (c2) at (-0.5,3) {};
    \node[hassegray] (c3) at (0.5,3) {};
    \node[hassegray] (c4) at (1.5,3) {};
    \draw  (1)--(a1) (1)--(a2) (a1)--(b1) (a1)--(b2) (a2)--(b1) (a2)--(b2) (b1)--(c1) (b2)--(c1);
    \draw[linegray] (1)--(a3) (a1)--(b3) (a2)--(b4) (a2)--(b5) (a3)--(b3) (a3)--(b4) (a3)--(b5) (b1)--(c2) (b2)--(c3) (b3)--(c2) (b3)--(c3) (b4)--(c2) (b4)--(c4) (b5)--(c3) (b5)--(c4);}  \right) = \mathrm{HS} \left( \smallhasse{
    \node[hasse] (1) at (0,0) {};
    \node[hassegray] (a1) at (-1,1) {};
    \node[hasse] (a2) at (0,1) {};
    \node[hasse] (a3) at (1,1) {};
    \node[hassegray] (b1) at (-2,2) {};
    \node[hassegray] (b2) at (-1,2) {};
    \node[hassegray] (b3) at (0,2) {};
    \node[hasse] (b4) at (1,2) {};
    \node[hasse] (b5) at (2,2) {};
    \node[hassegray] (c1) at (-1.5,3) {};
    \node[hassegray] (c2) at (-0.5,3) {};
    \node[hassegray] (c3) at (0.5,3) {};
    \node[hasse] (c4) at (1.5,3) {};
    \draw  (1)--(a3) (1)--(a2) (a3)--(b4) (a3)--(b5) (a2)--(b4) (a2)--(b5) (b4)--(c4) (b5)--(c4);
    \draw[linegray] (1)--(a1) (a1)--(b3) (a2)--(b1) (a2)--(b2) (a1)--(b3) (a1)--(b1) (a1)--(b2) (b1)--(c1) (b1)--(c2) (b2)--(c3) (b3)--(c2) (b3)--(c3) (b1)--(c2) (b2)--(c4) (b2)--(c1) (b2)--(c1);}  \right) $   &  $\frac{1 + 3 t^2 + 11 t^4 + 10 t^6 + 11 t^8 + 3 t^{10} + t^{12}}{(1 - t^2)^3 (1 - t^4)^3}$  &  
    $\frac{1 + 5 t^2 + 23 t^4 + 62 t^6 + 110 t^8 + 130 t^{10} +\pal + t^{20}}{(1-t^2)^{11} (1 - t^4)^3}$ \\
    $ \mathrm{HS} \left( \smallhasse{
    \node[hasse] (1) at (0,0) {};
    \node[hasse] (a1) at (-1,1) {};
    \node[hasse] (a2) at (0,1) {};
    \node[hasse] (a3) at (1,1) {};
    \node[hasse] (b1) at (-2,2) {};
    \node[hassegray] (b2) at (-1,2) {};
    \node[hasse] (b3) at (0,2) {};
    \node[hasse] (b4) at (1,2) {};
    \node[hassegray] (b5) at (2,2) {};
    \node[hassegray] (c1) at (-1.5,3) {};
    \node[hasse] (c2) at (-0.5,3) {};
    \node[hassegray] (c3) at (0.5,3) {};
    \node[hassegray] (c4) at (1.5,3) {};
    \draw  (1)--(a1) (1)--(a2) (1)--(a3) (a1)--(b1) (a1)--(b3) (a2)--(b1) (a2)--(b4) (a3)--(b3) (a3)--(b4) (b1)--(c2) (b3)--(c2) (b4)--(c2);
    \draw[linegray]  (a2)--(b2) (a1)--(b2) (a2)--(b5) (a3)--(b5) (b1)--(c1) (b2)--(c1) (b2)--(c3) (b3)--(c3) (b5)--(c3) (b5)--(c4) (b4)--(c4);}  \right) =  \mathrm{HS} \left( \smallhasse{
    \node[hasse] (1) at (0,0) {};
    \node[hasse] (a1) at (-1,1) {};
    \node[hasse] (a2) at (0,1) {};
    \node[hasse] (a3) at (1,1) {};
    \node[hassegray] (b1) at (-2,2) {};
    \node[hasse] (b2) at (-1,2) {};
    \node[hasse] (b3) at (0,2) {};
    \node[hassegray] (b4) at (1,2) {};
    \node[hasse] (b5) at (2,2) {};
    \node[hassegray] (c1) at (-1.5,3) {};
    \node[hassegray] (c2) at (-0.5,3) {};
    \node[hasse] (c3) at (0.5,3) {};
    \node[hassegray] (c4) at (1.5,3) {};
    \draw  (1)--(a1) (1)--(a2) (1)--(a3) (a1)--(b2) (a1)--(b3) (a2)--(b2) (a2)--(b5) (a3)--(b3) (a3)--(b5) (b2)--(c3) (b3)--(c3) (b5)--(c3);
    \draw[linegray] (a2)--(b1) (a1)--(b1) (a2)--(b4) (a3)--(b4) (b1)--(c1) (b1)--(c2) (b2)--(c1) (b3)--(c2) (b4)--(c2) (b4)--(c4) (b5)--(c4);}  \right) $ &  $\left( \frac{1+t^2}{(1-t^2)^2} \right)^3$  &  
    $ \frac{\scriptsize \left(
    \begin{array}{c} 1+11 t^2+88 t^4+460 t^6+1742 t^8+5044 t^{10}+11598 t^{12}+21699 t^{14}\\+33621 t^{16}+43544 t^{18}+47436 t^{20}+\pal +t^{40}\end{array}
    \right)}
    {(1-t^2)^{5} (1-t^4)^5  (1- t^6)^4} $ \\    
 \begin{tabular}{c}
 $  \mathrm{HS} \left( \smallhasse{
    \node[hasse] (1) at (0,0) {};
    \node[hasse] (a1) at (-1,1) {};
    \node[hasse] (a2) at (0,1) {};
    \node[hassegray] (a3) at (1,1) {};
    \node[hasse] (b1) at (-2,2) {};
    \node[hassegray] (b2) at (-1,2) {};
    \node[hassegray] (b3) at (0,2) {};
    \node[hassegray] (b4) at (1,2) {};
    \node[hassegray] (b5) at (2,2) {};
    \node[hassegray] (c1) at (-1.5,3) {};
    \node[hassegray] (c2) at (-0.5,3) {};
    \node[hassegray] (c3) at (0.5,3) {};
    \node[hassegray] (c4) at (1.5,3) {};
    \draw[black] (1)--(a1)--(b1)--(a2)--(1); 
    \draw[linegray] (1)--(a3) (a1)--(b3) (a2)--(b4) (a2)--(b5) (a3)--(b3) (a3)--(b4) (a3)--(b5) (b1)--(c2) (b2)--(c3) (b3)--(c2) (b3)--(c3) (b4)--(c2) (b4)--(c4) (b5)--(c3) (b5)--(c4) (a1)--(b2)  (a2)--(b2) (b1)--(c1) (b2)--(c1);} \right)  =  \mathrm{HS} \left( \smallhasse{
    \node[hasse] (1) at (0,0) {};
    \node[hassegray] (a1) at (-1,1) {};
    \node[hasse] (a2) at (0,1) {};
    \node[hasse] (a3) at (1,1) {};
    \node[hassegray] (b1) at (-2,2) {};
    \node[hassegray] (b2) at (-1,2) {};
    \node[hassegray] (b3) at (0,2) {};
    \node[hassegray] (b4) at (1,2) {};
    \node[hasse] (b5) at (2,2) {};
    \node[hassegray] (c1) at (-1.5,3) {};
    \node[hassegray] (c2) at (-0.5,3) {};
    \node[hassegray] (c3) at (0.5,3) {};
    \node[hassegray] (c4) at (1.5,3) {};
    \draw[black] (1)--(a2)--(b5)--(a3)--(1); 
    \draw[linegray] (a1)--(b3) (a2)--(b4) (a3)--(b3) (a3)--(b4) (b1)--(c2) (b2)--(c3) (b3)--(c2) (b3)--(c3) (b4)--(c2) (b4)--(c4) (b5)--(c3) (b5)--(c4) (1)--(a1) (a1)--(b1) (a1)--(b2) (a2)--(b1) (a2)--(b2) (b1)--(c1) (b2)--(c1);}
\right)  $ \\ $ =  \mathrm{HS} \left( \smallhasse{
    \node[hasse] (1) at (0,0) {};
    \node[hasse] (a1) at (-1,1) {};
    \node[hasse] (a2) at (0,1) {};
    \node[hassegray] (a3) at (1,1) {};
    \node[hassegray] (b1) at (-2,2) {};
    \node[hasse] (b2) at (-1,2) {};
    \node[hassegray] (b3) at (0,2) {};
    \node[hassegray] (b4) at (1,2) {};
    \node[hassegray] (b5) at (2,2) {};
    \node[hassegray] (c1) at (-1.5,3) {};
    \node[hassegray] (c2) at (-0.5,3) {};
    \node[hassegray] (c3) at (0.5,3) {};
    \node[hassegray] (c4) at (1.5,3) {};
    \draw (1)--(a1)--(b2)--(a2)--(1);
    \draw[linegray] (1)--(a3) (a1)--(b3) (a2)--(b4) (a2)--(b5) (a3)--(b3) (a3)--(b4) (a3)--(b5) (b1)--(c2) (b2)--(c3) (b3)--(c2) (b3)--(c3) (b4)--(c2) (b4)--(c4) (b5)--(c3) (b5)--(c4) (a1)--(b1)  (a2)--(b1) (b1)--(c1) (b2)--(c1);}
\right) =  \mathrm{HS} \left( \smallhasse{
    \node[hasse] (1) at (0,0) {};
    \node[hassegray] (a1) at (-1,1) {};
    \node[hasse] (a2) at (0,1) {};
    \node[hasse] (a3) at (1,1) {};
    \node[hassegray] (b1) at (-2,2) {};
    \node[hassegray] (b2) at (-1,2) {};
    \node[hassegray] (b3) at (0,2) {};
    \node[hasse] (b4) at (1,2) {};
    \node[hassegray] (b5) at (2,2) {};
    \node[hassegray] (c1) at (-1.5,3) {};
    \node[hassegray] (c2) at (-0.5,3) {};
    \node[hassegray] (c3) at (0.5,3) {};
    \node[hassegray] (c4) at (1.5,3) {};
    \draw (1)--(a2)--(b4)--(a3)--(1);
    \draw[linegray] (a1)--(b3) (a2)--(b5) (a3)--(b3) (a3)--(b5) (b1)--(c2) (b2)--(c3) (b3)--(c2) (b3)--(c3) (b4)--(c2) (b4)--(c4) (b5)--(c3) (b5)--(c4) (1)--(a1) (a1)--(b1) (a1)--(b2) (a2)--(b1) (a2)--(b2) (b1)--(c1) (b2)--(c1);} \right) $
\end{tabular}    &  $\left( \frac{1+t^2}{(1-t^2)^2} \right)^2$  &  
 $  \frac{1+8 t^2+43 t^4+128 t^6+238 t^8+288 t^{10}\pal+t^{20}}{(1-t^2)^{8} (1-t^4)^4}  $\\ 
    $  \mathrm{HS} \left( \smallhasse{
    \node[hasse] (1) at (0,0) {};
    \node[hasse] (a1) at (-1,1) {};
    \node[hassegray] (a2) at (0,1) {};
    \node[hasse] (a3) at (1,1) {};
    \node[hassegray] (b1) at (-2,2) {};
    \node[hassegray] (b2) at (-1,2) {};
    \node[hasse] (b3) at (0,2) {};
    \node[hassegray] (b4) at (1,2) {};
    \node[hassegray] (b5) at (2,2) {};
    \node[hassegray] (c1) at (-1.5,3) {};
    \node[hassegray] (c2) at (-0.5,3) {};
    \node[hassegray] (c3) at (0.5,3) {};
    \node[hassegray] (c4) at (1.5,3) {};
    \draw (1)--(a1)--(b3)--(a3)--(1);
    \draw[linegray]  (1)--(a2) (a2)--(b4) (a2)--(b5) (a3)--(b4) (a3)--(b5) (b1)--(c2) (b2)--(c3) (b3)--(c2) (b3)--(c3) (b4)--(c2) (b4)--(c4) (b5)--(c3) (b5)--(c4) (a1)--(b1) (a1)--(b2) (a2)--(b1) (a2)--(b2) (b1)--(c1) (b2)--(c1);} \right)  $ &  $\left( \frac{1+t^2}{(1-t^2)^2} \right)^2$  &  
    $ \frac{1+9 t^2+54 t^4 +194 t^6 +471 t^8 +771 t^{10}+916 t^{12}+\pal +t^{24}}{(1-t^2)^{6} (1-t^4)^6} $\\ 
    $ \mathrm{HS} \left( \smallhasse{
    \node[hasse] (1) at (0,0) {};
    \node[hasse] (a1) at (-1,1) {};
    \node[hassegray] (a2) at (0,1) {};
    \node[hassegray] (a3) at (1,1) {};
    \node[hassegray] (b1) at (-2,2) {};
    \node[hassegray] (b2) at (-1,2) {};
    \node[hassegray] (b3) at (0,2) {};
    \node[hassegray] (b4) at (1,2) {};
    \node[hassegray] (b5) at (2,2) {};
    \node[hassegray] (c1) at (-1.5,3) {};
    \node[hassegray] (c2) at (-0.5,3) {};
    \node[hassegray] (c3) at (0.5,3) {};
    \node[hassegray] (c4) at (1.5,3) {};
    \draw (a1)--(1);
    \draw[linegray] (1)--(a2) (1)--(a3) (a1)--(b3) (a2)--(b4) (a2)--(b5) (a3)--(b3) (a3)--(b4) (a3)--(b5) (b1)--(c2) (b2)--(c3) (b3)--(c2) (b3)--(c3) (b4)--(c2) (b4)--(c4) (b5)--(c3) (b5)--(c4) (a1)--(b1) (a1)--(b2) (a2)--(b1) (a2)--(b2) (b1)--(c1) (b2)--(c1);} \right)  =  \mathrm{HS} \left( \smallhasse{
    \node[hasse] (1) at (0,0) {};
    \node[hassegray] (a1) at (-1,1) {};
    \node[hassegray] (a2) at (0,1) {};
    \node[hasse] (a3) at (1,1) {};
    \node[hassegray] (b1) at (-2,2) {};
    \node[hassegray] (b2) at (-1,2) {};
    \node[hassegray] (b3) at (0,2) {};
    \node[hassegray] (b4) at (1,2) {};
    \node[hassegray] (b5) at (2,2) {};
    \node[hassegray] (c1) at (-1.5,3) {};
    \node[hassegray] (c2) at (-0.5,3) {};
    \node[hassegray] (c3) at (0.5,3) {};
    \node[hassegray] (c4) at (1.5,3) {};
    \draw (a3)--(1);
    \draw[linegray] (1)--(a2) (1)--(a1) (a1)--(b3) (a2)--(b4) (a2)--(b5) (a3)--(b3) (a3)--(b4) (a3)--(b5) (b1)--(c2) (b2)--(c3) (b3)--(c2) (b3)--(c3) (b4)--(c2) (b4)--(c4) (b5)--(c3) (b5)--(c4) (a1)--(b1) (a1)--(b2) (a2)--(b1) (a2)--(b2) (b1)--(c1) (b2)--(c1);} \right)  $& $  \frac{1+t^2}{(1-t^2)^2}   $ &  
    $ \frac{(1+t^2) (1-2 t+4 t^2-2 t^3+t^4) (1+2 t+4 t^2+2 t^3+t^4)}
    {(1-t^2)^{10}} $\\
    $\mathrm{HS} \left( \smallhasse{
    \node[hasse] (1) at (0,0) {};
    \node[hassegray] (a1) at (-1,1) {};
    \node[hasse] (a2) at (0,1) {};
    \node[hassegray] (a3) at (1,1) {};
    \node[hassegray] (b1) at (-2,2) {};
    \node[hassegray] (b2) at (-1,2) {};
    \node[hassegray] (b3) at (0,2) {};
    \node[hassegray] (b4) at (1,2) {};
    \node[hassegray] (b5) at (2,2) {};
    \node[hassegray] (c1) at (-1.5,3) {};
    \node[hassegray] (c2) at (-0.5,3) {};
    \node[hassegray] (c3) at (0.5,3) {};
    \node[hassegray] (c4) at (1.5,3) {};
    \draw (1)--(a2);
    \draw[linegray] (1)--(a3) (a1)--(b3) (a2)--(b4) (a2)--(b5) (a3)--(b3) (a3)--(b4) (a3)--(b5) (b1)--(c2) (b2)--(c3) (b3)--(c2) (b3)--(c3) (b4)--(c2) (b4)--(c4) (b5)--(c3) (b5)--(c4) (1)--(a1) (a1)--(b1) (a1)--(b2) (a2)--(b1) (a2)--(b2) (b1)--(c1) (b2)--(c1);} \right) $& $  \frac{1+t^2}{(1-t^2)^2}   $ &  
    $ \frac{1+11 t^2+57 t^4+170 t^6+324 t^8+398 t^{10}+\pal+ t^{20}}
    {(1-t^2)^{5} (1-t^4)^5} $ \\ 
    $\mathrm{HS} \left( \smallhasse{
    \node[hasse] (1) at (0,0) {};
    \node[hassegray] (a1) at (-1,1) {};
    \node[hassegray] (a2) at (0,1) {};
    \node[hassegray] (a3) at (1,1) {};
    \node[hassegray] (b1) at (-2,2) {};
    \node[hassegray] (b2) at (-1,2) {};
    \node[hassegray] (b3) at (0,2) {};
    \node[hassegray] (b4) at (1,2) {};
    \node[hassegray] (b5) at (2,2) {};
    \node[hassegray] (c1) at (-1.5,3) {};
    \node[hassegray] (c2) at (-0.5,3) {};
    \node[hassegray] (c3) at (0.5,3) {};
    \node[hassegray] (c4) at (1.5,3) {};
    \draw[linegray] (1)--(a2) (1)--(a3) (a1)--(b3) (a2)--(b4) (a2)--(b5) (a3)--(b3) (a3)--(b4) (a3)--(b5) (b1)--(c2) (b2)--(c3) (b3)--(c2) (b3)--(c3) (b4)--(c2) (b4)--(c4) (b5)--(c3) (b5)--(c4) (1)--(a1) (a1)--(b1) (a1)--(b2) (a2)--(b1) (a2)--(b2) (b1)--(c1) (b2)--(c1);} \right) $ & $1$  &  
    $\frac{(1+t^2)^2 (1+5 t^2+t^4)}{(1-t^2)^8} $ \\ \hline 
    \end{tabular}
\caption{Hilbert series for the different leaf closures in the Higgs branch of \eqref{electric:suspstart} with $p=2$ and $n=2,3$. }
\label{tab:HS}
 \end{table}
\end{landscape}
}

\subsection{3d Coulomb Branch and Full Moduli Space}
Consider the simplest non-trivial example, \eqref{eq:quiverSUn} with $n=2$. Inversion gives the following
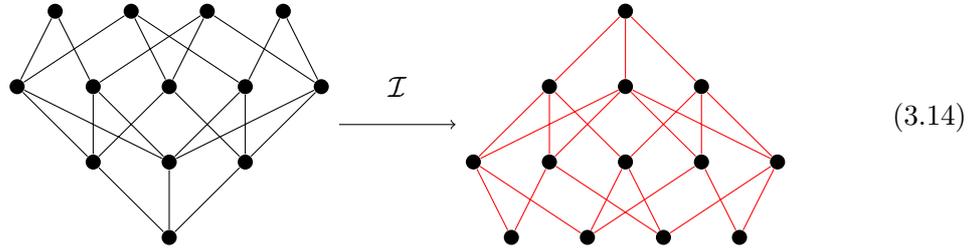
\begin{equation}
   \raisebox{-.5\height}{  \begin{tikzpicture}
        \node (cH) at (0,0) {$\begin{tikzpicture}
        \node[hasse] (1) at (0,0) {};
    \node[hasse] (a1) at (-1,1) {};
    \node[hasse] (a2) at (0,1) {};
    \node[hasse] (a3) at (1,1) {};
    \node[hasse] (b1) at (-2,2) {};
    \node[hasse] (b2) at (-1,2) {};
    \node[hasse] (b3) at (0,2) {};
    \node[hasse] (b4) at (1,2) {};
    \node[hasse] (b5) at (2,2) {};
    \node[hasse] (c1) at (-1.5,3) {};
    \node[hasse] (c2) at (-0.5,3) {};
    \node[hasse] (c3) at (0.5,3) {};
    \node[hasse] (c4) at (1.5,3) {};
    \draw (1)--(a1)--(b1)--(c1)--(b2)--(a1)--(b3)--(c2)--(b1)--(a2)--(b2)--(c3)--(b3)--(a3)--(1)--(a2)--(b4)--(c4)--(b5)--(a3)--(b4)--(c2) (c3)--(b5)--(a2);
    \end{tikzpicture}$};
    \node (qC) at (6,0) {$\begin{tikzpicture}
        \node[hasse] (1) at (0,0) {};
    \node[hasse] (a1) at (-1,-1) {};
    \node[hasse] (a2) at (0,-1) {};
    \node[hasse] (a3) at (1,-1) {};
    \node[hasse] (b1) at (-2,-2) {};
    \node[hasse] (b2) at (-1,-2) {};
    \node[hasse] (b3) at (0,-2) {};
    \node[hasse] (b4) at (1,-2) {};
    \node[hasse] (b5) at (2,-2) {};
    \node[hasse] (c1) at (-1.5,-3) {};
    \node[hasse] (c2) at (-0.5,-3) {};
    \node[hasse] (c3) at (0.5,-3) {};
    \node[hasse] (c4) at (1.5,-3) {};
    \draw[red] (1)--(a1)--(b1)--(c1)--(b2)--(a1)--(b3)--(c2)--(b1)--(a2)--(b2)--(c3)--(b3)--(a3)--(1)--(a2)--(b4)--(c4)--(b5)--(a3)--(b4)--(c2) (c3)--(b5)--(a2);
    \end{tikzpicture}$};
    \draw[->] (cH)--(qC);
    \node at (3,0.5) {$\mathcal{I}$};
    \end{tikzpicture}}
\end{equation}
where all elementary transitions are $A_1$. This indicates that there are 4 most singular points in the Coulomb branch. The physics at these 4 points, however, is not the same, as the magnetic quivers encoding the local Higgs directions have different Higgs and Coulomb branches.

\pagebreak

\section{5d Infinite Coupling}
\label{sec:5dinf}

We may interpret the gauge theories \eqref{electric:suspstart} as 5d $\mathcal{N}=1$ theories, and ask how their Higgs branches change when tuning specific couplings to infinity. Generically this is quite a complicated question, we therefore only study two examples to highlight the present phenomena. Furthermore we only study the Higgs branches as varieties, through their magnetic quivers and derived Hasse diagrams, and ignore any nilpotent operators (which arise for example classically or in the form of gaugino bilinears \cite{Cremonesi:2015lsa}).

Comparing the finite and infinite coupling, we observe three phenomena:
\begin{itemize}
    \item Cone enhancement: A cone in the classical Higgs branch grows in dimension, due to the contribution of instanton operators at infinite coupling \cite{Cremonesi:2015lsa}.
    \item Cone fusion: Two (or more) cones in the classical Higgs branch fuse into a single, bigger cone at infinite coupling. We observe this in combination with decoration.
    \item Decoration: Some cones in the infinite coupling Higgs branch have \emph{decorated} \cite{Bourget:2022ehw,Bourget:2022tmw} magnetic quivers. We observe this in combination with cone fusion.
\end{itemize}

\subsection{Tale of Two Cones Revisited -- Cone Enhancement}

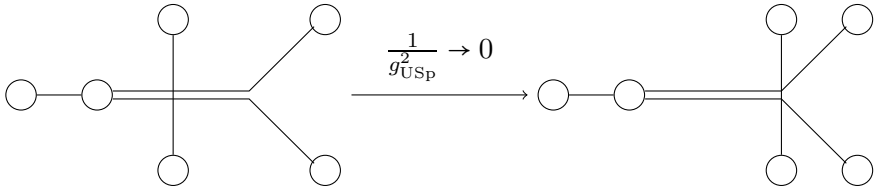
\begin{figure}
    \centering
    \begin{tikzpicture}
        \node (1) at (0,0) {$
        \begin{tikzpicture}
            \node[sev] (6) at (5,0) {};
            \node[sev] (7) at (6,0) {};
            \node[sev] (8a) at (7,1) {};
            \node[sev] (8b) at (7,-1) {};
            \node[sev] (10a) at (9,1) {};
            \node[sev] (10b) at (9,-1) {};
            \draw (6)--(7);
            \draw[transform canvas={yshift=1.5pt}] (7)--(8,0)--(10a);
            \draw[transform canvas={yshift=-1.5pt}] (7)--(8,0)--(10b);
            \draw (8a)--(8b);
        \end{tikzpicture}
        $};
        \node (2) at (7,0) {$
        \begin{tikzpicture}
            \node[sev] (6) at (5,0) {};
            \node[sev] (7) at (6,0) {};
            \node[sev] (8a) at (8,1) {};
            \node[sev] (8b) at (8,-1) {};
            \node[sev] (10a) at (9,1) {};
            \node[sev] (10b) at (9,-1) {};
            \draw (6)--(7);
            \draw[transform canvas={yshift=1.5pt}] (7)--(8,0)--(10a);
            \draw[transform canvas={yshift=-1.5pt}] (7)--(8,0)--(10b);
            \draw (8a)--(8b);
        \end{tikzpicture}
        $};
        \draw[->] (1)--(2);
        \node at (3.5,0.5) {$\frac{1}{g_{\mathrm{USp}}^2}\rightarrow0$};
    \end{tikzpicture}
    \caption{Brane webs for various choices of gauge coupling for the low energy theory \eqref{eq:5dinf2C}.}
    \label{fig:5dinf2Cones}
\end{figure}

It is again helpful to go back to the `tale of two cones'. For simplicity we consider the rank one case, i.e.\ the theory
\begin{equation}
\label{eq:5dinf2C}
     \raisebox{-.5\height}{\begin{tikzpicture}
    \node[gauge,label=below:{$\mathrm{USp}(2)$}] (2) at (2,0) {};
    \node[flavour,label=below:{$D_2$}] (3) at (4,0) {};
    \draw (2)--(3);
    \end{tikzpicture}}\;,
\end{equation}
viewed as an effective 5d $\mathcal{N}=1$ theory. There is a single coupling, $g_{\mathrm{USp}}$, which we may keep finite or take to infinity. The corresponding brane webs are depicted in Figure \ref{fig:5dinf2Cones}. When $g_{\mathrm{USp}}$ is finite, the Higgs branch is classical and as described in Section \ref{sec:2cones}. When $\frac{1}{g_{\mathrm{USp}}} \rightarrow 0$, the brane web has two maximal decompositions, shown below with the corresponding magnetic quivers:
\begin{subequations}
    \begin{equation}
        \raisebox{-.5\height}{\begin{tikzpicture}
            \node[sev] (6) at (6,0) {};
            \node[sev] (7) at (7,0) {};
            \node[sev] (9a) at (9,1) {};
            \node[sev] (9b) at (9,-1) {};
            \node[sev] (10a) at (10,1) {};
            \node[sev] (10b) at (10,-1) {};
            \draw[cyan] (6)--(7);
            \draw[cyan,transform canvas={yshift=1.5pt}] (7)--(9,0)--(10a);
            \draw[cyan,transform canvas={yshift=-2.5pt}] (7)--(9,0)--(10b);
            \draw[goodred] (9a)--(9b);
        \end{tikzpicture}}\qquad
        \raisebox{-.5\height}{ \begin{tikzpicture}
        \node[gaugec,label=left:{$1$}] (1) at (0,1) {};
        \node[gauger,label=right:{$1$}] (1u) at (1,1.5) {};
        \node (2) at (0,-1) {};
        \draw[transform canvas={yshift=-1.5pt}] (1)--(1u);
        \draw[transform canvas={yshift=1.5pt}] (1)--(1u);
        \draw[dashed] (1)--(-1,0)--(2);
    \end{tikzpicture}}
    \end{equation}
    \begin{equation}
        \raisebox{-.5\height}{  \begin{tikzpicture}
            \node[sev] (6) at (6,0) {};
            \node[sev] (7) at (7,0) {};
            \node[sev] (9a) at (9,1) {};
            \node[sev] (9b) at (9,-1) {};
            \node[sev] (10a) at (10,1) {};
            \node[sev] (10b) at (10,-1) {};
            \draw[goodmagenta] (6)--(7);
            \draw[goodgreen,transform canvas={yshift=1.5pt}] (7)--(9,0)--(10a);
            \draw[blue,transform canvas={yshift=-1.5pt}] (7)--(9,0)--(10b);
            \draw[goodgreen,transform canvas={xshift=1pt}] (9,2pt)--(9b);
            \draw[blue,transform canvas={xshift=-1pt}] (9a)--(9,-2pt);
        \end{tikzpicture}} \qquad
        \raisebox{-.5\height}{ \begin{tikzpicture}
        \node (1) at (0,1) {};
        \node[gaugem,label=left:{$1$}] (2) at (0,-1) {};
        \node[gaugeb,label=right:{$1$}] (2du) at (0.75,-1) {};
        \node[gaugeg,label=right:{$1$}] (2dd) at (1.25,-1.5) {};
        \draw (2)--(2du) (2)--(2dd) (2du)--(2dd);
        \draw[dashed] (1)--(-1,0)--(2);
    \end{tikzpicture}}
    \end{equation}
\end{subequations}
The dashed line is supposed to indicate the $D_2$ Dynkin diagram, and has no meaning in the quivers.

Comparing these magnetic quivers to \eqref{magnetic:twocones5d1} and \eqref{magnetic:twocones5d2} with $k=1$, we find that one cone is modified and changes in dimension. The Higgs branch Hasse diagram becomes
\begin{equation}
    \raisebox{-.5\height}{ \begin{tikzpicture}
    \node[hasse] (1) at (0,0) {};
    \node[hasse] (2) at (-0.5,1) {};
    \node[hasse] (3) at (0.5,1) {};
    \node at (0,2) {};
    \draw (1)--(2);
    \draw (1)--(3);
    \end{tikzpicture}}
    \raisebox{-.5\height}{ \begin{tikzpicture}
    \draw[->] (0,0)--(1,0);
    \node at (0.5,0.5) {$\frac{1}{g_{\mathrm{USp}}^2}\rightarrow0$};
    \end{tikzpicture}}
    \raisebox{-.5\height}{ \begin{tikzpicture}
    \node[hasse] (1) at (0,0) {};
    \node[hasse] (2) at (-0.5,1) {};
    \node[hasse] (3) at (0.5,2) {};
    \draw (1)--(2);
    \draw[blue] (1)--(3);
    \end{tikzpicture}}
\end{equation}
where black lines denote $A_1$ transitions and the blue line denotes an {\color{blue}$a_2$} transition. The height of a dot is proportional to the dimension of the corresponding leaf.

\subsection{Cone Fusion and Decorations}

\begin{figure}[p]
    \centering
    \begin{tikzpicture}
        \node (1) at (0,0) {$
        \begin{tikzpicture}
            \node[sev] (6) at (5,0) {};
            \node[sev] (7) at (6,0) {};
            \node[sev] (8a) at (7,1) {};
            \node[sev] (8b) at (7,-1) {};
            \node[sev] (9a) at (8,1) {};
            \node[sev] (9b) at (8,-1) {};
            \node[sev] (10a) at (10,1) {};
            \node[sev] (10b) at (10,-1) {};
            \draw[transform canvas={yshift=1.5pt}] (6)--(7);
            \draw[transform canvas={yshift=-1.5pt}] (6)--(7);
            \draw[transform canvas={yshift=3pt}] (7)--(9,0)--(10a);
            \draw[transform canvas={yshift=1pt}] (7)--(9,0)--(10a);
            \draw[transform canvas={yshift=-1pt}] (7)--(9,0)--(10b);
            \draw[transform canvas={yshift=-3pt}] (7)--(9,0)--(10b);
            \draw (8a)--(8b) (9a)--(9b);
        \end{tikzpicture}
        $};
        \node (2) at (-4,-5) {$
        \begin{tikzpicture}
            \node[sev] (6) at (5,0) {};
            \node[sev] (7) at (6,0) {};
            \node[sev] (8a) at (7.5,2) {};
            \node[sev] (8b) at (7.5,-2) {};
            \node[sev] (9a) at (7.5,1) {};
            \node[sev] (9b) at (7.5,-1) {};
            \node[sev] (10a) at (10,1) {};
            \node[sev] (10b) at (10,-1) {};
            \draw[transform canvas={yshift=1.5pt}] (6)--(7);
            \draw[transform canvas={yshift=-1.5pt}] (6)--(7);
            \draw[transform canvas={yshift=3pt}] (7)--(9,0)--(10a);
            \draw[transform canvas={yshift=1pt}] (7)--(9,0)--(10a);
            \draw[transform canvas={yshift=-1pt}] (7)--(9,0)--(10b);
            \draw[transform canvas={yshift=-3pt}] (7)--(9,0)--(10b);
            \draw[transform canvas={xshift=1.5pt}] (9a)--(9b);
            \draw[transform canvas={xshift=-1.5pt}] (9a)--(9b);
            \draw (8a)--(9a) (9b)--(8b);
        \end{tikzpicture}
        $};
        \node (3) at (4,-5) {$
        \begin{tikzpicture}
            \node[sev] (6) at (5,0) {};
            \node[sev] (7) at (6,0) {};
            \node[sev] (8a) at (7.5,1) {};
            \node[sev] (8b) at (7.5,-1) {};
            \node[sev] (9a) at (9,1) {};
            \node[sev] (9b) at (9,-1) {};
            \node[sev] (10a) at (10,1) {};
            \node[sev] (10b) at (10,-1) {};
            \draw[transform canvas={yshift=1.5pt}] (6)--(7);
            \draw[transform canvas={yshift=-1.5pt}] (6)--(7);
            \draw[transform canvas={yshift=3pt}] (7)--(9,0)--(10a);
            \draw[transform canvas={yshift=1pt}] (7)--(9,0)--(10a);
            \draw[transform canvas={yshift=-1pt}] (7)--(9,0)--(10b);
            \draw[transform canvas={yshift=-3pt}] (7)--(9,0)--(10b);
            \draw (9a)--(9b);
            \draw (8a)--(8b);
        \end{tikzpicture}
        $};
        \node (4) at (0,-10) {$
        \begin{tikzpicture}
            \node[sev] (6) at (5,0) {};
            \node[sev] (7) at (6,0) {};
            \node[sev] (8a) at (9,2) {};
            \node[sev] (8b) at (9,-2) {};
            \node[sev] (9a) at (9,1) {};
            \node[sev] (9b) at (9,-1) {};
            \node[sev] (10a) at (10,1) {};
            \node[sev] (10b) at (10,-1) {};
            \draw[transform canvas={yshift=1.5pt}] (6)--(7);
            \draw[transform canvas={yshift=-1.5pt}] (6)--(7);
            \draw[transform canvas={yshift=3pt}] (7)--(9,0)--(10a);
            \draw[transform canvas={yshift=1pt}] (7)--(9,0)--(10a);
            \draw[transform canvas={yshift=-1pt}] (7)--(9,0)--(10b);
            \draw[transform canvas={yshift=-3pt}] (7)--(9,0)--(10b);
            \draw[transform canvas={xshift=1.5pt}] (9a)--(9b);
            \draw[transform canvas={xshift=-1.5pt}] (9a)--(9b);
            \draw (8a)--(9a) (9b)--(8b);
        \end{tikzpicture}
        $};
        \draw[->] (1)--(2);
        \draw[->] (1)--(3);
        \draw[->] (2)--(4);
        \draw[->] (3)--(4);
        \node at (-2.5,-1.75) {$\frac{1}{g_{\mathrm{SU}}^2}\rightarrow0$};
        \node at (2.75,-2) {$\frac{1}{g_{\mathrm{USp}}^2}\rightarrow0$};
        \node at (3,-7.5) {$\frac{1}{g_{\mathrm{SU}}^2}\rightarrow0$};
        \node at (-2.75,-7.75) {$\frac{1}{g_{\mathrm{USp}}^2}\rightarrow0$};
    \end{tikzpicture}
    \caption{Brane webs for various choices of gauge coupling for the low energy theory \eqref{eq:5dInfQuiv}. In the top line both couplings are finite, in the middle line only one of the couplings is infinite, at the bottom both couplings are infinite.}
    \label{fig:5dinfBW}
\end{figure}
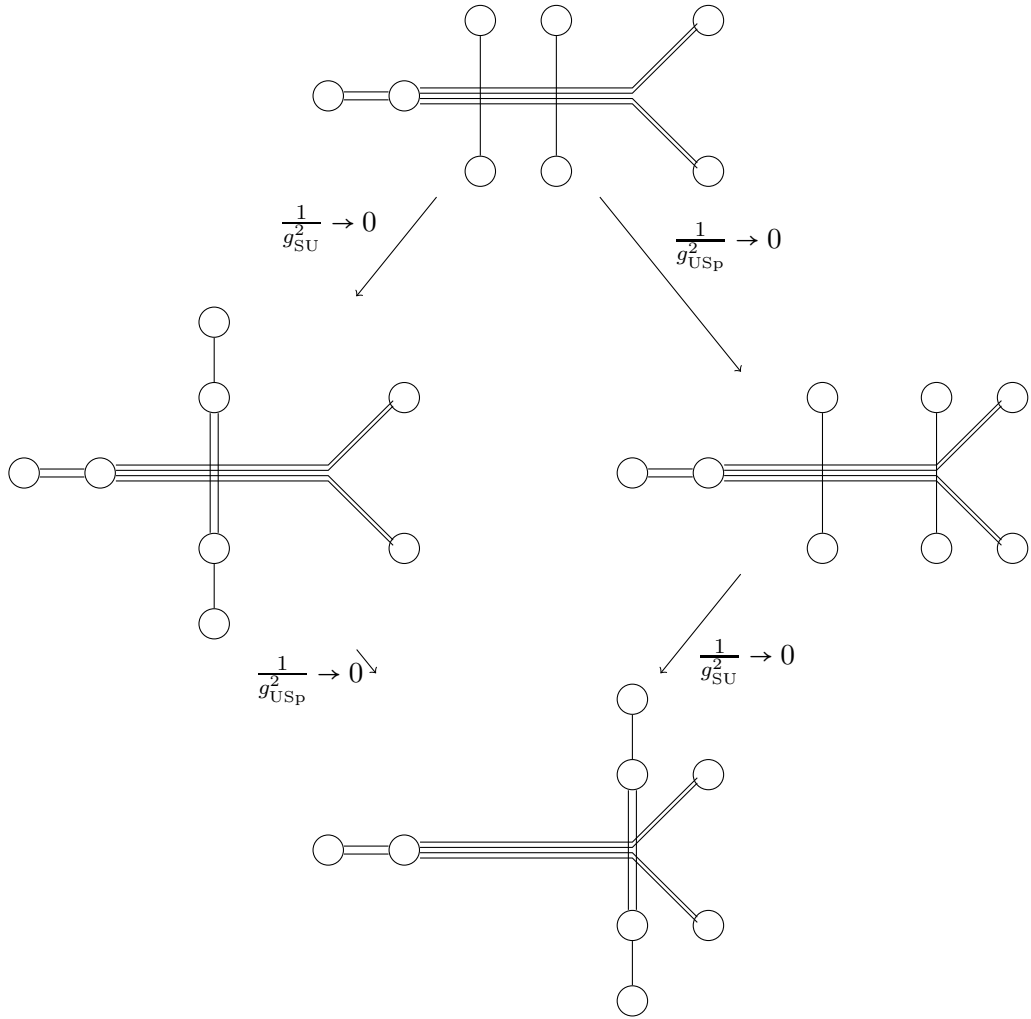

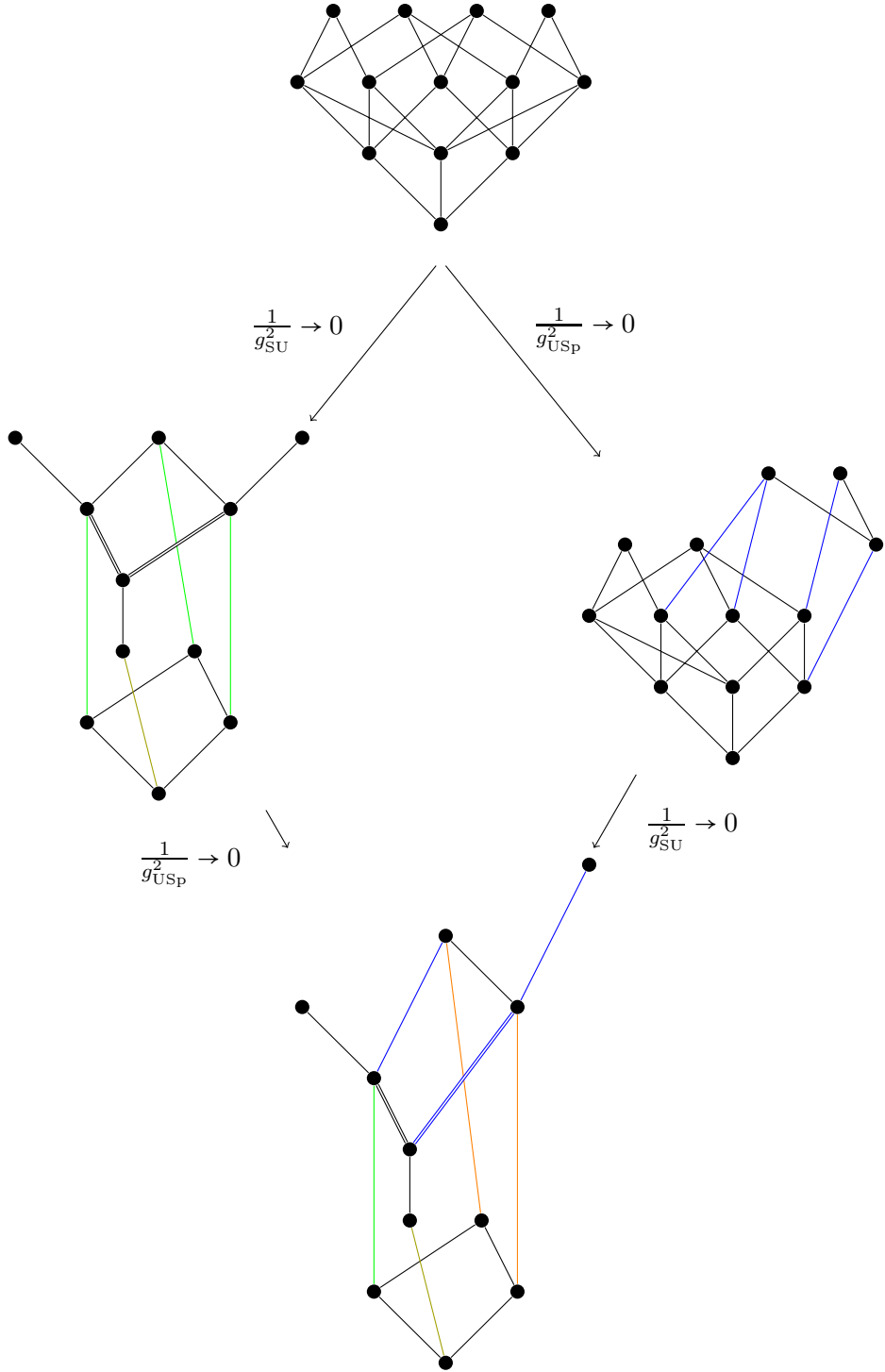
\begin{figure}[p]
    \centering
    \begin{tikzpicture}
        \node (1) at (0,2) {$
       \begin{tikzpicture}
        \node[hasse] (1) at (0,0) {};
    \node[hasse] (a1) at (-1,1) {};
    \node[hasse] (a2) at (0,1) {};
    \node[hasse] (a3) at (1,1) {};
    \node[hasse] (b1) at (-2,2) {};
    \node[hasse] (b2) at (-1,2) {};
    \node[hasse] (b3) at (0,2) {};
    \node[hasse] (b4) at (1,2) {};
    \node[hasse] (b5) at (2,2) {};
    \node[hasse] (c1) at (-1.5,3) {};
    \node[hasse] (c2) at (-0.5,3) {};
    \node[hasse] (c3) at (0.5,3) {};
    \node[hasse] (c4) at (1.5,3) {};
    \draw (1)--(a1)--(b1)--(c1)--(b2)--(a1)--(b3)--(c2)--(b1)--(a2)--(b2)--(c3)--(b3)--(a3)--(1)--(a2)--(b4)--(c4)--(b5)--(a3)--(b4)--(c2) (c3)--(b5)--(a2);
    \end{tikzpicture}
        $};
        \node (2) at (-4,-5) {$
        \raisebox{-.5\height}{ \begin{tikzpicture}
        \node[hasse] (a) at (0,0) {};
        \node[hasse] (b1) at (-1,1) {};
        \node[hasse] (b2) at (1,1) {};
        \node[hasse] (cd) at (-0.5,2) {};
        \node[hasse] (c) at (0.5,2) {};
        \node[hasse] (d) at (-0.5,3) {};
        \node[hasse] (e1) at (-1,4) {};
        \node[hasse] (e2) at (1,4) {};
        \node[hasse] (f1) at (-2,5) {};
        \node[hasse] (f0) at (0,5) {};
        \node[hasse] (f2) at (2,5) {};
        \draw (f1)--(e1)--(f0)--(e2)--(f2)
        (c)--(b1)--(a)--(b2)--(c);
        \draw[green] (e1)--(b1) (e2)--(b2) (f0)--(c);
        \draw[double] (e1)--(d)--(e2);
        \draw (d)--(cd);
        \draw[goodyellow] (cd)--(a);
    \end{tikzpicture}}
        $};
        \node (3) at (4,-5) {$
       \raisebox{-.5\height}{ \begin{tikzpicture}
    \node[hasse] (1) at (0,0) {};
    \node[hasse] (a1) at (-1,1) {};
    \node[hasse] (a2) at (0,1) {};
    \node[hasse] (a3) at (1,1) {};
    \node[hasse] (b1) at (-2,2) {};
    \node[hasse] (b2) at (-1,2) {};
    \node[hasse] (b3) at (0,2) {};
    \node[hasse] (b4) at (1,2) {};
    \node[hasse] (b5) at (2,3) {};
    \node[hasse] (c1) at (-1.5,3) {};
    \node[hasse] (c2) at (-0.5,3) {};
    \node[hasse] (c3) at (0.5,4) {};
    \node[hasse] (c4) at (1.5,4) {};
    \draw (1)--(a1)--(b1)--(c1)--(b2)--(a1)--(b3)--(c2)--(b1)--(a2)--(b2) (b3)--(a3)--(1)--(a2)--(b4) (c4)--(b5) (a3)--(b4)--(c2) (c3)--(b5);
    \draw[blue] (b2)--(c3)--(b3) (b4)--(c4) (b5)--(a3);
    \end{tikzpicture}}
        $};
        \node (4) at (0,-12) {$ \raisebox{-.5\height}{  \begin{tikzpicture}
        \node[hasse] (a) at (0,0) {};
        \node[hasse] (b1) at (-1,1) {};
        \node[hasse] (b2) at (1,1) {};
        \node[hasse] (cd) at (-0.5,2) {};
        \node[hasse] (c) at (0.5,2) {};
        \node[hasse] (d) at (-0.5,3) {};
        \node[hasse] (e1) at (-1,4) {};
        \node[hasse] (e2) at (1,5) {};
        \node[hasse] (f1) at (-2,5) {};
        \node[hasse] (f0) at (0,6) {};
        \node[hasse] (f2) at (2,7) {};
        \draw (f1)--(e1) (f0)--(e2)
        (c)--(b1)--(a)--(b2)--(c);
        \draw[blue] (e1)--(f0) (e2)--(f2);
        \draw[green] (e1)--(b1);
        \draw[orange] (e2)--(b2) (f0)--(c);
        \draw[double] (e1)--(d);
        \draw[double,blue] (d)--(e2);
        \draw (d)--(cd);
        \draw[goodyellow] (cd)--(a);
    \end{tikzpicture}}
        $};
        \draw[->] (1)--(2);
        \draw[->] (1)--(3);
        \draw[->] (2)--(4);
        \draw[->] (3)--(4);
        \node at (-2.0,-1) {$\frac{1}{g_{\mathrm{SU}}^2}\rightarrow0$};
        \node at (2,-1) {$\frac{1}{g_{\mathrm{USp}}^2}\rightarrow0$};
        \node at (3.5,-8.0) {$\frac{1}{g_{\mathrm{SU}}^2}\rightarrow0$};
        \node at (-3.5,-8.5) {$\frac{1}{g_{\mathrm{USp}}^2}\rightarrow0$};
    \end{tikzpicture}
    \caption{Hasse diagrams for various choices of gauge coupling for the low energy theory \eqref{eq:5dInfQuiv}. The black lines are $a_1$ transitions, the blue lines are {\color{blue}$a_2$} transitions, the double blue lines are {\color{blue}$a_2\cup a_2$} transitions, the green line is a {\color{green}$a_3$} transition, the orange lines are {\color{orange}$a_4$} transitions, and the yellow line is a {\color{goodyellow}$c_2$} transition. As can be seen from the Hasse diagrams: When sending a coupling to infinity there is always cone enhancement. When $g_{\mathrm{SU}}$ is sent to infinity there is also cone fusion (accompanied by decoration in the magnetic quiver).}
    \label{fig:5dinfHD}
\end{figure}

Let us consider the theory
\begin{equation}
   \raisebox{-.5\height}{  \begin{tikzpicture}
    \node[gauge,label=below:{$\mathrm{SU}(2)$}] (1) at (0,0) {};
    \node[gauge,label=below:{$\mathrm{USp}(4)$}] (2) at (2,0) {};
    \node[flavour,label=below:{$D_2$}] (3) at (4,0) {};
    \draw (1)--(2)--(3);
    \end{tikzpicture}}
    \label{eq:5dInfQuiv}
\end{equation}
as a 5d $\mathcal{N}=1$ theory. We have two gauge couplings, $g_{\mathrm{SU}}$ and $g_{\mathrm{USp}}$, and can send either of them, or both to infinity. The relevant brane webs are depicted in Figure \ref{fig:5dinfBW}. The Higgs branch Hasse diagrams are summarized in Figure \ref{fig:5dinfHD}, and we derive them individually in the following.

\subsubsection{\texorpdfstring{$\frac{1}{g_{\mathrm{USp}}^2}\rightarrow0$}{gUSp to infinity}}
The brane web has four maximal decompositions with associated magnetic quivers:
\begin{subequations}
    \begin{equation}
        \raisebox{-.5\height}{ \begin{tikzpicture}
            \node (00) at (9,1.2) {};
            \node (00) at (9,-1.2) {};
            \node[sev] (6) at (5,0) {};
            \node[sev] (7) at (6,0) {};
            \node[sev] (8a) at (7.5,1) {};
            \node[sev] (8b) at (7.5,-1) {};
            \node[sev] (9a) at (9,1) {};
            \node[sev] (9b) at (9,-1) {};
            \node[sev] (10a) at (10,1) {};
            \node[sev] (10b) at (10,-1) {};
            \draw[cyan,transform canvas={yshift=1.5pt}] (6)--(7);
            \draw[cyan,transform canvas={yshift=-1.5pt}] (6)--(7);
            \draw[cyan,transform canvas={yshift=3pt}] (7)--(9,0)--(10a);
            \draw[cyan,transform canvas={yshift=1pt}] (7)--(9,0)--(10a);
            \draw[cyan,transform canvas={yshift=-1pt}] (7)--(9,0)--(10b);
            \draw[cyan,transform canvas={yshift=-3pt}] (7)--(9,0)--(10b);
            \draw[goodred] (8a)--(8b);
            \draw[goodorange] (9a)--(9b);
        \end{tikzpicture}}
        \qquad \raisebox{-.5\height}{ \begin{tikzpicture}
        \node[gaugec,label=left:{$2$}] (1) at (0,1) {};
        \node[gauger,label=right:{$1$}] (1u) at (1,1.5) {};
        \node[gaugeo,label=right:{$1$}] (1d) at (1,0.5) {};
        \node (2) at (0,-1) {};
        \draw[transform canvas={yshift=-1.5pt}] (1)--(1u);
        \draw[transform canvas={yshift=1.5pt}] (1)--(1u);
        \draw[transform canvas={yshift=-1.5pt}] (1)--(1d);
        \draw[transform canvas={yshift=1.5pt}] (1)--(1d);
        \draw[dashed] (1)--(-1,0)--(2);
    \end{tikzpicture}}
    \end{equation}
    \begin{equation}
        \raisebox{-.5\height}{  \begin{tikzpicture}
            \node (00) at (9,1.2) {};
            \node (00) at (9,-1.2) {};
            \node[sev] (6) at (5,0) {};
            \node[sev] (7) at (6,0) {};
            \node[sev] (8a) at (7.5,1) {};
            \node[sev] (8b) at (7.5,-1) {};
            \node[sev] (9a) at (9,1) {};
            \node[sev] (9b) at (9,-1) {};
            \node[sev] (10a) at (10,1) {};
            \node[sev] (10b) at (10,-1) {};
            \draw[goodmagenta,transform canvas={yshift=1.5pt}] (6)--(7);
            \draw[cyan,transform canvas={yshift=-1.5pt}] (6)--(7);
            \draw[goodred,transform canvas={yshift=3pt}] (7)--(9,0)--(10a);
            \draw[cyan,transform canvas={yshift=1pt}] (7)--(9,0)--(10a);
            \draw[goodred,transform canvas={yshift=-1pt}] (7)--(9,0)--(10b);
            \draw[cyan,transform canvas={yshift=-3pt}] (7)--(9,0)--(10b);
            \draw[goodred] (8a)--(8b);
            \draw[goodorange] (9a)--(9b);
        \end{tikzpicture}}
        \qquad
        \raisebox{-.5\height}{ \begin{tikzpicture}
        \node[gaugec,label=left:{$1$}] (1) at (0,1) {};
        \node[gaugeo,label=right:{$1$}] (1d) at (1,0.5) {};
        \node[gaugem,label=left:{$1$}] (2) at (0,-1) {};
        \node[gauger,label=right:{$1$}] (2u) at (1,-0.5) {};
        \draw[transform canvas={yshift=-1.5pt}] (1)--(1d);
        \draw[transform canvas={yshift=1.5pt}] (1)--(1d);
        \draw[transform canvas={yshift=-1.5pt}] (2)--(2u);
        \draw[transform canvas={yshift=1.5pt}] (2)--(2u);
        \draw[transform canvas={xshift=1.5pt}] (1d)--(2u);
        \draw[transform canvas={xshift=-1.5pt}] (1d)--(2u);
        \draw[dashed] (1)--(-1,0)--(2);
    \end{tikzpicture}}
    \end{equation}
    \begin{equation}
        \raisebox{-.5\height}{\begin{tikzpicture}
            \node (00) at (9,1.3) {};
            \node (00) at (9,-1.3) {};
            \node[sev] (6) at (5,0) {};
            \node[sev] (7) at (6,0) {};
            \node[sev] (8a) at (7.5,1) {};
            \node[sev] (8b) at (7.5,-1) {};
            \node[sev] (9a) at (9,1) {};
            \node[sev] (9b) at (9,-1) {};
            \node[sev] (10a) at (10,1) {};
            \node[sev] (10b) at (10,-1) {};
            \draw[goodmagenta,transform canvas={yshift=1.5pt}] (6)--(7);
            \draw[cyan,transform canvas={yshift=-1.5pt}] (6)--(7);
            \draw[cyan,transform canvas={yshift=3pt}] (7)--(9,0)--(10a);
            \draw[goodgreen,transform canvas={yshift=1pt}] (7)--(9,0)--(10a);
            \draw[cyan,transform canvas={yshift=-1pt}] (7)--(9,0)--(10b);
            \draw[blue,transform canvas={yshift=-3pt}] (7)--(9,0)--(10b);
            \draw[goodred] (8a)--(8b);
            \draw[goodgreen,transform canvas={xshift=1pt}] (9,1.5pt)--(9b);
            \draw[blue,transform canvas={xshift=-1pt}] (9a)--(9,-3pt);
        \end{tikzpicture}} 
        \qquad
        \raisebox{-.5\height}{  \begin{tikzpicture}
        \node[gaugec,label=left:{$1$}] (1) at (0,1) {};
        \node[gauger,label=right:{$1$}] (1u) at (1,1.5) {};
        \node[gaugem,label=left:{$1$}] (2) at (0,-1) {};
        \node[gaugeb,label=right:{$1$}] (2du) at (0.75,-1) {};
        \node[gaugeg,label=right:{$1$}] (2dd) at (1.25,-1.5) {};
        \draw[transform canvas={yshift=-1.5pt}] (1)--(1u);
        \draw[transform canvas={yshift=1.5pt}] (1)--(1u);
        \draw (2)--(2du) (2)--(2dd) (2du)--(2dd);
        \draw (1u)--(2du) (1u)--(2dd);
        \draw[dashed] (1)--(-1,0)--(2);
    \end{tikzpicture}}
    \end{equation}
    \begin{equation}
        \raisebox{-.5\height}{\begin{tikzpicture}
            \node[sev] (6) at (5,0) {};
            \node[sev] (7) at (6,0) {};
            \node[sev] (8a) at (7.5,1) {};
            \node[sev] (8b) at (7.5,-1) {};
            \node[sev] (9a) at (9,1) {};
            \node[sev] (9b) at (9,-1) {};
            \node[sev] (10a) at (10,1) {};
            \node[sev] (10b) at (10,-1) {};
            \draw[goodmagenta,transform canvas={yshift=1.5pt}] (6)--(7);
            \draw[goodmagenta,transform canvas={yshift=-1.5pt}] (6)--(7);
            \draw[goodred,transform canvas={yshift=3pt}] (7)--(9,0)--(10a);
            \draw[goodgreen,transform canvas={yshift=1pt}] (7)--(9,0)--(10a);
            \draw[goodred,transform canvas={yshift=-1pt}] (7)--(9,0)--(10b);
            \draw[blue,transform canvas={yshift=-3pt}] (7)--(9,0)--(10b);
            \draw[goodred] (8a)--(8b);
            \draw[goodgreen,transform canvas={xshift=1pt}] (9,1.5pt)--(9b);
            \draw[blue,transform canvas={xshift=-1pt}] (9a)--(9,-3pt);
        \end{tikzpicture}}
        \qquad
        \raisebox{-.5\height}{ \begin{tikzpicture}
        \node (1) at (0,1) {};
        \node[gaugem,label=left:{$2$}] (2) at (0,-1) {};
        \node[gauger,label=right:{$1$}] (2u) at (1,-0.5) {};
        \node[gaugeb,label=right:{$1$}] (2du) at (0.75,-1) {};
        \node[gaugeg,label=right:{$1$}] (2dd) at (1.25,-1.5) {};
        \draw[transform canvas={yshift=-1.5pt}] (2)--(2u);
        \draw[transform canvas={yshift=1.5pt}] (2)--(2u);
        \draw (2)--(2du) (2)--(2dd) (2du)--(2dd);
        \draw[dashed] (1)--(-1,0)--(2);
    \end{tikzpicture}}\;.
    \end{equation}
\end{subequations}

Comparing these magnetic quivers to the ones for $n=2$ in \eqref{magnetic:fourcones5d1}-\eqref{magnetic:fourcones5d4} we find that two cones are modified, and change in dimension (cone enhancement). The Hasse diagram becomes
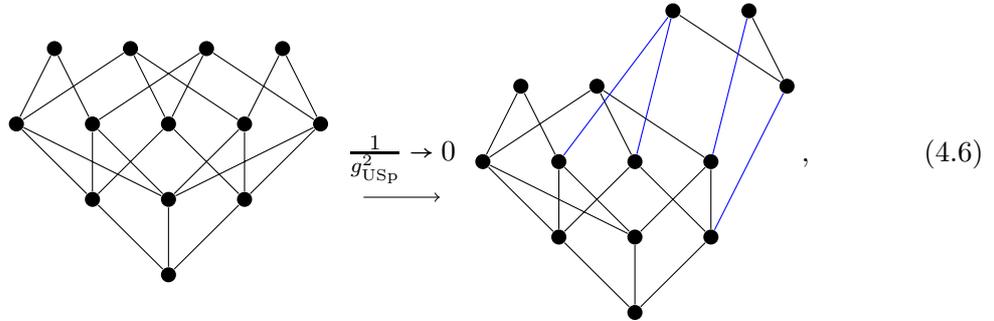
\begin{equation}
    \raisebox{-.5\height}{\begin{tikzpicture}
        \node[hasse] (1) at (0,0) {};
    \node[hasse] (a1) at (-1,1) {};
    \node[hasse] (a2) at (0,1) {};
    \node[hasse] (a3) at (1,1) {};
    \node[hasse] (b1) at (-2,2) {};
    \node[hasse] (b2) at (-1,2) {};
    \node[hasse] (b3) at (0,2) {};
    \node[hasse] (b4) at (1,2) {};
    \node[hasse] (b5) at (2,2) {};
    \node[hasse] (c1) at (-1.5,3) {};
    \node[hasse] (c2) at (-0.5,3) {};
    \node[hasse] (c3) at (0.5,3) {};
    \node[hasse] (c4) at (1.5,3) {};
    \draw (1)--(a1)--(b1)--(c1)--(b2)--(a1)--(b3)--(c2)--(b1)--(a2)--(b2)--(c3)--(b3)--(a3)--(1)--(a2)--(b4)--(c4)--(b5)--(a3)--(b4)--(c2) (c3)--(b5)--(a2);
    \end{tikzpicture}}
    \raisebox{-.5\height}{ \begin{tikzpicture}
    \draw[->] (0,0)--(1,0);
    \node at (0.5,0.5) {$\frac{1}{g_{\mathrm{USp}}^2}\rightarrow0$};
    \end{tikzpicture}}
    \raisebox{-.5\height}{ \begin{tikzpicture}
    \node[hasse] (1) at (0,0) {};
    \node[hasse] (a1) at (-1,1) {};
    \node[hasse] (a2) at (0,1) {};
    \node[hasse] (a3) at (1,1) {};
    \node[hasse] (b1) at (-2,2) {};
    \node[hasse] (b2) at (-1,2) {};
    \node[hasse] (b3) at (0,2) {};
    \node[hasse] (b4) at (1,2) {};
    \node[hasse] (b5) at (2,3) {};
    \node[hasse] (c1) at (-1.5,3) {};
    \node[hasse] (c2) at (-0.5,3) {};
    \node[hasse] (c3) at (0.5,4) {};
    \node[hasse] (c4) at (1.5,4) {};
    \draw (1)--(a1)--(b1)--(c1)--(b2)--(a1)--(b3)--(c2)--(b1)--(a2)--(b2) (b3)--(a3)--(1)--(a2)--(b4) (c4)--(b5) (a3)--(b4)--(c2) (c3)--(b5);
    \draw[blue] (b2)--(c3)--(b3) (b4)--(c4) (b5)--(a3);
    \end{tikzpicture}}\;,
    \label{eq:diagramCrownSUinf}
\end{equation}
where the black lines are $A_1$ transitions and the blue lines are {\color{blue}$a_2$} transitions. The height of a dot is proportional to the dimension of the corresponding leaf.

\subsubsection{\texorpdfstring{$\frac{1}{g_{\mathrm{SU}}^2}\rightarrow0$}{gSU to infinity}}

The brane web has only three maximal decompositions (cone fusion):
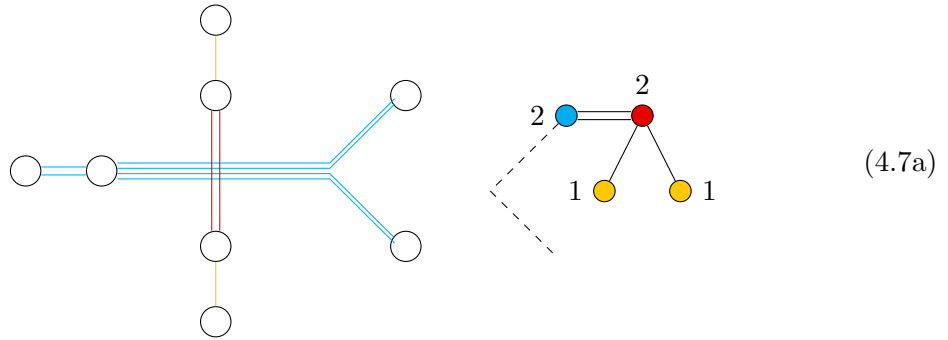
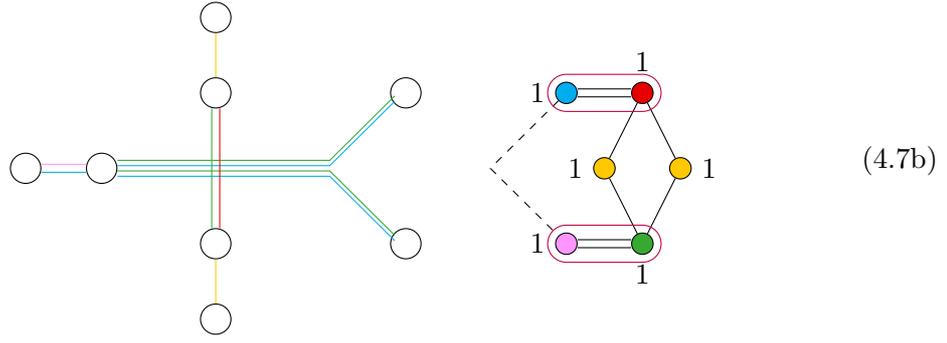
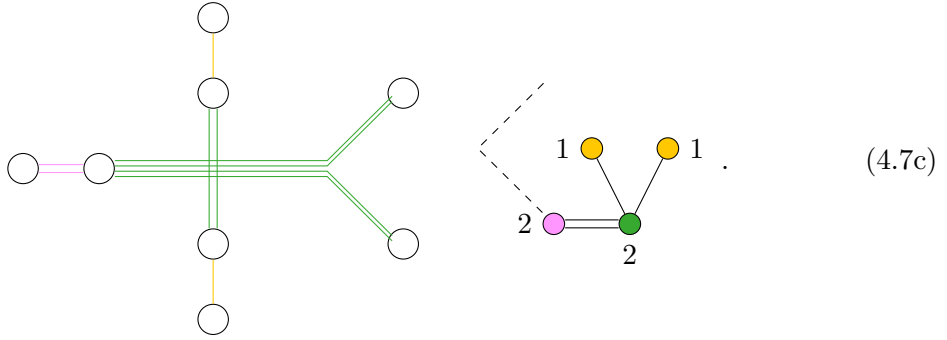
\begin{subequations}
    \begin{equation}
        \raisebox{-.5\height}{ \begin{tikzpicture}
   \node at (7,2.3) {};
   \node at (7,-2.3) {};
            \node[sev] (6) at (5,0) {};
            \node[sev] (7) at (6,0) {};
            \node[sev] (8a) at (7.5,2) {};
            \node[sev] (8b) at (7.5,-2) {};
            \node[sev] (9a) at (7.5,1) {};
            \node[sev] (9b) at (7.5,-1) {};
            \node[sev] (10a) at (10,1) {};
            \node[sev] (10b) at (10,-1) {};
            \draw[cyan,transform canvas={yshift=1.5pt}] (6)--(7);
            \draw[cyan,transform canvas={yshift=-1.5pt}] (6)--(7);
            \draw[cyan,transform canvas={yshift=3pt}] (7)--(9,0)--(10a);
            \draw[cyan,transform canvas={yshift=1pt}] (7)--(9,0)--(10a);
            \draw[cyan,transform canvas={yshift=-1pt}] (7)--(9,0)--(10b);
            \draw[cyan,transform canvas={yshift=-3pt}] (7)--(9,0)--(10b);
            \draw[goodred,transform canvas={xshift=1.5pt}] (9a)--(9b);
            \draw[goodred,transform canvas={xshift=-1.5pt}] (9a)--(9b);
            \draw[goodorange] (8a)--(9a);
            \draw[goodorange] (9b)--(8b);
        \end{tikzpicture}} 
        \qquad \raisebox{-.5\height}{ \begin{tikzpicture}
        \node[gaugec,label=left:{$2$}] (1) at (0,1) {};
        \node[gauger,label=above:{$2$}] (11) at (1,1) {};
        \node[gaugeo,label=left:{$1$}] (x) at (0.5,0) {};
        \node[gaugeo,label=right:{$1$}] (y) at (1.5,0) {};
        \node (2) at (0,-1) {};
        \draw[transform canvas={yshift=-1.5pt}] (1)--(11);
        \draw[transform canvas={yshift=1.5pt}] (1)--(11);
        \draw (x)--(11)--(y);
        \draw[dashed] (1)--(-1,0)--(2);
    \end{tikzpicture}}
    \end{equation}
    \begin{equation}
        \raisebox{-.5\height}{  \begin{tikzpicture}
   \node at (7,2.3) {};
   \node at (7,-2.3) {};
            \node[sev] (6) at (5,0) {};
            \node[sev] (7) at (6,0) {};
            \node[sev] (8a) at (7.5,2) {};
            \node[sev] (8b) at (7.5,-2) {};
            \node[sev] (9a) at (7.5,1) {};
            \node[sev] (9b) at (7.5,-1) {};
            \node[sev] (10a) at (10,1) {};
            \node[sev] (10b) at (10,-1) {};
            \draw[goodmagenta,transform canvas={yshift=1.5pt}] (6)--(7);
            \draw[cyan,transform canvas={yshift=-1.5pt}] (6)--(7);
            \draw[goodgreen,transform canvas={yshift=3pt}] (7)--(9,0)--(10a);
            \draw[cyan,transform canvas={yshift=1pt}] (7)--(9,0)--(10a);
            \draw[goodgreen,transform canvas={yshift=-1pt}] (7)--(9,0)--(10b);
            \draw[cyan,transform canvas={yshift=-3pt}] (7)--(9,0)--(10b);
            \draw[goodred,transform canvas={xshift=1.5pt}] (9a)--(9b);
            \draw[goodgreen,transform canvas={xshift=-1.5pt}] (9a)--(9b);
            \draw[goodorange] (8a)--(9a);
            \draw[goodorange] (9b)--(8b);
        \end{tikzpicture}} 
        \qquad \raisebox{-.5\height}{ \begin{tikzpicture}
        \node[gaugec,label=left:{$1$}] (1) at (0,1) {};
        \node[gauger,label=above:{$1$}] (11) at (1,1) {};
        \node[gaugeo,label=left:{$1$}] (x) at (0.5,0) {};
        \node[gaugeo,label=right:{$1$}] (y) at (1.5,0) {};
        \node[gaugem,label=left:{$1$}] (2) at (0,-1) {};
        \node[gaugeg,label=below:{$1$}] (22) at (1,-1) {};
        \draw[transform canvas={yshift=-1.5pt}] (1)--(11);
        \draw[transform canvas={yshift=1.5pt}] (1)--(11);
        \draw[transform canvas={yshift=-1.5pt}] (2)--(22);
        \draw[transform canvas={yshift=1.5pt}] (2)--(22);
        \draw (x)--(11)--(y);
        \draw (x)--(22)--(y);
        \draw[dashed] (1)--(-1,0)--(2);
        \draw[purple] \convexpath{1,11}{7pt};
        \draw[purple] \convexpath{2,22}{7pt};
    \end{tikzpicture}}
    \end{equation}
    \begin{equation}
        \raisebox{-.5\height}{ \begin{tikzpicture}
   \node at (7,2.3) {};
   \node at (7,-2.3) {};
            \node[sev] (6) at (5,0) {};
            \node[sev] (7) at (6,0) {};
            \node[sev] (8a) at (7.5,2) {};
            \node[sev] (8b) at (7.5,-2) {};
            \node[sev] (9a) at (7.5,1) {};
            \node[sev] (9b) at (7.5,-1) {};
            \node[sev] (10a) at (10,1) {};
            \node[sev] (10b) at (10,-1) {};
            \draw[goodmagenta,transform canvas={yshift=1.5pt}] (6)--(7);
            \draw[goodmagenta,transform canvas={yshift=-1.5pt}] (6)--(7);
            \draw[goodgreen,transform canvas={yshift=3pt}] (7)--(9,0)--(10a);
            \draw[goodgreen,transform canvas={yshift=1pt}] (7)--(9,0)--(10a);
            \draw[goodgreen,transform canvas={yshift=-1pt}] (7)--(9,0)--(10b);
            \draw[goodgreen,transform canvas={yshift=-3pt}] (7)--(9,0)--(10b);
            \draw[goodgreen,transform canvas={xshift=1.5pt}] (9a)--(9b);
            \draw[goodgreen,transform canvas={xshift=-1.5pt}] (9a)--(9b);
            \draw[goodorange] (8a)--(9a);
            \draw[goodorange] (9b)--(8b);
        \end{tikzpicture}} 
        \qquad \raisebox{-.5\height}{\begin{tikzpicture}
        \node (1) at (0,1) {};
        \node[gaugeo,label=left:{$1$}] (x) at (0.5,0) {};
        \node[gaugeo,label=right:{$1$}] (y) at (1.5,0) {};
        \node[gaugem,label=left:{$2$}] (2) at (0,-1) {};
        \node[gaugeg,label=below:{$2$}] (22) at (1,-1) {};
        \draw[transform canvas={yshift=-1.5pt}] (2)--(22);
        \draw[transform canvas={yshift=1.5pt}] (2)--(22);
        \draw (x)--(22)--(y);
        \draw[dashed] (1)--(-1,0)--(2);
    \end{tikzpicture}}\;.
    \end{equation}
\end{subequations}

Comparing these magnetic quivers to the ones for $n=2$ in \eqref{magnetic:fourcones5d1}-\eqref{magnetic:fourcones5d4} 
we see that all cones are modified and grow in dimension, and that the two middle cones are fused into one larger cone. Note also that the middle quiver comes with a decoration\footnote{See \cite[Appendix B]{Bourget:2022tmw} for explanation of a brane web decomposition corresponding to a decorated quiver.} in the sense of \cite{Bourget:2022ehw,Bourget:2022tmw}. The Hasse diagram becomes
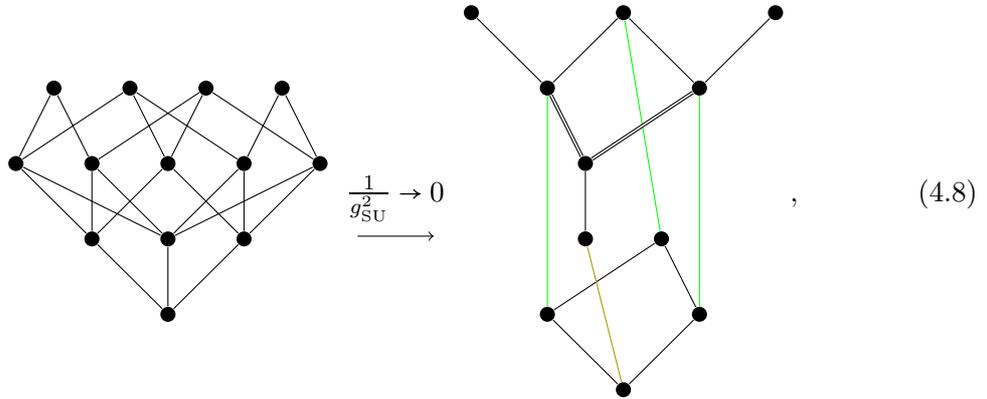
\begin{equation}
    \raisebox{-.5\height}{\begin{tikzpicture}
        \node[hasse] (1) at (0,0) {};
    \node[hasse] (a1) at (-1,1) {};
    \node[hasse] (a2) at (0,1) {};
    \node[hasse] (a3) at (1,1) {};
    \node[hasse] (b1) at (-2,2) {};
    \node[hasse] (b2) at (-1,2) {};
    \node[hasse] (b3) at (0,2) {};
    \node[hasse] (b4) at (1,2) {};
    \node[hasse] (b5) at (2,2) {};
    \node[hasse] (c1) at (-1.5,3) {};
    \node[hasse] (c2) at (-0.5,3) {};
    \node[hasse] (c3) at (0.5,3) {};
    \node[hasse] (c4) at (1.5,3) {};
    \draw (1)--(a1)--(b1)--(c1)--(b2)--(a1)--(b3)--(c2)--(b1)--(a2)--(b2)--(c3)--(b3)--(a3)--(1)--(a2)--(b4)--(c4)--(b5)--(a3)--(b4)--(c2) (c3)--(b5)--(a2);
    \end{tikzpicture}}
    \raisebox{-.5\height}{ \begin{tikzpicture}
    \draw[->] (0,0)--(1,0);
    \node at (0.5,0.5) {$\frac{1}{g_{\mathrm{SU}}^2}\rightarrow0$};
    \end{tikzpicture}}
    \raisebox{-.5\height}{ \begin{tikzpicture}
        \node[hasse] (a) at (0,0) {};
        \node[hasse] (b1) at (-1,1) {};
        \node[hasse] (b2) at (1,1) {};
        \node[hasse] (cd) at (-0.5,2) {};
        \node[hasse] (c) at (0.5,2) {};
        \node[hasse] (d) at (-0.5,3) {};
        \node[hasse] (e1) at (-1,4) {};
        \node[hasse] (e2) at (1,4) {};
        \node[hasse] (f1) at (-2,5) {};
        \node[hasse] (f0) at (0,5) {};
        \node[hasse] (f2) at (2,5) {};
        \draw (f1)--(e1)--(f0)--(e2)--(f2)
        (c)--(b1)--(a)--(b2)--(c);
        \draw[green] (e1)--(b1) (e2)--(b2) (f0)--(c);
        \draw[double] (e1)--(d)--(e2);
        \draw (d)--(cd);
        \draw[goodyellow] (cd)--(a);
    \end{tikzpicture}}\;,
\end{equation}
where the black lines are $A_1$ transitions, the double black lines are $a_1\cup a_1$ transitions,\footnote{Elementary transitions which are unions of cones show up e.g.\ in the nilcone of $\mathrm{SO}(2r+1)$ for $r>2$ \cite{Kraft1982,Cabrera:2017njm}, as well as in symmetric products of Kleinian singularities and in the moduli space of instantons \cite[Sections 2\&3]{Bourget:2022tmw}.} the green lines are {\color{green}$a_3$} transitions and the yellow line is a {\color{goodyellow}$c_2$} transition. The height of a dot is proportional to the dimension of the corresponding leaf. We refer to \cite[Section 2]{Bourget:2022tmw} for details about how the diagram is obtained from quiver subtraction.

\subsubsection{\texorpdfstring{$\frac{1}{g_{\mathrm{USp}}^2}\rightarrow0$ and $\frac{1}{g_{\mathrm{SU}}^2}\rightarrow0$}{gUSP and gSU to infinity}}

The brane web has three maximal decompositions and associated magnetic quivers:

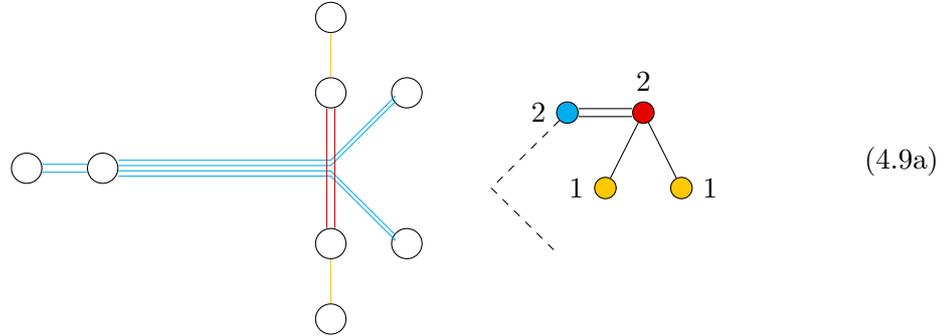
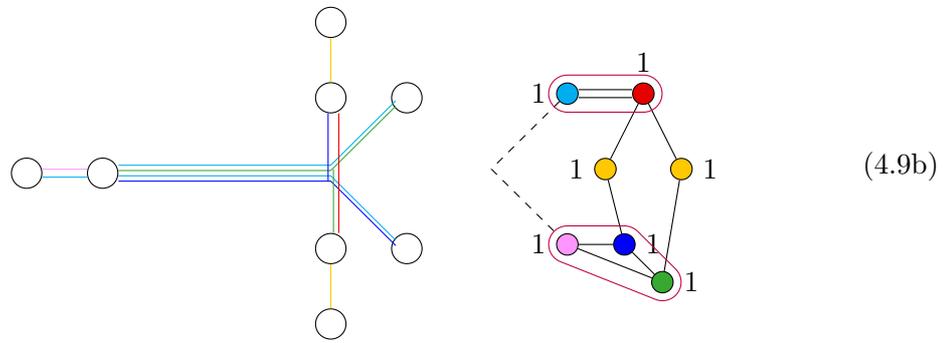
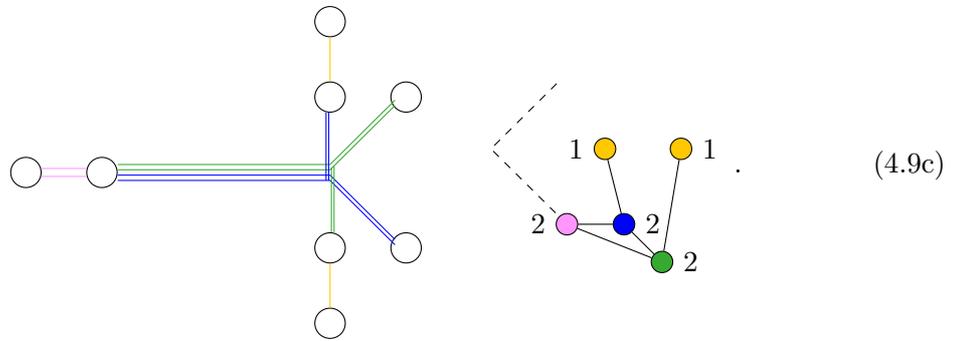
\begin{subequations}
    \begin{equation}
        \raisebox{-.5\height}{  \begin{tikzpicture}
   \node at (9,2.3) {};
   \node at (9,-2.3) {};
        \node[sev] (6) at (5,0) {};
        \node[sev] (7) at (6,0) {};
        \node[sev] (8a) at (9,2) {};
        \node[sev] (8b) at (9,-2) {};
        \node[sev] (9a) at (9,1) {};
        \node[sev] (9b) at (9,-1) {};
        \node[sev] (10a) at (10,1) {};
        \node[sev] (10b) at (10,-1) {};
        \draw[cyan,transform canvas={yshift=1.5pt}] (6)--(7);
        \draw[cyan,transform canvas={yshift=-1.5pt}] (6)--(7);
        \draw[cyan,transform canvas={yshift=3pt}] (7)--(9,0)--(10a);
        \draw[cyan,transform canvas={yshift=1pt}] (7)--(9,0)--(10a);
        \draw[cyan,transform canvas={yshift=-1pt}] (7)--(9,0)--(10b);
        \draw[cyan,transform canvas={yshift=-3pt}] (7)--(9,0)--(10b);
        \draw[goodred,transform canvas={xshift=1.5pt}] (9a)--(9b);
        \draw[goodred,transform canvas={xshift=-1.5pt}] (9a)--(9b);
        \draw[goodorange] (8a)--(9a) (9b)--(8b);
    \end{tikzpicture}}
    \qquad
    \raisebox{-.5\height}{  \begin{tikzpicture}
        \node[gaugec,label=left:{$2$}] (1) at (0,1) {};
        \node[gauger,label=above:{$2$}] (11) at (1,1) {};
        \node[gaugeo,label=left:{$1$}] (x) at (0.5,0) {};
        \node[gaugeo,label=right:{$1$}] (y) at (1.5,0) {};
        \node (2) at (0,-1) {};
        \draw[transform canvas={yshift=-1.5pt}] (1)--(11);
        \draw[transform canvas={yshift=1.5pt}] (1)--(11);
        \draw (x)--(11)--(y);
        \draw[dashed] (1)--(-1,0)--(2);
    \end{tikzpicture}}
    \end{equation}
    \begin{equation}
        \raisebox{-.5\height}{  \begin{tikzpicture}
   \node at (9,2.3) {};
   \node at (9,-2.3) {};
        \node[sev] (6) at (5,0) {};
        \node[sev] (7) at (6,0) {};
        \node[sev] (8a) at (9,2) {};
        \node[sev] (8b) at (9,-2) {};
        \node[sev] (9a) at (9,1) {};
        \node[sev] (9b) at (9,-1) {};
        \node[sev] (10a) at (10,1) {};
        \node[sev] (10b) at (10,-1) {};
        \draw[goodmagenta,transform canvas={yshift=1.5pt}] (6)--(7);
        \draw[cyan,transform canvas={yshift=-1.5pt}] (6)--(7);
        \draw[cyan,transform canvas={yshift=3pt}] (7)--(9,0)--(10a);
        \draw[goodgreen,transform canvas={yshift=1pt}] (7)--(9,0)--(10a);
        \draw[cyan,transform canvas={yshift=-1pt}] (7)--(9,0)--(10b);
        \draw[blue,transform canvas={yshift=-3pt}] (7)--(9,0)--(10b);
        \draw[goodred,transform canvas={xshift=3pt}] (9a)--(9b);
            \draw[goodgreen,transform canvas={xshift=1pt}] (9,1.5pt)--(9b);
            \draw[blue,transform canvas={xshift=-1pt}] (9a)--(9,-3pt);
        \draw[goodorange] (8a)--(9a) (9b)--(8b);
    \end{tikzpicture}} 
    \qquad
    \raisebox{-.5\height}{  \begin{tikzpicture}
        \node[gaugec,label=left:{$1$}] (1) at (0,1) {};
        \node[gauger,label=above:{$1$}] (11) at (1,1) {};
        \node[gaugeo,label=left:{$1$}] (x) at (0.5,0) {};
        \node[gaugeo,label=right:{$1$}] (y) at (1.5,0) {};
        \node[gaugem,label=left:{$1$}] (2) at (0,-1) {};
        \node[gaugeb,label=right:{$1$}] (2du) at (0.75,-1) {};
        \node[gaugeg,label=right:{$1$}] (2dd) at (1.25,-1.5) {};
        \draw[transform canvas={yshift=-1.5pt}] (1)--(11);
        \draw[transform canvas={yshift=1.5pt}] (1)--(11);
        \draw (x)--(11)--(y);
        \draw (x)--(2du) (2dd)--(y);
        \draw (2)--(2du)--(2dd)--(2);
        \draw[dashed] (1)--(-1,0)--(2);
        \draw[purple] \convexpath{1,11}{7pt};
        \draw[purple] \convexpath{2,2du,2dd}{7pt};
    \end{tikzpicture}}
    \end{equation}
    \begin{equation}
        \raisebox{-.5\height}{ \begin{tikzpicture}
   \node at (9,2.3) {};
   \node at (9,-2.3) {};
        \node[sev] (6) at (5,0) {};
        \node[sev] (7) at (6,0) {};
        \node[sev] (8a) at (9,2) {};
        \node[sev] (8b) at (9,-2) {};
        \node[sev] (9a) at (9,1) {};
        \node[sev] (9b) at (9,-1) {};
        \node[sev] (10a) at (10,1) {};
        \node[sev] (10b) at (10,-1) {};
        \draw[goodmagenta,transform canvas={yshift=1.5pt}] (6)--(7);
        \draw[goodmagenta,transform canvas={yshift=-1.5pt}] (6)--(7);
        \draw[goodgreen,transform canvas={yshift=3pt}] (7)--(9,0)--(10a);
        \draw[goodgreen,transform canvas={yshift=1pt}] (7)--(9,0)--(10a);
        \draw[blue,transform canvas={yshift=-1pt}] (7)--(9,0)--(10b);
        \draw[blue,transform canvas={yshift=-3pt}] (7)--(9,0)--(10b);
            \draw[goodgreen,transform canvas={xshift=0.5pt}] (9,2pt)--(9b);
            \draw[goodgreen,transform canvas={xshift=1.5pt}] (9,2.5pt)--(9b);
            \draw[blue,transform canvas={xshift=-1.5pt}] (9a)--(9,-3pt);
            \draw[blue,transform canvas={xshift=-0.5pt}] (9a)--(9,-3pt);
        \draw[goodorange] (8a)--(9a) (9b)--(8b);
    \end{tikzpicture}} 
    \qquad
    \raisebox{-.5\height}{ \begin{tikzpicture}
        \node (1) at (0,1) {};
        \node[gaugeo,label=left:{$1$}] (x) at (0.5,0) {};
        \node[gaugeo,label=right:{$1$}] (y) at (1.5,0) {};
        \node[gaugem,label=left:{$2$}] (2) at (0,-1) {};
        \node[gaugeb,label=right:{$2$}] (2du) at (0.75,-1) {};
        \node[gaugeg,label=right:{$2$}] (2dd) at (1.25,-1.5) {};
        \draw (x)--(2du) (2dd)--(y);
        \draw (2)--(2du)--(2dd)--(2);
        \draw[dashed] (1)--(-1,0)--(2);
    \end{tikzpicture}}\;.
    \end{equation}
\end{subequations}

Comparing these magnetic quivers to the ones for $n=2$ in \eqref{magnetic:fourcones5d1}-\eqref{magnetic:fourcones5d4} 
we see that all cones are modified and grow in dimension, and that the two middle cones are fused into one larger cone whose magnetic quiver is decorated in the sense of \cite{Bourget:2022ehw,Bourget:2022tmw}. The Hasse diagram becomes
\begin{equation}
    \raisebox{-.5\height}{\begin{tikzpicture}
        \node[hasse] (1) at (0,0) {};
    \node[hasse] (a1) at (-1,1) {};
    \node[hasse] (a2) at (0,1) {};
    \node[hasse] (a3) at (1,1) {};
    \node[hasse] (b1) at (-2,2) {};
    \node[hasse] (b2) at (-1,2) {};
    \node[hasse] (b3) at (0,2) {};
    \node[hasse] (b4) at (1,2) {};
    \node[hasse] (b5) at (2,2) {};
    \node[hasse] (c1) at (-1.5,3) {};
    \node[hasse] (c2) at (-0.5,3) {};
    \node[hasse] (c3) at (0.5,3) {};
    \node[hasse] (c4) at (1.5,3) {};
    \draw (1)--(a1)--(b1)--(c1)--(b2)--(a1)--(b3)--(c2)--(b1)--(a2)--(b2)--(c3)--(b3)--(a3)--(1)--(a2)--(b4)--(c4)--(b5)--(a3)--(b4)--(c2) (c3)--(b5)--(a2);
    \end{tikzpicture}}
    \raisebox{-.5\height}{ \begin{tikzpicture}
    \draw[->] (0,0)--(1,0);
    \node at (0.5,0.5) {$\frac{1}{g_{\mathrm{USp}}^2},\frac{1}{g_{\mathrm{SU}}^2}\rightarrow0$};
    \end{tikzpicture}}
   \raisebox{-.5\height}{  \begin{tikzpicture}
        \node[hasse] (a) at (0,0) {};
        \node[hasse] (b1) at (-1,1) {};
        \node[hasse] (b2) at (1,1) {};
        \node[hasse] (cd) at (-0.5,2) {};
        \node[hasse] (c) at (0.5,2) {};
        \node[hasse] (d) at (-0.5,3) {};
        \node[hasse] (e1) at (-1,4) {};
        \node[hasse] (e2) at (1,5) {};
        \node[hasse] (f1) at (-2,5) {};
        \node[hasse] (f0) at (0,6) {};
        \node[hasse] (f2) at (2,7) {};
        \draw (f1)--(e1) (f0)--(e2)
        (c)--(b1)--(a)--(b2)--(c);
        \draw[blue] (e1)--(f0) (e2)--(f2);
        \draw[green] (e1)--(b1);
        \draw[orange] (e2)--(b2) (f0)--(c);
        \draw[double] (e1)--(d);
        \draw[double,blue] (d)--(e2);
        \draw (d)--(cd);
        \draw[goodyellow] (cd)--(a);
    \end{tikzpicture}}\;,
\end{equation}
where the black lines are $A_1$ transitions, the double black lines are $a_1\cup a_1$ transitions, the blue lines are {\color{blue}$a_2$} transitions, the double blue lines are {\color{blue}$a_2\cup a_2$} transitions, the green line is a {\color{green}$a_3$} transition, the orange lines are {\color{orange}$a_4$} transitions, and the yellow line is a {\color{goodyellow}$c_2$} transition. The height of a dot is proportional to the dimension of the corresponding leaf.

\clearpage 

\section{Outlook}

In this paper we show the importance of a novel brane configuration of NS5-D3-D5-O$5^-$ setups. This leads us to explore theories whose Higgs branches consist of many cones using magnetic quivers. The double T-dual brane web hosts theories whose Higgs branches consist of many more cones, which are modified by taking gauge couplings to infinity.

Based on the principle of inversion, Coulomb branch Hasse diagrams can be conjectured from the Higgs branch Hasse diagrams which are computable from the magnetic quivers. The theories we study are bad, and hence a direct computation of their Coulomb branches is difficult. It remains a challenge to test our conjectures with explicit computations.

One feature of the present theories is that their Higgs branch chiral ring may contain nilpotent operators which are difficult to study. We leave several questions about them for future work:
\begin{itemize}
    \item How can we identify nilpotent operators from the brane system?
    \item What is the fate of nilpotent operators in the quantum moduli space?
    \item What is the physics of a nilpotent operator?
\end{itemize}

\section*{Acknowledgements}

We are indebted to Santiago Cabrera for many insightful discussions and collaboration at an early stage of this work. AH would like to thank Hirotaka Hayashi, Sung-Soo Kim, Kimyeong Lee and Futoshi Yagi for discussions.
AB is supported by the ERC Consolidator Grant 772408-Stringlandscape, and by the LabEx ENS-ICFP: ANR-10-LABX-0010/ANR-10-IDEX-0001-02 PSL*. The work of AB, JFG, AH, RK and ZZ is partially supported by STFC grant ST/T000791/1. The work of MS was in part supported by the Yau Mathematical Sciences Center at Tsinghua University, the National Natural Science Foundation of China (grant no.\ 11950410497), and the China Postdoctoral Science Foundation (grant no.\ 2019M650616).
ZZ is partially supported by the ERC Consolidator Grant \# 864828 “Algebraic Foundations of Supersymmetric Quantum Field Theory” (SCFTAlg).

\appendix

\section{Bad Coulomb Branches from Inversion, and the full Moduli Space}
\label{app:badCB}

Coulomb branches of bad \cite{Gaiotto:2008ak} 3d $\mathcal{N}=4$ theories are difficult to study. The monopole formula \cite{Cremonesi:2013lqa}, for example, diverges due to the non-conical nature of these Coulomb branches. In some special cases one can use the Hall-Littlewood formula \cite{Cremonesi:2014kwa} to obtain a Hilbert series which captures aspects of, but does not completely describe the Coulomb branch of the bad theory in question. Framed unitary Dynkin quivers were described as generalized affine Grassmannian slices in \cite{2016arXiv160403625B} via the BFN construction \cite{Nakajima:2015txa,Nakajima2015,Braverman:2016wma}, but no description for more general theories was given so far.\footnote{Coulomb branches of orthosymplectic quivers for example have only been given a BFN type description fairly recently \cite{Braverman:2022zei}.} In principle one can use abelianisation \cite{Bullimore:2015lsa} to study bad Coulomb branches of any gauge theory, as was successfully done in \cite{Assel:2017jgo,Assel:2018exy} for theories with a single gauge node, but this is difficult to do for more complicated theories.

It is therefore desirable to find indirect ways to describe the structure of Coulomb branches of bad theories. Magnetic quivers have proven a useful tool to study Higgs branches of good, ugly, and bad theories, as is demonstrated for example in the present paper. Let $\mathsf{Q}_e$ be a (good, ugly, or bad)  3d $\mathcal{N}=4$ `electric' theory and $\{\mathsf{Q}_m^i\}$ its set of magnetic quivers, which we assume to be good.\footnote{It sometimes happens that one finds a magnetic quiver which is bad, and in such cases it is not clear whether one could find a good magnetic theory, which may not be a quiver, instead.} The Coulomb branches of $\mathsf{Q}_m^i$ describe the different cones which make up the Higgs branch of $\mathsf{Q}_e$.

We can view an individual magnetic quiver $\mathsf{Q}_m^i$ as a theory in its own right. This theory has a Higgs branch, which by 3d mirror symmetry\footnote{Since the magnetic quiver is good there is a point (origin) in its moduli space where it flows to a fully interacting SCFT. For SCFTs there is a notion of 3d mirror symmetry \cite{Intriligator:1996ex}.} is the Coulomb branch of some (possibly non-Lagrangian) SCFT $\mathcal{T}^\vee{}_m^i$. It is natural to conjecture that this SCFT $\mathcal{T}^\vee{}_m^i$ is realized as a low energy theory on a point (or symplectic leaf) in the Coulomb branch of our original electric theory $\mathsf{Q}_e$, and that the Higgs branch of a magnetic quiver $\mathsf{Q}_m^i$ describes the local singular geometry to a specific locus in the Coulomb branch of our original electric theory $\mathsf{Q}_e$.

Furthermore, if all elementary slices in the Coulomb and Higgs branch of our electric theory $\mathsf{Q}_e$ are either Kleinian singularities or ADE minimal nilpotent orbit closures, then we propose that the Hasse diagram of the Coulomb branch of $\mathsf{Q}_e$ can be obtained through \emph{inversion}\footnote{If $\mathfrak{H}$ is a Hasse diagram where all elementary slices are Kleinian singularities or $ADE$ minimal nilpotent orbit closures, then the \emph{inversion} $\mathcal{I}(\mathfrak{H})$ is obtained by reversing the partial order and exchanging Kleinian singularities and minimal nilpotent orbit closures associated to the same $ADE$ algebra \cite{Grimminger:2020dmg}.} \cite{Grimminger:2020dmg} from the Hasse diagram of the classical Higgs branch of $\mathsf{Q}_e$, which in turn can be obtained from the Hasse diagrams of the Coulomb branches of the magnetic quivers $\mathsf{Q}_m^i$, computed e.g.\ via quiver subtraction.

We test these conjectures on two examples which were studied in detail in the literature, namely U$(k)$ SQCD in Section \ref{subsec:appU} and USp$(2k)$ SQCD in Section \ref{subsec:appSp}. We then make new predictions about the Coulomb branch and full moduli space of bad SU$(k)$ SQCD in Section \ref{subsec:appSU}.

\subsection{\texorpdfstring{$\mathrm{U}(k)$ SQCD}{U(k) SQCD}}
\label{subsec:appU}
Consider the electric theory
\begin{equation}
\label{eq:USQCD}
\raisebox{-.5\height}{
    \begin{tikzpicture}
        \node at (-1,0.5) {$\mathsf{Q}_e=$};
        \node[gauge,label=right:{U$(k)$}] (1) at (0,0) {};
        \node[flavour,label=right:{$N$}] (2) at (0,1) {};
        \draw (1)--(2);
    \end{tikzpicture}
    }
\end{equation}
which is good for $N\geq 2k$, ugly for $N=2k-1$, and bad for $N\leq 2k-2$. We discuss the good and non-good cases separately.

\paragraph{$N\geq2k$.} The theory is good and there is complete Higgsing. The magnetic quiver for good \eqref{eq:USQCD} is its 3d mirror
\begin{equation}
\label{eq:USQCDmagGOOD}
\raisebox{-.5\height}{
    \begin{tikzpicture}
        \node at (0,0.5) {$\mathsf{Q}_m=$};
        \node[gauge,label=below:{$1$}] (1) at (1,0) {};
        \node[gauge,label=below:{$2$}] (2) at (2,0) {};
        \node (3) at (3,0) {$\cdots$};
        \node[gauge,label=below:{$k$}] (4) at (4,0) {};
        \node[flavour,label=left:{$1$}] (4f) at (4,1) {};
        \node (5) at (5,0) {$\cdots$};
        \node[gauge,label=below:{$k$}] (6) at (6,0) {};
        \node[flavour,label=right:{$1$}] (6f) at (6,1) {};
        \node (7) at (7,0) {$\cdots$};
        \node[gauge,label=below:{$2$}] (8) at (8,0) {};
        \node[gauge,label=below:{$1$}] (9) at (9,0) {};
        \draw (1)--(2)--(3)--(4)--(5)--(6)--(7)--(8)--(9) (4)--(4f) (6)--(6f);
        \draw [decorate,decoration={brace,amplitude=5pt}]
    (9.3,-0.6)--(0.7,-0.6) node [black,midway,yshift=-10] {$N-1$};
    \end{tikzpicture}
    }\;.
\end{equation}
It is straightforward to obtain the Coulomb branch Hasse diagram of \eqref{eq:USQCDmagGOOD} from quiver subtraction, which is the Higgs branch Hasse diagram of \eqref{eq:USQCD}. As discussed in \cite{Grimminger:2020dmg} the inversion of the Coulomb branch Hasse diagram of \eqref{eq:USQCDmagGOOD} is the Higgs branch Hasse diagram of \eqref{eq:USQCDmagGOOD} and hence the Coulomb branch Hasse diagram of \eqref{eq:USQCD}.
\begin{equation}
    \mathfrak{H}\left(\mathcal{C}\eqref{eq:USQCDmagGOOD}\right)=
    \mathfrak{H}\left(\mathcal{H}\eqref{eq:USQCD}\right)=\qquad
        \raisebox{-.5\height}{\begin{tikzpicture}
        \node[hasse] (0) at (0,0) {};
        \node[hasse] (1) at (0,1) {};
        \node[hasse] (2) at (0,2) {};
        \node (3) at (0,3) {$\vdots$};
        \node[hasse] (4) at (0,4) {};
        \node[hasse] (5) at (0,5) {};
        \node[hasse] (6) at (0,6) {};
        \draw (0)--(1)--(2)--(3)--(4)--(5)--(6);
        \node at (0.5,0.5) {$a_{N-1}$};
        \node at (0.5,1.5) {$a_{N-3}$};
        \node at (1,4.5) {$a_{N-2k+3}$};
        \node at (1,5.5) {$a_{N-2k+1}$};
    \end{tikzpicture}}
    \raisebox{-.5\height}{\begin{tikzpicture}
        \draw[->] (0,0)--(1,0);
        \node at (0.5,0.5) {$\mathcal{I}$};
    \end{tikzpicture}}\qquad
    \raisebox{-.5\height}{\begin{tikzpicture}
        \node[hasse] (0) at (0,0) {};
        \node[hasse] (1) at (0,1) {};
        \node[hasse] (2) at (0,2) {};
        \node (3) at (0,3) {$\vdots$};
        \node[hasse] (4) at (0,4) {};
        \node[hasse] (5) at (0,5) {};
        \node[hasse] (6) at (0,6) {};
        \draw (0)--(1)--(2)--(3)--(4)--(5)--(6);
        \node at (0.5,5.5) {$A_{N-1}$};
        \node at (0.5,4.5) {$A_{N-3}$};
        \node at (1,1.5) {$A_{N-2k+3}$};
        \node at (1,0.5) {$A_{N-2k+1}$};
    \end{tikzpicture}}\quad=
    \mathfrak{H}\left(\mathcal{H}\eqref{eq:USQCDmagGOOD}\right)=
    \mathfrak{H}\left(\mathcal{C}\eqref{eq:USQCD}\right)\;.
\end{equation}
Applying quiver subtraction directly to \eqref{eq:USQCD} yields the right hand side, confirming the inversion conjecture.

\paragraph{$N<2k$.} The theory \eqref{eq:USQCD} is not good and there is incomplete Higgsing. On a generic point of the Higgs branch of \eqref{eq:USQCD} there is an unbroken
\begin{equation}
\label{eq:USQCDrem}
    \raisebox{-.5\height}{\begin{tikzpicture}
        \node[gauge,label=right:{U$(k-\frac{N-\epsilon}{2})$}] (1) at (0,0) {};
        \node[flavour,label=right:{$\epsilon$}] (2) at (0,1) {};
        \draw (1)--(2);
    \end{tikzpicture}}\textnormal{     with  }\epsilon=\begin{cases}
    1\textnormal{ , }N\textnormal{ odd}\\
    0\textnormal{ , }N\textnormal{ even}
    \end{cases}
\end{equation}
remaining. This is easy to see e.g.\ from the brane system.

The magnetic quiver for \eqref{eq:USQCD} is (independent of $k<2N$)\footnote{Here we only consider the case of zero FI parameter. For $N\geq k$ one may turn on an FI parameter, which changes the magnetic quiver as discussed e.g.\ in \cite[Appendix B]{Bourget:2021jwo}.}
\begin{equation}
\label{eq:USQCDmagBAD}
    \mathsf{Q}_m=\begin{cases}
        \begin{tikzpicture}
        \node[gauge,label=below:{$1$}] (1) at (1,0) {};
        \node[gauge,label=below:{$2$}] (2) at (2,0) {};
        \node (3) at (3,0) {$\cdots$};
        \node[gauge,label=below:{$\frac{N-1}{2}$}] (4) at (4,0) {};
        \node[flavour,label=left:{$1$}] (4f) at (4,1) {};
        \node[gauge,label=below:{$\frac{N-1}{2}$}] (5) at (5,0) {};
        \node[flavour,label=right:{$1$}] (5f) at (5,1) {};
        \node (6) at (6,0) {$\cdots$};
        \node[gauge,label=below:{$2$}] (7) at (7,0) {};
        \node[gauge,label=below:{$1$}] (8) at (8,0) {};
        \draw (1)--(2)--(3)--(4)--(5)--(6)--(7)--(8) (4)--(4f) (5)--(5f);
        \draw [decorate,decoration={brace,amplitude=5pt}]
    (8.3,-0.8)--(0.7,-0.8) node [black,midway,yshift=-10] {$N-1$};
        \node at (10,0) {, $N$ odd};
    \end{tikzpicture}\\
    \begin{tikzpicture}
        \node[gauge,label=below:{$1$}] (1) at (1,0) {};
        \node[gauge,label=below:{$2$}] (2) at (2,0) {};
        \node (3) at (3,0) {$\cdots$};
        \node[gauge,label=below:{$\frac{N}{2}$}] (4) at (4,0) {};
        \node[flavour,label=left:{$2$}] (4f) at (4,1) {};
        \node (5) at (5,0) {$\cdots$};
        \node[gauge,label=below:{$2$}] (6) at (6,0) {};
        \node[gauge,label=below:{$1$}] (7) at (7,0) {};
        \draw (1)--(2)--(3)--(4)--(5)--(6)--(7) (4)--(4f);
        \draw [decorate,decoration={brace,amplitude=5pt}]
    (7.3,-0.8)--(0.7,-0.8) node [black,midway,yshift=-10] {$N-1$};
        \node at (10,0) {, $N$ even};
    \end{tikzpicture}
    \end{cases}\;.
\end{equation}
Applying quiver subtraction to \eqref{eq:USQCDmagBAD} we obtain its Coulomb branch Hasse diagram, and hence the Higgs branch Hasse diagram of \eqref{eq:USQCD}. The inversion of this Hasse diagram is the Higgs branch Hasse diagram of \eqref{eq:USQCDmagBAD}, as discussed in the previous case. We can conjecture that it is also the Coulomb branch Hasse diagram of \eqref{eq:USQCD}. This is a non-trivial statement, as \eqref{eq:USQCDmagBAD} is only a magnetic quiver for \eqref{eq:USQCD} and not its 3d mirror.\footnote{3d mirror symmetry is a duality of SCFTs. We are currently studying gauge theories which do not flow to a fully interacting SCFT in the IR (i.e.\ they are ugly or bad). While the Coulomb branch of the magnetic quiver is the Higgs branch of the electric quiver, the Higgs branch of the magnetic quiver is not the Coulomb branch of the electric quiver.}
\begin{equation}
\label{eq:USQCDhasseBAD}
    \mathfrak{H}\left(\mathcal{C}\eqref{eq:USQCDmagBAD}\right)=
    \mathfrak{H}\left(\mathcal{H}\eqref{eq:USQCD}\right)=\qquad
        \raisebox{-.5\height}{\begin{tikzpicture}
        \node[hasse] (0) at (0,0) {};
        \node[hasse] (1) at (0,1) {};
        \node[hasse] (2) at (0,2) {};
        \node (3) at (0,3) {$\vdots$};
        \node[hasse] (4) at (0,4) {};
        \node[hasse] (5) at (0,5) {};
        \node[hasse] (6) at (0,6) {};
        \draw (0)--(1)--(2)--(3)--(4)--(5)--(6);
        \node at (0.5,0.5) {$a_{N-1}$};
        \node at (0.5,1.5) {$a_{N-3}$};
        \node at (0.5,4.5) {$a_{3+\epsilon}$};
        \node at (0.5,5.5) {$a_{1+\epsilon}$};
    \end{tikzpicture}}
    \raisebox{-.5\height}{\begin{tikzpicture}
        \draw[->] (0,0)--(1,0);
        \node at (0.5,0.5) {$\mathcal{I}$};
    \end{tikzpicture}}\qquad
    \raisebox{-.5\height}{\begin{tikzpicture}
        \node[hasse] (0) at (0,0) {};
        \node[hasse] (1) at (0,1) {};
        \node[hasse] (2) at (0,2) {};
        \node (3) at (0,3) {$\vdots$};
        \node[hasse] (4) at (0,4) {};
        \node[hasse] (5) at (0,5) {};
        \node[hasse] (6) at (0,6) {};
        \draw (0)--(1)--(2)--(3)--(4)--(5)--(6);
        \node at (0.5,5.5) {$A_{N-1}$};
        \node at (0.5,4.5) {$A_{N-3}$};
        \node at (0.5,1.5) {$A_{3+\epsilon}$};
        \node at (0.5,0.5) {$A_{1+\epsilon}$};
    \end{tikzpicture}}\quad=
    \mathfrak{H}\left(\mathcal{H}\eqref{eq:USQCDmagBAD}\right)\overset{\textnormal{conjecture}}{=}
    \mathfrak{H}\left(\mathcal{C}\eqref{eq:USQCD}\right)\;.
\end{equation}
Naively applying the quiver subtraction algorithm to \eqref{eq:USQCD} indeed yields the Hasse diagram on the right hand side of \eqref{eq:USQCDhasseBAD}. When the quiver subtraction algorithm terminates one is left with \eqref{eq:USQCDrem}, the theory which remains on a generic point of the Higgs branch of \eqref{eq:USQCD}. 

This has a simple interpretation: A symplectic singularity described by the Hasse diagram on the right hand side of \eqref{eq:USQCDhasseBAD} has quaternionic co-dimension\footnote{The dimension of the transverse slice from the bottom to the top leaf. All dimensions given in what follows are quaternionic dimensions. } $\frac{N-\epsilon}{2}$. The Coulomb branch of \eqref{eq:USQCD} has dimension $k$. If the right hand side of \eqref{eq:USQCDhasseBAD} is the Hasse diagram of the Coulomb branch of \eqref{eq:USQCD}, then the bottom leaf should have dimension $k-\frac{N-\epsilon}{2}$, which is non-zero as for bottom leaves of cones. A natural candidate for this leaf is the Coulomb branch of \eqref{eq:USQCDrem}, which is a smooth space with precisely this dimension. We will henceforth label bottom leaves of non-zero dimension pink in the Coulomb branch Hasse diagram, denoting the dimension of the leaf. We can propose the following Hasse diagram for the Coulomb branch of \eqref{eq:USQCD}:
\begin{equation}
    \raisebox{-.5\height}{\begin{tikzpicture}
        \node[hassebr,label=left:{\color{red}$k-\frac{N-\epsilon}{2}$}] (0) at (0,0) {};
        \node[hasse] (1) at (0,1) {};
        \node[hasse] (2) at (0,2) {};
        \node (3) at (0,3) {$\vdots$};
        \node[hasse] (4) at (0,4) {};
        \node[hasse] (5) at (0,5) {};
        \node[hasse] (6) at (0,6) {};
        \draw[red] (0)--(1)--(2)--(3)--(4)--(5)--(6);
        \node at (0.5,5.5) {$A_{N-1}$};
        \node at (0.5,4.5) {$A_{N-3}$};
        \node at (0.5,1.5) {$A_{3+\epsilon}$};
        \node at (0.5,0.5) {$A_{1+\epsilon}$};
    \end{tikzpicture}}
\end{equation}
This is consistent with the description of the Coulomb branch of \eqref{eq:USQCD} computed in \cite{2016arXiv160403625B} via the BFN construction and given in terms of generalized affine Grassmannian slices, and computed in \cite{Assel:2017jgo} via abelianisation.

The Hasse diagram of the full moduli space can also be obtained from inversion, as explained in \cite{Grimminger:2020dmg}. For $k\geq2$ and $N=4$ in \eqref{eq:USQCD} we propose the following Hasse diagram of the full moduli space $\mathcal{M}$:
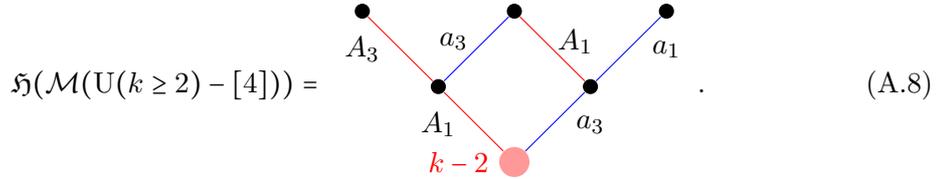
\begin{equation}
    \mathfrak{H}(\mathcal{M}(\mathrm{U}(k\geq2)-[4]))=
    \raisebox{-.5\height}{
    \begin{tikzpicture}
        \node[hassebr,label=left:{\color{red}$k-2$}] (0) at (0,0) {};
        \node[hasse] (1) at (-1,1) {};
        \node[hasse] (2) at (-2,2) {};
        \node[hasse] (a) at (1,1) {};
        \node[hasse] (b) at (2,2) {};
        \node[hasse] (1a) at (0,2) {};
        \draw[red] (0)--(1)--(2) (a)--(1a);
        \draw[blue] (0)--(a)--(b) (1)--(1a);
        \node at (-1,0.5) {$A_1$};
        \node at (-2,1.5) {$A_3$};
        \node at (1,0.5) {$a_3$};
        \node at (2,1.5) {$a_1$};
        \node at (0.8,1.6) {$A_1$};
        \node at (-0.8,1.6) {$a_3$};
    \end{tikzpicture}}\;.
\end{equation}

\subsection{\texorpdfstring{$\mathrm{USp}(2k)$ SQCD}{USp(2k) SQCD}}
\label{subsec:appSp}
Consider the electric theory
\begin{equation}
\label{eq:SpSQCD}
\raisebox{-.5\height}{
    \begin{tikzpicture}
        \node at (-1,0.5) {$\mathsf{Q}_e=$};
        \node[gauge,label=right:{USp$(2k)$}] (1) at (0,0) {};
        \node[flavour,label=right:{$D_{N}$}] (2) at (0,1) {};
        \draw (1)--(2);
    \end{tikzpicture}
    }
\end{equation}
which is good if $N>2k$, it is bad if $N\leq 2k$. The good case works completely analogously to the discussion for the unitary gauge theory. Let us turn to the more interesting case where \eqref{eq:SpSQCD} is bad, i.e.\ $N\leq 2k$. This theory was studied extensively in \cite{Assel:2018exy}.

\paragraph{$N$ odd.}
If $N$ is odd, there is one magnetic quiver
\begin{equation}
\label{eq:SpSQCDmagODD}
    \mathsf{Q}_m=\raisebox{-.5\height}{ \begin{tikzpicture}
        \node[gauge,label=below:{\small$1$}] (1) at (0,0) {};
        \node[gauge,label=below:{\small$2$}] (2) at (1,0) {};
        \node (3) at (2,0) {$\cdots$};
        \node[gauge,label=above:{\small$N-2$}] (4) at (3,0) {};
        \node[gauge,label=above:{\small$\frac{N-1}{2}$}] (5a) at (4,0.5) {};
        \node[gauge,label=below:{\small$\frac{N-1}{2}$}] (5b) at (4,-0.5) {};
        \node[flavour,label=right:{\small$1$}] (6a) at (5,0.5) {};
        \node[flavour,label=right:{\small$1$}] (6b) at (5,-0.5) {};
        \draw (1)--(2)--(3)--(4)--(5a)--(6a) (6b)--(5b)--(4);
    \end{tikzpicture}}\;.
\end{equation}
Applying quiver subtraction to \eqref{eq:SpSQCDmagODD} gives its Coulomb branch Hasse diagram, which we can identify with the Higgs branch Hasse diagram of \eqref{eq:SpSQCD}. We can use inversion to obtain the Higgs branch Hasse diagram of \eqref{eq:SpSQCDmagODD}.
\begin{equation}
    \mathfrak{H}\left(\mathcal{C}\eqref{eq:SpSQCDmagODD}\right)=
    \mathfrak{H}\left(\mathcal{H}\eqref{eq:SpSQCD}\right)
    =\qquad\raisebox{-.5\height}{\begin{tikzpicture}
        \node (l) at (0,0) {$\begin{tikzpicture}
        \node[hasse] (0) at (0,0) {};
        \node[hasse] (1) at (0,1) {};
        \node[hasse] (2) at (0,2) {};
        \node (3) at (0,3) {$\vdots$};
        \node[hasse] (4) at (0,4) {};
        \node[hasse] (5) at (0,5) {};
        \node[hasse] (6) at (0,6) {};
        \draw (0)--(1)--(2)--(3)--(4)--(5)--(6);
        \node at (0.5,0.5) {$d_{N}$};
        \node at (0.5,1.5) {$d_{N-2}$};
        \node at (0.5,4.5) {$d_{5}$};
        \node at (0.5,5.5) {$d_{3}$};
    \end{tikzpicture}$};
    \node (r) at (3,0) {$\begin{tikzpicture}
        \node[hasse] (0) at (0,0) {};
        \node[hasse] (1) at (0,1) {};
        \node[hasse] (2) at (0,2) {};
        \node (3) at (0,3) {$\vdots$};
        \node[hasse] (4) at (0,4) {};
        \node[hasse] (5) at (0,5) {};
        \node[hasse] (6) at (0,6) {};
        \draw (0)--(1)--(2)--(3)--(4)--(5)--(6);
        \node at (0.5,0.5) {$D_{3}$};
        \node at (0.5,1.5) {$D_{5}$};
        \node at (0.5,4.5) {$D_{N-2}$};
        \node at (0.5,5.5) {$D_{N}$};
    \end{tikzpicture}$};
    \draw[->] (l)--(r);
    \node at (1.5,0.5) {$\mathcal{I}$};
    \end{tikzpicture}}\quad=\mathfrak{H}\left(\mathcal{H}\eqref{eq:SpSQCDmagODD}\right)\;.
\end{equation}
Following the reasoning of before, we can conjecture the Coulomb branch Hasse diagram of \eqref{eq:SpSQCD}:
\begin{equation}
    \mathfrak{H}\left(\mathcal{C}\eqref{eq:SpSQCD}\right)=
    \raisebox{-0.5\height}{\begin{tikzpicture}
        \node[hassebr,label=left:{\color{red}$k-\frac{N-1}{2}$}] (0) at (0,0) {};
        \node[hasse] (1) at (0,1) {};
        \node[hasse] (2) at (0,2) {};
        \node (3) at (0,3) {$\vdots$};
        \node[hasse] (4) at (0,4) {};
        \node[hasse] (5) at (0,5) {};
        \node[hasse] (6) at (0,6) {};
        \draw[red] (0)--(1)--(2)--(3)--(4)--(5)--(6);
        \node at (0.5,0.5) {$D_{3}$};
        \node at (0.5,1.5) {$D_{5}$};
        \node at (0.5,4.5) {$D_{N-2}$};
        \node at (0.5,5.5) {$D_{N}$};
    \end{tikzpicture}}
\end{equation}
where the bottom smooth leaf is the Coulomb branch of the theory 
\begin{equation}
\raisebox{-.5\height}{
    \begin{tikzpicture}
        \node[gauge,label=right:{USp$(2k-N-1)$}] (1) at (0,0) {};
        \node[flavour,label=right:{$D_{1}$}] (2) at (0,1) {};
        \draw (1)--(2);
    \end{tikzpicture}\;.
    }
\end{equation}
This Hasse diagram is consistent with the analysis of \cite{Assel:2018exy}.

For $k\geq2$ and $N=5$ in \eqref{eq:SpSQCD} the proposed Hasse diagram of the full moduli space is
\begin{equation}
    \mathfrak{H}(\mathcal{M}(\mathrm{USp}(k\geq2)-[D_5]))=
    \raisebox{-.5\height}{
    \begin{tikzpicture}
        \node[hassebr,label=left:{\color{red}$k-2$}] (0) at (0,0) {};
        \node[hasse] (1) at (-1,1) {};
        \node[hasse] (2) at (-2,2) {};
        \node[hasse] (a) at (1,1) {};
        \node[hasse] (b) at (2,2) {};
        \node[hasse] (1a) at (0,2) {};
        \draw[red] (0)--(1)--(2) (a)--(1a);
        \draw[blue] (0)--(a)--(b) (1)--(1a);
        \node at (-1,0.5) {$D_3$};
        \node at (-2,1.5) {$D_5$};
        \node at (1,0.5) {$d_5$};
        \node at (2,1.5) {$D_3$};
        \node at (0.5,1.5) {$D_3$};
        \node at (-0.5,1.5) {$d_5$};
    \end{tikzpicture}}\;.
\end{equation}

\paragraph{$N$ even.}
If $N$ is even, there are two magnetic quivers (see Section \ref{sec:3dd})
\begin{subequations}
    \begin{equation}
    \label{eq:SpSQCDmagEVEN1}
        \mathsf{Q}_m^1=\raisebox{-.5\height}{ \begin{tikzpicture}
            \node[gauge,label=below:{\small$1$}] (1) at (0,0) {};
            \node[gauge,label=below:{\small$2$}] (2) at (1,0) {};
            \node (3) at (2,0) {$\cdots$};
            \node[gauge,label=above:{\small$N-2$}] (4) at (3,0) {};
            \node[gauge,label=above:{\small$\frac{N}{2}$}] (5a) at (4,0.5) {};
            \node[gauge,label=below:{\small$\frac{N}{2}-1$}] (5b) at (4,-0.5) {};
            \node[flavour,label=right:{\small$2$}] (6a) at (5,0.5) {};
            \draw (1)--(2)--(3)--(4)--(5a)--(6a) (5b)--(4);
        \end{tikzpicture}}\;,
    \end{equation}
    \begin{equation}
    \label{eq:SpSQCDmagEVEN2}
        \mathsf{Q}_m^2=\raisebox{-.5\height}{ \begin{tikzpicture}
            \node[gauge,label=below:{\small$1$}] (1) at (0,0) {};
            \node[gauge,label=below:{\small$2$}] (2) at (1,0) {};
            \node (3) at (2,0) {$\cdots$};
            \node[gauge,label=above:{\small$N-2$}] (4) at (3,0) {};
            \node[gauge,label=above:{\small$\frac{N}{2}-1$}] (5a) at (4,0.5) {};
            \node[gauge,label=below:{\small$\frac{N}{2}$}] (5b) at (4,-0.5) {};
            \node[flavour,label=right:{\small$2$}] (6b) at (5,-0.5) {};
            \draw (1)--(2)--(3)--(4)--(5a) (6b)--(5b)--(4);
        \end{tikzpicture}}\;.
    \end{equation}
\end{subequations}
One for each cone in the classical Higgs branch of \eqref{eq:SpSQCD}. The intersection of the two cones in the classical Higgs branch is given by
\begin{equation}
\label{eq:SpSQCDmagEVEN12}
    \mathsf{Q}_m^{12}=\raisebox{-.5\height}{ \begin{tikzpicture}
            \node[gauge,label=below:{\small$1$}] (1) at (0,0) {};
            \node[gauge,label=below:{\small$2$}] (2) at (1,0) {};
            \node (3) at (2,0) {$\cdots$};
            \node[gauge,label=above:{\small$N-2$}] (4) at (3,0) {};
            \node[gauge,label=above:{\small$\frac{N}{2}-1$}] (5a) at (4,0.5) {};
            \node[gauge,label=below:{\small$\frac{N}{2}-1$}] (5b) at (4,-0.5) {};
            \node[flavour,label=left:{\small$1$}] (4f) at (3,-1) {};
            \draw (1)--(2)--(3)--(4)--(5a) (5b)--(4)--(4f);
        \end{tikzpicture}}\;.
\end{equation}
Applying quiver subtraction to \eqref{eq:SpSQCDmagEVEN1} and \eqref{eq:SpSQCDmagEVEN2}, with \eqref{eq:SpSQCDmagEVEN12} their intersection, we get the Hasse diagram for the classical Higgs branch of \eqref{eq:SpSQCD}. Following \cite{Grimminger:2020dmg} we can conjecture its inversion to be the Hasse diagram of the quantum Coulomb branch of \eqref{eq:SpSQCD}:
\begin{equation}
    \mathfrak{H}(\mathcal{H}_{\textnormal{classical}}\eqref{eq:SpSQCD})=\qquad
    \raisebox{-.5\height}{
        \begin{tikzpicture}
            \node[hasse,label={[shift={(0.5,0)}]$d_{N}$}] (1) at (0,0) {};
            \node[hasse,label={[shift={(0.5,0)}]$d_{N-2}$}] (2) at (0,1) {};
            \node[hasse] (3) at (0,2) {};
            \node (4) at (0,3) {$\vdots$};
            \node[hasse,label={[shift={(0.5,0)}]$d_{4}$}] (5) at (0,4) {};
            \node[hasse] (6) at (0,5) {};
            \node[hasse,label={[shift={(0,-0.8)}]$a_1$}] (7a) at (-0.5,6) {};
            \node[hasse,label={[shift={(0,-0.8)}]$a_1$}] (7b) at (0.5,6) {};
            \draw (1)--(2)--(3)--(4)--(5)--(6)--(7a) (7b)--(6);
        \end{tikzpicture}
    }
    \raisebox{-.5\height}{\begin{tikzpicture}
        \draw[->] (0,0)--(1,0);
        \node at (0.5,0.5) {$\mathcal{I}$};
    \end{tikzpicture}}\quad\raisebox{-.5\height}{
        \begin{tikzpicture}
            \node[hasse,label={[shift={(0.5,-1)}]$D_{N}$}] (1) at (0,6) {};
            \node[hasse,label={[shift={(0.5,-1)}]$D_{N-2}$}] (2) at (0,5) {};
            \node[hasse] (3) at (0,4) {};
            \node (4) at (0,3) {$\vdots$};
            \node[hasse,label={[shift={(0.5,-1)}]$D_{4}$}] (5) at (0,2) {};
            \node[hasse] (6) at (0,1) {};
            \node[hassebr,label=left:{$k-\frac{N}{2}$},label={[shift={(0,0.2)}]$a_1$}] (7a) at (-0.5,0) {};
            \node[hassebr,label=right:{$k-\frac{N}{2}$},label={[shift={(0,0.2)}]$a_1$}] (7b) at (0.5,0) {};
            \draw[red] (1)--(2)--(3)--(4)--(5)--(6)--(7a) (7b)--(6);
        \end{tikzpicture}
    }\overset{\textnormal{conjecture}}{=}\mathfrak{H}(\mathcal{C}\eqref{eq:SpSQCD})\;.
\end{equation}
In this case, there are two most singular leaves in the quantum Coulomb branch of \eqref{eq:SpSQCD}). Each leaf is the Coulomb branch of pure USp$(2k-N)$.
From one lowest leaf there is an emanating Higgs branch which is described by the Coulomb branch of the first magnetic quiver. From the other leaf there is an emanating Higgs branch which is described by the Coulomb branch of the second magnetic quiver. This is also consistent with the analysis of \cite{Assel:2018exy}.

For $k\geq2$ and $N=4$ in \eqref{eq:SpSQCD} the proposed Hasse diagram of the full moduli space is
\begin{equation}
    \mathfrak{H}(\mathcal{M}(\mathrm{USp}(k\geq2)-[D_4]))=
    \raisebox{-.5\height}{
    \begin{tikzpicture}
        \node[hassebr,label=left:{\color{red}$k-2$}] (0x) at (-0.5,0) {};
        \node[hassebr,label=right:{\color{red}$k-2$}] (0y) at (0.5,0) {};
        \node[hasse] (1) at (0,1) {};
        \node[hasse] (2) at (-0.5,2) {};
        \node[hasse] (ax) at (-1.5,1) {};
        \node[hasse] (bx) at (-2.5,2) {};
        \node[hasse] (ay) at (1.5,1) {};
        \node[hasse] (by) at (2.5,2) {};
        \node[hasse] (1a) at (1,2) {};
        \draw[red] (0x)--(1)--(2) (0y)--(1) (ax)--(1a) (ay)--(1a);
        \draw[blue] (0x)--(ax)--(bx) (0y)--(ay)--(by) (1)--(1a);
        \node at ($(0x)!0.5!(1)$) {\scriptsize$A_1$};
        \node at ($(0y)!0.5!(1)$) {\scriptsize$A_1$};
        \node at ($(1)!0.3!(2)$) {\scriptsize$D_4$};
        \node at ($(ax)!0.3!(1a)$) {\scriptsize$A_1$};
        \node at ($(0x)!0.5!(ax)$) {\scriptsize$d_4$};
        \node at ($(ax)!0.5!(bx)$) {\scriptsize$a_1$};
        \node at ($(0y)!0.5!(ay)$) {\scriptsize$a_3$};
        \node at ($(ay)!0.5!(by)$) {\scriptsize$a_1$};
        \node at ($(ay)!0.5!(1a)$) {\scriptsize$A_1$};
        \node at ($(1)!0.5!(1a)$) {\scriptsize$d_4$};
        \node at ($(ay)!0.5!(1a)$) {\scriptsize$A_1$};
    \end{tikzpicture}}\;.
\end{equation}

\subsection{\texorpdfstring{$\mathrm{SU}(k)$ SQCD}{SU(k) SQCD}}
\label{subsec:appSU}
Let us now consider the electric theory
\begin{equation}
\label{eq:SUSQCD}
\raisebox{-.5\height}{
    \begin{tikzpicture}
        \node at (-1,0.5) {$\mathsf{Q}_e=$};
        \node[gauge,label=right:{SU$(k)$}] (1) at (0,0) {};
        \node[flavour,label=right:{$N$}] (2) at (0,1) {};
        \draw (1)--(2);
    \end{tikzpicture}
    }
\end{equation}
 which is good for $N\geq2k-1$ and bad for $N<2k-1$. The classical Higgs branch of this theory was studied in detail in \cite{Argyres:1996eh,Bourget:2019rtl}. We only discuss $k\leq N\leq 2k-2$, as this is the most interesting case. There are two cones in the classical Higgs branch, the baryonic cone and the mesonic cone. At a generic point on the baryonic cone there is complete Higgsing (this can be seen e.g.\ from the brane web construction). At a generic point on the mesonic cone there is incomplete Higgsing. The magnetic quivers are \cite{Bourget:2019rtl}
\begin{subequations}
    \begin{align}
    \label{eq:SUSQCDmagB}
        \mathsf{Q}_m^{\mathrm{B}}&=\raisebox{-.5\height}{\begin{tikzpicture}
        \node[gauge,label=below:{$1$}] (1) at (1,0) {};
        \node[gauge,label=below:{$2$}] (2) at (2,0) {};
        \node (3) at (3,0) {$\cdots$};
        \node[gauge,label=below:{$N-k$}] (4) at (4,0) {};
        \node[gauge,label=left:{$1$}] (4u) at (4,1) {};
        \node (5) at (5,0) {$\cdots$};
        \node[gauge,label=below:{$N-k$}] (6) at (6,0) {};
        \node[gauge,label=right:{$1$}] (6u) at (6,1) {};
        \node (7) at (7,0) {$\cdots$};
        \node[gauge,label=below:{$2$}] (8) at (8,0) {};
        \node[gauge,label=below:{$1$}] (9) at (9,0) {};
        \draw (1)--(2)--(3)--(4)--(5)--(6)--(7)--(8)--(9) (4)--(4u) (6)--(6u);
        \draw[thick,double] (4u)--(6u);
        \node at (5,1.3) {$2k-N$};
        \draw [decorate,decoration={brace,amplitude=5pt}]
    (9.3,-0.6)--(0.7,-0.6) node [black,midway,yshift=-10] {$N-1$};
    \end{tikzpicture}}\;, \\
    \label{eq:SUSQCDmagM}
        \mathsf{Q}_m^{\mathrm{M}}&=\begin{cases}
        \begin{tikzpicture}
        \node[gauge,label=below:{$1$}] (1) at (1,0) {};
        \node[gauge,label=below:{$2$}] (2) at (2,0) {};
        \node (3) at (3,0) {$\cdots$};
        \node[gauge,label=below:{$\frac{N-1}{2}$}] (4) at (4,0) {};
        \node[flavour,label=left:{$1$}] (4f) at (4,1) {};
        \node[gauge,label=below:{$\frac{N-1}{2}$}] (5) at (5,0) {};
        \node[flavour,label=right:{$1$}] (5f) at (5,1) {};
        \node (6) at (6,0) {$\cdots$};
        \node[gauge,label=below:{$2$}] (7) at (7,0) {};
        \node[gauge,label=below:{$1$}] (8) at (8,0) {};
        \draw (1)--(2)--(3)--(4)--(5)--(6)--(7)--(8) (4)--(4f) (5)--(5f);
        \draw [decorate,decoration={brace,amplitude=5pt}]
    (8.3,-0.8)--(0.7,-0.8) node [black,midway,yshift=-10] {$N-1$};
        \node at (10,0) {, if $N$ odd,};
    \end{tikzpicture}\\
    \begin{tikzpicture}
        \node[gauge,label=below:{$1$}] (1) at (1,0) {};
        \node[gauge,label=below:{$2$}] (2) at (2,0) {};
        \node (3) at (3,0) {$\cdots$};
        \node[gauge,label=below:{$\frac{N}{2}$}] (4) at (4,0) {};
        \node[flavour,label=left:{$2$}] (4f) at (4,1) {};
        \node (5) at (5,0) {$\cdots$};
        \node[gauge,label=below:{$2$}] (6) at (6,0) {};
        \node[gauge,label=below:{$1$}] (7) at (7,0) {};
        \draw (1)--(2)--(3)--(4)--(5)--(6)--(7) (4)--(4f);
        \draw [decorate,decoration={brace,amplitude=5pt}]
    (7.3,-0.8)--(0.7,-0.8) node [black,midway,yshift=-10] {$N-1$};
        \node at (10,0) {, if $N$ even.};
    \end{tikzpicture}
    \end{cases}
    \end{align}
\end{subequations}
The intersection of the two cones in the classical Higgs branch is given by
\begin{equation}
\label{eq:SUSQCDmagBM}
    \mathsf{Q}_m^{\mathrm{BM}}=\raisebox{-.5\height}{\begin{tikzpicture}
        \node[gauge,label=below:{$1$}] (1) at (1,0) {};
        \node[gauge,label=below:{$2$}] (2) at (2,0) {};
        \node (3) at (3,0) {$\cdots$};
        \node[gauge,label=below:{$N-k$}] (4) at (4,0) {};
        \node[gauge,label=left:{$1$}] (5u) at (5,1) {};
        \node (5) at (5,0) {$\cdots$};
        \node[gauge,label=below:{$N-k$}] (6) at (6,0) {};
        \node (7) at (7,0) {$\cdots$};
        \node[gauge,label=below:{$2$}] (8) at (8,0) {};
        \node[gauge,label=below:{$1$}] (9) at (9,0) {};
        \draw (1)--(2)--(3)--(4)--(5)--(6)--(7)--(8)--(9) (4)--(5u)--(6);
        \draw [decorate,decoration={brace,amplitude=5pt}]
    (9.3,-0.6)--(0.7,-0.6) node [black,midway,yshift=-10] {$N-1$};
    \end{tikzpicture}}\;.
\end{equation}
Applying quiver subtraction to \eqref{eq:SUSQCDmagB} and \eqref{eq:SUSQCDmagM} with \eqref{eq:SUSQCDmagBM} their intersection yields the Hasse diagram of the classical Higgs branch of \eqref{eq:SUSQCD} \cite{Bourget:2019rtl}:
\begin{equation}
    \label{eq:pre-InvertSU}
    \mathfrak{H}(\mathcal{H}_{\textnormal{classical}}\eqref{eq:SUSQCD})=\qquad
    \raisebox{-.5\height}{
        \begin{tikzpicture}
            \node[hasse] (1) at (0,0) {};
            \node[hasse] (2) at (0,1) {};
            \node[hasse] (3) at (0,2) {};
            \node (4) at (0,3) {$\vdots$};
            \node[hasse] (5) at (0,4) {};
            \node[hasse] (6) at (0,5) {};
            \node[hasse] (B) at (-1,6) {};
            \node[hasse] (7) at (1,6) {};
            \node (8) at (2,7) {$\udots$};
            \node[hasse] (9) at (3,8) {};
            \node[hasse] (10) at (4,9) {};
            \draw (1)--(2)--(3)--(4)--(5)--(6)--(7)--(8)--(9)--(10) (6)--(B);
            \node at (0.5,0.5) {$a_{N-1}$};
            \node at (0.5,1.5) {$a_{N-3}$};
            \node at (0.8,4.5) {$a_{2k-N+1}$};
            \node at (1.3,5.5) {$a_{2k-N-1}$};
            \node at (4,8.5) {$a_{1+\epsilon}$};
            \node at (4,9.5) {M};
            \node at (-1.3,5.5) {$A_{2k-N-1}$};
            \node at (-1,6.5) {B};
        \end{tikzpicture}
    }\textnormal{with }\epsilon=\begin{cases}
        1\textnormal{ , if $N$ odd}\\
        0\textnormal{ , if $N$ even}
    \end{cases}\;,
\end{equation}
which we can invert to obtain the Hasse diagram of the quantum Coulomb branch:
\begin{equation}
    \eqref{eq:pre-InvertSU}\quad\raisebox{-.5\height}{\begin{tikzpicture}
        \draw[->] (0,0)--(1,0);
        \node at (0.5,0.5) {$\mathcal{I}$};
    \end{tikzpicture}}\quad\raisebox{-.5\height}{
        \scalebox{0.8}{\begin{tikzpicture}
            \node[hasse] (1) at (0,0) {};
            \node[hasse] (2) at (0,-1) {};
            \node[hasse] (3) at (0,-2) {};
            \node (4) at (0,-3) {$\vdots$};
            \node[hasse] (5) at (0,-4) {};
            \node[hasse] (6) at (0,-5) {};
            \node[hasse] (7) at (1,-6) {};
            \node (8) at (2,-7) {$\ddots$};
            \node[hasse] (9) at (3,-8) {};
            \node[circle,fill=red!40!white,draw,inner sep=0.3cm,label=right:{\color{red}$k-\frac{N-\epsilon}{2}-1$}] (10) at (4.4,-9.4) {};
            \node[hasse,label=left:{\color{red}$0$}] (B) at (-1,-11) {};
            \draw[red] (1)--(2)--(3)--(4)--(5)--(6)--(7)--(8)--(9)--(10) (6)--(B);
            \node at (0.5,-0.5) {$A_{N-1}$};
            \node at (0.5,-1.5) {$A_{N-3}$};
            \node at (0.8,-4.5) {$A_{2k-N+1}$};
            \node at (1.3,-5.5) {$A_{2k-N-1}$};
            \node at (4,-8.5) {$A_{1+\epsilon}$};
            \node at (4.4,-10.1) {M${}^\vee$};
            \node at (-1.3,-8) {$a_{2k-N-1}$};
            \node at (-1,-11.5) {B${}^\vee$};
        \end{tikzpicture}}
    }\overset{\textnormal{conjecture}}{=}\mathfrak{H}(\mathcal{C}\eqref{eq:SUSQCD})\;.
\end{equation}
The leaf in the Coulomb branch of \eqref{eq:SUSQCD} denoted B${}^\vee$ is where the baryonic Higgs branch (the Coulomb branch of \eqref{eq:SUSQCDmagB}) emanates. It has co-dimension $k-1$, which is the dimension of the Coulomb branch, and hence this leaf is of zero dimension. This is consistent with the fact, that the theory is completely Higgsed on the baryonic branch. The local geometry at this point B${}^\vee$ in the Coulomb branch of \eqref{eq:SUSQCD} is the Higgs branch of \eqref{eq:SUSQCDmagB}.

The leaf in the Coulomb branch of \eqref{eq:SUSQCD} denoted M${}^\vee$ is where the mesonic Higgs branch (the Coulomb branch of \eqref{eq:SUSQCDmagM}) emanates. It has co-dimension $\frac{N-\epsilon}{2}$, and hence this leaf is of dimension $k-\frac{N-\epsilon}{2}-1$. This is consistent with the fact, that the theory is incompletely Higgsed on the mesonic branch. The local geometry transverse to a point on M${}^\vee$ in the Coulomb branch of \eqref{eq:SUSQCD} is the Higgs branch of \eqref{eq:SUSQCDmagM}.

The quantum moduli space of \eqref{eq:SUSQCD} as a 4d $\mathcal{N}=2$ theory was explored in \cite{Argyres:1996eh}. Their description is consistent with our results.

For $k=3$ and $N=4$ in \eqref{eq:SUSQCD} the proposed Hasse diagram of the full moduli space is
\begin{equation}
\label{eq:fullMSSU3with4}
    \mathfrak{H}(\mathcal{M}(\mathrm{SU}(3)-[4]))=
    \raisebox{-.5\height}{
    \begin{tikzpicture}
        \node[hasse,label=left:{\color{red}0}] (0x) at (-0.5,0) {};
        \node[hasse,label=right:{\color{red}0}] (0y) at (0.5,0) {};
        \node[hasse] (1) at (0,1) {};
        \node[hasse] (2) at (-0.5,2) {};
        \node[hasse] (ax) at (-1.5,1) {};
        \node[hasse] (bx) at (-2.5,2) {};
        \node[hasse] (ay) at (1.5,1) {};
        \node[hasse] (by) at (2.5,2) {};
        \node[hasse] (1a) at (1,2) {};
        \draw[red] (0x)--(1)--(2) (0y)--(1) (ax)--(1a) (ay)--(1a);
        \draw[blue] (0x)--(ax)--(bx) (0y)--(ay)--(by) (1)--(1a);
        \node at ($(0x)!0.5!(1)$) {\scriptsize$a_1$};
        \node at ($(0y)!0.5!(1)$) {\scriptsize$A_1$};
        \node at ($(1)!0.3!(2)$) {\scriptsize$A_3$};
        \node at ($(ax)!0.3!(1a)$) {\scriptsize$a_1$};
        \node at ($(0x)!0.5!(ax)$) {\scriptsize$a_3$};
        \node at ($(ax)!0.5!(bx)$) {\scriptsize$A_1$};
        \node at ($(0y)!0.5!(ay)$) {\scriptsize$a_3$};
        \node at ($(ay)!0.5!(by)$) {\scriptsize$a_1$};
        \node at ($(1)!0.5!(1a)$) {\scriptsize$d_4$};
    \end{tikzpicture}}\;.
\end{equation}
Since $A_1=a_1$ this Hasse diagram has a $\mathbb{Z}_2$ symmetry. However there is no such $\mathbb{Z}_2$ symmetry in the moduli space. The two most singular points (co-dimension 2) in the Coulomb branch do not have the same local geometry, which is easily seen from the corresponding magnetic quivers \eqref{eq:SUSQCDmagB} and \eqref{eq:SUSQCDmagM}.

For $k=6$ and $N=7$ in \eqref{eq:SUSQCD} the proposed Hasse diagram of the full moduli space is
\begin{equation}
    \raisebox{-.5\height}{
    \begin{tikzpicture}
        \node at (-4,7) {$\mathfrak{H}(\mathcal{M}(\mathrm{SU}(6)-[7]))=$};
        \node[hasse] (3) at (0,0) {};
        \node[hasse] (2) at (0,-1) {};
        \node[hasse] (1) at (1,-2) {};
        \node[hassebr,label=right:{\color{red}$2$}] (0) at (2,-3) {};
        \node[hasse] (2a) at (3,5) {};
        \node[hasse] (1a) at (4,4) {};
        \node[hasse] (a) at (5,3) {};
        \node[hasse] (1b) at (6,8) {};
        \node[hasse] (b) at (7,7) {};
        \node[hasse] (c) at (8,9) {};
        \node[hasse,label=left:{\color{red}$0$}] (B) at (-2,-5) {};
        \node[hasse] (B1) at (-5,1) {};
        \node[hasse] (B2) at (-5.5,2) {};
        \draw[red] (B)--(2) (0)--(1)--(2)--(3) (B1)--(2a)--(1a)--(a) (1b)--(b);
        \draw[blue] (B)--(B1)--(B2) (0)--(a)--(b)--(c) (1)--(1a)--(1b) (2)--(2a);
        \node at (-2,-5.5) {B${}^\vee$};
        \node at (2,-3.5) {M${}^\vee$};
        \node at (-5.5,2.5) {B};
        \node at (0,0.5) {$\mathcal{C}$};
        \node at (8,9.5) {M};
        \node at (6,8.5) {mixed};
        \node at (3,5.5) {mixed};
        \node at ($(B)!0.5!(B1)$) {$a_6$};
        \node at ($(B1)!0.5!(B2)$) {$A_4$};
        \node at ($(B)!0.5!(2)$) {$a_4$};
        \node at ($(B1)!0.5!(2a)$) {$a_4$};
        \node at ($(0)!0.5!(1)$) {$A_2$};
        \node at ($(1)!0.5!(2)$) {$A_4$};
        \node at ($(2)!0.5!(3)$) {$A_6$};
        \node at ($(0)!0.5!(a)$) {$a_6$};
        \node at ($(a)!0.5!(b)$) {$a_4$};
        \node at ($(b)!0.5!(c)$) {$a_2$};
        \node at ($(1)!0.5!(1a)$) {$a_6$};
        \node at ($(1a)!0.5!(1b)$) {$a_4$};
        \node at ($(2)!0.5!(2a)$) {$a_6$};
        \node at ($(a)!0.5!(1a)$) {$A_2$};
        \node at ($(1a)!0.5!(2a)$) {$A_4$};
        \node at ($(b)!0.5!(1b)$) {$A_2$};
    \end{tikzpicture}}\;,
\end{equation}
Here B (M) denotes the baryonic (mesonic) branch in the classical Higgs branch, emanating from the leaf B${}^\vee$ (M${}^\vee$) in the quantum Coulomb branch. $\mathcal{C}$ denotes the Coulomb branch, and mixed stands for either mixed branch.

\pagebreak

\section{\texorpdfstring{Details on the Eight Cones of $\mathrm{SU}(2)-\mathrm{SU}(4)-\mathrm{USp}(6)-[D_2]$}{Details on the Eight Cones of SU(2)-SU(4)-USp(6)-[D2]}}
\label{app:eightcones}

In this Appendix, we consider the quiver \eqref{electric:suspstart} with $n=2$ and $p=3$. For this case, there are $2^p = 8$ cones to consider.  This study is relegated to the appendix as it becomes very technical, and at first sight does not contain new conceptual difficulties. However it is a good illustration of the fact that new technical tools and notations become necessary to keep track of the combinatorial complexity. In addition, we make an important point regarding multiplicities in the corresponding Higgs scheme. 

The brane system is as follows: 
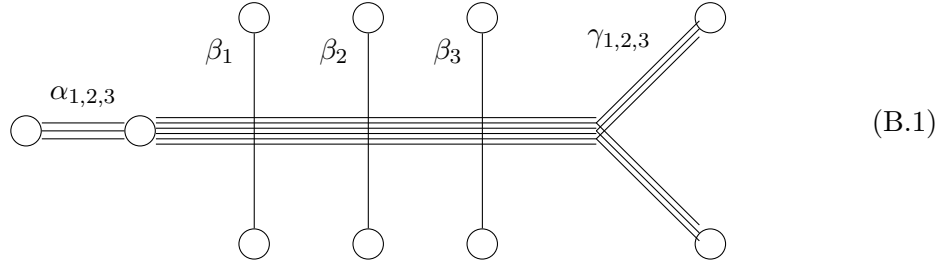
\begin{equation}
   \raisebox{-.5\height}{  \begin{tikzpicture}[xscale=1.5,yscale=1.5]
    \node[sev] (6) at (4,0) {};
    \node[sev] (7) at (5,0) {};
    \node[sev] (8a) at (6,1) {};
    \node[sev] (8b) at (6,-1) {};
    \node[sev] (9a) at (7,1) {};
    \node[sev] (9b) at (7,-1) {};
    \node[sev] (11a) at (8,1) {};
    \node[sev] (11b) at (8,-1) {};
    \node[sev] (10a) at (10,1) {};
    \node[sev] (10b) at (10,-1) {};
    \draw[transform canvas={yshift=-5pt}] (7)--(9,0);
    \draw[transform canvas={yshift=-3pt}] (7)--(9,0);
    \draw[transform canvas={yshift=-1pt}] (7)--(9,0);
    \draw[transform canvas={yshift=1pt}] (7)--(9,0);
    \draw[transform canvas={yshift=3pt}] (7)--(9,0);
    \draw[transform canvas={yshift=5pt}] (7)--(9,0);
    \draw[transform canvas={yshift=0pt}] (6)--(7) (9,0)--(10a) (10b)--(9,0);
    \draw[transform canvas={yshift=-3pt}] (6)--(7) (9,0)--(10a) (10b)--(9,0);
    \draw[transform canvas={yshift=3pt}] (6)--(7) (9,0)--(10a) (10b)--(9,0);
    \draw (8a)--(8b) (9a)--(9b) (11a)--(11b);
    \node at (5.7,.7) {$\beta_1$};
    \node at (6.7,.7) {$\beta_2$};
    \node at (7.7,.7) {$\beta_3$};
    \node at (4.5,.3) {$\alpha_{1,2,3}$};
    \node at (9.2,.8) {$\gamma_{1,2,3}$};
    \end{tikzpicture}}
    \label{branes:fourcones5d}
\end{equation}
We have decomposed the brane web into nine pieces that individually satisfy charge conservation, but not the s-rule: $\alpha_{1,2,3}$, $\beta_{1,2,3}$, $\gamma_{1,2,3}$. In order to satisfy the s-rule, one can pair up an $\gamma$ piece with either an $\alpha$ piece or with a $\beta$ piece. 

In a given phase of the theory, certain subsets of these nine pieces lie at the same position in the directions transverse to the drawing above, and we can read a magnetic quiver for this phase. Maximally Higgsed phases correspond to maximal decompositions, while partially Higgsed phases correspond to non-maximal decompositions. It is important to keep track of which pieces contributes to each phase, so we adopt a somewhat exotic notation for the resulting quivers. For instance, the maximal decomposition 
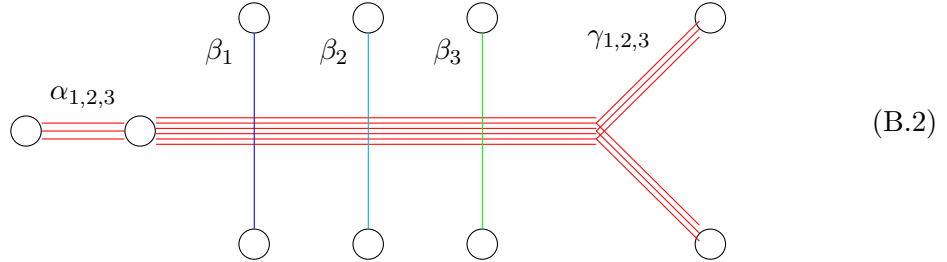
\begin{equation}
   \raisebox{-.5\height}{  \begin{tikzpicture}[xscale=1.5,yscale=1.5]
    \node[sev] (6) at (4,0) {};
    \node[sev] (7) at (5,0) {};
    \node[sev] (8a) at (6,1) {};
    \node[sev] (8b) at (6,-1) {};
    \node[sev] (9a) at (7,1) {};
    \node[sev] (9b) at (7,-1) {};
    \node[sev] (11a) at (8,1) {};
    \node[sev] (11b) at (8,-1) {};
    \node[sev] (10a) at (10,1) {};
    \node[sev] (10b) at (10,-1) {};
    \draw[red,transform canvas={yshift=-5pt}] (7)--(9,0);
    \draw[red,transform canvas={yshift=-3pt}] (7)--(9,0);
    \draw[red,transform canvas={yshift=-1pt}] (7)--(9,0);
    \draw[red,transform canvas={yshift=1pt}] (7)--(9,0);
    \draw[red,transform canvas={yshift=3pt}] (7)--(9,0);
    \draw[red,transform canvas={yshift=5pt}] (7)--(9,0);
    \draw[red,transform canvas={yshift=0pt}] (6)--(7) (9,0)--(10a) (10b)--(9,0);
    \draw[red,transform canvas={yshift=-3pt}] (6)--(7) (9,0)--(10a) (10b)--(9,0);
    \draw[red,transform canvas={yshift=3pt}] (6)--(7) (9,0)--(10a) (10b)--(9,0);
    \draw[blue] (8a)--(8b) ;
    \draw[cyan] (9a)--(9b);
    \draw[green] (11a)--(11b);
    \node at (5.7,.7) {$\beta_1$};
    \node at (6.7,.7) {$\beta_2$};
    \node at (7.7,.7) {$\beta_3$};
    \node at (4.5,.3) {$\alpha_{1,2,3}$};
    \node at (9.2,.8) {$\gamma_{1,2,3}$};
    \end{tikzpicture}}
\end{equation}
has magnetic quiver 
\begin{equation}
    \raisebox{-.5\height}
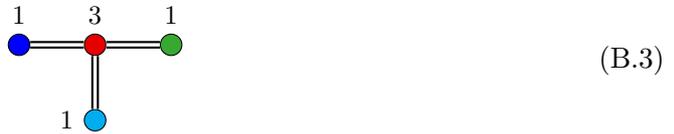
{ \begin{tikzpicture}
\node[gaugeb,label=above:{\small$1$}] (1) at (-1,0) {};
\node[gauger,label=above:{\small$3$}] (2) at (0,0) {};
\node[gaugeg,label=above:{\small$1$}] (3) at (1,0) {};
\node[gaugec,label=left:{\small$1$}] (4) at (0,-1) {};
\draw[doublea] (1)--(2)--(3) (2)--(4);
    \end{tikzpicture}}
\end{equation}
which will be denoted 
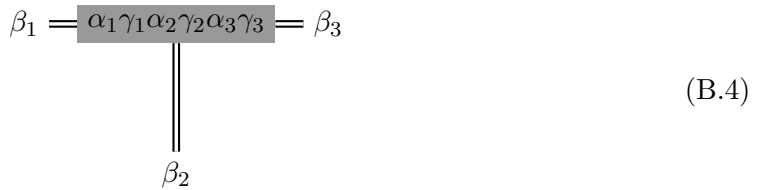
\begin{equation}
 \raisebox{-.5\height}{\begin{tikzpicture}[x=2cm,y=2cm]
\node[fill=black!40] (1) at (0,0) {$\alpha_1 \gamma_1 \alpha_2 \gamma_2 \alpha_3  \gamma_3$};
\node (2) at (-1,0) {$ \beta_1$};
\node (3) at (0,-1) {$ \beta_2$};
\node (4) at (1,0) {$\beta_3$};
\draw[doublea] (1)--(2) (1)--(3) (1)--(4);
\end{tikzpicture}}
\end{equation}
The central node $\alpha_1 \gamma_1 \alpha_2 \gamma_2 \alpha_3  \gamma_3$ is three copies of the same subweb, and as such it gives rise to a U(3) gauge node. For better readability, the dark shaded vertices mark $\mathrm{U}(3)$ gauge nodes, light shaded vertices mark $\mathrm{U}(2)$ gauge nodes while unshaded vertices are $\mathrm{U}(1)$ gauge nodes. The edge multiplicities are denoted by parallel lines as usual. In our examples the edge multiplicity is either 2 or 4.

Using this coding system, it is possible to draw the full Higgs branch Hasse diagram shown in Figure \ref{fig:hasseBig}. In this figure, each box represents a certain number of phases / leaves, according to the index structure: 1 for the white boxes, 3 for the orange boxes and 6 for the red boxes. For instance the 8 cones of the Higgs branch, which are represented in the top row of Figure \ref{fig:hasseBig}, divide into two singlets and two triplets of the $S_3$ permutation group (permuting the $\beta_{i=1,2,3}$ in \eqref{branes:fourcones5d}), as shown by the two white boxes and the two orange boxes. 

\paragraph{Higgs Scheme. } The Higgs Scheme Hilbert series can be computed from the hyper-K\"ahler quotient, and one finds 
\begin{equation}
H = \frac{\left(  \begin{array}{c}
  1+4 t^2+20 t^4+98 t^6+315 t^8+865 t^{10} +1860 t^{12} +3124 t^{14}+3918 t^{16} \\ +3671 t^{18}+1996 t^{20}  +264 t^{22}-819 t^{24}-510 t^{26}-106 t^{28} \\ +280 t^{30}  +29 t^{32}+3 t^{34}-50
   t^{36}+10 t^{38}+4 t^{40}-t^{42}
\end{array} \right)}{(1 - t^2)^4 (1 - t^4)^2 (1 - t^6)^4} \, . 
\label{eq:HKresultBIG}
\end{equation}
In order to reproduce this from Figure \ref{fig:hasseBig}, one needs to evaluate the Coulomb branch Hilbert series for all the quivers, and then to find the multiplicities, if they are non-trivial polynomials in $t^2$. We don't know of a general method for identifying these polynomials, so we look for solutions that match with the result \eqref{eq:HKresultBIG}. Namely, there are 17 inequivalent quivers in Figure \ref{fig:hasseBig}, that form a collection $\mathcal{Q}$, and we look for a coefficient $c_{\mathsf{Q}} \in \mathbb{Z}[t^2]$ for each $\mathsf{Q} \in \mathcal{Q}$ such that 
\begin{equation}
    \sum\limits_{\mathsf{Q} \in \mathcal{Q}}  c_{\mathsf{Q}}  \mathrm{HS} (\mathcal{C} (\mathsf{Q})) = H \, . 
\end{equation}
Expanding this equation in $t$ and limiting the search for coefficients of degrees $\leq d$ gives infinitely many equations for finitely many unknowns. For $d=0$ we find no solution, meaning that there are nilpotent operators in the scheme. For $d=1$ we do find solutions. Perhaps surprisingly, there is a 12-dimensional space of solutions: this simply echoes the non-uniqueness of primary decomposition of ideals. The multiplicities of the top two rows of Figure \ref{fig:hasseBig} are however uniquely fixed, and the $c_{\mathsf{Q}}$ for these rows are degree 0 polynomials. This suggests that there are no nilpotent operators on the top leaves, but that they do appear on the lower leaves. It remains a challenge to give an explicit description of the nilpotent operators.

\begin{landscape}

\begin{figure}
\vspace*{-2.3cm}\hspace*{-2.5cm}\scalebox{.5}{
\begin{tikzpicture}[x=3cm,y=6cm]
\node[draw,fill=c1] (n1) at (-5,0) {\begin{tikzpicture}[x=2cm,y=2cm]
\node[fill=black!40] (1) at (0,0) {$\gamma_1 \gamma_2 \gamma_3$};
\node (2) at (-1,0) {$\alpha_1 \beta_1$};
\node (3) at (0,-1) {$\alpha_2 \beta_2$};
\node (4) at (1,0) {$\alpha_3 \beta_3$};
\draw[doublea] (1)--(2) (1)--(3) (1)--(4);
\end{tikzpicture}};
\node[draw,fill=c3] (n2)  at (-2,0) {\begin{tikzpicture}[x=2cm,y=2cm]
\node[fill=black!40] (1) at (0,0) {$\gamma_i \gamma_j$};
\node (2) at (1,0) {$\alpha_i \beta_i$};
\node (3) at (0,-1) {$\alpha_j \beta_j$};
\node (4) at (1,-1) {$\beta_k$};
\node (5) at (2,-1) {$\alpha_k \gamma_k$};
\draw[doublea] (1)--(2)--(4)--(3)--(1) (4)--(5);
\end{tikzpicture}};
\node[draw,fill=c3] (n3)  at (2,0) {\begin{tikzpicture}[x=2cm,y=2cm]
\node[fill=black!40] (1) at (0,0) {$\alpha_i  \gamma_i \alpha_j  \gamma_j$};
\node (2) at (1,0) {$\beta_i$};
\node (3) at (0,-1) {$\beta_j$};
\node (4) at (1,-1) {$\alpha_k  \beta_k$};
\node (5) at (2,-1) {$\gamma_k$};
\draw[doublea] (1)--(2)--(4)--(3)--(1) (4)--(5);
\end{tikzpicture}};
\node[draw,fill=c1] (n4)  at (5,0) {\begin{tikzpicture}[x=2cm,y=2cm]
\node[fill=black!40] (1) at (0,0) {$\alpha_1 \gamma_1 \alpha_2 \gamma_2 \alpha_3  \gamma_3$};
\node (2) at (-1,0) {$ \beta_1$};
\node (3) at (0,-1) {$ \beta_2$};
\node (4) at (1,0) {$\beta_3$};
\draw[doublea] (1)--(2) (1)--(3) (1)--(4);
\end{tikzpicture}};
\node[draw,fill=c3] (n5)  at (-6,-1) {\begin{tikzpicture}[x=2cm,y=2cm]
\node[fill=black!20] (1) at (0,0) {$\gamma_j \gamma_k$};
\node (2) at (1,0) {$\alpha_k \beta_k$};
\node (3) at (0,-1) {$\alpha_j \beta_j$};
\node (4) at (1,-1) {$\alpha_i \beta_i \gamma_i$};
\draw[doublea] (1)--(2)--(4)--(3)--(1);
\end{tikzpicture}};
\node[draw,fill=c6] (n6)  at (-3,-1) {\begin{tikzpicture}[x=2cm,y=2cm]
\node[fill=black!20] (1) at (0,0) {$\gamma_i \gamma_j$};
\node (2) at (1,0) {$\alpha_i \beta_i \beta_k$};
\node (3) at (0,-1) {$\alpha_j \beta_j$};
\node (4) at (1,-1) {$\alpha_k \gamma_k$};
\draw[doublea] (1)--(2)--(3)--(1) (2)--(4);
\end{tikzpicture}};
\node[draw,fill=c6] (n7)  at (0,-1) {\begin{tikzpicture}[x=2cm,y=2cm]
\node (1) at (0,0) {$ \gamma_j$};
\node (2) at (1,0) {$\alpha_i \beta_i \gamma_i$};
\node (3) at (0,-1) {$\alpha_j \beta_j$};
\node (4) at (1,-1) {$\beta_k$};
\node (5) at (2,-1) {$\alpha_k \gamma_k$};
\draw[doublea] (1)--(3)--(2)--(4)--(3) (4)--(5);
\end{tikzpicture}};
\node[draw,fill=c6] (n8)  at (3,-1) {\begin{tikzpicture}[x=2cm,y=2cm]
\node[fill=black!20] (1) at (0,0) {$\alpha_i  \gamma_i \alpha_j  \gamma_j$};
\node (2) at (1,0) {$\beta_i \alpha_k  \beta_k$};
\node (3) at (0,-1) {$\beta_j$};
\node (4) at (1,-1) {$\gamma_k$};
\draw[doublea] (1)--(2)--(3)--(1) (2)--(4);
\end{tikzpicture}};
\node[draw,fill=c3] (n9)  at (6,-1) {\begin{tikzpicture}[x=2cm,y=2cm]
\node[fill=black!20] (1) at (0,0) {$\alpha_j \gamma_j \alpha_k  \gamma_k$};
\node (2) at (1,0) {$\beta_k$};
\node (3) at (0,-1) {$ \beta_j$};
\node (4) at (1,-1) {$\alpha_i \beta_i \gamma_i$};
\draw[doublea] (1)--(2)--(4)--(3)--(1);
\end{tikzpicture}};
\node[draw,fill=c6] (n11)  at (-8,-2) {\begin{tikzpicture}[x=2cm,y=2cm]
\node[fill=black!20] (1) at (0,0) {$\gamma_j \gamma_k$};
\node (2) at (1,0) {$\alpha_k \beta_k$};
\node (4) at (1,-1) {$\alpha_i \beta_i \gamma_i \alpha_j \beta_j$};
\draw[doublea] (1)--(2)--(4)--(1);
\end{tikzpicture}};
\node[draw,fill=c3] (n12)  at (-6,-2) {\begin{tikzpicture}[x=2cm,y=2cm]
\node[fill=black!20] (1) at (0,0) {$\gamma_i \gamma_j$};
\node (2) at (1,0) {$\alpha_i \beta_i \alpha_j \beta_j \beta_k$};
\node (4) at (1,-1) {$\alpha_k \gamma_k$};
\draw[doublea] (2)--(4);
\draw[quadruple] (2)--(1);
\end{tikzpicture}};
\node[draw,fill=c3] (n10)  at (-4,-2) {\begin{tikzpicture}[x=2cm,y=2cm]
\node (1) at (0,0) {$ \gamma_k$};
\node (2) at (1,0) {$\alpha_k \beta_k$};
\node (3) at (0,-1) {$\alpha_j \beta_j \gamma_j$};
\node (4) at (1,-1) {$\alpha_i \beta_i \gamma_i$};
\draw[doublea] (1)--(2)--(3)--(4)--(2);
\end{tikzpicture}};
\node[draw,fill=c6] (n14)  at (-2,-2) {\begin{tikzpicture}[x=2cm,y=2cm]
\node (1) at (0,0) {$ \gamma_j$};
\node (2) at (1,0) {$\alpha_i \beta_i \gamma_i \beta_k$};
\node (3) at (0,-1) {$\alpha_j \beta_j$};
\node (4) at (1,-1) {$\alpha_k \gamma_k$};
\draw[doublea] (1)--(3) (2)--(4);
\draw[quadruple] (2)--(3);
\end{tikzpicture}};
\node[draw,fill=c6] (n13)  at (-0,-2) {\begin{tikzpicture}[x=2cm,y=2cm]
\node (1) at (0,0) {$\gamma_i $};
\node (2) at (1,0) {$\alpha_i \beta_i \beta_k$};
\node (3) at (0,-1) {$\alpha_j \beta_j \gamma_j$};
\node (4) at (1,-1) {$\alpha_k \gamma_k$};
\draw[doublea] (1)--(2)--(4);
\draw[quadruple] (2)--(3);
\end{tikzpicture}};
\node[draw,fill=c6] (n15)  at (2,-2) {\begin{tikzpicture}[x=2cm,y=2cm]
\node (1) at (0,0) {$ \gamma_j$};
\node (2) at (1,0) {$\alpha_i \beta_i \gamma_i \alpha_j \beta_j$};
\node (4) at (0,-1) {$\beta_k$};
\node (5) at (1,-1) {$\alpha_k \gamma_k$};
\draw[doublea] (1)--(2) (4)--(5);
\draw[quadruple] (2)--(4);
\end{tikzpicture}};
\node[draw,fill=c3] (n16)  at (4,-2) {\begin{tikzpicture}[x=2cm,y=2cm]
\node (1) at (0,0) {$ \alpha_j \beta_j\gamma_j$};
\node (2) at (1,0) {$\alpha_i \beta_i \gamma_i$};
\node (4) at (0,-1) {$\beta_k$};
\node (5) at (1,-1) {$\alpha_k \gamma_k$};
\draw[doublea] (4)--(1)--(2)--(4)--(5);
\end{tikzpicture}};
\node[draw,fill=c3] (n18)  at (6,-2) {\begin{tikzpicture}[x=2cm,y=2cm]
\node[fill=black!20] (1) at (0,0) {$\alpha_i \gamma_i \alpha_j  \gamma_j$};
\node (2) at (1,0) {$ \beta_i \beta_j \alpha_k \beta_k$};
\node (4) at (1,-1) {$ \gamma_k$};
\draw[doublea] (2)--(4);
\draw[quadruple] (2)--(1);
\end{tikzpicture}};
\node[draw,fill=c6] (n17)  at (8,-2) {\begin{tikzpicture}[x=2cm,y=2cm]
\node[fill=black!20] (1) at (0,0) {$\alpha_j \gamma_j \alpha_k  \gamma_k$};
\node (2) at (1,0) {$\beta_k$};
\node (4) at (1,-1) {$\alpha_i \beta_i \gamma_i  \beta_j$};
\draw[doublea] (1)--(2)--(4)--(1);
\end{tikzpicture}};
\node[draw,fill=c3] (n22)  at (-7,-3.5) {\begin{tikzpicture}[x=2cm,y=2cm]
\node[fill=black!20] (1) at (0,0) {$\gamma_j \gamma_k$};
\node (4) at (1,-1) {$\alpha_i \beta_i \gamma_i \alpha_j \beta_j \alpha_k \beta_k$};
\draw[quadruple] (1)--(4);
\end{tikzpicture}};
\node[draw,fill=c3] (n21)  at (-5,-3.5) {\begin{tikzpicture}[x=2cm,y=2cm]
\node (1) at (0,0) {$ \gamma_k$};
\node (2) at (1,0) {$\alpha_k \beta_k$};
\node (4) at (1,-1) {$\alpha_i \beta_i \gamma_i \alpha_j \beta_j \gamma_j$};
\draw[doublea] (1)--(2);
\draw[quadruple] (2)--(4);
\end{tikzpicture}};
\node[draw,fill=c6] (n20)  at (-3,-3.5) {\begin{tikzpicture}[x=2cm,y=2cm]
\node (1) at (0,0) {$\gamma_j $};
\node (2) at (1,0) {$\alpha_k \beta_k \gamma_k$};
\node (4) at (1,-1) {$\alpha_i \beta_i \gamma_i \alpha_j \beta_j$};
\draw[doublea] (1)--(4);
\draw[quadruple] (4)--(2);
\end{tikzpicture}};
\node[draw,fill=c1] (n19)  at (-1,-3.5) {\begin{tikzpicture}[x=2cm,y=2cm]
\node (2) at (1,0) {$\alpha_1 \beta_1 \gamma_1$};
\node (3) at (0,-1) {$\alpha_2 \beta_2 \gamma_2$};
\node (4) at (1,-1) {$\alpha_3 \beta_3 \gamma_3$};
\draw[doublea] (2)--(3)--(4)--(2);
\end{tikzpicture}};
\node[draw,fill=c6] (n23)  at (1,-3.5) {\begin{tikzpicture}[x=2cm,y=2cm]
\node (1) at (0,0) {$\gamma_i $};
\node (2) at (1,0) {$\alpha_i \beta_i \alpha_j \beta_j \gamma_j \beta_k$};
\node (4) at (1,-1) {$\alpha_k \gamma_k$};
\draw[doublea] (2)--(4);
\draw[doublea] (2)--(1);
\end{tikzpicture}};
\node[draw,fill=c6] (n24)  at (3,-3.5) {\begin{tikzpicture}[x=2cm,y=2cm]
\node (2) at (1,0) {$\alpha_i \beta_i \gamma_i \beta_k$};
\node (3) at (0,-1) {$\alpha_j \beta_j \gamma_j$};
\node (4) at (1,-1) {$\alpha_k \gamma_k$};
\draw[doublea] (2)--(4);
\draw[quadruple] (2)--(3);
\end{tikzpicture}};
\node[draw,fill=c3] (n26)  at (5,-3.5) {\begin{tikzpicture}[x=2cm,y=2cm]
\node (1) at (0,0) {$ \alpha_k \gamma_k$};
\node (2) at (1,0) {$\beta_k$};
\node (4) at (1,-1) {$\alpha_i \beta_i \gamma_i \alpha_j \beta_j \gamma_j$};
\draw[doublea] (1)--(2);
\draw[quadruple] (2)--(4);
\end{tikzpicture}};
\node[draw,fill=c3] (n25)  at (7,-3.5) {\begin{tikzpicture}[x=2cm,y=2cm]
\node[fill=black!20] (1) at (0,0) {$\alpha_j \gamma_j \alpha_k \gamma_k$};
\node (4) at (1,-1) {$\alpha_i \beta_i \gamma_i  \beta_j  \beta_k$};
\draw[quadruple] (1)--(4);
\end{tikzpicture}};
\node[draw,fill=c3] (n28)  at (-4,-4.5) {\begin{tikzpicture}[x=2cm,y=2cm]
\node (1) at (1,0) {$\gamma_k $};
\node (4) at (1,-1) {$\alpha_i \beta_i \gamma_i \alpha_j \beta_j \gamma_j \alpha_k \beta_k $};
\draw[doublea] (1)--(4);
\end{tikzpicture}};
\node[draw,fill=c3] (n27)  at (-0,-4.5) {\begin{tikzpicture}[x=2cm,y=2cm]
\node (2) at (1,0) {$\alpha_k \beta_k \gamma_k$};
\node (4) at (1,-1) {$\alpha_i \beta_i \gamma_i \alpha_j \gamma_j  \beta_j$};
\draw[quadruple] (4)--(2);
\end{tikzpicture}};
\node[draw,fill=c3] (n29)  at (4,-4.5) {\begin{tikzpicture}[x=2cm,y=2cm]
\node (1) at (1,0) {$\alpha_k \gamma_k $};
\node (4) at (1,-1) {$\alpha_i \beta_i \gamma_i \alpha_j \beta_j \gamma_j  \beta_k $};
\draw[doublea] (1)--(4);
\end{tikzpicture}};
\node[draw,fill=c1] (n30)  at (0,-5.5) {\begin{tikzpicture}[x=2cm,y=2cm]
\node (4) at (1,-1) {$\alpha_1 \beta_1 \gamma_1 \alpha_2 \beta_2 \gamma_2 \alpha_3 \beta_3 \gamma_3 $};
\end{tikzpicture}};
\draw (n1)--(n5);
\draw (n2)--(n5);
\draw (n2)--(n6);
\draw (n2)--(n7);
\draw (n3)--(n7);
\draw (n3)--(n8);
\draw (n3)--(n9);
\draw (n4)--(n9);
\draw (n5)--(n10);
\draw (n5)--(n11);
\draw (n6)--(n11);
\draw (n6)--(n12);
\draw (n6)--(n13);
\draw (n6)--(n14);
\draw (n7)--(n10);
\draw (n7)--(n13);
\draw (n7)--(n14);
\draw (n7)--(n15);
\draw (n7)--(n16);
\draw (n8)--(n13);
\draw (n8)--(n15);
\draw (n8)--(n17);
\draw (n8)--(n18);
\draw (n9)--(n16);
\draw (n9)--(n17);
\draw (n10)--(n19);
\draw (n10)--(n20);
\draw (n10)--(n21);
\draw (n11)--(n20);
\draw (n11)--(n21);
\draw (n11)--(n22);
\draw (n12)--(n22);
\draw (n12)--(n23);
\draw (n13)--(n20);
\draw (n13)--(n23);
\draw (n13)--(n24);
\draw (n14)--(n21);
\draw (n14)--(n23);
\draw (n14)--(n24);
\draw (n15)--(n20);
\draw (n15)--(n23);
\draw (n15)--(n26);
\draw (n16)--(n19);
\draw (n16)--(n24);
\draw (n16)--(n26);
\draw (n17)--(n24);
\draw (n17)--(n25);
\draw (n17)--(n26);
\draw (n18)--(n23);
\draw (n18)--(n25);
\draw (n19)--(n27);
\draw (n20)--(n27);
\draw (n20)--(n28);
\draw (n21)--(n27);
\draw (n21)--(n28);
\draw (n22)--(n28);
\draw (n23)--(n28);
\draw (n23)--(n29);
\draw (n24)--(n27);
\draw (n24)--(n29);
\draw (n25)--(n29);
\draw (n26)--(n27);
\draw (n26)--(n29);
\draw (n27)--(n30);
\draw (n28)--(n30);
\draw (n29)--(n30);
\end{tikzpicture}
}
    \caption{Diagram for theory \eqref{electric:suspstart} with $p=3$ and $n=2$. See the text for the explanation of colors and notations.  }
    \label{fig:hasseBig}
\end{figure}

\end{landscape}

\providecommand{\href}[2]{#2}\begingroup\raggedright\endgroup

\end{document}